\documentclass[11pt, a4paper, twosides]{Thesis} 
\usepackage{transparent}
\graphicspath{{./}{Figures/}}
\usepackage{framed}
\usepackage{tabularx}
\usepackage[usenames,dvipsnames]{color}
\usepackage{calc}
\usepackage{fancyhdr}
\usepackage{ifthen}
\usepackage{setspace}
\usepackage{moaccopyright}
\usepackage{fancybox}
\usepackage{lipsum}
\usepackage{ upgreek }
\usepackage{ mathrsfs }
\usepackage{booktabs}
\usepackage{multirow}
\usepackage{upgreek}
\usepackage{rotating}
\usepackage{pdflscape}
\usepackage{afterpage}
\usepackage{capt-of}
\usepackage{lipsum}
\usepackage{setspace}
\usepackage{umoline}
\usepackage{soul}
\usepackage{amsmath}
\usepackage{ marvosym }
\usepackage{epigraph}
\usepackage{quotchap}
\usepackage[T1]{fontenc}
\usepackage{subfigure}
\usepackage{lettrine}
\usepackage{amsmath}

\allowdisplaybreaks
\usepackage[font=small,format=plain,labelfont=bf ,textfont=normal, justification=justified,singlelinecheck=false]{caption}

 \newcommand{\be}{\begin{equation}}
 \newcommand{\ee}{\end{equation}}
  \newcommand{\bea}{\begin{eqnarray}}
 \newcommand{\eea}{\end{eqnarray}}
 \newcommand{\define}{\equiv}
\newcommand{\gothg}{\mathfrak{g}}
\newcommand{\half}{\frac{1}{2}}

\newcommand{\GR}{{\mbox{\tiny GR}}}

\newcommand{\bc}{\begin{center}}
\newcommand{\ec}{\end{center}}
\newcommand{\vonevtwo}{{\bf v}_1\cdot {\bf v}_2}
\newcommand{\vonen}{{\bf v}_1\cdot {\bf n}}
\newcommand{\vtwon}{{\bf v}_2\cdot {\bf n}}
\def\nva{{\bf v}_1 \cdot {\bf n}}
\def\nvb{{\bf v}_2 \cdot {\bf n}}
\newcommand{\ppE}{{\mbox{\tiny ppE}}}
\newcommand{\f}[2]{\frac{#1}{#2}}

\newcommand{\beq}{\begin{equation}}
\newcommand{\eeq}{\end{equation}}
\newcommand{\beqa}{\begin{eqnarray}}
\newcommand{\eeqa}{\end{eqnarray}}

\def\xone{\mathbf{x}_1}
\def\xtwo{\mathbf{x}_2}
\def\n{\mathbf{n}}
\def\a{\mathbf{a}}
\def\b{\mathbf{b}}
\def\c{\mathbf{c}}
\def\x{\mathbf{x}}
\def\vone{\mathbf{v}_1}
\def\vtwo{\mathbf{v}_2}
\def\aone{\mathbf{a}_1}
\def\atwo{\mathbf{a}_2}
\def\datwo{\dot{\mathbf{a}}_2}
\def\Nvone{(\mathbf{v}_1\cdot\mathbf{n})}
\def\Nvtwo{(\mathbf{v}_2\cdot\mathbf{n})}
\def\vonevtwo{(\mathbf{v}_1\cdot\mathbf{v}_2)}
\def\ddatwo{\ddot{\mathbf{a}}_2}
\def\Naone{(\mathbf{a}_1\cdot\mathbf{n})}
\def\Natwo{(\mathbf{a}_2\cdot\mathbf{n})}

\usepackage[titletoc]{appendix}

\fancypagestyle{myplain}
{ 
 \fancyhf{}

  \fancyfoot[C]{\thepage}
}
\fancypagestyle{myfancy}
{ \fancyhf{}

}
\usepackage[T1]{fontenc}
\usepackage[ansinew]{inputenc}
\usepackage{xcolor,graphicx,wallpaper}

\makeatletter
\let\orig@chapter\@chapter
\def\@chapter[#1]#2{\ifnum \c@secnumdepth >\m@ne
                       \if@mainmatter
                         \refstepcounter{chapter}%
                         \typeout{\@chapapp\space\thechapter.}%
                         \addcontentsline{toc}{chapter}%
                                   {Chapter~\protect\numberline{\thechapter.} #1}%
                       \else
                         \addcontentsline{toc}{chapter}{#1}%
                       \fi
                    \else
                      \addcontentsline{toc}{chapter}{#1}%
                    \fi
                    \chaptermark{#1}%
                    \addtocontents{lof}{\protect\addvspace{10\p@}}%
                    \addtocontents{lot}{\protect\addvspace{10\p@}}%
                    \if@twocolumn
                      \@topnewpage[\@makechapterhead{#2}]%
                    \else
                      \@makechapterhead{#2}%
                      \@afterheading
                    \fi}
\makeatother

\def\vec#1{
{\bf #1}
}

\def\mypart#1#2#3#4{
\par 
\pagestyle{plain}
\newpage\clearpage
\ClearWallPaper\TileWallPaper{220mm}{297.4mm}{#4} 
\vspace*{2cm} 
\refstepcounter{part}
{\centering \textbf{\huge PART \thepart}\par}%
\vspace{1cm}
{\centering \linespread{2}\selectfont \textbf{\Huge #1}\par \linespread{1}\selectfont}%
\thispagestyle{empty}
\vspace{2cm}
\null
#3
#2
\addcontentsline{toc}{part}{Part \Roman{part}:\; \textbf{#1}}
\pagestyle{fancy}
}

\def\myapp#1#2#3#4{
\pagestyle{plain}
\newpage\clearpage
\ClearWallPaper\TileWallPaper{220mm}{297.4mm}{#4} 
\vspace*{2cm} 
\refstepcounter{part}
\vspace{1cm}
{\centering \linespread{2}\selectfont \textbf{\Huge #1}\par \linespread{1}\selectfont}%
\thispagestyle{empty}
\vspace{2cm}
\null
#3
\vfill
#2
\addcontentsline{toc}{part}{\textbf{#1}}
\pagestyle{fancy}
}
\hypersetup{
  colorlinks,
  citecolor=OliveGreen,
  linkcolor=NavyBlue,
  urlcolor=Magenta}

\usepackage{moactitlepage}

\usepackage[Bjornstrup]{fncychap}

\usepackage{xcolor}
\colorlet{partbgcolor}{gray!30}
\colorlet{partnumcolor}{gray}
\colorlet{chapbgcolor}{gray!30}
\colorlet{chapnumcolor}{red}

\makeatletter
\renewcommand*{\@part}{}
\def\@part[#1]#2{%
  \ifnum \c@secnumdepth >-2\relax
    \refstepcounter{part}%
    \@maybeautodot\thepart%
    \addparttocentry{\thepart}{#1}%
  \else
    \addparttocentry{}{#1}%
  \fi
  \begingroup
    \setparsizes{\z@}{\z@}{\z@\@plus 1fil}\par@updaterelative
    \raggedpart
    \interlinepenalty \@M
    \normalfont\sectfont\nobreak
    \setlength\fboxsep{0pt}
    \colorbox{partbgcolor}{\rule{0pt}{40pt}%
    \makebox[\linewidth]{%
    \begin{minipage}{\dimexpr\linewidth+20pt\relax}
      \ifnum \c@secnumdepth >-2\relax
        \vskip-25pt
        \size@partnumber{\partformat}%
      \fi      %
      \vskip\baselineskip
      \hspace*{\dimexpr\myhi+10pt\relax}%
      \parbox{\dimexpr\linewidth-2\myhi-20pt\relax}{\raggedleft\LARGE#2\strut}%
      \hspace*{\myhi}\par\medskip%
    \end{minipage}%
      }%
    }%
    \partmark{#1}\par
  \endgroup
  \@endpart
}

\renewcommand\DOCH{%
  \settowidth{\py}{\CNoV\thechapter}
  \addtolength{\py}{-10pt}
  \fboxsep=0pt%
  \colorbox{chapbgcolor}{\rule{0pt}{40pt}\parbox[b]{\textwidth}{\hfill}}%
  \kern-\py\raise20pt%
  \hbox{\color{chapnumcolor}\CNoV\thechapter}\\%
}

\makeatother

\usepackage[square, numbers, comma, sort&compress]{natbib} 
\hypersetup{urlcolor=blue, colorlinks=true} 
\title{\ttitle} 

\begin{document}








\pagenumbering{gobble}

\renewcommand{\moacdepartment}{Department of Physics}

\renewcommand{\moacdate}{August 2013}

\renewcommand{\moacsupervisors}{Prof. Clifford M. Will}

\renewcommand{\usemoaclogo}{false}






\moactitlepages



\frontmatter 

\setstretch{1.3} 

\fancyhead{} 
\lhead[]{\thechapter}      
\rhead[\thechapter]{}

\newcommand{\HRule}{\rule{\linewidth}{0.5mm}} 

\hypersetup{pdftitle={\ttitle}}
\hypersetup{pdfsubject=\subjectname}
\hypersetup{pdfauthor=\authornames}
\hypersetup{pdfkeywords=\keywordnames}

\clearpage
\pagestyle{empty}
\null\vfill
\noindent
\begin{center}
{\textcopyright 2013,
\moacauthor\\
All Rights Reserved}
\end{center}

\setcounter{tocdepth}{2}

\pagestyle{myfancy} 
\thispagestyle{myplain}
\lhead[\thepage]{Contents}      
\rhead[Contents]{\thepage}

\begingroup
\hypersetup{linkcolor=blue}
\tableofcontents
\endgroup 

\lhead[\thepage]{List of Figures}      
\rhead[List of Figures]{\thepage}
\listoffigures 

\lhead[\thepage]{List of Tables}      
\rhead[List of Tables]{\thepage}
\listoftables 

\pagestyle{myplain}
\setstretch{1.3} 

\acknowledgements{\addtocontents{toc}{} 
{Since I started my journey in the field of theoretical physics until today, there have been many people from several universities and institutes who have had undeniable effects on my career and have made all the work presented in this dissertation possible. I want to take this opportunity to acknowledge some of them. I would like to thank the Physics Department at {\em University of Tehran} and {\em Washington University in St. Louis} for their great undergraduate and graduate programs, specially for their advanced courses on gravitation and astrophysics. I would also like to thank {\em McDonnell Center for the Space Sciences} in Washington University for their strong support. I am very thankful to the {\em Massachusetts Institute of Technology}, {\em Institute d'astrophysique de Paris}, and the {\em University of Florida} for their hospitality during working on parts of this dissertation. This research was supported in part by the {\em National Science Foundation}, Grant Nos. PHY 09-65133 and 12-60995.

First and foremost, I would like to thank my PhD advisor for critically reading this dissertation and for his extremely helpful comments. It was a great pleasure for me to have {\em Clifford Will} as my advisor during my doctoral studies since Spring 2009. Here I could ---and probably should--- acknowledge him for all those many unique and precious lessons he gave me, but I keep it short by saying that people like Cliff, make me have a strong faith in a brighter future for science and humanity. I look at him as a role model and I hope some day I could contribute to the scientific community just as he does. I know this won't be easy, but that is my goal.

I also have to greatly thank {\em Nicolas Yunes}, who has been a collaborator on a part of this dissertation. His advices and comments were always constructive and to the point. I would also like to thank {\em Frances Ferrer} and {\em K. G. Arun} for helpful discussions and acknowledge {\em Leo Stein} for his help in streamlining our Mathematica code. My thanks and appreciations also go to {\em Emanuele Berti} and {\em Michael Horbatsch} for their essential argument on a part of this work.

Alongside my advisor, who taught me a great deal of what I know in the field of gravitation, I would like to thank all of my teachers and professors. My first thanks go to my high-school physics teacher {\em H. Doroodian} for an excellent first impression he gave me into physics. I still keep the Halliday-Resnick book translated in Farsi which I got from him as an award in one of his classes. The same gratitude goes to all my teachers and professors in University of Tehran and Washington University in St. Louis, which I was very fortunate to be a student of them; I would like to thank {\em Amir~M. Abbassi}, {\em Mark Alford}, {\em Carl Bender}, {\em Claude Bernard}, {\em Anders Carlsson}, {\em Ram Cowsik}, {\em Jonathan Katz}, {\em Hamid~R. Moshfegh}, and {\em Wai-Mo Suen}, to name a few.

As a graduate student in Washington University in St. Louis I had the chance to visit several other universities and research institutes and interact with many excellent professors and researchers in my field of research. I would like to take this opportunity to thank {\em Scott Hughes} for his hospitality and the great conversation that we had during my visit at MIT in Spring 2011. I need to thank {\em Luc Blanchet} at Institut d'astrophysique de Paris for his hospitality and helpful discussions. From the same institute I thank {\em Roya Mohayaee}, and {\em Jacques Colin} for making me feel at home during my stay in Paris in the summer of 2012. I am thankful for invitations of {\em Reza Mansouri} at IPM, {\em Mohammad Nouri-Zonoz} at University of Tehran, and {\em Sohrab Rahvar} at Sharif University of Technology that gave me the opportunity to present parts of this work and interact with the experts in their institutes. I also have to thank {\em Steven Detweiler} for his support and hospitality at the University of Florida.

During my doctoral studies in Dr. Will's research group I have been glad to work with several postdocs and graduate students in our group including {\em K.~G. Arun}, {\em Adamantios Stavridis}, {\em Ryan Lang}, {\em Dimitris Manolidis}, {\em Laleh Sadeghian}, {\em Alexandre Le Tiec}, and {\em Pierre Fromholz}. I would like to thank all of them for creating such a good working environment. I would also like to thank all the people whom I have learnt ``new'' things from them during my PhD time period inside and outside of the working environment, including {\em James Bendert}, {\em Benjamin Burch}, {\em Steven Dorsher}, {\em Lauren Edge}, {\em Daniel Flanagan}, {\em Daniel Hunter}, {\em Joben Lewis}, {\em Matthew Lightman}, {\em Faraz Monifi}, {\em Ryan Murphy}, {\em Danial Sabri}, {\em Sarah Thibadeau}, {\em Kaveh Vejdani}, {\em Kasey Wagoner}, and {\em Shannon Kian Zare}, to name a few. A special thanks goes to {\em Sina Mossahebi}, {\em Morvarid Karimi}, {\em Javad Komijani}, {\em Morteza Shahriari-Nia}, {\em Mehdi Saremi}, and {\em Moojan Daneshmand} for their warm hospitalities.


Last but not least, I am grateful to my family and friends for all the support and positive energy that they always offer. In particular, I would like to thank my wife, {\em Laleh Sadeghian}, my parents, {\em Gholam-Abbas Mirshekari} and {\em Goli Abedi}, and my siblings, {\em Fatemeh}, {\em Masoumeh}, and {\em Soroush}. I have been very lucky to have so many valuable friends wherever I have lived so far: Tehran, St. Louis, Paris, and Gainesville. It is impossible to list all the names here but I would like them to know that I won't forget their support, help, friendship, brotherhood, and love.
}

}
\clearpage 


\pagestyle{empty} 

\setstretch{1.3} 

\dedicatory{to Laleh} 

\addtocontents{toc}{} 


\pagestyle{myplain} 

\addtotoc{Abstract} 

\setstretch{1.2} 

\abstract{\addtocontents{toc}{} 
{
This dissertation consists of four parts. In Part~\ref{part:1}, we briefly review fundamental theories of gravity, performed experimental tests, and gravitational waves. The framework and the methods that we use in our calculations are discussed in Part~\ref{part:2}. This part includes reviewing the methods of the Parametrized Post-Newtonian (PPN) framework, Direct Integration of Relaxed Einstein Equations (DIRE), and Matched Filtering.

In Part~\ref{part:3}, we calculate the explicit equations of motion for non-spinning compact objects (neutron stars or black holes) to 2.5 post-Newtonian order, or $O(v/c)^5$ beyond Newtonian gravity, in a general class of alternative theories to general relativity known as scalar-tensor theories. For the conservative part of the motion, we obtain the two-body Lagrangian and conserved energy and momentum through second post-Newtonian order. We find the contributions to gravitational radiation reaction to 1.5 post-Newtonian and 2.5 post-Newtonian orders, the former corresponding to the effects of dipole gravitational radiation. For binary black holes we show that the motion through 2.5 post-Newtonian order is observationally identical to that predicted by general relativity.\footnote{ \href{http://prd.aps.org/abstract/PRD/v87/i8/e084070}{S. Mirshekari and C.~M. Will,  {\bf Physical Review D 87, 8}, (Apr. 2013)}}

In Part~\ref{part:4}, we construct a parametrized dispersion relation that can produce a range of predictions of alternative theories of gravity for violations of Lorentz invariance in gravitation, and investigate their impact on the propagation of gravitational waves. We show how such corrections map to the waveform observable by a gravitational-wave detector, and to the ``parametrized post-Einsteinian framework'', proposed to model a range of deviations from General Relativity. Given a gravitational-wave detection, the lack of evidence for such corrections could then be used to place a constraint on Lorentz violation.\footnote{ \href{http://prd.aps.org/abstract/PRD/v85/i2/e024041}{S. Mirshekari, N. Yunes, and C.~M. Will, {\bf Physical Review D 85, 2}, (Jan. 2012)}}
}
}

\clearpage 

\mainmatter 
\pagestyle{myfancy} 


\mypart{Foundations}{{\vfill \small{\em This part includes introductory materials to the rest of the dissertation. In \cref{chapter1} the fundamental theory of gravity from its early days up to date is reviewed briefly. The status of tests of theories of gravity specially general relativity is discussed in \cref{chapter3}. \cref{chapter4} is an introduction to gravitational waves.}
} }
{\begin{framed}
\begin{itemize}
\item \cref{chapter1}--- \nameref{chapter1}
\item \cref{chapter3}--- \nameref{chapter3}
\item \cref{chapter4}--- \nameref{chapter4}
\end{itemize}
\end{framed}
}{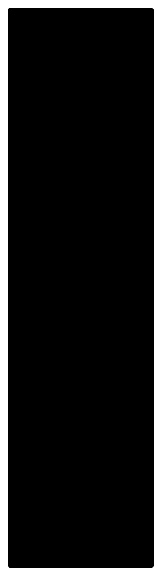}\label{part:1}

\begin{savequote}[0.55\linewidth]
{\scriptsize ``The fact that we live at the bottom of a deep gravity well, on the surface of a gas covered planet going around a nuclear fireball 90 million miles away and think this to be normal is obviously some indication of how skewed our perspective tends to be.''}
\qauthor{\scriptsize---Douglas Adams}
\end{savequote}

\chapter{Fundamental Theory of Gravity}
\thispagestyle{myplain}
\label{chapter1} 
\lhead[\thepage]{Chapter 1. \emph{Fundamental Theory of Gravity}}      
\rhead[Chapter 1. \emph{Fundamental Theory of Gravity}]{\thepage}
\section{From Newtonian Gravity to Einstein's General Relativity} \ClearWallPaper
\label{sec:fromNewtontoEinstein}

It is not a long time in the history of humanity that we know where we are in the Universe. Since ancient thinkers until the development of the heliocentric model by Nicolaus Copernicus in the 16th century, the accepted view about the Universe was that the Earth is at the center and the Sun and other planets orbit around it~\footnote{In the early 20th century, the discovery of other galaxies and the development of the Big Bang theory led to the development of cosmological models of a homogeneous, isotropic Universe (which lacks a central point) that is expanding at all points.}. This popular belief was based on the Ptolemaic geocentric system. The publication of Copernicus' book proposing a heliocentric system, just before his death in 1543, is considered a major event in the history of science. Tycho Brahe (1546--1601) performed the most accurate and comprehensive astronomical and planetary observations until his time. Brahe's observational data helped his young colleague, Johannes Kepler (1571--1630), to develop his laws of planetary motion. These works eventually led to the first well-stablished theory of gravitation by Isaac Newton in 1679. We refer the interested readers to \cite{sleepwalkers} for an interesting detailed history of astronomical science before Newton. 

Newtonian gravity was the dominant theory of gravity in celestial mechanics for almost two centuries. The first observed deviation from Newtonian gravity in the solar system was recognized in 1859 in the motion of Mercury~\cite{verrier}. Analysis of the best available timed observations of transits of Mercury over the Sun's disk shows that the actual rate of the precession of Mercury's perihelion (point of closest approach to the Sun) disagrees with that predicted from Newton's theory by 43" (arc seconds) per century. All attempts failed to explain this deviation by Newtonian gravity until Einstein's theory of gravity in 1916~\cite{1916SPAW.......688E}. The basic concepts of this theory are briefly summarized in the next section. Einstein showed that general relativity agrees closely with the observed amount of perihelion shift of Mercury. This was a powerful factor motivating the further tests of general relativity.

Although general relativity has successfully passed all the performed tests (see \cref{chapter3}), we are still interested to continue testing general relativity and studying alternative theories, for three reasons: (1) Gravity is a fundamental interaction of nature; deeper understanding of gravity leads to deeper understanding of the Universe. (2) All attempts to quantize gravity and to unify it with other types of interaction (i.e. electroweak and strong interactions) suggest that standard general relativity is not likely to be the last word. (3) Since general relativity contains no adjustable parameter, its predictions are fixed and therefore every test of the theory is either a potentially deadly test or a possible probe for new physics~\cite{lrr-2006-3}.

\section{General Relativity in a Nutshell}
\label{subsec:GR-Intro}
The way general relativity describes the cause of motion is quite different from the Newtonian explanation. In general relativity, there is no need to define gravitational forces, as Newton did, to describe the motion of massive objects in gravitational fields. In general relativity, the distribution of matter (massive particles) changes the geometry of spacetime such that massive objects just follow their optimum natural paths through the spacetime (geodesics). Paraphrasing John Wheeler (1911-2008), spacetime tells matter how to move and matter tells spacetime how to curve. 

To briefly review the basic concepts of the theory of general relativity, we start from the key concept of the invariant, differential line element $ds$ at spacetime point $\vec{x}$ as
\be
ds^2=g_{\mu\nu} (\vec{x})\;dx^\mu dx^\nu,
\ee
where $g_{\mu\nu}$ is a $4\times 4$ symmetric tensor ({\em metric tensor}), and repeated indices imply summation. Two examples are (1) the Minkowski metric in a Cartesian-coordinate system i.e. $(t, x, y, z)$ as $g_{\mu\nu}= \text{diag} (-1, 1, 1, 1)$
which has fixed values for its components and describes the {\em flat} spacetime in the absence of matter (or at very far distances from the gravitational source where the gravitational field is negligible), and (2) the Schwarzschild metric in a spherical-coordinate system i.e. $(t, r, \theta, \phi)$ which at a distance $r$ from the source mass $M$ is given by
\be
g_{\mu\nu} (\vec{x})=
\begin{pmatrix} 
-(1-2GM/c^2 r)  & 0 & 0 & 0\\ 
0  & (1-2GM/c^2 r) ^{-1} & 0 & 0\\
0  & 0 & r^2 & 0\\
0  & 0 & 0 & r^2 \sin^2\theta  
\end{pmatrix},
\ee
describing the {\em curved} spacetime around a static, spherically symmetric mass distribution of total mass $M$, where $G$ is Newton's gravitational constant and $c$ is the speed of light.

The geodesic equation of motion for a test particle is given by
\be\label{geodesics}
\frac{d^2 x^\mu}{d\tau^2}=-\Gamma^\mu_{\alpha\beta} \frac{dx^\alpha}{d\tau} \frac{dx^\beta}{d\tau},
\ee
where $\tau$ is the {\em proper} time measured by a clock traveling with the particle, and $\Gamma^\mu_{\alpha\beta}$ are the {\em Christoffel symbols} (also known as connection coefficients) defined by
\be
\label{christoffel}
 \Gamma^\mu_{\alpha\beta}=\frac{1}{2} g^{\mu\nu} (g_{\nu\alpha, \beta}+g_{\nu\beta, \alpha}-g_{\alpha\beta,\nu}),
\ee
where the comma followed by a subscript denotes a partial derivative with respect to that coordinate, and from which we can define the {\em Riemann curvature tensor} as
\be
R^\alpha_{\beta\mu\nu}=\Gamma^\alpha_{\beta\nu,\mu}-\Gamma^\alpha_{\beta\mu,\nu}+\Gamma^\alpha_{\mu\gamma} \Gamma^\gamma_{\beta\nu}-\Gamma^\alpha_{\nu\gamma} \Gamma^\gamma_{\beta\mu}.
\ee
The {\em Ricci tensor} and Ricci scalar can be defined by contracting two of the indices of the Riemann tensor, and then contracting again,
\bea
R_{\mu\nu}=R^{\alpha}_{\mu\alpha\nu},&&  R = g^{\mu\nu} R_{\mu\nu} ;
\eea
the Ricci tensor and scalar appears in the famous Einstein's field equations in general relativity:
\be
\label{GR:fieldequations}
R_{\mu\nu}-\frac{1}{2} g_{\mu\nu} R = \frac{8\pi G}{c^4} T_{\mu\nu}.
\ee
where $T_{\mu\nu}$ is the energy-momentum tensor for the matter. The Einstein--Hilbert action in general relativity is the action that yields the Einstein field equations, given by \eref{GR:fieldequations}, through the principle of least action. It is given by
\be
\displaystyle S=\frac{c^4}{16\pi G} \int R \sqrt{-g} d^4x\,.
\ee

The stress-energy tensor can be regarded as having the following qualitative form:

\be
T_{\mu\nu} (\vec{x})=\left(
\begin{array}{c| ccc} 
\text{Energy Density} &  & \frac{1}{c}\text{Energy Flux} & \\
 \hline 
  &  &  &\\
  \frac{1}{c} \text{(Momentum} &  &  \text{Stress Tensor} &\\
   \text{Density)} &  & (3\times 3) & \\
   &&&
   \end{array}\right).
\ee
Specifically, $T^{00}(\vec{x})$ is the local energy density, $T^{0i}(\vec{x})$ and $T^{i0}(\vec{x})$ are, respectively, the flux of energy and the density of momentum both in the direction of $x^i$ (note $T^{0i}=T^{i0}$). $T^{ij}$ is the $i$th component of the force per unit area exerted across a surface with normal in direction $x^j$. The diagonal elements $T^{ii}$ (no summation over $i$) represent pressure components, and the off-diagonal elements represent shear stresses. 

For more details about general relativity, we refer the interested readers to many published textbooks on this topic including those written by Wald~\cite{Wald:1984cw}, Weinberg~\cite{weinbergGravity}, Misner, Thorne, and Wheeler ~\cite{mtw}, Schutz~\cite{schutzbook}, Hughston and Tod~\cite{Hughston}, Stephani~\cite{Stephani}, d'Inverno~\cite{Inverno}, Carroll~\cite{Carroll}, Kopeikin, Efroimsky, and Kaplan~\cite{2011rcms.book.....K}, and Poisson and Will~\cite{poissonwill}.


\section{Alternative Theories of Gravity}
Alternative theories of gravity are interesting because although general relativity has successfully passed all the tests performed to date, but there are some issues in which general relativity is not quite promising, such as to quantize and unify gravity. An alternative theory might be the solution such that it is compatible with general relativity in certain limits and also can explain the ambiguous sectors like quantum gravity and unifying gravity with other forces. However, so far no alternative theory has been completely successful. The space of possible alternative theories is infinite but the most desirable theories of gravity are those which satisfy a certain number of properties including~\cite{nicoreview}:

\begin{itemize}
\item {\bf Precision tests.} The predictions of the gravitational theory must be consistent with the Solar system, binary pulsar, and experimental tests that have been performed so far. Namely, {\bf (a)} there must exist some limit in which the predictions of the theory are consistent with those of general relativity within experimental precision ({\em general relativity limit}), {\bf (b)} the theory must admit solutions that correspond to observed phenomena, including but not limited to (nearly) flat spacetime, (nearly) Newtonian stars, and cosmological solutions ({\em existence of known solutions}), and {\bf (c)} the special solutions described in (b) must be stable to small perturbations ({\em stability of solutions}). Of course, these properties are not all necessarily independent. For example, the existence of a weak-field limit usually also implies the existence of known solutions.

\item{\bf Well-motivated from fundamental physics.} There must be some fundamental theory or principle from which the alternative theory derives. This fundamental theory would solve some fundamental problem such as the incompatibility between quantum mechanics and general relativity.
\end{itemize}

Since Einstein (1916) many various feasible and unfeasible alternative theories of gravity have been proposed to modify or replace general relativity. In this section we shall introduce two classes out of many: (1) scalar-tensor theories and (2) massive graviton theories. In this dissertation, we only focus on these specific classes (see Part~\ref{part:3},\ref{part:4}). To have a review of alternative theories of gravity specially those are testable via gravitational wave observations we refer the interested reader to \cite{2013IJMPD..2241012A, nicoreview, tegp, poissonwill, 2006PhDT.......281C}.

\subsection{Scalar-Tensor Theories}
\label{sec:ST}
One of the cornerstones of every theory of gravity is its action. 
Although the {\em Einstein frame}~\cite{1999astro.ph.10176F, fujiimaeda} gives the simplest presentation of the scalar-tensor theory, the metric used in this frame is not the same as the physical metric $g_{\mu\nu}$ that governs clocks and rods. Through a conformal transformation
one can recast the theory into the {\em Jordan frame} in which clocks and rods measure the physical values of time and distance. The action in the Jordan frame is given by 
\begin{equation}
S = \frac{1}{16\pi} \int \left [ \phi \;R - \phi^{-1}\; \omega(\phi)\; g^{\alpha\beta}\; \partial_\alpha \phi\; \partial_\beta \phi \right ] \sqrt{-g}\; d^4x + S_{NG} ( \mathfrak{m}, g_{\alpha\beta}) \,,
\label{STaction1}
\end{equation}
where the non-gravitational, matter action $S_{NG}$ involves the matter fields $\mathfrak{m}$ and the metric only. Applying the principle of the least action to \eref{STaction1} leads to the following field equations 
\begin{subequations}\label{STfieldEqs1}
\begin{eqnarray}
G_{\mu\nu} &=& \frac{8\pi}{\phi} T_{\mu\nu} + \frac{\omega(\phi)}{\phi^2} \left ( \phi_{,\mu} \phi_{,\nu} - \frac{1}{2} g_{\mu\nu} \phi_{,\lambda} \phi^{,\lambda} \right ) + \frac{1}{\phi} \left ( \phi_{;\mu\nu} - g_{\mu\nu} \Box_g \phi \right ) \,, 
\label{fieldeq1}\\
\Box_g \phi  &=& \frac{1}{3 + 2\omega(\phi)} \left ( 8\pi T - 16\pi \phi \frac{\partial T}{\partial \phi} - \frac{d\omega}{d\phi} \phi_{,\lambda} \phi^{,\lambda} \right ) \,.
\label{fieldeq2}
\end{eqnarray}
\end{subequations}

If the coupling $\omega(\phi)=\omega_{\text{\tiny BD}}$ is constant, then the general scalar-tensor theory in Eqs.~(\ref{STaction1}) reduces to the massless Brans-Dicke theory~\cite{Brans:1961sx} which is the simplest scalar-tensor theory that one could construct. For more details and more complicated versions of this theory we refer the interested reader to~\cite{tegp, lrr-2006-3, fujiimaeda, DamourEsposito92}.

Like general relativity, scalar-tensor theories are among metric theories of gravity and predict gravitational waves. But they predict an extra scalar (spin-0) mode of polarization in addition to the two transverse-traceless (spin-2) modes of general relativity. The emission of dipolar radiation in scalar-tensor theories is not predicted by general relativity. 

The form of the action in Eqs.~(\ref{STaction1}) suggests that in the weak-field limit one may consider scalar-tensor theories as modifying Newton's gravitational constant via $G\rightarrow G(\phi)=G/\phi$. Scalar-tensor theories have a continuous limit to Einstein's theory such that in the limit of $\omega\rightarrow \infty$ one recovers general relativity. Because of this, scalar-tensor theories have passed all the performed precision tests. The massless Brans-Dicke theory agrees with all known experimental tests provided $\omega_{\text{\tiny BD}}>4\times 10^4$, given by measurements of the time delay in tracking signals to the Cassini spacecraft, while observations of the Nordtvedt effect with Lunar Laser Ranging and observations of the orbital period derivative of white-dwarf/neutron-star binaries yield looser constraints~\cite{2003Natur.425..374B}. Massive Brans-Dicke theory has been recently constrained to $\omega_{\text{\tiny BD}}>4\times 10^4$ and $m_{\text{\tiny s}}<2.5\times 10^{-20}$ eV, with $m_{\text{\tiny s}}$ the mass of the scalar field, through the observations of Shapiro time delay~\cite{alsing}.

Scalar-tensor theories have not only passed the precision tests but also are very well-motivated by fundamental physics. Specially, they can be derived from the low-energy limit of certain string theories. The integration of string quantum fluctuations leads to a higher-dimensional string theoretical action that reduces locally to a field theory similar to a scalar-tensor one~\cite{1985NuPhB.261....1F, 1993NuPhB.400..416G}. In addition, scalar-tensor theories can be mapped to the general class of $f(R)$ theories which have been proposed as a way to account for the acceleration of the universe without resorting to dark energy. (see~\cite{tsujikawa, 2006CQGra..23.5117S, 2010RvMP...82..451S} for a review of $f(R)$ theories and their correspondence to scalar-tensor theories). 

Black holes and stars continue to exist in scalar-tensor theories. Stellar configurations are modified from their general relativistic profile~\cite{willzaglauer, 2011JCAP...08..027H, dimitri}, while black holes are not. Hawking~\cite{hawking} has shown that {\em stationary} black holes in Brans-Dicke theory are identical to those in general relativity. Many extensions of Hawking's theorem have been carried out since then, including \cite{PhysRevD.51.R6608, 2013arXiv1304.2836B, 2012PhRvL.109h1102B, 2013arXiv1304.2836B}. In particular, Sotiriou and Faraoni~\cite{sotirioufaraoni} have generalized Hawking's proof from pure Branse-Dicke theory to a general class of scalar-tensor theories. Recently, Hawking's result has been extended even further to {\em quasi-stationary} black holes. These extensions have been done in general scalar-tensor theories, through the study of post-Newtonian comparable-mass inspirals~\cite{mir13}, extreme-mass ratio inspirals~\cite{yunespanicardoso} and numerical simulations of comparable-mass black hole mergers~\cite{healy}. Post-Newtonian calculations, accurate to $(v/c)^5$ order beyond Newtonian limit, predict no measurable difference between the equations of motion of binary black holes in general relativity and in general scalar-tensor theories of gravity~\cite{mir13}.

\subsection{Massive Graviton Theories and Lorentz Violation}

Einstein's theory of general relativity predicts massless gauge bosons i.e. gravitons for gravitational propagation which travel with the speed of light. In the other hand, in massive graviton theories, the gravitational interaction is propagated by a massive gauge boson i.e. a graviton with mass $m_g\neq 0$. The corresponding Compton wavelength is $\lambda_g\define h/(m_g c)< \infty$. For a detailed review of massive graviton theories see e.g.~\cite{2012RvMP...84..671H}.

Like scalar-tensor theories, massive graviton theories are somewhat well-motivated by fundamental physics, especially by theories of quantum gravity. In the cosmological extension of loop quantum gravity i.e. loop quantum cosmology~\cite{2005LRR.....8...11B, 2003gr.qc.....4074A}, the graviton dispersion relation predicts massive gravitons~\cite{2008PhRvD..77b3508B}. Massive graviton models also arise in some alternative theories inspired by string theory such as Dvali's compact, extra-dimensional theory~\cite{2000PhLB..485..208D}. Other modified theories that imply massive gravitons include Rosen bimetric theory~\cite{rosen1, rosen2}, Visser's theory~\cite{Visser:1997hd}, TeVez~\cite{teves}, and Bigravity~\cite{2012JHEP...09..002P}.

Massive graviton theories have a theoretical issue, the van Dam-Veltman-Zakharov (vDVZ) discontinuity~\cite{vanDam1970397, zakharov}. They do not quite satisfy the precision tests. In particular, certain predictions of massive graviton theories do not reduce to those of general relativity in the $m_g\rightarrow 0$ limit. Roughly speaking, this discontinuity is due to the fact that, in this limit the scalar mode in spin states does not decouple~\cite{nicoreview}. The vDVZ discontinuity, however, can be evaded by carefully including non-linearities in massive graviton theories ~\cite{doi:10.1142/9789814374552_0470, 2011PhRvL.106w1101D, 2012CQGra..29w5026E}.

Although the absence of any particular well-accepted action for massive graviton theories makes it very difficult to ascertain many of the properties of these theories, we can still consider certain phenomenological effects~\cite{nicoreview}. The two main consequences of massive graviton theories are modifications to (1) the Newtonian limit, and (2) gravitational wave propagation. 

The first class of modifications corresponds to the replacement of the Newtonian potential by a Yukawa-type potential. In the non-radiative, near-zone of mass $M$, the Yukawa potential is given by $V=(M/r) \exp(-r/\lambda_g)$, where $r$ is the distance to the massive body~\cite{Will:1997bb}. The proposed tests of Yukawa interactions include the observations of bound clusters, tidal interactions between galaxies~\cite{1974PhRvD...9.1119G}, and weak lensing~\cite{2004APh....21..559C}. These proposed tests are all model-dependent. 

The second class of modifications can be clearly seen in a modified gravitational wave dispersion relation~\cite{mir12, Will:1997bb}. Explicit forms of modifications are given in Eqs.~(\ref{SR}, \ref{SRvelocity}). Either modification to the dispersion relation has the net effect of slowing gravitons down, such that for the same observable event the arrival times of photons and gravitons are different (see \fref{fig:redshift}). We will discuss this issue in more detail in \cref{chapter13}.

Although it is extremely difficult (if not impossible) to measure the mass of a single graviton~\cite{freemandyson}, many authors have tried to put an upper limit on the graviton's mass via different methods including the data analysis of binary pulsars and gravitational waves~\cite{Will:1997bb, mir12, Finn:2002fk, Cutler:2002ef, 2008ApJ...684..870K}. Table~\ref{summary} shows a list of obtained upper limits on the mass of the graviton by a recent matched filtering analysis.

Although massive graviton theories unavoidably lead to a modification to the graviton dispersion relation, the converse is not necessarily true. A modification of the dispersion relation is usually accompanied by a modification to either the Lorentz group or its action in real or momentum space~\cite{nicoreview}. Such Lorentz-violating effects are commonly found in quantum gravitational theories, including loop quantum gravity~\cite{2008PhRvD..77b3508B} and string theory~\cite{2005hep.th....8119C, 2010GReGr..42....1S}, as well as other effective models~\cite{Berezhiani:2008nr, Berezhiani:2007zf}. In Doubly Special Relativity~\cite{2001PhLB..510..255A, AmelinoCamelia:2002wr, 2010arXiv1003.3942A, 2002PhRvL..88s0403M}, the graviton dispersion relation is modified at high energies by modifying the law of transformation of inertial observers. Modified graviton dispersion relations have also been shown to arise in generic extra-dimensional models~\cite{2011PhLB..696..119S}, in Ho\v{r}ava-Lifshitz theory~\cite{Horava:2008ih, Horava:2009uw, Blas:2011zd, 2010arXiv1010.5457V} and in theories with non-commutative geometries~\cite{2011arXiv1102.0117G, Garattini:2011kp, Garattini:2011hy}. None of these theories necessarily requires a massive graviton, but rather the modification to the dispersion relation is introduced due to Lorentz-violating effects.

\section{Parametrized Post-Newtonian Theory as a Powerful Tool}

In the 1970's, Nordtvedt and Will~\cite{nor68b, 1972ApJ...177..757W, 1972ApJ...177..775N, tegp} developed a general Parametrized Post-Newtonian theory (PPN) of gravity in which general relativity and many viable alternative theories of gravity such as scalar-tensor theories can be described by choosing proper values for 10 independent parameters. The PPN parameters and their physical significance are shown in Table~\ref{tab:PPNmeaning}. The values of the PPN parameters differ for different theories (e.g. see Table~\ref{tab:ppnvalues}). In general relativity, all the PPN parameters vanish except $\gamma=\beta=1$. In the next chapter we use the PPN framework as a powerful tool to study the tests of gravitational theories, and leave more details until Chapter~\ref{chapter5} where we discuss the PPN framework. 
\begin {table}[h]
\begin{center}
\begin{tabular}{ | c| l| ccc|  }
\hline
&  & Value in& Value in& Value in\\
PPN& What this parameter measures& general& semiconservative& fully conservative\\
Parameter&relative to general relativity & relativity &theories &theories \\ 
\hline \hline
\multirow{3}{*} &{How much space-curvature} &  & & \\
$\gamma$& is produced by unit& 1 & $\gamma$ & $\gamma$\\ 
& rest mass?&  &  & \\ 
\hline

\multirow{3}{*}&{How much {\em nonlinearity}} &  & & \\
 $\beta$&is there in the superposition& 1 & $\beta$& $\beta$\\
 &law for gravity?&  & & \\ \hline
 
 $\xi$ &{Are there preferred-location} & 0 & $\xi$ & $\xi$\\
&effects?& & & \\\hline
 
 $\alpha_1$ &  & 0 & $\alpha_1$& 0\\
 $\alpha_2$&Are there preferred-frame& 0 & $\alpha_2$& 0\\
$\alpha_3$ &effects?& 0 & 0& 0 \\ \hline
 
$\alpha_3$ &  & 0 & 0 & 0\\
$\zeta_1$ & Is there  & 0 & 0 & 0\\
$\zeta_2$ & violation of  conservation& 0 & 0& 0\\
$\zeta_3$ & of total momentum? & 0 & 0 & 0\\
$\zeta_4$ & & 0 & 0& 0 \\ \hline

\end{tabular}
\end{center}
\caption {The PPN parameters and their physical significance are shown. {\em Semi-conservative} and {\em fully-conservative} theories of gravity are two different classes of theories. In the fully-conservative class the four-angular-momentum $J^{\mu\nu}$ and the four-linear-momentum $P^{\mu}$ are both conserved while in semi-conservative theories only $P^{\mu}$ is conserved. In fully-conservative theories of gravity $\gamma$, $\beta$, and $\xi$ are the only PN parameters of the theory.~\cite{tegp}} \label{tab:PPNmeaning} 
\end {table}

\begin{savequote}[0.55\linewidth]
{\scriptsize ``No amount of experimentation can ever prove me right; a single experiment can prove me wrong.''}
\qauthor{\scriptsize---Albert Einstein}
\end{savequote}

\chapter{Tests of Gravitational Theories} 
\thispagestyle{myplain}
\label{chapter3} 

This chapter is devoted to reviewing tests of gravitation theory. It is important to know that what kind of experiments and observations have been done so far primarily in the weak-field slow-motion regime, the regime covered by the PPN framework. Our results in Part~\ref{part:3} and Part~\ref{part:4} are among the next steps toward providing new tools and abilities to test alternative theories of gravity, using future observations by gravitational-wave detectors. For a review of possible tests of gravitational theories with gravitational-wave detectors see~\cite{mishra:2010tp, 2012arXiv1212.5575G, 2013IJMPD..2241012A}. In this chapter we briefly review the classical tests and tests of the Strong Equivalence Principle. Then we discuss the gravitational-wave's properties that we can use to put alternative theories of gravity to the test. We finish up this chapter with a list of performed tests and a summary of all the obtained bounds on the PPN parameters via various tests.
\lhead[\thepage]{Chapter 2. \emph{Tests of Gravitational Theories}}      
\rhead[Chapter 2. \emph{Tests of Gravitational Theories}]{\thepage}
\section{The Classical and SEP Tests}
\label{sec:classical}
In this section we focus on three ket tests of relativistic gravity, including: (1) the perihelion advance of Mercury, (2) the deflection of light, and (3) the time delay of light. Strong Equivalence Principle (SEP) tests make up another class of tests for gravitational theories, that we discuss in this section.
\subsection{The Classical Tests}
\subsubsection{The Perihelion Advance of Mercury's Orbit}
An anomalous rate of precession of the perihelion of Mercury's orbit had been a puzzle since 1859 \cite{verrier}. taking all the possible Newtonian effects into account, the observational results still showed a deviation as big as $43''$ per century in the perihelion shift of Mercury. This remaining precession can be explained accurately by Einstein's general relativity, and the predicted value agrees closely with the observed amount of perihelion shift. This was a powerful factor motivating the adoption of general relativity. Based on recent measurements of the perihelion advance of Mercury's orbit and using the PPN formalism for fully conservative theories of gravity ($\alpha_1\equiv\alpha_2\equiv\alpha_3\equiv\zeta_2\equiv0$) it is possible to place a bound on the PPN parameters $\gamma$ and $\beta$ (Eq.7.55 in TEGP). The results agree with general relativity. Using 24 years of observing the perihelion shift of Mercury (1966-1990), Shapiro and his collaborators have estimated the following constrains on the PPN parameter combination~\cite{shapiro, 1990grg..conf..313S}:
\be
\mid 2\gamma-\beta-1\mid <3\times 10^{-3}.
\ee
Analysis of data taken since 1990 could improve the accuracy.
\subsubsection{The Deflection of Light}
\begin{figure}[t]
\centering
\includegraphics[width=0.6\textwidth]{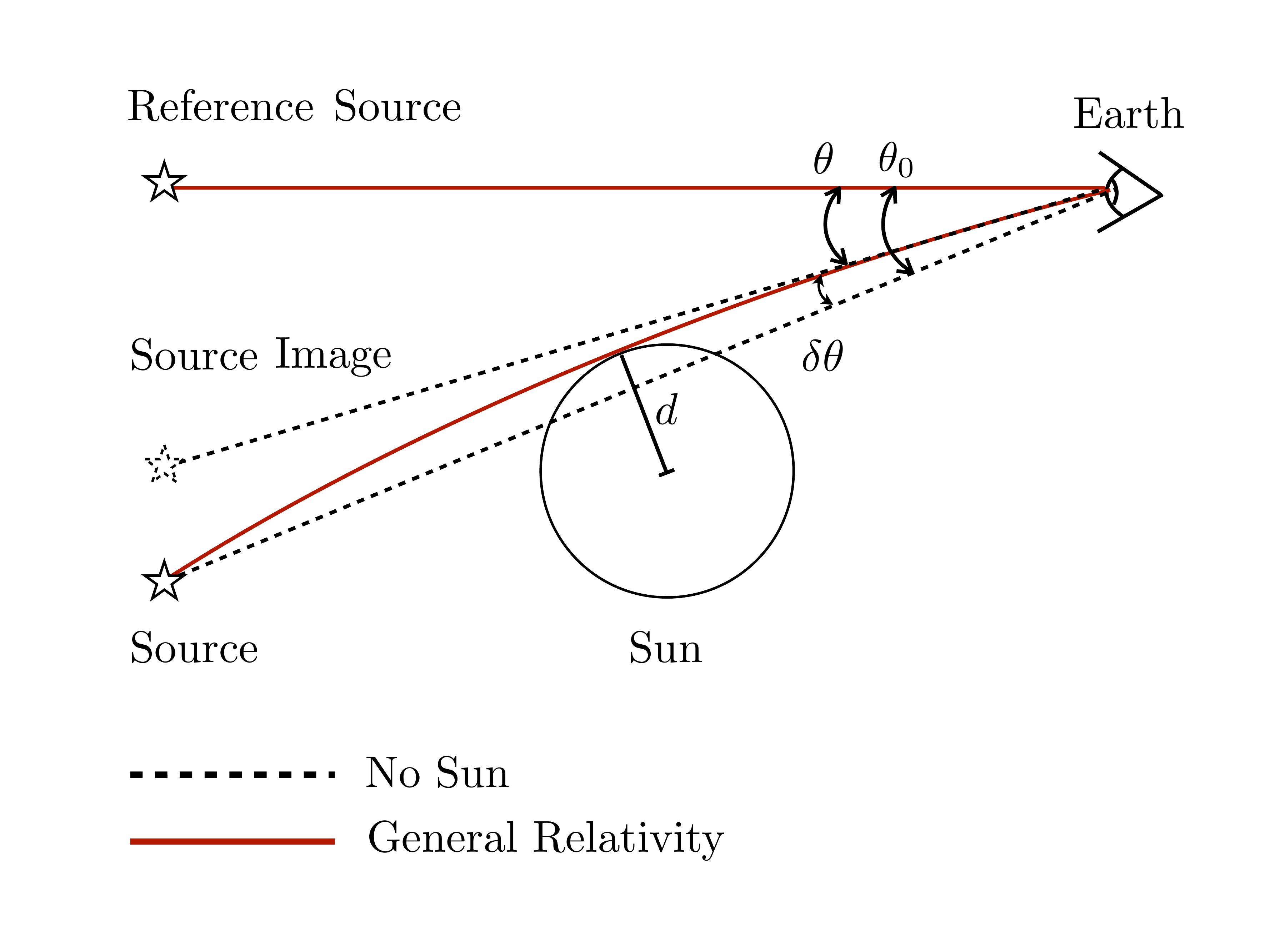}
\caption{Geometry of light-deflection measurements.}
\label{fig:deflection}
\end{figure}

Accurate measurements of the deflection of light near massive bodies like our Sun can test gravitational theories in the PPN formalism by bounding the value of the PPN parameter $\gamma$. A straightforward calculation in the PPN formalism, based on the equations of motion for photons i.e. Eqs.~(6.14, 6.15) of \cite{tegp}, shows that the deflection angle of a light ray coming from a very distant source which is passing nearby a massive object on its way to our detectors on the Earth is given by
\be
\delta\theta=(\frac{1+\gamma}{2}) \frac{4\,m}{d} (\frac{1+\cos\theta_0}{2}),
\ee
where $\gamma$ is the PPN parameter, $m$ is the mass of the body which causes the deflection, $d$ is the closest distance between the light ray and the mass $m$, and $\theta_0$ indicates the angle between the undeflected ray and the direction to the source star (see \fref{fig:deflection}). 

For the Sun, the deflection is maximum for a grazing ray i.e. for $\theta_0\simeq0$, $d\simeq R_\odot\simeq6.96\times10^5$ km, $m=m_\odot=1.476$ km. For light in the visible band, the effect is detectable from the Earth only at the time of total solar eclipses. In this case
\be
\delta\theta_{max}=\frac{1}{2} (1+\gamma) 1''.75.
\ee

\begin{figure}[t]
\centering
\includegraphics[width=0.6\textwidth]{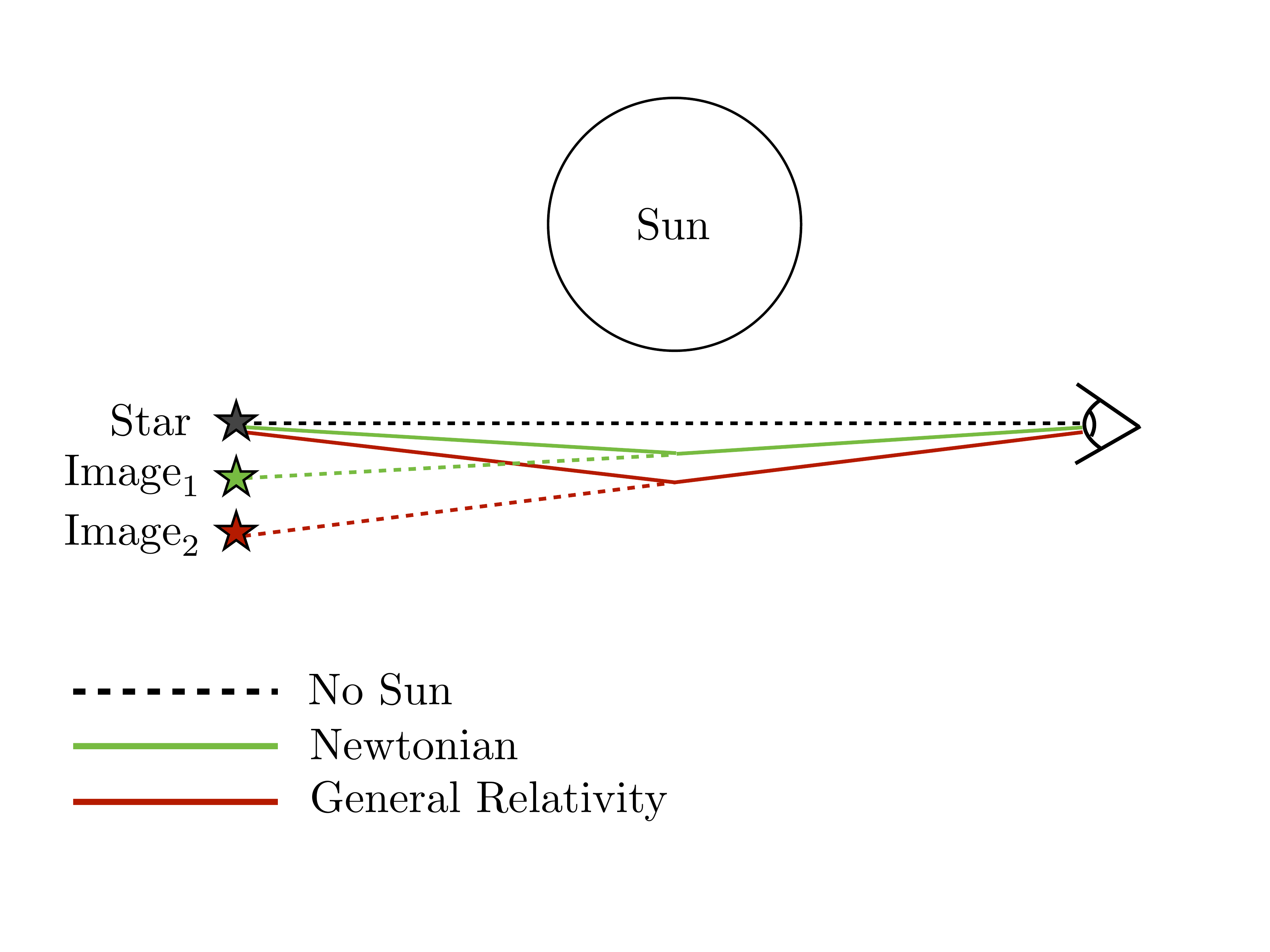}
\caption{Path of a light signal from a far source star to Earth in presence of the Sun's gravitational field, predicted by Newtonian gravity and general relativity. The deflection of light in general relativity is twice what Newtonian gravity predicts.}
\label{fig:deflectionNGR}
\end{figure}

The light deflection phenomenon had been predicted as a Newtonian effect \cite{cavendish, sol1804} many years before Einstein's general relativity in 1915. The first observational test to measure this effect was performed by Arthur Eddington in 1919~\cite{eddington}. The level of accuracy was not very high in the first experiment but clearly enough to reject the Newtonian prediction for the deflection angle which is half of what general relativity predicts. Figure~\ref{fig:deflectionNGR} illustrates the prediction of these theories for the path of a light ray from a far star passing near the Sun.
After Eddington, many other groups measured the deflection of light via different methods and techniques such as very-long-baseline radio-interferometric techniques (VLBI). A complete list of performed measurements of light deflection has been presented in Fig. 7.2 of TEGP~\cite{tegp}. A recent VLBI analysis~\cite{VLBI} yieldes
\be
\frac{1}{2} (1+\gamma)=0.99992\pm 0.00023,
\ee
which is much more accurate that earlier measurements in the 1970's (see~\cite{texas}, for example).
\subsubsection{The Time-Delay of Light}
The spacetime path of a light ray is affected by the gravitational field that it travels through, in two ways: (1) non-uniform gravitational fields cause the optimal path of the light rays to be curved, not straight (2) for a given distance, general relativity predicts a longer time travel for  photon compared to what Newtonian gravity predicts. Here we concentrate on the second aspect i.e. the time delay of light. 

\begin{figure}[t]
\centering
\includegraphics[width=0.6\textwidth]{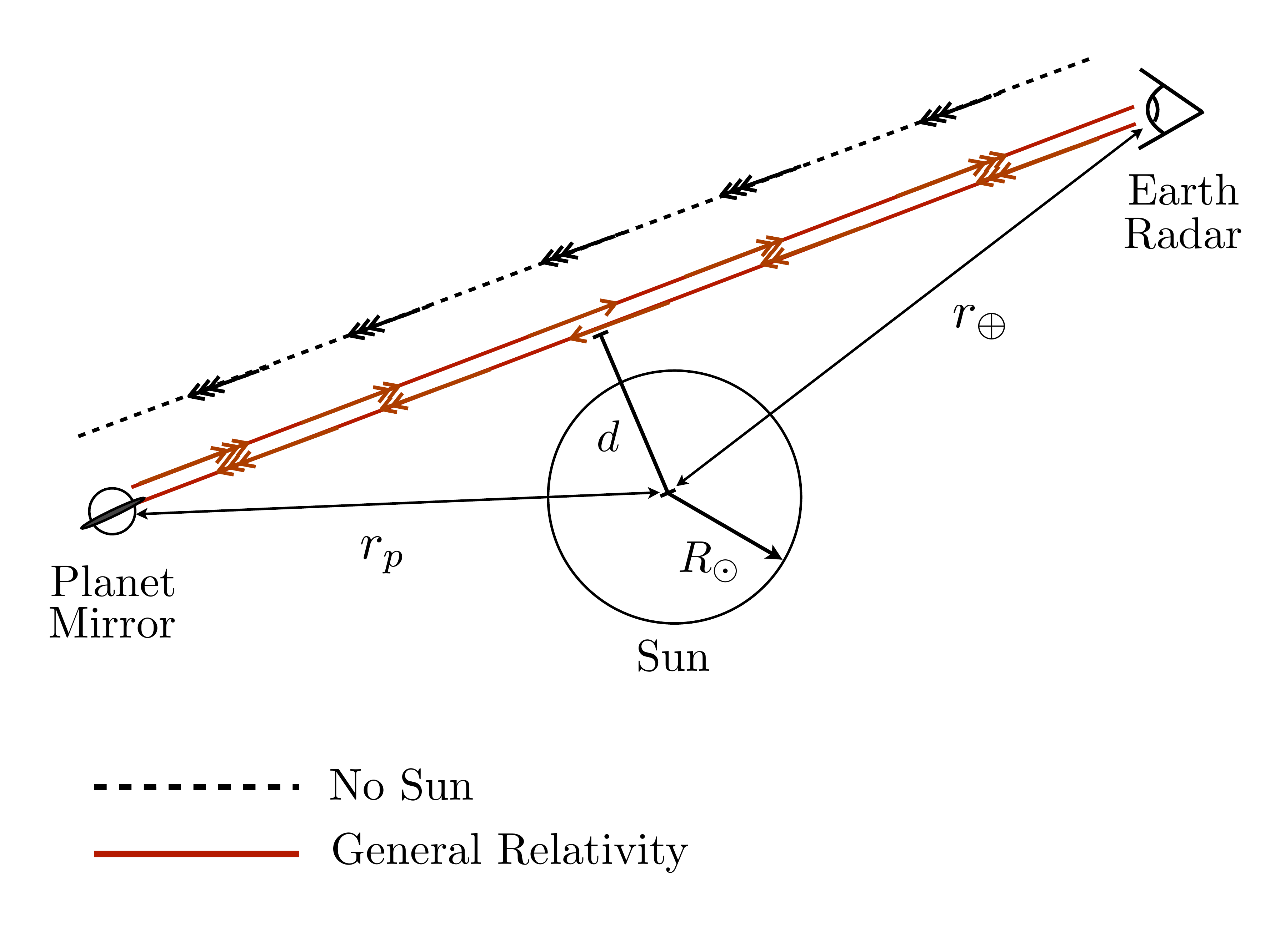}
\caption{Schematic configuration for measuring the time delay effect. It takes longer for a light signal to do a round trip from Earth to another planet in the presence of the gravitational field of the Sun. The number of arrows is inversely proportional to the local time delay of light.}
\label{fig:timedelay}
\end{figure}

For a radar signal, we can measure the time travel of a round trip by sending it toward a far planet such that it passes close to the Sun. The additional time delay $\delta t$ caused by the gravitational field of the Sun is a maximum when the reflector planet is on the far side of the Sun from the earth (superior conjunction); Fig.~\ref{fig:timedelay} shows this configuration. It is straightforward to show that \cite{tegp}
\bea
\delta t &=& 2 (1+\gamma) m \ln (\frac{4 r_\oplus r_p}{d^2})\nonumber\\
&=& \frac{1}{2} (1+\gamma) \biggl\{ 240 \; \mu s -20\; \mu s \ln \biggl[ (\frac{d}{R_\odot})^2 (\frac{a}{r_p})\biggr]\biggr\}\
\eea
where $R_\odot$ is the radius of the Sun, $d$ is the closest distance between the radar beam and the Sun, $r_p$ is the distance between the Sun and the target planet, and $a$ is an astronomical unit.

Many different tests have been done so far to measure the time delay of light. With a high level of accuracy all of the tests confirm general relativity. A complete list of the performed radar time-delay experiments is presented in Fig.7.3 of TEGP~\cite{tegp}. Compared to earlier experiment in the 70's, such as Viking experiment~\cite{1979ApJ...234L.219R}, a significant improvement was reported in 2003 in measuring the parameter $\gamma$ using Doppler tracking data for the Cassini spacecraft~\cite{Cassinni}. 

Most of the theories shown in Table.5.1 of TEGP can select their adjustable parameters or cosmological boundary conditions with sufficient freedom to meet this constraint. From the results of the Cassini experiment, we can conclude that the coefficient $\displaystyle\frac{1}{2} (1+\gamma)$ must be within at most 0.0012 percent of unity. Scalar-tensor theories must have $\omega>40,000$ to be compatible with this constraint. 

\subsection{Tests of the Strong Equivalence Principle}
\subsubsection{Weak, Strong, and Einstein Equivalence Principles}
Besides the classical tests of gravity, there is another class of solar-system experiments that tests the Strong Equivalence Principle (SEP). SEP contains the Einstein Equivalence principle (EEP) as a special case in which local gravitational forces are ignored. EEP is the cornerstone of all {\em metric theories of gravity} including general relativity, scalar-tensor gravity, etc. In metric theories of gravity, matter and non-gravitational fields respond only to the spacetime metric $g_{\mu\nu}$. The only theories of gravity that have a hope of being viable are metric theories, or possibly theories that are metric apart from very weak or short-range non-metric couplings (such as {\em string theory}). In all metric theories of gravity:
\begin{itemize}
\item There exists a symmetric metric tensor.
\item Test bodies move along the {\em geodesics} of the metric.
\item The non-gravitational laws of physics are equivalent to the special relativistic laws in local Lorentz frames.
\end{itemize}

Here we list all the conditions (sub-principles) that are required for a gravitational theory to satisfy {\bf EEP}:
\begin{itemize}
\item \label{wep} WEP (Weak Equivalence Principle) which states that the trajectory of a freely falling {\em test body} (one not acted upon by such forces as electromagnetism and too small to be affected by tidal gravitational forces) is independent of its internal structure and composition. 
\item \label{lli} LLI (Local Lorentz Invariance) which states that the outcome of any local non-gravitational experiment is independent of the velocity of the freely-falling reference frame in which it is performed.
\item \label{lpi} LPI (Local Position Invariance) which states that the outcome of any local non-gravitational experiment is independent of where and when in the universe it is performed.
\end{itemize}

Every metric theory of gravitation satisfies the conditions of EEP, yet does not necessarily satisfy SEP. SEP contains the same principles as EEP but with stronger conditions. {\bf SEP} is satisfied if and only if the following conditions are satisfied:
\begin{itemize}
\item \label{gwep} GWEP (General Weak Equivalence Principle) which states that WEP is valid for {\em self-gravitating} bodies as well as for test bodies.
\item \label{glli} GLLI (General Local Lorentz Invariance) and GLPI (General Local Position Invariance) which respectively state that LLI and LPI are valid not only for local non-gravitational experiments but also for local gravitational experiments too.
\end{itemize}

\subsubsection{Nordtvedt Effect and Other SEP tests}
It has been pointed out~\cite{tegp} that many metric theories of gravity (perhaps all except general relativity) can be expected to violate one or more aspects of SEP (for example see the following equations in TEGP: 6.33, 6.40, 6.75, 6.88). The breakdown in SEP has some observable consequences that many experiments have tested. The Lunar E\"{o}tv\"{o}s experiment to test the Nordtvedt effect is one in which the breakdown in GWEP is the target. The Nordtvedt effect is a prediction of many gravitational theories in which the Earth and the Moon fall toward the Sun with different accelerations. Considering the inertial mass as $m_i$ and passive gravitational mass as $m_p$ we have $m_i a = m_p \nabla U$ and from~\cite{tegp}, we find that many theories predict
\be
\frac{m_p}{m_i}=1+\eta_N\frac{E_g}{m_i},
\ee
where $\eta_N$ (Nordtvedt parameter) is a linear combination of PPN parameters as
\be
\eta_N=4\beta-\gamma-3-\frac{10}{3} \zeta-\alpha_1+\frac{2}{3} \alpha_2-\frac{2}{3}\zeta_1-\frac{1}{3}\zeta_2,
\ee
and $E_g$ is the gravitational self-energy of the body. Since for laboratory-sized objects the value of $E_g/m_i$ is extremely small ($E_g/m_i\leq 10^{-27}$) the existence of the Nordtvedt effect does not violate the results of laboratory E\"{o}tv\"{o}s experiments~\cite{eotvos}. This is far below the sensitivity of current and future E\"{o}tv\"{o}s-type experiments. On the other hand, for the Sun, Earth, and the Moon, $E_g/m_i$ is respectively $3.6\times 10^{-6}$, $4.6\times 10^{-10}$, and $0.2\times 10^{-10}$. Measuring the Nordtvedt effect for the Earth-Moon-Sun system via Lunar Laser Ranging gives
\be\label{nordtvedt}
\eta_N = \left\{
    \begin{array}{ll}
        0.00\pm 0.03 & \mbox{\cite{williams}}\\
        0.001\pm 0.015 & \mbox{\cite{shapiro76}}\\
	0.00044\pm 0.00045  & \mbox{\cite{Nordvedt1, Nordvedt2}}
    \end{array}
\right.
\ee

General relativity does not violate SEP and therefore there is no Nordtvedt effect in general relativity ($\eta_N=0$), but this effect is certainly present in general scalar-tensor theories \footnote{except for a particular choice of the function $\omega(\phi)$, i.e. Barker's constant $G$ Theory, in which $\omega(\phi)=\displaystyle\frac{4-3\phi}{2\phi-2}$.} such that $\eta_N=1/(1+2\omega)+4\zeta\lambda_1$ where $\lambda_1$ and $\zeta$ are defined in Eqs.~~(\ref{deflambdaone}, \ref{defzeta}). In scalar-tensor theories of gravity, the internal structure of bodies clearly affects the dynamics of motion and therefore violates the SEP.

Besides the Nordtvedt effect and Lunar E\"{o}tv\"{o}s experiments there are many other SEP experiments that test preferred-frame effects, preferred-location effects, and constancy of the Newtonian gravitational constant. Preferred-frame and preferred-location effects can be tested via two type of experiments: (1) geophysical tests (2) orbital tests. Interested readers can see lots of details in sections 8.1-8.4 of TEGP~\cite{tegp}. These SEP  experiments can measure the PPN parameters and therefore put additional bounds on some of them. 



\section{Gravitational-Wave Tests}
In the previous section we showed that a variety of tests of gravity in the solar system confirm general relativity. However the post-Newtonian limit of any other alternative metric theories of gravity, within a small margin of error (ranging from $1\%$ to parts in $10^{-7}$) must agree with that in general relativity. Most currently viable theories of gravity, such as scalar-tensor theories, can accommodate these constraints by choosing appropriate values for their arbitrary, intrinsic parameters and functions. Of course, no such adjustments are needed for general relativity. This fact makes general relativity the simplest and the most favorable one. In the other hand, because general relativity contains no adjustable parameter, any deviation from the fixed general relativistic predictions would kill the theory.

In addition to the post-Newtonian tests that we discussed in \sref{sec:classical}, new testing grounds where the differences among competing theories may appear in observable ways are also possible. Measuring the properties of gravitational waves, observing binary pulsars, and cosmological tests are new arenas for testing theories of gravity besides the classical and SEP tests. In this section we focus on gravitational radiation as a tool for testing relativistic gravity. Although Einstein's theory of relativity had predicted the existence of gravitational waves as ripples of spacetime, Eddington~\cite{edi23} suggested that they might represent merely ripples of the coordinates of spacetime and as such would not be observable. Forty years later, Bondi and his collaborators \cite{bondi62} showed in invariant, coordinate-free terms that gravitational radiation is physically observable. They explicitly showed that gravitational waves carry energy and momentum away from systems, and that the mass of systems that radiate gravitational waves must decrease. 

The {\em existence} of gravitational radiation is not particularly strong evidence for or against any proposed theories of gravity, because almost all viable alternative metric theories of gravity predict gravitational waves as well as general relativity. Therefore it is not the existence of gravitational waves that will concern us here to test gravity but the detailed properties of these waves, including speed, polarization, and radiation back-reaction.

In the weak-field, slow-motion, and far-zone limit, the predictions of various viable metric theories of gravity might be different from each other and from the predictions of general relativity at least in three important ways. They may predict: (1) different values for the speed of radiated gravitational waves which might not be necessarily equal to the speed of light, (2) different polarization states for generic gravitational waves, and (3) different multi-polarities (monopole, dipole, quadrupole, etc.) of gravitational radiation. Although the detection of gravitational waves is required for tests of speed and polarization, the tests of multi-polarities do not necessarily require direct gravitational-wave detection. The multi-polarities of gravitational waves can be studied by analyzing the back influence of the emission of radiation on the source (radiation reaction) for different multipoles. For instance, the emission of gravitational radiation changes the period of a two-body orbit, such as a binary pulsar. This is because the system loses energy via radiation of gravitational waves. 
\subsection{Speed of Gravitational Waves}
General relativity and scalar-tensor theories of gravity both predict that gravitational waves  propagate along null geodesics with a speed equal to the speed of light, $v_g=c$ (in the limit in which the wavelength of gravitational waves is small compared to radius of curvature of the background spacetime). On the other hand, if gravitation propagates by a massive field (a massive graviton), the speed of gravitational waves could differ from $c$ (see more details in \sref{speedofGW}). Vector-tensor theories~\cite{wil72, helling}, Rosen's bimetric theories~\cite{rosen1, rosen2}, and Rastall's theory~\cite{1972PhRvD...6.3357R} predict different values for the speed of gravitational radiation depending on the parameters of the theory (see section 10.1 of \cite{tegp} for details).

The most obvious way to measure (or bound) the speed of gravitational waves is by comparing the arrival times of a gravitational-wave signal and of an electromagnetic-wave signal from the same event, for example a supernova. For an event at a distance $D$ from our detector, the speed of gravitational radiation can be bounded by measuring the time interval between emission and arrival of an electromagnetic and gravitational signal from the same source. According to \cite{lrr-2006-3}
\be\label{vgbound}
1-\frac{v_g}{c}=5\times 10^{-17} (\frac{200 \text{Mpc}}{D}) (\frac{\Delta t}{1 s}),
\ee
where $\Delta t\equiv \Delta t_a-(1+Z) \Delta t_e$ is the {\em time difference}, where $\Delta t_a$ and $\Delta t_e$ are the differences in arrival time and emission time of the two signals, respectively, and $Z$ is the redshift of the source. The value of $\Delta t_e$ is considered to be unknown in many cases, so that the best one can do is to employ an upper bound on $\Delta t_e$ based on observation or modeling. 

If the frequency of the gravitational-waves is such that $h f \gg m_g c^2$, where $h$ is Planck's constant, then
\be
v_g/c \approx 1-\frac{1}{2} \biggl(\frac{c}{\lambda_g f}\biggr)^2,
\ee
where $\lambda_g=h/(m_g c)$ is the graviton Compton wavelength, and the bound on $v_g$ at \eref{vgbound} can be converted to a bound on $\lambda_g$ as
\be\label{lambdagbound}
\lambda_g > 3\times 10^{12} \text{km}\; \biggl(\frac{D}{200 \text{Mpc}} \frac{100 \text{Hz}}{f}\biggr)^{1/2} \biggl(\frac{1}{f \Delta t}\biggr)^{1/2}.
\ee

In the above analysis we have assumed that the source emits {\em both} gravitational and electromagnetic signals and we are able to detect them accurately enough. We have also assumed that the relative time of emission, $\Delta t_e$, is either very small or measurable to sufficient accuracy.

Instead of using both electromagnetic and gravitational signals from the same source, Will \cite{Will:1997bb} proposed a method in which a bound on the graviton mass can be set by studying gravitational radiation alone. This has been shown specifically in the case of inspiralling compact binary systems. Roughly speaking, by using Will's method the {\em phase interval} $f \Delta t$ in Eq. (\ref{lambdagbound}) can be measured to an accuracy $1/\rho$, where $\rho$ is the signal-to-noise ratio. Thus, one can estimate the bounds on $\lambda_g$ achievable for various compact inspiral systems, and for various detectors. In part \ref{part:4} we will discuss this method and a generalized version of it in detail. Other possible gravitational-wave based methods include (1) using binary pulsar data to bound modifications of gravitational radiation damping by a massive graviton~\cite{Finn:2002fk}, and (2) using LISA-like observations of the phasing of waves from compact white-dwarf binaries, eccentric galactic binaries, and eccentric inspiral binaries~\cite{Cutler:2002ef, Jones:2004uy}.

\subsection{Polarization of Gravitational Waves}
\begin{figure}[t]
\centering
\includegraphics[width=1\textwidth]{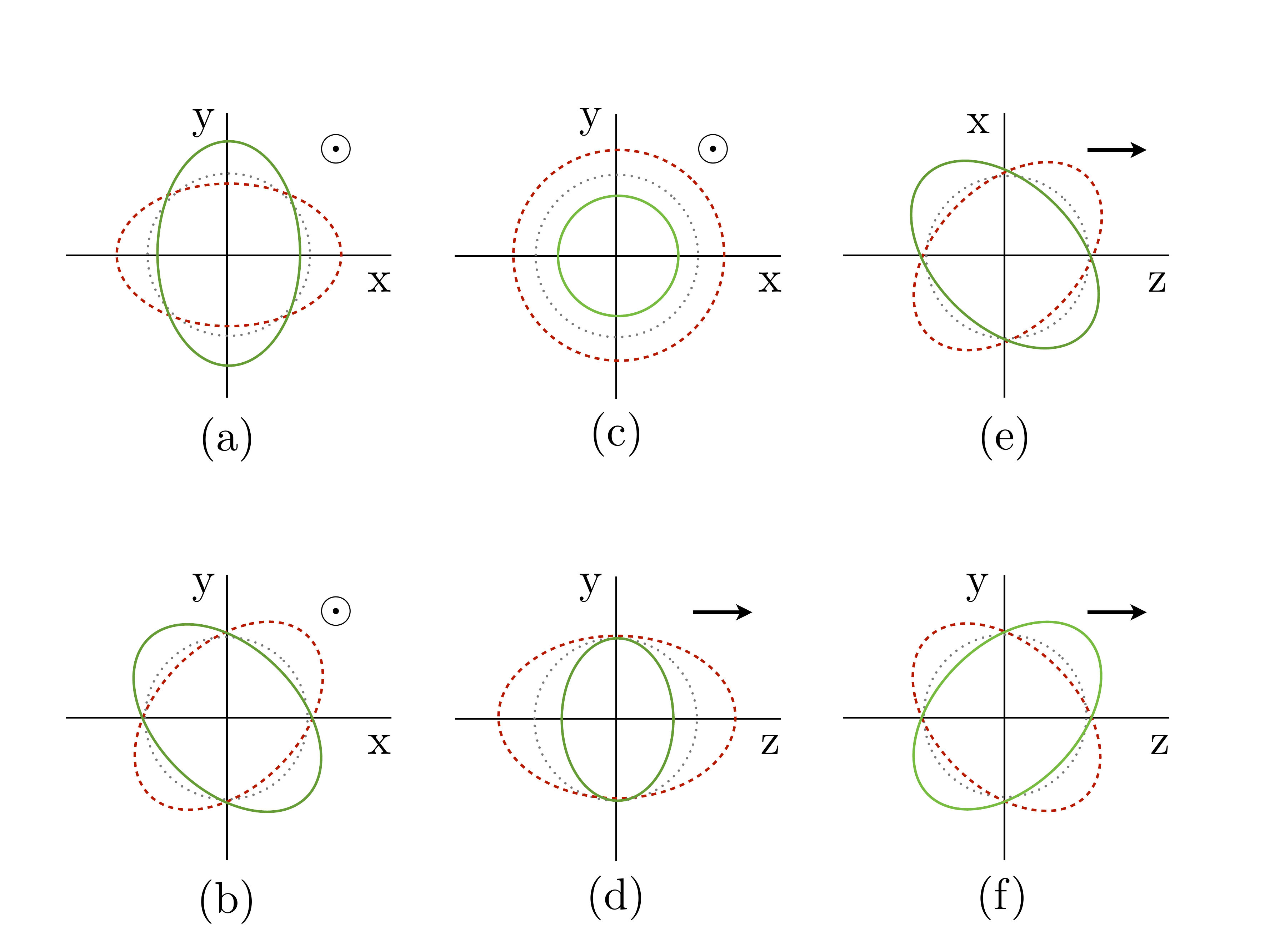}
\caption{The six polarization modes of a weak, plane gravitational wave permitted in any metric theory of gravity. Shown is the displacement that each mode induces on a ring of test particles (gray, dotted circle). We assume that the wave always propagates in the $+z$ direction and has time dependence $\cos{\omega t}$. In (a), (b), (c) the wave propagates out of the plane; in (d), (e), (f), the wave propagates in the plane. The red, solid line is a snapshot at $\omega t=0$ while the gray, dotted line and the green, dashed line are two snapshots at $\omega t=\pi/2$ and $\omega t=\pi$, respectively. There is no displacement perpendicular to the plane of the figure. In general relativity, (a) and (b) are the only possible polarizations; in massless scalar-tensor gravity, (c) may also be present \cite{tegp}.}
\label{fig:polarization}
\end{figure}

In principle, a well-designed gravitational-wave antenna, for example AdLIGO, can measure the local components of a symmetric $3\times 3$ tensor which is composed of the {\em electric} components of the Riemann curvature tensor, $R_{0i0j}$, via the equation of geodesic deviation~\cite{Eardleyprl, Eardleyprd}. If we show the spatial separation distance between two freely falling test masses by $x^i$, based on general relativity, the equation of geodesic deviation is $\ddot{x}^i=-R_{0i0j} x^j$. The symmetric $R_{0i0j}$ has six independent components, which can be expressed in terms of six modes of polarization. Figure~\ref{fig:polarization} shows these six possible independent polarization modes. This figure indicates how a ring of freely falling test particles can be distorted due to each of these polarization modes. Three of these six generic polarization modes represent transverse waves and the other three represent longitudinal waves. Four of them (a), (b), (e), (f) are quadruple modes in different planes while there is one monopolar {\em breathing} mode (c) and one axially symmetric {\em stretching} mode in the propagation direction (d).

In general relativity only two transverse quadrupole modes (a), (b) are present, independent of the source. Modes (a) and (b) correspond to the waveforms $h_+$ and $h_\times$, respectively. A suitable array of gravitational-wave antennas could describe or limit the number of polarization modes present in a given wave. Any observational evidence for other modes, besides (a) and (b), will be disastrous for general relativity. {\em Massless} scalar-tensor theories differ from general relativity by prediction of an extra polarization mode beside the general-relativistic polarization modes, namely a monopolar breathing mode (c). Notice that the absence of a breathing mode in future observational data would not necessarily rule out scalar-tensor gravity, because the strength of that mode depends on the nature of the source. In {\em massive} scalar-tensor theories the longitudinal stretching mode (d) is also possible, in addition to (a), (b), and (c), but it is suppressed relative to breathing mode (c) by a factor of $(\lambda/\lambda_c)^2$, where $\lambda$ is wavelength of the radiation, and $\lambda_c$ is the Compton wavelength of the massive scalar. More general metric theories predict additional longitudinal modes, up to the full complement of six (see chapter 10 of \cite{tegp} for details).

Implementing polarization observations has been studied in detail~\cite{tegp, 1998grwa.conf..168L, 1997rggr.conf..419W}. One important question is whether the current and future interferometric gravitational-wave detectors (ground-based and space-based, or a combination of both types) could perform interesting polarization measurements~\cite{1999PhRvD..59d4027B, 1999PhLB..470...67G, 2000PhRvD..62b4004M, 1997rggr.conf..433W, 2005CQGra..22S1321W}. The two LIGO observatories (in Washington and Louisiana states) have been constructed to have their arms as parallel as possible, apart from the curvature of Earth. Although this maximizes the joint sensitivity of the two detectors to gravitational-waves, unfortunately it minimizes their ability to detect the two modes of polarizations. Installing the INDIGO detector~\cite{indigo} in India will be a major help in this regard.

\subsection{Gravitational Radiation Back-Reaction}
In addition to measuring the speed and polarization of gravitational-waves, gravitational-wave-based tests of gravity are also possible via studying radiation reaction effects in compact binary sources. In the case of binary pulsars, the first derivative of the binary frequency $\dot{f}_b$ is measured using radio signals from the orbiting pulsar to measure the orbit precisely, while in the case of inspiralling compact binaries, we are able to measure the full nonlinear variation of $f_b$ as a function of time via gravitational-wave signals. 

Broad-band laser interferometers are especially sensitive to the phase evolution of the gravitational waves.  To extract gravitational-wave signals from noisy outputs of the detectors, we need to have an ensemble of theoretical {\em template waveforms} which depend on the intrinsic parameters of the inspiralling binary, such as the component masses, spins, and so on, and on its inspiral evolution. Data analysis involves some matched filtering of the noisy detector output against this ensemble of templates. For this purpose we need templates, accurate to an appropriate post-Newtonian order.

The evolution of the gravitational-wave frequency $f=2 f_b$ has been calculated up to the accuracy of 3.5PN order (see \cite{2004PhRvL..93i1101B} for a review). To avoid lengthy expressions at higher orders, here we only show the expression until 2PN order, calculated by Blanchet and his collaborators \cite{1995PhRvD..51.5360B, 1995PhRvL..74.3515B, wil96}:
\bea\label{fdot}
\dot{f}&=&\frac{96\pi}{5} f^2 (\pi \mathcal{M} f)^{5/3} \biggl[ 1-\biggl(\frac{743}{336}+\frac{11}{4}\eta\biggr) (\pi m f)^{2/3}+4\pi (\pi m f)\nonumber\\
&&\quad+\biggl(\frac{34103}{18144}+\frac{13661}{2016}\eta+\frac{59}{18}\eta^2\biggr) (\pi m f)^{4/3}+\mathcal{O}[(\pi m f)^{5/3}]\biggr],
\eea
where $m$, $\mathcal{M}$, $\eta$ are total mass, chirp mass, and mass-ratio parameters, respectively, given by Eqs.~(\ref{massdefinitions}, \ref{massdefinitions2}). This rate of change in the frequency is related to the rate of orbital energy loss by Kepler's third law via
\be\label{bridge}
\frac{\dot{f}}{f}=\frac{3}{2 E} \frac{dE}{dt}.
\ee

In a generic metric theory of gravity the rate of energy loss from an inspiralling compact binary system can be parametrized to leading order in a post-Newtonian expansion, as \cite{tegp}:
\be\label{Edot}
\frac{dE}{dt}=-\biggl\langle \frac{\mu^2 m^2}{r^4} \biggl[\frac{8}{15} (\kappa_1 v^2-\kappa_2 \dot{r}^2)+\frac{1}{3} \kappa_D \mathfrak{S}^2 \biggr] \biggr\rangle,
\ee
where $r$ is orbital separation, and $v$ is relative velocity. $\mathfrak{S}$ is the difference in the self-gravitational binding energy per unit mass between the two bodies. $\kappa_1$ and $\kappa_2$ are known as {\em PM parameters}, because of the pioneering work of Peters and Mathews \cite{1963PhRv..131..435P}, and their values depend on the theory (see Table \ref{tab:multipole}). While $\kappa_1$ and $\kappa_2$ represent quadruple radiation, $\kappa_D$ represents dipole radiation. There is no dipole radiation in general relativity and therefore $\kappa_D=0$, but scalar-tensor theories predict a dipolar contribution in the energy rate.  In general relativity ($\kappa_1=12$, $\kappa_2 =11$), the orbital frequency change induced by Eq. (\ref{Edot}) corresponds to the leading term ---the factor unity in the square brackets--- in  Eq. (\ref{fdot}).  
\setlength{\tabcolsep}{5pt} 
\setlength{\extrarowheight}{1.5pt}
\begin{table}[t]
\centering
\begin{tabular}{| l |  c c c| }
\hline
Theory & $\kappa_1$ & $\kappa_2$ & $\kappa_D$ \\
\hline\hline
General Relativity & 12 & 11 & 0\\
Brans-Dicke & $12-\displaystyle\frac{5}{2+\omega}$&$11-\displaystyle\frac{45}{8+4\omega}$&$\displaystyle\frac{2}{2+\omega}$\\
&&&\\
\hline
\end{tabular}
\caption{\label{tab:multipole} Multipole gravitational radiation parameters in general relativity and Brans-Dicke theory. A complete list of these parameters in other alternative theories of gravity can be found in Table 10.2 of TEGP~\cite{tegp}}
\end{table}

Based on above discussion, there are three possibilities that can be suggested to use radiation reaction effects to test gravity:
\begin{enumerate}
\item Performing accurate observations with sensitive detectors, we might be able to measure the coefficients of different powers of frequency in Eq.~(\ref{fdot}), leading to a possible test of general relativity. Blanchet and Sathyaprakash have shown that an interesting test of the so-called {\em tail-effects} (the third term in \eref{fdot}) could be possible by observing a source with a sufficiently strong signal~\cite{1994CQGra..11.2807B, 1995PhRvL..74.1067B}.
\item Another possibility is studying radiated gravitational-waves from a system of a small mass orbiting and inspiralling into a spinning black-hole. According to general relativity the spacetime around this spinning black-hole must be a Kerr spacetime which can be uniquely described by its mass and angular momentum (no-hair theorem), and consequently, observation of the waves could test this fundamental hypothesis~\cite{Poisson:1996tc,Ryan:1995wh}.
\item As we pointed out earlier, prediction of an additional dipole-radiation contribution in the energy lost formula, i.e. Eq.~(\ref{Edot}), can be used as a test of the gravitational theory. For example, any observational evidence for a dipolar contribution to the orbital evolution will be disastrous for general relativity in which no dipole radiation is predicted. Many authors have worked on the capabilities of both ground-based and space-based detectors to distinguish between general relativity and alternative theories of gravity and have shown that observing gravitational-waves even in the best case i.e. from mixed neutron-star/black-hole inspirals (which are the most promising type of binary sources among others such as black-hole/black-hole, and neutron-star/neutron-star  to observe any difference between general-relativity and scalar-tensor theories, see Chapter~\ref{chapter12}) is not likely to bound scalar-tensor gravity at a level competitive with the Cassini bound or with future solar-system improvements \cite{BBW, BBW2, kro95, scharrewill, willST, Will:2004xi}. On the other hand, such observations would be testing these theories in the radiative regime, as opposed to the non-radiative regime of the PPN framework.
\end{enumerate}

\section{Other Tests and Summarizing the Experimental Results}

In addition to classical tests, tests of SEP, and 
gravitational-wave based tests, there remains a number of tests of post-Newtonian gravitational effects that do not fit into any of these mentioned categories. In some cases, the prior constrains on the parameters are tighter than the best limit these experiments could hope to achieve. Obviously, one might ask why we should bother performing any other test when we already have obtained stronger bounds on the PPN parameters? The answer is that in spite of previous tests, for the following reasons it is important to carry out such experiments: (1) each new test provides independent, though potentially weaker, checks of the values of the PPN parameters and therefore is an independent test of gravitation theory,  (2) we should not treat the PPN formalism in a prejudicial way; it reduces the importance of experiments that have independent, compelling justifications for their performance, (3) any result which shows any disagreement with general relativity would be very interesting. 

Remaining tests of general relativity and alternative theories of gravity include: the Gravity Probe-B gyroscope experiment~\cite{tegp, schiff-b, schiff-c, 1974exgr.conf..331E, 1974exgr.conf..361L, Everitt:2011hp, 2011PhyOJ...4...43W}, laboratory tests of post-Newtonian gravity~\cite{kreuzer, 1972PhRvL..28.1665G, morrison, 1977PhRvD..15.2047B}, tests of post-Newtonian conservation laws~\cite{1976ApJ...205..861W, tegp}, stellar system tests which include: internal structure dependance~\cite{tegp, Lasky:2008fs, dimitri} and the binary pulsars~\cite{tegp, 2003LRR.....6....5S, 2006Sci...314...97K, 2007arXiv0704.0749D, 2009PhRvD..80d2004Y}, cosmological tests~\cite{tegp, 2005PhRvD..71l3526C, 2006AIPC..847...25C, 1999PhRvD..59l3502D, 1997PhRvD..56.7627S}. Table~\ref{ppnbounds} summarizes the tightest bounds on the PPN parameters, obtained by different experiments. Notice that no feasible experiment or observation has ever been proposed that would set direct limits on the parameters $\zeta_1$ or $\zeta_4$. However, these parameters do appear in combination with other PPN parameters in observable effects, for example in the Nordtvedt effect. 

A resource letter by Will~\cite{2010AmJPh..78.1240W} provides an introduction to some of the main current topics in experimental tests of general relativity as well as to some of the historical literature.

\setlength{\tabcolsep}{5pt} 
\setlength{\extrarowheight}{1.5pt}
\begin{table}[h]
\centering
\begin{tabular}{| c|  l r l| }
\hline
Parameter& Effects & Limit & Remarks\\
\hline\hline
$\gamma-1$&	 time delay&	 $2.3\times 10^{-5}$&	Cassini trackimg\\
	&	light deflecttion&	$4\times 10^{-4}$& VLBI\\
\hline
$\beta-1$&	perihelion shift&	$3\times 10^{-3}$&	$J_2=10^{-7}$ from helioseismology\\
	&	Nordtvedt effect&	$2.3\times 10^{-4}$&	 $\eta_N=4\beta-\gamma-3$ assumed\\
\hline
$\zeta$&	Earth tides&	$10^{-3}$&	gravimeter data\\
\hline
$\alpha_1$& orbital polarization&	$10^{-4}$&	Lunar laser ranging\\
	&	&	$2\times 10^{-4}$&	PSR J2317+1439\\
\hline
$\alpha_2$&	spin precession&	$4\times 10^{-7}$&	solar alignment with ecliptic\\
\hline
$\alpha_3$&	pulsar acceleration&	$4\times 10^{-20}$&	pulsar $\dot{P}$ statistics\\
\hline
$\eta_N$&	Nordtvedt effect&	$9\times 10^{-4}$&	lunar laser ranging\\
\hline
$\zeta_1$&	--- &	$2\times 10^{-2}$&	combined PPN bounds\\
\hline
$\zeta_2$&	binary acceleration&	$4\times 10^{-5}$&	$\ddot{P}_p$ for PSR 1913+16\\
\hline
$\zeta_3$&	Newton's 3rd law&	$10^{-8}$&	lunar acceleration\\
\hline
$\zeta_4$&	---&	--- &	not independent (see equation (58) of \cite{lrr-2006-3})\\
\hline
\end{tabular}
\caption{\label{ppnbounds}Current limits on the PPN parameters. Here $\eta_N$ is a combination of the PPN parameters as given in Eqs.~(\ref{nordtvedt}) \cite{lrr-2006-3}.}
\end{table}

\begin{savequote}[0.55\linewidth]
{\scriptsize ``It doesn't matter how beautiful your theory is, it doesn't matter how smart you are. If it doesn't agree with experiment, it's wrong.''}
\qauthor{\scriptsize---Richard P. Feynman}
\end{savequote}
\chapter{Gravitational Waves: Sources and Detection} 
\label{chapter4} 
\thispagestyle{myplain}

The existence of gravitational waves is one of the direct predictions of general relativity (and of almost all other alternative theories of gravity), produced by the acceleration of mass. No gravitational-wave signal has been detected directly to date. What are gravitational waves? What sources can generate these waves? How do they propagate and how can we detect them? These issues will be discussed in this chapter. 

In summary, gravitational-waves can be thought of as ripples in the curvature of spacetime. Why are we interested in their direct detection? First, that would be another verification of general relativity and it would be a major upset if gravitational waves did not exist! Second, and more importantly, the detection of gravitational waves will open a new window to the Universe, as a new branch of astronomy. Gravitational-wave astronomy will provide powerful tools for looking into the heart of some of the most violent events in the Universe in a way that is totally different from electromagnetic astronomy. 

It is believed that the reason for not detecting any gravitational-wave signal so far with the first generation of detectors such as initial-LIGO/VIRGO is the lack of strong-enough astronomical sources in the sensitive range of the detectors. 

In addition to the references cited in this chapter for specific topics on gravitational-waves, there exist many informative books and review articles including those written by Saulson~\cite{saulson}, Maggiore~\cite{maggoire}, Creighton and Anderson~\cite{anderson}, Jaranowski and Krol\'{a}k~\cite{lrr-2005-3}, Hartle~\cite{hartle2003}, Collins~\cite{collinsbook} Misner, Thorne, and Wheeler~\cite{mtw}, Schutz~\cite{schutzbook}, Sathyaprakash and Schutz (2009)~\cite{sat09}, Freise and Strain (2010)~\cite{lrr-2010-1}, Pitkin {\em et al.} (2011)~\cite{lrr-2011-5}, and more recently by Blair {\em et al.} (2012)~\cite{blairbook}, and Riles (2013)~\cite{Riles:2012fk}. 

\lhead[\thepage]{Chapter 3. \emph{Gravitational Waves}}      
\rhead[Chapter 3. \emph{Gravitational Waves}]{\thepage}

\section{Generation and Propagation}
According to general relativity, the presence of any matter will curve the spacetime around it. The proper distance between two neighboring points is given by $ds^2=g_{\mu\nu} dx^\mu dx^\nu$ where $g_{\mu\nu}$ is the metric tensor. In the absence of matter (or at very far distances from the matter), spacetime is flat (asymptotically flat) and the metric tensor is the Minkowski metric i.e. $\eta_{\mu\nu}=(-1, +1, +1, +1)$ in a Cartesian coordinate system.

The origin of gravitational waves is implicit in the tensorial field equations of the theory (for general relativity and scalar-tensor theories of gravity see \eref{GR:fieldequations} and \eref{STfieldEqs1}, respectively). To see why, consider a region far from a source, a nearly-flat region where the gravitational-wave perturbs a flat Cartesian metric $\eta_{\mu\nu}$ by only a small amount $h_{\mu\nu}$, i.e. $g_{\mu\nu}=\eta_{\mu\nu}+h_{\mu\nu}$ where $h_{\mu\nu}\ll 1$. Choosing an appropriate gauge condition, it  can be shown that this {\em linearized gravity} yields simple wave equations for the components of tensor $h$ such that in vacuum we have $\Box h_{\mu\nu}=0$. The amplitude of the wave $h$ is related to the perturbation of the metric which is in turn related to the curvature of spacetime. In addition, $h$ can be interpreted as a physical strain in space or more precisely $h\sim\delta L/L$ where $\delta L$ is the change in separation of two masses a distance $L$ apart.

From elementary electrodynamics, we know that the acceleration of charged particles generates electromagnetic waves. In the same way, we expect accelerating gravitationally charged particles (masses) to generate waves. However, the existence of only one sign of mass (not two positive/negative types of charge as in electrodynamics) together with the conservation law of linear momentum implies that there is neither monopolar nor dipolar gravitational radiation. Gravitational radiation starts from quadrupole radiation and continues up to higher multipoles. 

\begin{figure}[t] 
\centering
\includegraphics[width=0.6\textwidth]{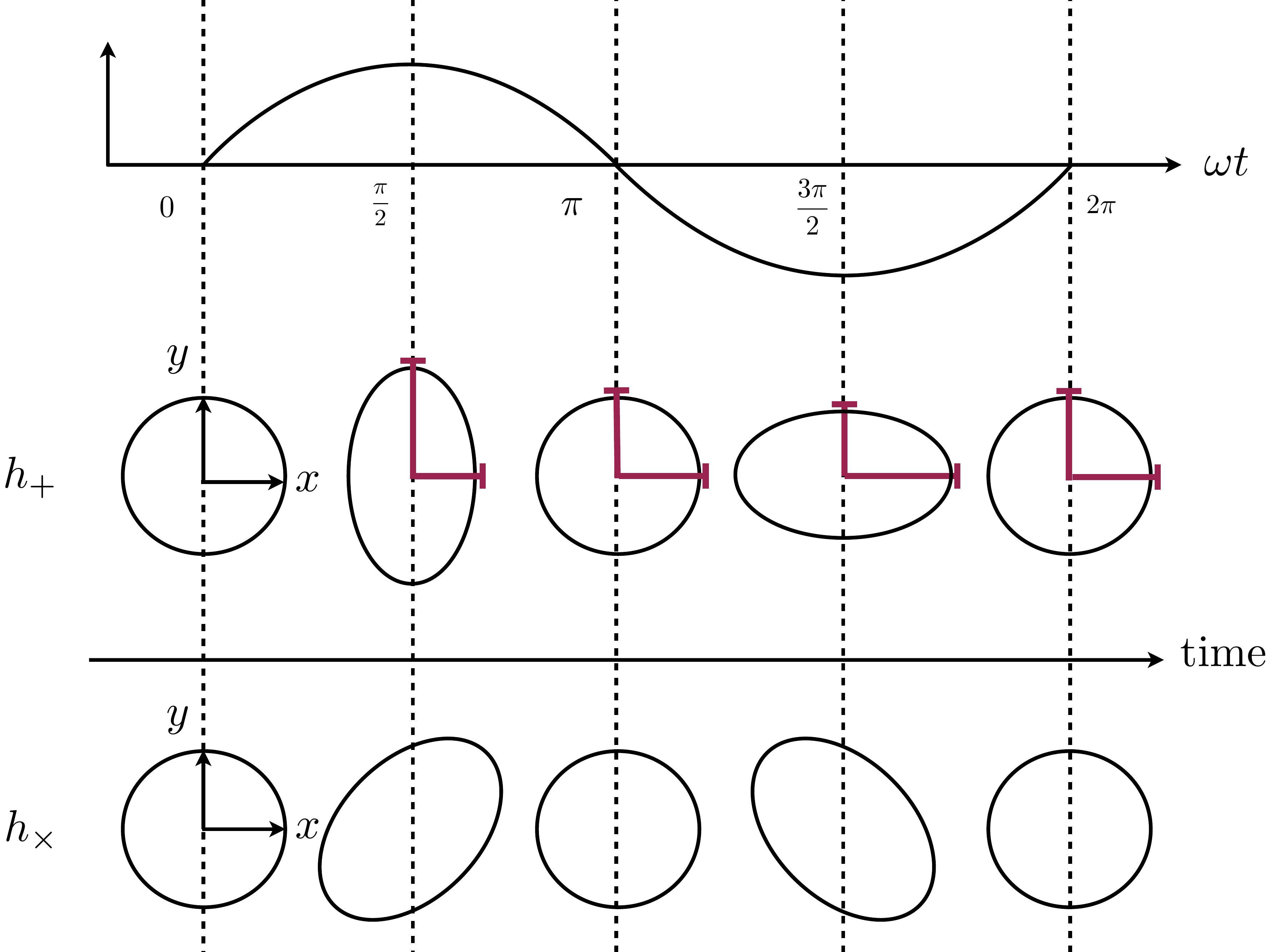}
\caption{Schematic diagram of how gravitational waves interact with test particles on a ring. The quadrupole nature of the interaction can be clearly seen. The direction of the gravitational wave is perpendicular to the page. In the middle panel, the arms of an interferometer are shown.}
\label{fig:pluscross}
\end{figure}

In general relativity, gravitational waves propagate with the speed of light and there are two possible polarization modes: $h_+$ and $h_\times$. The effect of these polarization modes on a ring of test particles is shown in Fig.~\ref{fig:pluscross}.  From this, the principle of most gravitational-wave detectors --- looking for changes in the length of mechanical systems such as bars of aluminum or the arms of Michelson-Morley-type interferometric detectors --- can be clearly seen. We will discuss more details about interferometric detection in \sref{sec:interferometric}.

The magnitude of the components of a perturbing gravitational signal $h^{ij}$ produced at a distance $r$ from a source at time $t$ is proportional to the second time derivative of the quadrupole moment of the source (at earlier time $t-r/c$) and inversely proportional to $r$ \cite{hartle2003}, 
\be
\label{amplitude}
h^{ij}(t,\vec{x})\approx \frac{2 G}{r c^4} \frac{d^2}{dt^2} [I^{ij}(t-r/c)].
\ee
Notice that in the above formula the extremely small value of coefficient $G/c^4$ clearly shows why the gravitational-wave signals are very hard to detect. The energy luminosity of the source is proportional to the square of the third time derivative of the quadrupole moment \cite{hartle2003} i.e.
\be
\mathcal{L}=\frac{G}{5 c^5}\langle \dddot{\text{ \Midline{I}}}_{ij} \dddot{\text{\Midline{I}}}^{ij}\rangle, 
\ee
where $\langle \; \rangle$ represents an average over several cycles; $I^{ij}$ is the moment of inertia defined as 
\be
I^{ij}=\int \rho(t,\vec{x}) x^i x^j d^3x,
\ee 
and $\text{\Midline{I}}^{ij}$ is the symmetric trace-free (STF) moment of inertia or {\em quadrupole tensor}:
\be
\text{\Midline{I}}^{ij}\equiv I^{ij}-\frac{1}{3} \delta^{ij} I^k_k.
\ee

The energy flux of gravitational waves can be very large. For example, the energy flux of a sinusoidal, linearly polarized wave of amplitude $h_+$ and angular frequency $\omega$ is~\cite{hartle2003}
\be
\mathcal{F}=\frac{1}{32 \pi} \frac{c^3}{G} h_+^2 \omega^2,
\ee
which for a $100$-Hz sinusoidal wave of amplitude $h_+=10^{-21}$, one obtains a flux of $1.6$ mW.m$^{-2}$. A simple comparison shows that during a short time when the waves of a coalescing binary neutron-star system in Virgo cluster pass the Earth, the implicit energy flux is more than a millionth that from the Sun! As we will see, however, detecting the passage of this energy flux is a very difficult task. In a sense, spacetime is extremely stiff, in that the ``ripples'' may be exceedingly small, yet can transmit considerable energy.

Before moving on to likely sources of detectable gravitational waves, it is useful to make a comparison between gravitational and electromagnetic waves:
\begin{itemize}

\item Detectable gravitational-wave signals ($<$ a few kHz) reflect {\em coherent} motion of extremely massive celestial objects, while in contrast, electromagnetic radiation generally arises from an {\em incoherent} superposition of the motions of charges.

\item Unlike photon detection, the detection of individual quanta of gravitation (gravitons) is impossible with any foreseeable tool~\cite{freemandyson}.

\item Compared to electromagnetic-waves, gravitational radiation suffers no more than a tiny absorption or scattering (although, like light, it is subject to deflection near massive objects). Gravitational-wave astronomy provides excellent tools to carry information about violent processes, for example deep within stars or behind dust clouds.

\item Astrophysical events, where there are potentially huge masses accelerating very strongly, are the only detectable sources of gravitational radiation by the current (and proposed) detectors. The gravitational-waves emitted from the best possible manmade sources are utterly undetectable with current technology. For example, imagine a dumbbell consisting of two 1-ton compact masses with their centers separated by 2 meters and spinning at 1kHz about a line bisecting and orthogonal to their symmetry axis.  For an observer 300-km away (in the radiation-zone) the amplitude of $h\thicksim\delta L/L$ is $10^{-38}$  \cite{saulson}, 14 orders of magnitude smaller than the best sensitivity of Advanced LIGO. For a LIGO-scale detector (4-km arms), it means measuring a distortion as small as Planck's length!

\end{itemize}

\section{Sources of Gravitational Waves}
\begin{figure}[t] 
\centering
\includegraphics[width=0.6\textwidth]{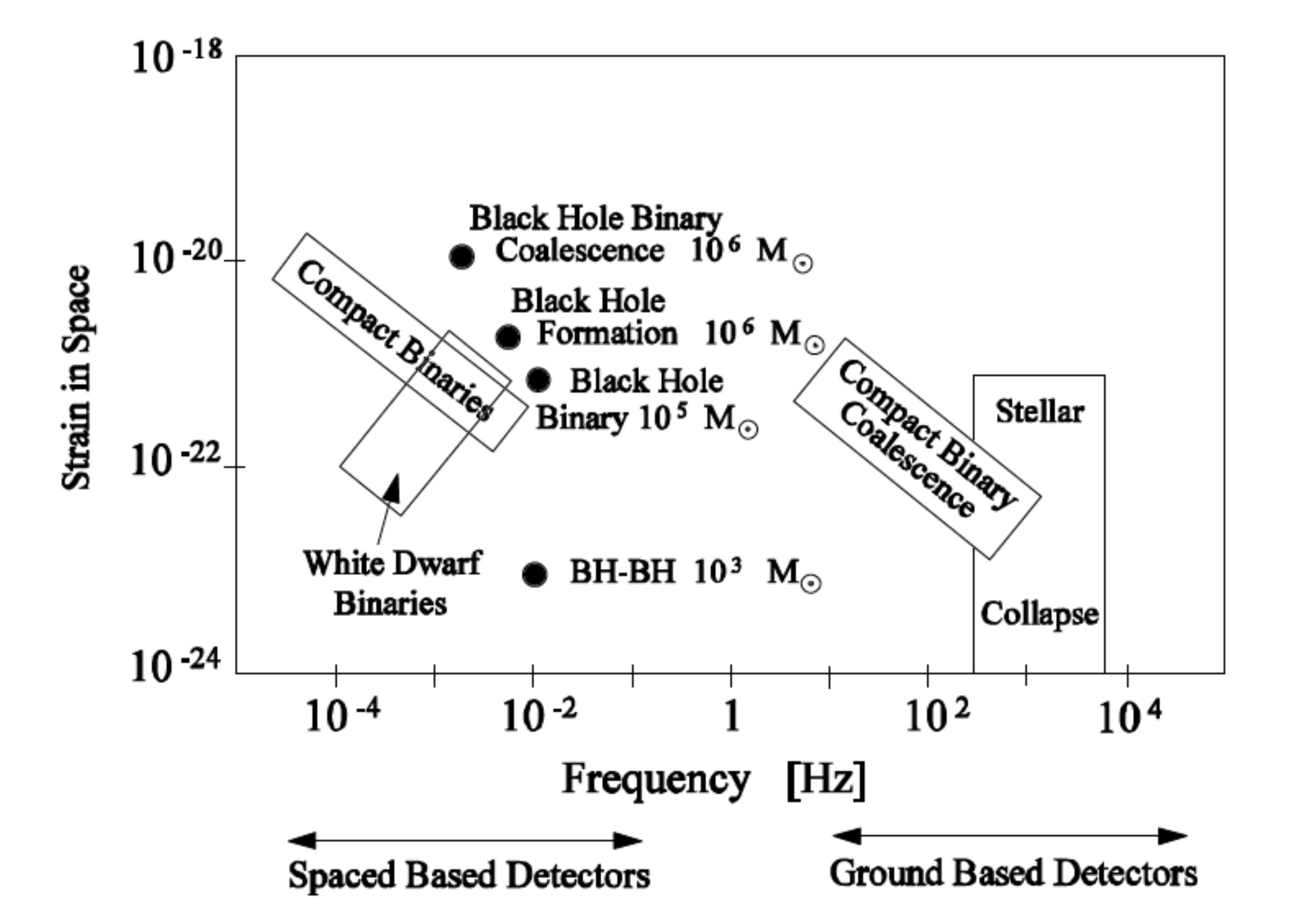}
\caption{Predicted signal strengths for a number of possible gravitational-wave sources. Different sources emit gravitational radiation in different ranges of frequency. Ground-based detectors are unable to detect low frequency waves (< a few Hz) while space-based detectors have this ability~\cite{2005JPhB...38S.497H}.}
\label{fig:frequency-sources}
\end{figure}

Studying sources of gravitational waves by current and future detectors will uncover dark sectors of the Universe in extreme physical conditions including strong, non-linear gravity in relativistic motion and extremely high density, temperature and magnetic fields. Sources of gravitational waves are expected to emit in a wide range of frequency, from $10^{-7}$ Hz in the case of ripples in the cosmological background to $10^3$ Hz for the birth of neutron-stars in supernova explosions. Fig.~\ref{fig:frequency-sources} shows the signal strength at the Earth, integrated over appropriate time intervals, for a number of sources. This figure also illustrates the estimated frequency range for different types of sources. The ground- and space-based detectors are sensitive to high and low frequency ranges, respectively. In section~\ref{sec:detectors} we will discuss different types of detectors together with their abilities and limitations.

There are many sources of great astrophysical interest including the interaction and coalescences of black-holes and neutron-stars, low-mass X-ray binaries such as Sco-X1~\footnote{Scorpius X-1 was the first extrasolar X-ray source discovered, and, aside from the Sun, it is the strongest source of X-rays in the sky.}, supernova explosions, rotating asymmetric neutron stars such as pulsars, and processes in the early Universe. In this dissertation we focus on the inspiralling compact binary sources, which are crucial for our work in parts~\ref{part:3} and \ref{part:4}, and refer the reader to recent reviews \cite{2004autt.book..281G, Cutler:uq, 1999CQGra..16A.131S, sat09} for further reading on other types of sources. 
\subsection{Compact Binary Systems and Prospects for Detection}
\begin{figure}[h] 
\centering
\includegraphics[width=0.6\textwidth]{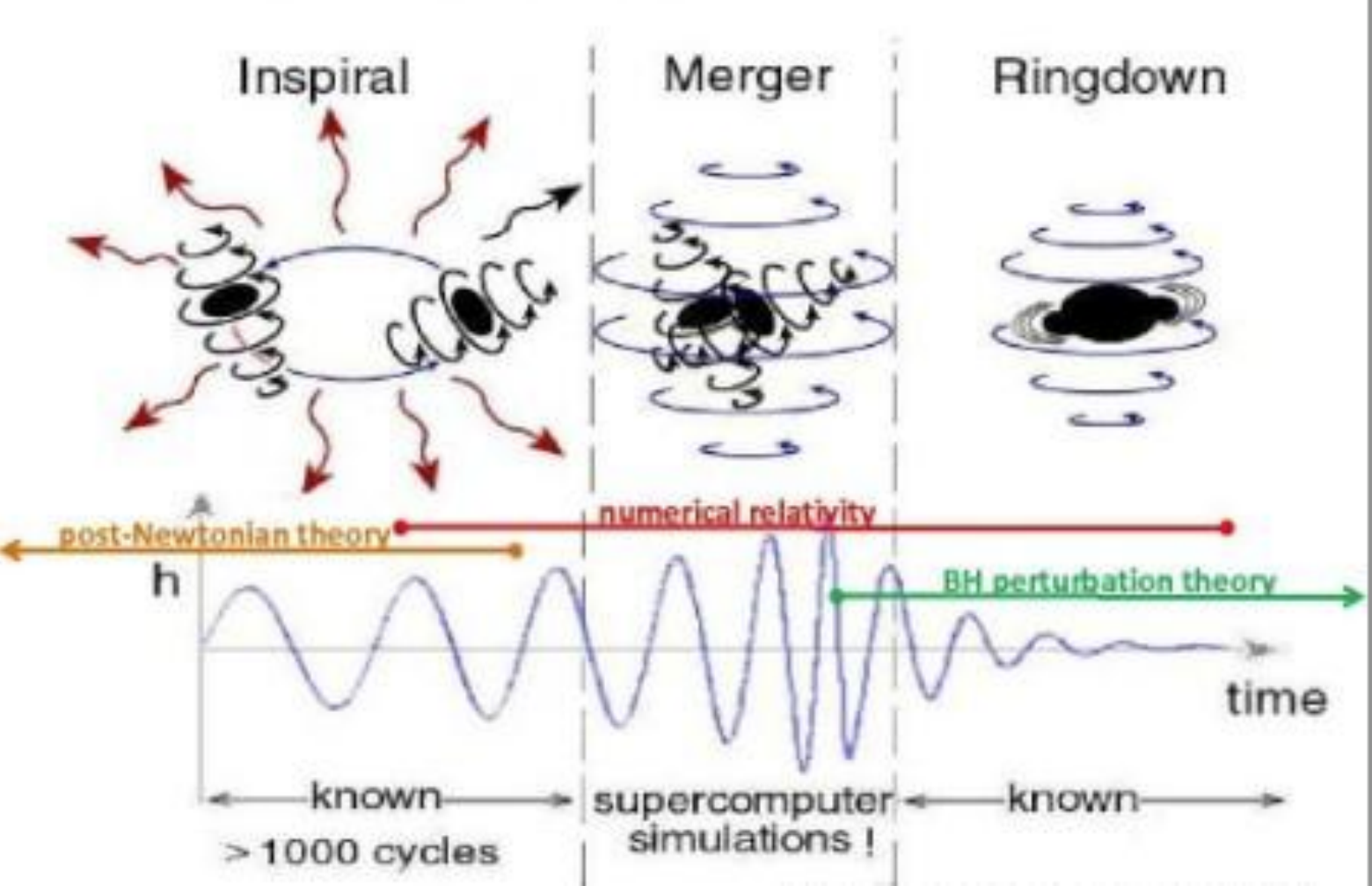}
\caption{Phase evolution of a compact binary system: inspiral, merger, and ringdown. [picture credit: Kip Thorne]}
\label{fig:binaryphases}
\end{figure}

The coalescence of a compact binary produces short-lived and well defined signals of gravitational-waves and therefore belongs to the most promising category of sources for detection. A compact binary system has two companions, which could be a neutron star (NS) or a black hole (BH), orbiting around the center of mass of the system. The system loses energy and angular momentum by emitting gravitational radiation. This leads to an inspiral of the two bodies toward each other and consequently an increase in rotational frequency of the system. The dynamics of every inspiralling compact binary have three phases which are illustrated in Fig.~\ref{fig:binaryphases} including:
 
\begin{enumerate}
\item The early {\em inspiral phase}, in which the separation distance is large and therefore the gravitational field strength at each body due to the other one is weak. Systems could spend hundreds of million years in this phase with a low gravitational radiation power. The emitted gravitational signal in this phase has a characteristic shape with slowly increasing amplitude and frequency. It is often called a {\em chirp} waveform. A chirping binary could be considered as an {\em ideal standard candle} in the sky~\cite{1986Natur.323..310S}; we can measure the luminosity distance by observing gravitational radiation from a chirping binary. The post-Newtonian approximation is valid in this weak-field, slow-motion regime.
\item The {\em merger phase}, when the two bodies get very close to each other and gravitational fields get extremely strong. The post-Newtonian approximation breaks down and a full numerical calculation is the only possible tool to obtain the motion in this regime. Studying strong, non-linear gravity and violent phenomena such as tidal deformation and disruption in the merger phase of coalescing binaries has been the domain of {\em Numerical Relativity} in the last two decades (for example NINJA project~\cite{ninja2}). For a review of numerical relativity see~\cite{2010nure.book.....B, 2001CQGra..18R..25L}. 
\item The {\em ringdown phase} (also called {\em late merger}), when the two compact bodies have merged to form either a single black hole or neutron star. The final, compact object in this phase could still radiate gravitational waves because of its asymmetries. It can be considered as a perturbed, rotating compact object and therefore perturbation theory can be applied to obtain the quasi-normal modes in this phase.
\end{enumerate}

Three types of compact binaries are possible:
\begin{itemize}
\item {\em NS-NS binaries} --- Based on observational data of three NS binaries containing one or more pulsars in our own Galaxy, detected by radio telescopes, it can be estimated that Galactic coalescence rate of NS-NS binaries is $\sim 9\times 10^{-5} \text{ yr}^{-1}$ ~\cite{2003Natur.426..531B}. Any NS binary within the range of $300$ Mpc from the Earth should be seeable by advanced ground-based detectors such as Advanced-LIGO which would imply an event rate between $0.1$ and $500 \text{ yr}^{-1}$ for NS-NS coalescences.
\item {\em NS-BH binaries} --- Since no astrophysical NS-BH binary has been observed to date, the estimation of their population is not as certain as for NS binaries. But, there still exist methods that we can use to estimate the event rate of NS-BH binaries in the detectable band of advanced detectors. Population synthesis models \cite{Grishchuk:2000gh} give an event rate between $1$ and $1500 \text{ yr}^{-1}$ within $650$ Mpc from the Earth (NS-BH sensitive distance for Advanced LIGO, see \sref{sec:adligo}).
\item {\em BH-BH binaries} --- Population synthesis models are highly uncertain about the Galactic rate of BH-BH coalesces. Nevertheless BH mergers may be promising candidate sources for a first direct detection of gravitational waves because the signal is significantly stronger than for BH-NS and NS-NS binaries. 
\end{itemize}



\section{Detection and Data Analysis}
\label{sec:detectors}
There have been many attempts to detect gravitational-waves beginning with the pioneering work by Joseph Weber in the  1960's. He reported in 1970 coincident excitations of two resonant-bar detectors in widely separate laboratories~\cite{1969PhRvL..22.1320W, 1970PhRvL..25..180W}. However, subsequent experiments by other groups (either with the same level of accuracy or better) failed to confirm the reported detections~\cite{1978ARA&amp;A..16..521T}. The first gravitational-wave detectors were metal cylinders  and the way that they were supposed to detect gravitational waves was quite simple. If the characteristic frequency of the incident wave is near the resonance frequency of the bar, the response to the wave is magnified and sudden changes in the amplitude of nominally thermal motion of the bar are expected. This effect is similar to an RLC antenna circuit's response to an electromagnetic-wave and we could measure it via piezoelectric transducers.

In the late 1990's, before the first generation of gravitational-wave interferometers came online, there were five major bar detectors operating cooperatively in the International Gravitational Event Collaboration (IGEC)~\cite{2007PhRvD..76j2001A}. These bars achieved impressive strain amplitude spectral noise densities near $10^{-21}/\sqrt{\text{Hz}}$, but only in narrow bands of $\sim 1-30$ Hz~\cite{2003PhRvL..91k1101A} near their resonant frequencies (ranging from $\sim 700$ Hz to $\sim 900$ Hz). Today, narrowband resonance bar detectors are almost completely phased out while the broadband interferometer detectors such as LIGO/VIRGO are leading the effort. Almost all of the current operating gravitational-wave detectors and all the  proposed ones use interferometry techniques for detection. In the next section we briefly introduce the basics of interferometric detection.

\subsection{Basics of Interferometer Detectors}
\label{sec:interferometric}

\begin{figure}[t]
\begin{center}$
\begin{array}{cc}
\includegraphics[width=0.45\textwidth]{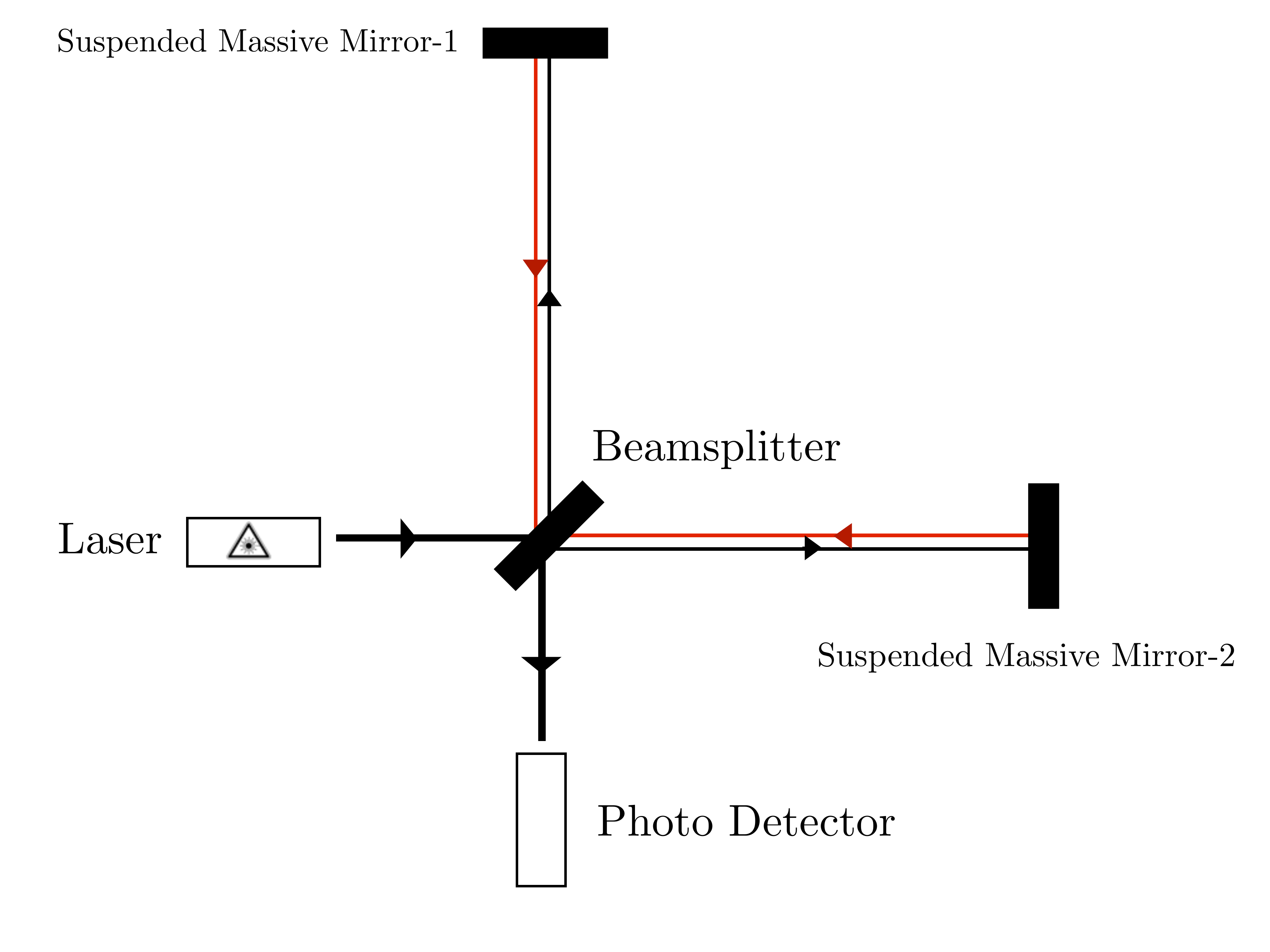} &
\includegraphics[width=0.45\textwidth]{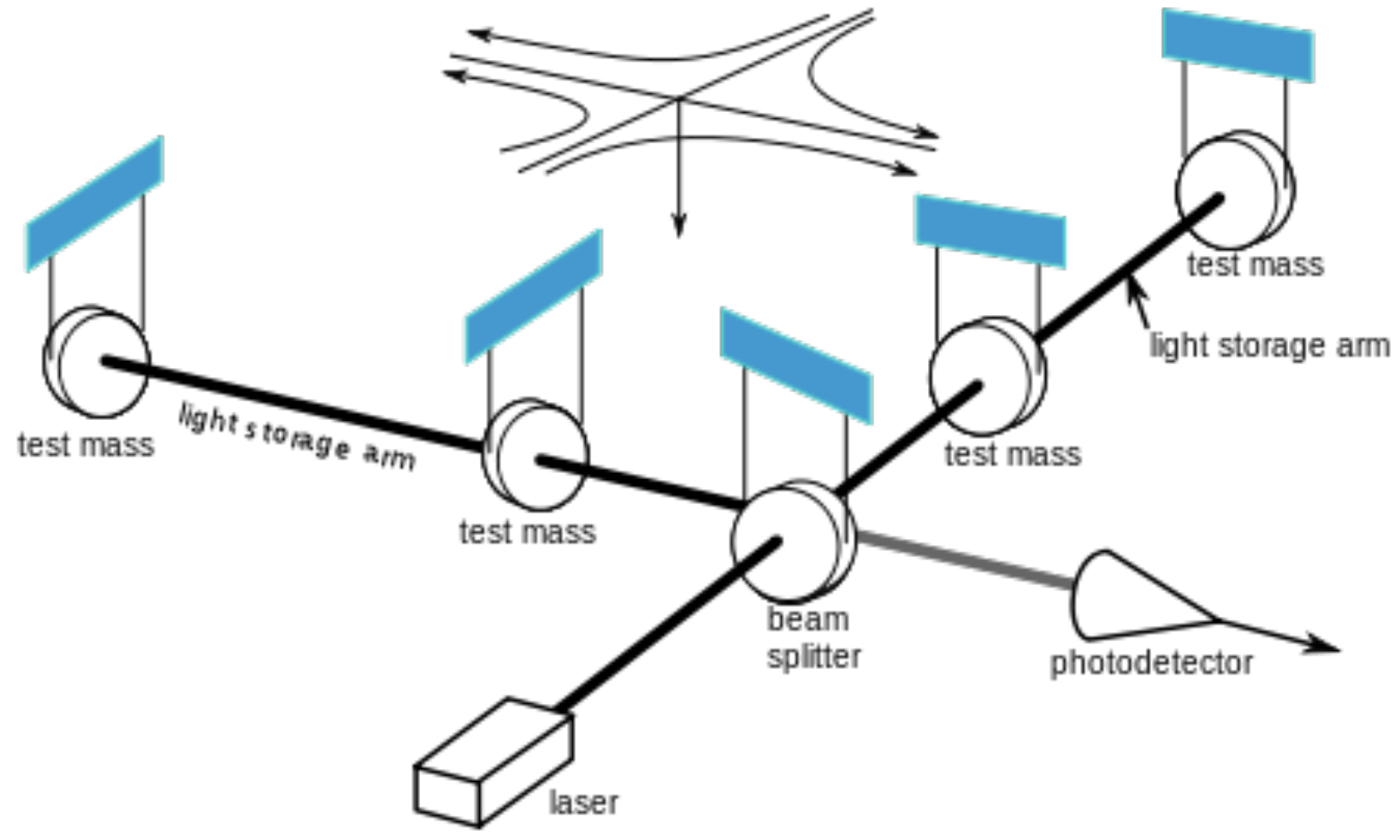}
\end{array}$
\end{center}
\caption{Schematic of an interferometer gravitational-wave detector such as LIGO. Left panel shows a top-view of a simple interferometer while the right panel shows a side-view with more details in an actual interferometric gravitational-wave detector.}
\label{fig:interferometer}\end{figure}

Interferometric gravitational-wave detectors are very similar to the classic 1887 Michelson-Morley interferometer. A simple illustration is shown in Fig.~\ref{fig:interferometer}. The apparatus is composed of two straight, equal-length arms in orthogonal directions. There is a beam splitter at the intersection of the arms which splits the coherent laser beam into two beams directed along each arm. There is a suspended massive mirror at the end of each arm which reflects the beams back to the beam splitter. The returning electromagnetic-wave signals will interfere constructively at the beam splitter, if the lengths of the arms are equal. Studying the interference pattern can show tiny changes in the lengths of the arms due to gravitational waves. The real apparatus is, of course, more sophisticated (see \cite{saulson}).

\begin{figure}[t] 
\centering
\includegraphics[width=0.8\textwidth]{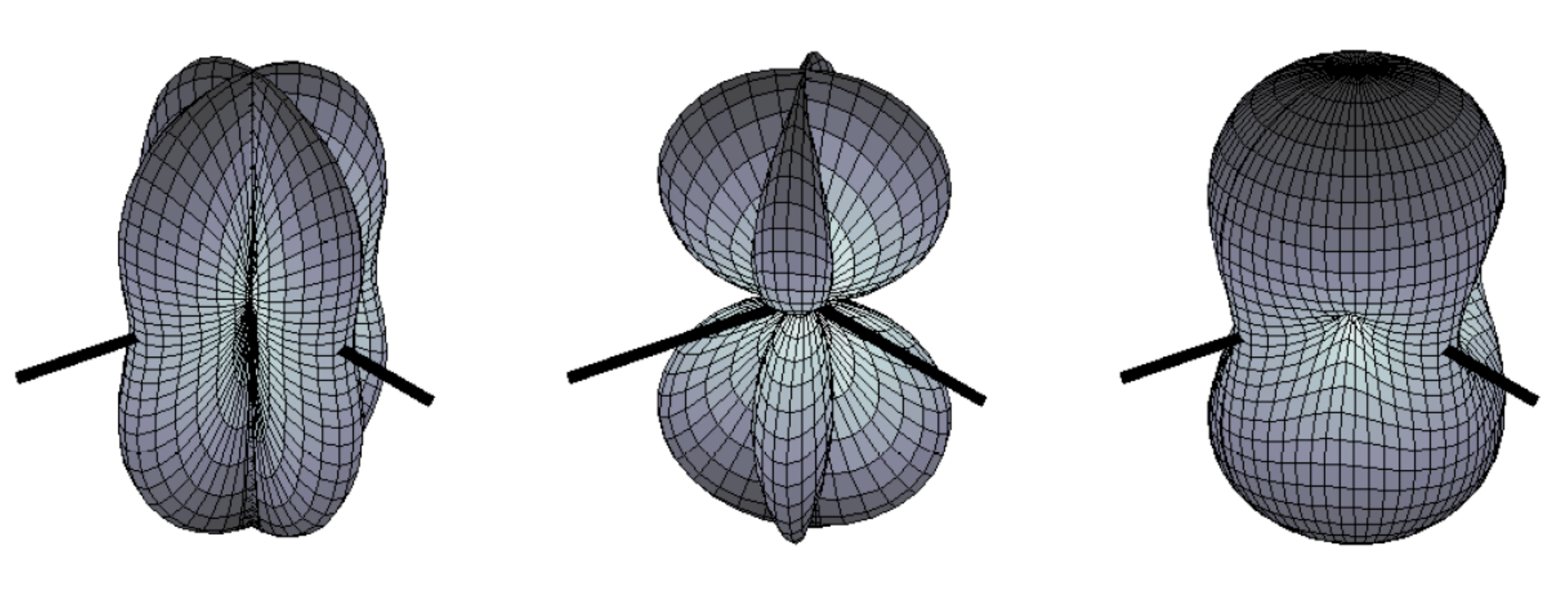}
\caption{Antenna response pattern for a Michelson interferometer in the long-wavelength approximation. The interferometer beamsplitter is located at the center of each pattern, and the thick black lines indicate the orientation of the interferometer arms. The distance from a point of the plot surface to the center of the pattern is a measure of the gravitational wave sensitivity in this direction. The pattern on the left is for $+$ polarization, the middle pattern is for $\times$ polarization, and the right-most one is for unpolarized waves~\cite{Abbott:2007kv}.}
\label{fig:antennapattern}
\end{figure}

Gravitational-wave detectors are better thought of as antennae than as telescopes, because their sizes are small compared to the wavelengths they are meant to detect. For example, the LIGO detectors when searching at 4 kHz have $L/\lambda$ of only about 0.05. This small ratio imply broad antenna lobes. Figure~\ref{fig:antennapattern} shows the antenna lobes for $+$, $\times$ linear polarizations and unpolarized case vs. incident direction for a Michelson interferometer in the long-wavelength limit. As a result, a single interferometer observing a transient event has very poor directionality.

\subsection{Interferometric Detection on the Earth}
\label{sec:adligo}
One can think of the ground-based gravitational-waves detectors as having three generations.
\subsubsection{First Generation}
Prototypes of gravitational-wave detectors since Weber (1960) led eventually to the building of major interferometric detectors on the Earth including:
\begin{itemize}
\item {\em LIGO}~\cite{ligo} which consists of three independent interferometric detectors: (1\&2) a 2-km and a 4-km length detector at Hanford, WA, USA, and (3) a 4-km length detector at Livingston, LA, USA. They all use the same laser type (Nd:YAG) with the same wavelength ($\lambda=1064$ nm) and the same test mass mirrors (10.7 kg). The major interferometers share many design characteristic, but also display significant differences. LIGO is sensitive to loud-enough gravitational-waves in the frequency range of 30-7000 Hz, roughly speaking.

Since the first operation in 1999, LIGO has had three phases so far: Initial-, Enhanced-, and Advanced-LIGO during while significant improvements have been made. According to NSF (2008) LIGO is the largest single enterprise undertaken by NSF, with capital investments of nearly \$300 million and operating costs of more than \$30 million/year~\cite{nsf}.
\item {\em VIRGO}~\cite{virgo} which has been operating since May 2007 in Europe, Italy. The VIRGO interferometer has quite similar design to that of LIGO and comparable performance. The primary differences are in the arm lengths (3 km vs 4 km) and laser power (17 W vs 10 W). The Italian/French VIRGO collaboration has also put lots of effort to ultra-stable lasers, high reflectivity mirrors, active seismic isolation and position and alignment control. 

While not as sensitive as LIGO in the most sensitive band near 150 Hz, VIRGO is more sensitive at low frequencies (below 40 Hz), because of aggressive seismic isolation. This lower reach offers the potential to  detect low-frequency spinning neutron-stars that are inaccessible to LIGO. VIRGO's sensitivity range of frequency is from 10 to 10,000 Hz. The VIRGO project is founded by CNRS and INFN on an annual \EUR
10 million budget~\cite{virgotalk}.

\item {\em GEO 600}~\cite{geo} is a smaller scale interferometer with 600-meter, folded arms (non-Fabry-Perot), built in 1995 at Sarstedt, Germany with a relatively small budget. Although this detector has a lower sensitivity compared to LIGO and VIRGO, it plays an important rule as a testbed for Advanced LIGO technology. It has pioneered several innovations to be used in Advanced LIGO: multi-pendulum suspension, signal recycling, rod-laser amplification, and photon squeezing. Meanwhile, it can serve (1) as an observatory keeping watch on the nearby galaxy when LIGO and VIRGO are down, (2) as a potential confirmation instrument for very loud signals. The sensitive frequency range of GEO-600 is from 50 to 1500 Hz.

\item {\em TAMA} ~\cite{tama} was a 300-meter interferometer, similar to the LIGO detectors with Fabry-Perot arms and using power recycling, located at the Mitaka campus of the National Astronomical Observatory of Japan. It operated at comparable sensitivity to LIGO in LIGO's early runs. It was an initial project by the gravitational-wave studies group at the Institute for Cosmic Ray Research (ICRR) of the University of Tokyo. The goal of the project was to develop advanced techniques needed for a future kilometer sized interferometer and to detect gravitational waves that may occur by chance within our local group of galaxies. The Japanese collaboration that built TAMA is now building the 2nd-generation KAGRA detector (formerly known as LCGT).

\end{itemize}
\subsubsection{Second Generation}

The LIGO and VIRGO detectors are now undergoing major upgrades to become Advanced LIGO~\cite{adligo, 2010CQGra..27h4006H} and Advanced VIRGO~\cite{advirgo}. These upgrades are expected to improve their broadband strain sensitivities by an order of magnitude, thereby increasing their effective ranges by the same amount. Since the volume of accessible space grows as the cube of the range, one can expect the advanced detectors to probe roughly 1000 times more volume and therefore have expected transient event rates O(1000) times higher than for the 1st-generation detectors.

In parallel, a primarily Japanese collaboration is proceeding to build an underground 3-km interferometer (KAGRA)~\cite{Kuroda:2010zzb} in a set of new tunnels in the Kamiokande mountain near the famous Super-Kamiokande neutrino detector. Placing the interferometer underground dramatically suppresses noise due to ambient seismic disturbances. 

In addition, INDIGO~\cite{indigo} ---which is a planned LIGO-type observatory in India--- has recently received initial approvals by the U.S.A. and Indian governments. The LIGO instrumentation that was initially scheduled to be installed at the 2-km interferometer at Hanford will be transported to India to add to the global network of gravitational-wave detectors, providing better source localization and better sensitivity to the polarization of gravitational-waves. Novel types of interferometers including AGIS~\cite{PhysRevD.78.122002} and TOBA~\cite{PhysRevLett.105.161101} have been also proposed recently.

\subsubsection{Third Generation}
With construction of second-generation interferometers well under way, the gravitational wave community has started looking ahead to third-generation underground detectors, for which KAGRA will provide a path finding demonstration. A European consortium is in the conceptual design stages of a 10-km cryogenic underground trio of triangular interferometers called Einstein Telescope~\cite{et}, which would use a 500-W laser and aggressive squeezing, yielding a design sensitivity an order of magnitude better than the 2nd-generation advanced detectors now under construction. With such capability, the era of precision gravitational wave astronomy and cosmology would open. Large statistics for detections and immense reaches ($\sim$Gpc) would allow new distributional analyses and cosmological probes. LIGO scientists too are starting to consider a 3rd-generation cryogenic detector, with a possible location in the proposed DUSEL underground facility~\cite{dusel, Riles:2012fk}.

\begin{figure}[t] 
\centering
\includegraphics[width=.6\textwidth]{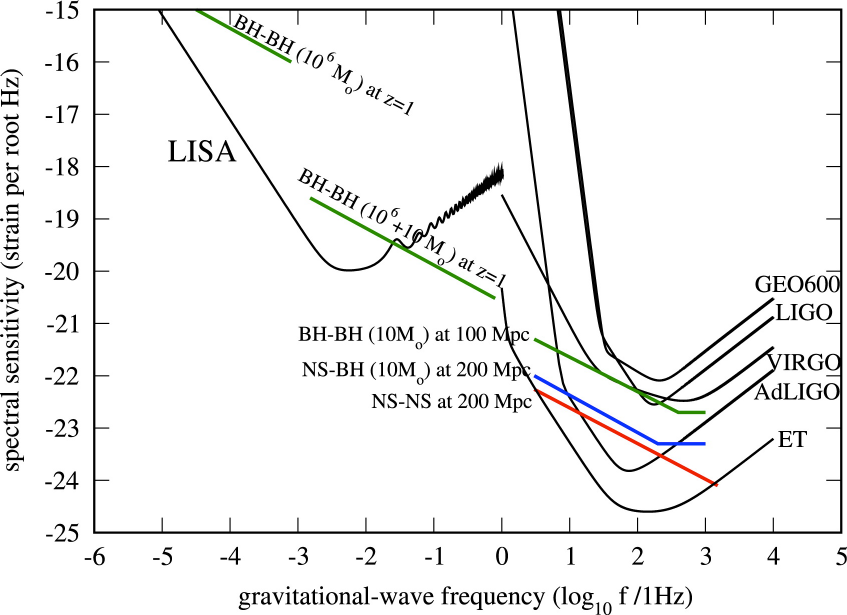}
\caption{Sensitivity curves of different generations of interferometric ground-based detectors. The sensitivity curve of space-based LISA in low frequencies is also shown for comparison~\cite{2011PrPNP..66..239A}.}
\label{fig:spectrum}
\end{figure}

The sensitivity curves of these detectors with different types of coalescence binary sources are shown in \fref{fig:spectrum}. Space-based detectors are needed to detect low-frequency gravitational-waves.

\subsection{Space-Based Detectors}
Some of the most interesting gravitational wave signals, such as those resulting from the formation and coalescence of black holes in the range $10^3$ to $10^6$ solar masses, will lie in the region of $10^{-4}$ to $10^{-1}$ Hz. To search for these requires a detector whose strain sensitivity is approximately $10^{-23}$ over relevant timescales. It has been pointed out that the most promising way of looking for such signals is to fly a laser interferometer in space, i.e. to launch a number of drag free spacecraft into orbit and to compare the distances between test masses in these craft using laser interferometry. The sensitivity curve of LISA is shown in \fref{fig:spectrum}.

\begin{figure}[t] 
\centering
\includegraphics[width=0.6\textwidth]{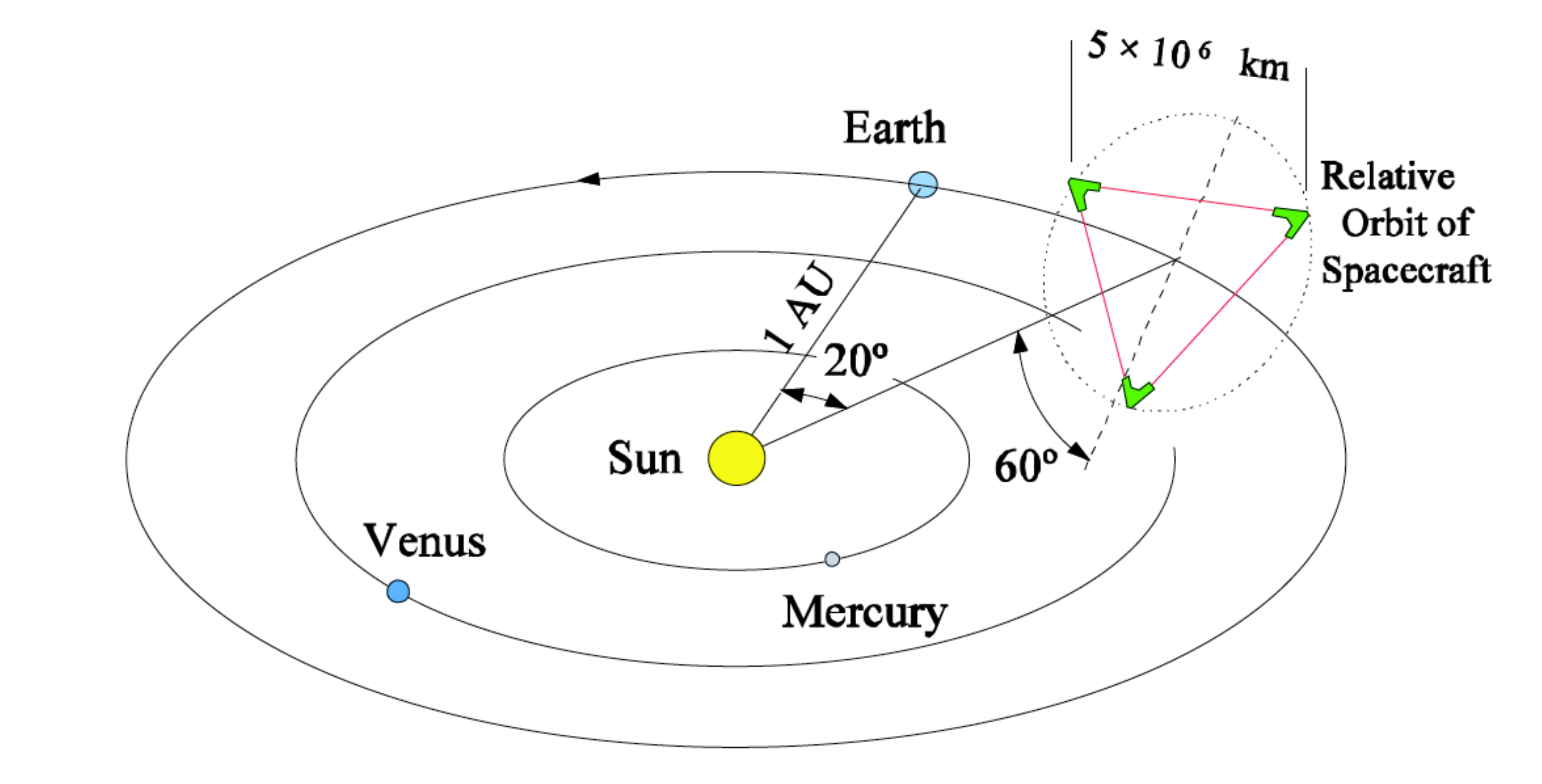}
\caption{Schematic diagram of proposed initial-LISA and its orbit about the sun. LISA was consisting of an array of three drag free spacecraft at the vertices of an equilateral triangle of length of side $5\times10^6$ km. This cluster is placed in an Earth-like orbit at a distance of 1 AU from the Sun, and 20 degrees behind the Earth as shown.}
\label{fig:lisa}
\end{figure}

An ambitious and long-studied proposed joint NASA-ESA project called LISA (Laser Interferometer Space Antenna) envisioned a triangular configuration (roughly equilateral with sides of $5\times10^6$ km) of three satellites (Fig.~\ref{fig:lisa}). As discussed above, there are many low-frequency gravitational wave sources expected to be detectable with LISA, and the proposed project has received very favorable review by a number of American and European scientific panels. Nonetheless, primarily for budgetary reasons, the project has been turned down by NASA (2012). Subsequently, NASA and ESA have solicited separate and significantly descoped new proposals. The funding prospects for these new proposals are quite uncertain, with ESA having recently passed over a descoped version of LISA called NGO (New Gravitational-wave Observer) in favor of a mission to Jupiter. Beside LISA-like missions, DECIGO~\cite{2009JPhCS.154a2040S} and BBO~\cite{PhysRevD.72.083005} are other existing possibilities for future spaced-based observatories that have been proposed recently.

\subsection{Pulsar Timing Arrays}
Detection of stochastic gravitational waves, potentially, can be done by performing precise pulsar timing via radio astronomy. This could be thought of as an entirely different method compared to the interferometry method in LIGO/VIRGO, for instance. Very-low-frequency (VLF) waves ($\sim$ several nHz) in the vicinity of the Earth could lead to a quadrupolar pattern in the timing residuals from a large number of pulsars observed at different directions on the sky~\cite{1978AZh....55...65S, 1979ApJ...234.1100D, 1983ApJ...265L..39H}. Three collaborations have formed in recent years to carry out the precise observations required: (1) The Parkes Pulsar Timing Array (PPTA-{Australia)~\cite{mtw}, (2) the European Pulsar Timing Array (EPTA-{UK, France, Netherlands, Italy)~\cite{mtw}, and (3) the North American NanoHertz Observatory for Gravitational Waves (NANOGrav {USA and Canada)~\cite{2009arXiv0909.1058J}. 
\subsection{Data Analysis}
The most challenging task for gravitational-wave detectors is extracting the signal from noisy data. This issue is less challenging for LISA-like detectors where data is signal-dominated compared to the ground-based detectors such as LIGO which are noise-dominated. Different sources of noise are involved, including seismic noise, thermal noise, photoelectron shot noise. A number of data analysis methods have been derived, which provide useful tools to do this task. The goal of any data analysis method include detection of gravitational waves, inferring the nature of the source from the detailed properties of the wave signal, and testing general relativity. We will discuss this topic in more detail in \cref{chapter7}, focusing on the Matched Filtering method.


\blankpage
\mypart{Methods}{{\vfill  \small{\em The framework and methods that will be used in the following parts are introduced in this part, including the methods of the Parametrized Post-Newtonian (PPN) framework, Direct Integration of Relaxed Einstein Equations (DIRE), and Matched Filtering.}}}
{\begin{framed}
\begin{itemize}
\item \cref{chapter5}--- \nameref{chapter5}
\item \cref{chapter6}--- \nameref{chapter6}
\item \cref{chapter7}--- \nameref{chapter7}
\end{itemize}
\end{framed}
} {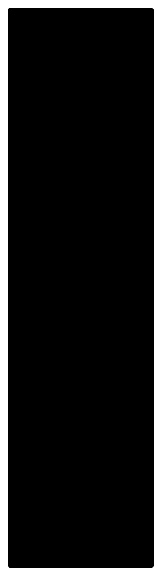}\label{part:2}
\begin{savequote}[0.55\linewidth]
{\scriptsize ``There are in fact two things, science and opinion; the former begets knowledge, the latter ignorance.''}
\qauthor{\scriptsize---Hippocrates}
\end{savequote}

\chapter{Parametrized Post-Newtonian Theory} 
\label{chapter5} 
\thispagestyle{myplain}
\lhead[\thepage]{Chapter 4. \emph{Parametrized Post-Newtonian Theory}}      
\rhead[Chapter 4. \emph{Parametrized Post-Newtonian Theory}]{\thepage}
\ClearWallPaper
To compare various theories of gravity and also to analyze the significance of various experiments to test the fundamental theory of gravity, two theoretical frameworks have been postulated: the Dicke framework and the Parametrized Post-Newtonian (PPN) framework. The Dicke framework, suggested by Robert Dicke, is particularly powerful for discussing null experiments, for delineating the qualitative nature of gravity, and for devising new covariant theories of gravity. The Dicke formalism has been discussed in more detail in~\cite{wil74a}.

The PPN framework starts where the Dicke framework leaves off: By analyzing a number of experiments within the Dicke framework one arrives at (among others) two {\em fair-confidence} conclusions about the nature of gravity. These are (i) that gravity is associated, at least in part, with a symmetric tensor field, the {\em metric}; and (ii) that the response of matter and fields to gravity is described by $\nabla\cdot T=0$, where $\nabla\cdot$ is the divergence with respect to the metric, and $T$ is the stress-energy tensor for all matter and non-gravitational fields. These two conclusions in the Dicke framework become the postulates upon which the PPN framework is built.

In this chapter, we briefly review the PPN formalism because we will need some part of it in our future calculations and also because it will help us to a better understanding of Part \ref{part:3} of this dissertation. This formalism provides a framework which is extremely useful for discussing specific alternative metric theories of gravity including scalar-tensor theories and for analyzing the solar system tests of gravitational effects. We will refer to this chapter when we study the equations of motion for compact binary systems in alternative theories of gravity in Part \ref{part:3}. This chapter is mostly based on Will's work in \cite{wil74a, tegp}.

The main advantage of working in a {\em parametrized} post-Newtonian framework is that, in principle, a wide range of metric theories of gravity can be accurately described in this framework only by tuning the values of the PPN parameters for each theory. The PPN formalism provides a useful framework in which comparing the theories and testing gravitational effects are easier to do with very few {\em a priori} assumptions about the nature of gravity. The PPN framework is a very practical tool to test alternative theories of gravity in solar system and beyond. Information given by future gravitational wave detection will also provide lots of useful data that can be applied to test alternative theories of gravity, although the PPN framework is less useful for those types of test.

\section{The Newtonian Limit} 
Classic Newtonian mechanics works well on solar system scales. The gravitational field is weak enough and characteristic velocities are such small compared to the speed of light that any general relativistic effect will be extremely small. These two conditions are called weak-field and slow-motion conditions, respectively. Nothing prevents using the Post-Newtonian theory even beyond solar-system scales as long as the weak-field and slow-motion conditions are satisfied. In the solar system, to an accuracy of better than part in $10^5$, light rays travel on straight lines at constant speed, and test bodies move according to
\be
{\bf a}=\nabla U,
\ee
where ${\bf a}$ is the acceleration of moving body, and $U$ is the Newtonian gravitational potential produced by rest-mass density $\rho$ according to
\bea
\nabla^2 U&=&-4\pi \rho,\\
U({\bf x},t)&=&\int \frac{\rho(\vec{x'},t)}{\mid \vec{x}-\vec{x'}\mid }d^3x'.
\eea
Note that we have assumed $c=G=1$. Considering perfect fluids with no viscosity, the Eulerian equations of hydrodynamics are 
\begin{subequations}\label{euler-equations}
\bea\frac{\partial\rho}{\partial t}+\nabla\cdot(\rho\vec{v})&=&0\\
\rho\frac{d\vec{v}}{dt}&=&\rho\nabla U-\nabla p\\
\frac{d}{dt}&\define&\frac{\partial}{\partial t}+\vec{v}\cdot\nabla,
\eea
\end{subequations}
where $\vec{v}$ is the velocity of an element of the fluid, $\rho$ is the rest-mass density of matter, $p$ is the pressure.
Considering a test body momentarily at rest in a static external gravitational field, the body's acceleration $a^k$ in a static $(t,\vec{x})$ coordinate system reduces from \ref{geodesic} to
\be\label{newtonianacc}
a^k=-\Gamma^k_{00}=\frac{1}{2} g^{kl} g_{00,l}.
\ee
We expect general relativity (or any other alternative theory of gravity) to be the same as Newtonian gravity very far away from the gravitational sources. In another words, we expect the metric in an appropriately chosen coordinate system to reduce to the flat Minkowski metric i.e.
\be
g_{\mu\nu}\rightarrow\eta_{\mu\nu}=\text{diag}(-1, +1, +1, +1).
\ee

To keep everything self-consistent in the Newtonian limit the only choice for the metric components including gravity are to be
\be\label{newtonian-metric}
g^{jk}\simeq\delta^{jk},\qquad g_{00}\simeq -1+2U.
\ee

Given the stress-energy tensor for perfect fluids as 
\be
T^{00}=\rho,\qquad T^{0j}=\rho v^j,\qquad T^{jk}=\rho v^j v^k+ p\delta^{jk},
\ee
this is straightforward to show that the Eulerian equations of motions in \eref{euler-equations} are equivalent to 
\be
T^{\mu\nu}_{;\nu}\simeq T^{\mu\nu}_{,\nu}+\Gamma^\mu_{00} T^{00}=0,
\ee
where we retain only terms of lowest order in $v^2\sim U\sim p/\rho$. 

Beyond the Newtonian limit when we begin to take into account the accuracies greater than a part in $10^5$, we need a more accurate approximation to the spacetime metric that goes beyond or {\em post} Newtonian theory (and this is why we called this theory as post-Newtonian theory). For example, for Mercury's additional perihelion shift of $\sim 5\times 10^{-7}$ radians per orbit, the accuracy of the Newtonian gravity is no longer enough, we have to consider the post-Newtonian limits of this problem as well.

\section{Post-Newtonian Bookkeeping}
For future use, it is very helpful to first develop a {\em bookkeeping} system for keeping track of {\em small quantities} in our post-Newtonian calculations. Because in the post-Newtonian formalism we often do an expansion in terms of small quantity $v/c$, it would be useful to compare the order of magnitude of the other quantities with $v/c$. The Virial theorem in its general form i.e. $2\times\langle\text{Kinetic Energy}\rangle_t=\langle\text{potential energy}\rangle_t$ in the effective one-body problem immediately yields $\mu v^2\sim \mu/r$ which clearly means
\be
v^2\lessapprox U.
\ee

The matter making up the Sun and planets is under pressure $p$, but this pressure is generally smaller than the matter's gravitational energy density $\rho\, U$ i.e.
\be
\frac{p}{\rho}\lessapprox U.
\ee
For instance, in the Sun $p/\rho\sim10^{-5}$ and in the Earth $p/\rho\sim 10^{-10}$. Other than gravitational energy $U$, one can also think about other forms of energy such as compressional energy, radiation, and thermal energy. But they are also very small compared to $\rho$. Defining $\Pi$ as the specific energy density (ratio of energy density to rest-mass density), $\Pi$ is $\sim 10^{-5}$ in the Sun and $\sim 10^{-10}$ in the Earth. We can think of the order of magnitude of $\Pi$ as
\be
\Pi \lessapprox U.
\ee
We assign to these above mentioned small quantities a bookkeeping label that denotes their {\em order of smallness}:
\be\label{bookkeeping}
U\sim v^2\sim \frac{p}{\rho}\sim \Pi \sim \mathcal{O}(\epsilon).
\ee
Later in this dissertation we will neglect the effect of non-gravitational energy density $\Pi$ in our calculation but we keep it for now to be able to describe all the parameters in the complete PPN formalism. Based on \eref{bookkeeping}, we can conclude that single powers of velocity $v$ are $\mathcal{O}(\epsilon^{1/2})$, $U^2$ is $\mathcal{O}(\epsilon^2)$, $U v$ is $\mathcal{O}(\epsilon^{3/2})$, and so on. Also since the time evolution of the solar system is governed by the motion of its constituents, we have
\be
\partial/\partial t \sim \vec{v}\cdot\vec{\nabla},
\ee
and thus,
\be
\frac{\mid \partial/\partial t\mid }{\mid \partial/\partial x\mid }\sim \mathcal{O}(\epsilon^{1/2}).
\ee

Now, we are ready to analyze the {\em post-Newtonian} metric using this bookkeeping system. The action for the motion of a point particle in any metric theory of gravity can be written as
\bea\nonumber
I_0&=&-m_0 \int (-g_{\mu\nu} \frac{dx^\mu}{dt}\frac{dx^\nu}{dt})^{1/2} dt\\
&=&-m_0\int(-g_{00}-2 g_{0j} v^j-g_{jk} v^j v^k)^{1/2} dt.
\label{NGaction}
\eea

The integrand in Eq. (\ref{NGaction}) can be considered as a Lagrangian $L$ for a single particle in a metric gravitational field. In the Newtonian limit we can substitute the metric components from Eq. (\ref{newtonian-metric}) to get
\be
L=(1-2U-v^2)^{1/2}.
\ee
It is straightforward to confirm that this Lagrangian yields the equations of motion by using the Euler-Lagrange equations. In other words, Newtonian physics can be recovered by using an approximation for the Lagrangian correct to $\mathcal{O}(\epsilon)$. Therefore $L$ to $\mathcal{O}(\epsilon^2)$ must give post-Newtonian physics.

Since half-integer-order terms, such as $\mathcal{O}(\epsilon^{1/2})$ and $\mathcal{O}(\epsilon^{3/2})$, contain an odd number of factors of velocity $\vec{v}$ or of time derivatives $\partial/\partial t$, and these factors are not symmetric under the time reversal operator, half-integer-order terms must be representing energy dissipation or absorption by the system. But what happened to half-integer-order terms, $\mathcal{O}(\epsilon^{1/2})$ or $\mathcal{O}(\epsilon^{3/2})$, in the Newtonian Lagrangian?  Because of the conservation of rest mass, terms of $\mathcal{O}(\epsilon^{1/2})$ don't appear and conservation of energy in the Newtonian limit prevents terms of $\mathcal{O}(\epsilon^{3/2})$. Beyond $\mathcal{O}(\epsilon^2)$, different theories may treat things differently. General relativity predicts that the first odd-order terms appear at $\mathcal{O}(\epsilon^{7/2})$, which represents energy lost from the system by gravitational radiation. Terms of $\mathcal{O}(\epsilon^{5/2})$ are prohibited by the conservation of post-Newtonian energy in general relativity.

Going one step beyond the Newtonian limit i.e. to first post-Newtonian order (1PN), we have to express $L$ to $\mathcal{O}(\epsilon^2)$. To do so we have to know the various metric components to an appropriate order as shown in the following,
\be
L=\{1-2U-v^2-g_{00}[\mathcal{O}(\epsilon^2)]-2 g_{0j}[\mathcal{O}(\epsilon^{3/2})] v^j-g_{jk} [\mathcal{O}(\epsilon)] v^j v^k\}^{1/2}.
\ee
Thus the first post-Newtonian limit of any metric theory of gravity requires a knowledge of 
\begin{subequations}
\bea
g_{00}\quad &to& \quad \mathcal{O}(\epsilon^2),\\
g_{0j}\quad &to& \quad \mathcal{O}(\epsilon^{3/2}),\\
g_{jk}\quad &to& \quad \mathcal{O}(\epsilon).
\eea
\end{subequations}
For calculation in the second post-Newtonian limit (2PN) we need to know each metric component to an additional power of $\epsilon$ higher that what has been shown above for 1PN.

Similarly, it can be verified that if one takes the perfect fluid stress-energy tensor which is given by
\be
T^{\mu\nu}=(\rho+\rho\,\Pi+p) u^\mu u^\nu+ p g^{\mu\nu},
\ee
and expand it through the following orders of accuracy:
\begin{subequations}
\bea
T^{00}\quad &to& \quad \rho \mathcal{O}(\epsilon),\\
T^{0j}\quad &to& \quad \rho\mathcal{O}(\epsilon^{3/2}),\\
T^{jk}\quad &to& \quad \rho\mathcal{O}(\epsilon^2),
\eea
\end{subequations}
and combine it with the post-Newtonian metric, then the equations of motion $T^{\mu\nu}_{;\nu}=0$ will yield consistent {\em post-Eulerian} equations of hydrodynamics. 

\section{The Most General Post-Newtonian Metric}
\label{subset:The-Most-General-Post-Newtonian-Metric}
The most general post-Newtonian metric can be found by simply writing down metric terms composed of all possible post-Newtonian functions of matter variables, each multiplied by an arbitrary coefficient that may depend on the cosmological matching conditions and on other constants, and adding these terms to the Minkowski metric to obtain the physical metric. Unfortunately, there is an infinite number of such functionals, so that in order to obtain a formalism that is both useful and manageable, we must impose some restrictions on the possible terms to be considered, guided in part by a subjective notation of {\em reasonableness} and in part by evidence obtained from known gravitation theories. A list of the restrictions is given in section 4.1d of TEGP, specially:
\begin{itemize}
\item The deviations of the metric from flat space are all of Newtonian or post-Newtonian order; no post-post-Newtonian or higher-order deviations are included (see \cite{chandra65} for a discussion on distinction between Newtonian, post-Newtonian, and post-post-Newtonian terms).
\item For the field points at very far distances from the matter source where $\mid \vec{x}-\vec{x}'\mid $ is extremely large, the metric is flat (asymptoticly flat condition). This condition prevents the appearance of terms such as $\int v(\vec{x}')^2 \Pi (\vec{x}') d^3x'$ or $\int \Pi(\vec{x}') [p(\vec{x}')/\rho(\vec{x}')] d^3x'$ in $g_{00}$, for example.
\item The gradients of small order quantities related to matter including rest mass, energy, velocity, and pressure are not allowed in the metric. Terms involving gradients, such as $\int v_j(\vec{x}') (x_j-x'_j) \, [p(\vec{x}')/\rho(\vec{x}')]_{,i} d^3x'$ in $g_{0i}$, for instance, are prohibited by this condition. 
\end{itemize}


We now can construct a very general form for the post-Newtonian perfect-fluid metric in any metric theory of gravity, expressed in a local, quasi-Cartesian coordinate system moving with respect to the universe rest frame, and in a standard gauge as shown in \eref{PPNmetric}. The only way that that the metric of any one theory can differ from that of any other theory is in the coefficients that multiply each term in the metric. By replacing each coefficient by an arbitrary parameter we obtain a {\em super metric theory of gravity} whose special cases (particular values of the parameters) are the post-Newtonian metrics of particular theories of gravity. This {\em super metric} is called the parametrized post-Newtonian (PPN) metric, and the parameters are called PPN parameters.

The most mature version of the post-Newtonian metric in its most general form is given in \cite{tegp} as
\begin{subequations}\label{PPNmetric}
\bea
g_{00}&=&-1+2U-2(\psi-\beta U^2)+\Phi^{PF}\\
g_{0i}&=&-[2(1+\gamma)+\frac{1}{2}\alpha_1] U_j-\frac{1}{2} [1+\alpha_2-\zeta_1+2\xi] \partial_{tj} X+\Phi^{PF}_j\\
g_{ij}&=&(1+2\gamma U) \delta_{ij}\\
\label{psi-ppn-metric} \psi&:=& \frac{1}{2} (2\gamma+1+\alpha_3+\zeta_1-2\xi) \Phi_1-(2\beta-1-\zeta_2-\xi) \Phi_2+(1+\zeta_3) \Phi_3\nonumber \\
&& +(3\gamma+3\zeta_4-2\xi) \Phi_4-\frac{1}{2} (\zeta_1-2\xi) \Phi_6-\xi \Phi_W.
\eea
\end{subequations}
where $\gamma$, $\beta$, $\zeta$, $\alpha_1$, $\alpha_2$, $\alpha_3$, $\zeta_1$, $\zeta_2$, $\zeta_3$, $\zeta_4$ are 10 PPN parameters and the post-Newtonian potentials are defined to be functions of matter properties as
\begin{subequations}
\label{PPNpotentials}
\bea
\label{U} U&\define&\int\frac{{\rho^*}'}{\mid \vec{x}-\vec{x}'\mid }d^3x',\\
\Phi_1&\define&\int\frac{{\rho^*}' v'^2}{\mid \vec{x}-\vec{x}'\mid }d^3x',\\
\Phi_2&\define&\int\frac{{\rho^*}' U'}{\mid \vec{x}-\vec{x}'\mid }d^3x',\\
\Phi_3&\define&\int\frac{{\rho^*}' \Pi'}{\mid \vec{x}-\vec{x}'\mid }d^3x',\\
\Phi_4&\define&\int\frac{p'}{\mid \vec{x}-\vec{x}'\mid }d^3x',\\
\Phi_6&\define&\int\frac{{\rho^*}' [\vec{v}'\cdot(\vec{x}-\vec{x}')]^2}{\mid \vec{x}-\vec{x}'\mid ^3}d^3x',\\
\Phi_W&\define&\int \frac{{\rho^*}' {\rho^*}'' (\vec{x}-\vec{x}')}{\mid \vec{x}-\vec{x}'\mid ^3}\cdot \biggl( \frac{\vec{x}'-\vec{x}''}{\mid \vec{x}-\vec{x}''\mid }-\frac{\vec{x}-\vec{x}''}{\mid \vec{x}'-\vec{x}''\mid }\biggr) d^3x' d^3x'',\\
U^j&\define&\int\frac{{\rho^*}'v'^j}{\mid \vec{x}-\vec{x}'\mid }d^3x',\\
X&\define&\int {\rho^*}'\mid \vec{x}-\vec{x}'\mid d^3x',
\eea
\end{subequations}
and the preferred-frame potentials are
\begin{subequations}\label{PF-potentials}
\bea
\Phi^{PF}&\define& (\alpha_3-\alpha_1) w^2 U+\alpha_2 w^j w^k \partial_{jk} X+ (2\alpha_3-\alpha_1) w^j U_j,\\
\Phi_j^{PF}&\define&-\frac{1}{2} \alpha_1 w_j U+\alpha_2 w^k \partial_{jk} X.
\eea
\end{subequations}
where all above potentials are functions of $(\vec{x},t)$ while primed functions show the same functions evaluated at $(\vec{x}',t)$. For example, $\rho'$ and $\vec{v}'$ stand for $\rho (\vec{x}',t)$ and $\vec{v} (\vec{x}',t)$, respectively. Notice that $w^i$ in \ref{PF-potentials} indicates the coordinate velocity of the PPN coordinate system relative to the mean rest frame of the universe; $v^i$ is the coordinate velocity of matter i.e. $dx^i/dt$; $\rho$ and $p$ are the density and pressure of the matter both measured in a local freely falling frame momentarily co-moving with the matter; $\Pi$ represents internal energy per unit rest mass. It includes  all non-rest mass and non gravitational energy, for instance thermal energy and energy of compression.

In Eq. (\ref{PPNmetric}) we are in a nearly globally Lorentz coordinate system in which the coordinates are $(t, x^1, x^2, x^3)$. All coordinate arbitrariness ({\em gauge freedom}) has been removed by specialization of the coordinates to the standard PPN gauge. For more details about applying Lorentz transformations to the coordinate system and also about the standard PPN gauge see section 4.2 and 4.3 of TEGP. 

\section{The PPN Parameters and Their Significance}
As we explained in \sref{subset:The-Most-General-Post-Newtonian-Metric}, the use of parameters to describes the post-Newtonian limit of metric theories of gravity is called the {\em Parametrized Post-Newtonian (PPN) Formalism}. A primitive version of such a formalism was devised and studied by Eddington (1922)
, Robertson~\cite{rob62}, and Schiff~\cite{sch67}. In this formalism, which was developed for solar system tests of general relativity, the Sun is considered to be a non-rotating, spherical, massive object, and planets are modeled as test bodies moving on geodesics of the spacetime metric. The metric in this version of the formalism reads 
\begin{subequations}
\be\label{ERSmetric}
ds^2=-\biggl[1-2\frac{M}{r}+2\beta(\frac{M}{r})^2\biggr] dt^2+\biggl[1+2\gamma\frac{M}{r}\biggr] (dx^2+dy^2+dz^2),\\
\ee
\end{subequations}
where $M$ is the mass of the Sun, and $\beta$ and $\gamma$ are the only PPN parameters in this version. In standard PPN gauge, the parameter $\beta$ measures the amount of nonlinearity of a theory in $g_{00}$ while the parameter $\gamma$ represents the curvature of spacetime produced by the Sun at radius $r$.

Schiff~\cite{sch60b} generalized the metric in \eref{ERSmetric} to incorporate rotation (Lense-Thirring effect), and Baierlein~\cite{bai67} developed a primitive perfect-fluid PPN metric. But the pioneering development of the full PPN formalism was initiated by Kenneth Nordtvedt, Jr.~\cite{nor68b}, who studied the post-Newtonian metric of a system of gravitating point masses. Will~\cite{wil71a} generalized the formalism to incorporate matter described by a perfect-fluid. A unified version of the PPN formalism was then presented by Will and Nordtvedt~\cite{wil72} and summarized by Will in \cite{wil74a} (hereafter TTEG). The Whitehead term $\Phi_W$ was added by Will~\cite{wil73}. 

Although linear combinations of PPN parameters have been used in \eref{PPNmetric}, it can be seen quite easily that a given set of numerical coefficients for the post-Newtonian terms will yield a unique set of values for the parameters. The linear combinations were chosen in such a way that the parameters $\alpha_1$, $\alpha_2$, $\alpha_3$, $\zeta_1$, $\zeta_2$, $\zeta_3$, and $\zeta_4$ will have special physical significance. Evaluating every PPN parameter in a theory of gravitation is equivalent to measuring some specific properties of the theory. 
\section{Post-Newtonian Limits of Alternative Metric Theories}
The PPN formalism is sufficiently general that a wide range of theories of gravity can be described by this formalism with some specific values for the PPN parameters. The interested reader might refer to TEGP~\cite{tegp} which presents a {\em cookbook} for calculating the post-Newtonian limits of many metric theories of gravity. However, in this section we only focus on two major classes of gravitational theories i.e general relativity and scalar-tensor theories of gravity. We show the final post-Newtonian form of the metric tensor in terms of the constants and variables of each theory and read the PPN parameters from that.

The field equations in general relativity are given by [see \sref{subsec:GR-Intro}]
\be
R_{\mu\nu}-\frac{1}{2} g_{\mu\nu} R = 8\pi T_{\mu\nu}.
\ee
Considering the stress-energy tensor of matter in the form of a perfect fluid and following the cookbook steps in TEGP, the final form of the metric in general relativity is
\bea
g_{00}&=&-1+2U+3 \Phi_1-2\Phi_2+2\Phi_3+6\Phi_4- U^2,\\
g_{0j}&=&-4 U_j-\frac{1}{2} \partial_{tj} X,\\
g_{jk}&=&(1+2U) \delta_{jk}.
\eea
Keeping all the calculations in the standard PPN gauge, the PPN parameters can be read off immediately 
\bea
\gamma=\beta=1,\quad\xi=0,\\
\alpha_1=\alpha_2=\alpha_3=\zeta_1=\zeta_2=\zeta_3=\zeta_4=0.
\eea
Based on table~\ref{tab:PPNmeaning} and the values of PPN parameters in general relativity one can confirm that this theory is a fully conservative theory of gravity ($\alpha_3=\zeta_i=0$) and predicts no preferred-frame effects ($\alpha_i=0$) as we expect.

In general scalar-tensor theories of gravity, a dynamical scalar field $\phi$ is introduced in addition to the metric tensor $g_{\mu\nu}$.  The interaction between $\phi$ and $g_{\mu\nu}$ is governed by a coupling function $\omega(\phi)$. If $\omega=constant$ the scalar-tensor theory reduces to its specific form of Brans-Dicke theory~\cite{Brans:1961sx}. The field equations in scalar-tensor theories are derived from the action 
\be
I=\frac{1}{16\pi} \int \sqrt{-g}  \biggl[\phi R-\frac{\omega(\phi)}{\phi} g^{\mu\nu} \phi_{,\mu} \phi_{,\nu}\biggr] d^4x+I_{NG},
\ee
where the matter action $I_{NG}$ is a function only of matter variables and $g_{\mu\nu}$. It does {\em not} depend on the scalar field $\phi$.
\bea
R_{\mu\nu}-\frac{1}{2} g_{\mu\nu} R &=& \frac{8\pi}{\phi}  T_{\mu\nu}+ \frac{\omega{\phi}}{\phi^2} \biggl(\phi_{,\mu}\phi_{,\nu}-\frac{1}{2}g_{\mu\nu} \phi_{,\lambda} \phi^{,\lambda}\biggr)+\frac{1}{\phi} (\phi_{;\mu\nu}-g_{\mu\nu} \Box_g \phi),\\
 \Box_g \phi&=&\frac{1}{3+2\omega(\phi)} \biggl(8\pi T -\frac{d\omega}{d\phi} \phi_{,\lambda} \phi^{,\lambda}\biggr).
\eea
We choose coordinates (local quasi-Cartesian) in which the metric is asymptotically flat and $\phi$ takes the asymptotic value $\phi_0$. Defining
\bea
\omega&\define\omega(\phi_0),\\
 \omega'&\define\displaystyle\frac{d\omega}{d\phi}\mid _{\phi_0},\\
\zeta&\define\displaystyle\frac{1}{4+2\omega},\label{defzeta}\\
\lambda_1&\define\displaystyle\frac{\omega'\xi}{3+2\omega},\label{st-para}\label{deflambdaone}
\eea
and following the TEGP method we obtain the post-Newtonian metric of general scalar-tensor gravity as
\begin{subequations}
\bea\label{st-ppn-metric}
g_{00}&=&-1+2U+2 [\psi-(1+\xi\lambda_1) U^2]\\
g_{0j}&=&-4(1-\zeta) U_j-\frac{1}{2} \partial_{tj} X,\\
g_{jk}&=& [1+2(1-2\xi) U ] \;\delta_{jk}.
\eea
\end{subequations}
where
\be
\psi=\frac{1}{2} (3-4\xi) \Phi_1 -(1+2\xi\lambda_1) \Phi_2+\Phi_3+3 (1-2\xi) \Phi_4.
\ee
Notice that in going to geometrized units, we have set
\be\label{today}
G_{today}\define \frac{1}{\phi_0} \frac{4+2\omega}{3+2\omega}=1. 
\ee
Comparing \eref{st-ppn-metric} with \eref{PPNmetric}, the PPN parameters in scalar-tensor gravity are \cite{nut69a, nor70b} 
\begin{subequations}
\label{ST-PPN-paras}
\bea
\gamma=1-2\xi=\frac{1+\omega}{2+\omega}, \quad \beta=1+\xi\lambda_1, \quad \xi=0,\\
\alpha_1=\alpha_2=\alpha_3=\zeta_1=\zeta_2=\zeta_3=\zeta_4=0.
\eea
\end{subequations}
Again, $\alpha_3=\zeta_i=0$ and $\alpha_i=0$ confirms that scalar-tensor theories are fully conservative theories with no preferred-frame effects. In the limit of $\omega\rightarrow\infty$, the PPN parameters $\gamma$ and $\beta$ reduce to their general relativistic values i.e. unity. Table~\ref{tab:ppnvalues} summerizes the PPN parameters of general relativity and one of the most popular alternative class of theories i.e. general scalar-tensor theories including Brans-Dicke theory.

\begin{table}[t]
\begin{center}
\begin{tabular}{| l| | c| c | ccccc| }
\multicolumn{8}{r}{PPN Parameter\;\;\;\;\;\;\;\;\;} \\
\cline{4-8}
Theory    & Arbitrary Functions & Matching Parameters&$\gamma$&$\beta$&$\zeta$&$\alpha_1$&$\alpha_2$  \\
\hline\hline
General Relativity      & none   & none & 1& 1& 0& 0& 0     \\
\hline
BD Theory       & $\omega_{BD}$     & $\phi_0$  & $\displaystyle \frac{1+\omega_{BD}}{2+\omega_{BD}}$& 1& 0& 0& 0    \\
General ST& $A(\phi),\; V(\phi)$  & $\phi_0$   & $\displaystyle\frac{1+\omega}{2+\omega}$ &$1+\zeta\lambda_1$& 0& 0& 0    \\
\hline
\end{tabular}
\caption{\label{tab:ppnvalues}The values of the PPN parameters for general relativity and scalar-tensor theories including Brans-Dicke theory.}
\end{center}
\end{table}
\section{Equations of Motion in the PPN Formalism}
We define a {\em conserved density} $\rho^*$ by
\be
\rho^*\define \sqrt{-g} u^0 \rho,
\ee
where $u^0$ is the time component of the fluid element's four velocity, and $\rho$ is the locally measured mass density (see \sref{sec:conserved-density} for details). Using the general form of PPN in \eref{PPNmetric}, up to the first post-Newtonian order we find
\be
\rho^*= \biggl[1+\frac{1}{2} v^2+3\gamma U + \mathcal{O}(\epsilon^2)\biggr] \rho.
\ee
The components of the stress-energy tensor are given to the required order by
\begin{subequations}
\bea
T^{00}&=& \rho^* \biggl[1+\biggl(\frac{1}{2} v^2-(3\gamma-2) U+\Pi\biggr)\biggr]+ \mathcal{O}(\epsilon)\\
T^{0j}&=&\rho^* \biggl[1+\biggl(\frac{1}{2} v^2-(3\gamma-2) U+\Pi+\frac{p}{\rho^*}\biggr)\biggr]+ \mathcal{O}(\epsilon^{3/2})\\
T^{jk}&=&\rho^* v^j v^k \biggl[1+\biggl(\frac{1}{2} v^2-(3\gamma-2) U+\Pi+\frac{p}{\rho^*}\biggr)\biggr] + p\biggl(1-2\gamma U\biggr) \delta^{jk}+ \mathcal{O}(\epsilon^2).
\eea
\end{subequations}
It is straightforward to calculate the Christoffel symbols from the PPN metric in \eref{PPNmetric}. Having the Christoffel symbols and stress-energy tensor components up to appropriate order, one can substitute them into the equations of motion $T^{\mu\nu}_{;\nu}=0$ and obtain the PPN equations of hydrodynamics as
\begin{subequations}
\label{PPNeom}
\bea
\rho^*\frac{dv^j}{dt}&=&-\partial_j p+\rho^* \partial_j U+\biggl[\biggl(\frac{1}{2} v^2+(2-\gamma) U+\Pi+\frac{p}{\rho^*}\biggr)\partial_j p- v^j\partial_t p\biggr]\\
&&+\rho^* \biggl[\biggl(\gamma v^2-2(\gamma+\beta) U\biggr) \partial_j U-v^j \biggl( (2\gamma+1)\partial_t U +2 (\gamma+1) v^k \partial_k U\biggr)\\
&&\quad +\frac{1}{2} (4\gamma+4+\alpha_1) \; \biggl( \partial_t U_j+v^k (\partial_k U_j-\partial_j U_k)\biggr ) + \partial_j \Psi\biggr]\\
&&+\rho^* \biggl[\frac{1}{2} \partial_j \Phi^{PF}-\partial_t \Phi_j^{PF}-v^k (\partial_k \Phi_j^{PF}-\partial_j \Phi_k^{PF}) \biggr]+ \mathcal{O}(\epsilon^2),
\eea
\end{subequations}
where
\be
\Psi=\psi+\frac{1}{2} (1+\alpha_2-\zeta_1+2\xi)\; \ddot{X},
\ee
where $\psi$, $\Phi^{PF}$, and $\Phi_j^{PF}$ are given in \eref{psi-ppn-metric} and \eref{PF-potentials}. Note that $\ddot{X}=\Phi_1+2\Phi_4-\Phi_5-\Phi_6$.
\begin{savequote}[0.55\linewidth]
{\scriptsize ``If we knew what it was we were doing, it would not be called research, would it?''}
\qauthor{\scriptsize---Albert Einstein}
\end{savequote}

\chapter{DIRE: Direct Integration of Relaxed Einstein Equations} 

\label{chapter6} 
\thispagestyle{myplain}
\lhead[\thepage]{Chapter 5. \emph{DIRE: Direct Integration of Relaxed Einstein Equations}}      
\rhead[Chapter 5. \emph{DIRE: Direct Integration of Relaxed Einstein Equations}]{\thepage}

Direct Integration of the Relaxed Einstein Equations (DIRE) is one of three well-developed approaches to compute analytic, approximate solutions of the nonlinear field equations in general relativity via post-Newtonian methods (the other two methods includes the Blanchet-Damour-Iyer (BDI) approach~\cite{bla86, bla88, bla89, dam91, bla92, bla95} and the Effective Field Theory (EFT) approach~\cite{PhysRevD.73.104029}). The DIRE approach has been developed by Will and Pati~\cite{pat00, pat02} built upon earlier work by Epstein, Wagoner, Will and Wiseman~\cite{eps75, wag76, wis91, wis92, wis93, wil96}. 

In this chapter we introduce this approach and show, step by step, how it can be applied to solve the Einstein field equations and obtain the explicit general relativistic equations of motion for non-spinning compact binary systems, including black holes and neutron stars. Here we review what has been done in \cite{pat00, pat02} only up to the lowest post-Newtonian order because of two main reasons: First, showing more details of the calculations and technics that the authors in \cite{pat00, pat02} have used. Second, to provide a well-defined, reference framework in which we can compare our new results with, in the next part of this dissertation. In addition, having the structure of DIRE method in GR will avoid repeating many similar, lengthy steps in some future calculations in this dissertation. In the next part, we will generalize DIRE method from GR to a well-motivated, general class of alternative theories of gravity namely scalar-tensor theories. 

\section{Foundations of DIRE}
\subsection{The Relaxed Einstein Equations}
\label{sec:relaxedeq}
The method of DIRE is based on a reformation of the field equations of general relativity into a form known as the {\em relaxed Einstein equations}. The main idea is to recast Einstein's field equations from their regular form,
\be
R^{\mu\nu}-\frac{1}{2} g^{\mu\nu} R = 8 \pi T^{\mu\nu},
\label{einstein}
\ee
to their ``relaxed'' form,
\be
\Box_\eta h^{\mu\nu} = -16 \pi \tau^{\mu\nu}.
\label{relaxed}
\ee

We choose a particular coordinate system and stick with it hereafter in which
\be
h^{\mu\nu}_{,\nu}=0. 
\label{harmonic}
\ee
This combined with the definition of $h^{\mu\nu}$ in Eq. (\ref{h}) is called the \emph{De Donder} gauge condition in the literature. We also can call this specific coordinate system as {\em harmonic} coordinates, simply because Eq. (\ref{harmonic}) requires all the four coordinates to satisfy the curved spacetime scalar wave equation i.e. $\Box_g x^\mu=0$.

In Eq. (\ref{relaxed}) the box operator is the flat d'Alambertian, $\Box_\eta=\eta^{\mu\nu}\partial_\mu\partial_\nu$, and $h^{\mu\nu}$, referred to  as {\em gravitational field}, defined as
\be
h^{\mu\nu} \define \eta^{\mu\nu} -\gothg^{\mu\nu},
\label{h}
\ee
where
\be\label{gotg}
\gothg^{\mu\nu}\define \sqrt{-g} g^{\mu\nu}.
\ee

Equation (\ref{relaxed}) is in the form of a flat spacetime \emph{wave equation} and therefore its solution can be treated via well-known \emph{Green's functions}. The equation is called ``relaxed'' because it can be solved formally as a functional of source variables without specifying the motion of the source. 

Here we have to emphasize that $h^{\mu\nu}$ plays an important role in gravitational-wave calculations. The spatial components of $h^{\mu\nu}$, evaluated far from the source, describe the gravitational waveform and are directly related to the signal which a gravitational-wave detector measures. 

The source term in \eref{relaxed}, $\tau^{\mu\nu}$, is defined to be an {\em effective} stress-energy pseudotensor as the sum of a matter part ($T^{\mu\nu}$) and a gravitational part ($\Lambda^{\mu\nu}$):
\be
16\pi \tau^{\mu\nu} = 16\pi (-g) T^{\mu\nu} + \Lambda^{\mu\nu},
\label{tau}
\ee
where $T^{\mu\nu}$ is the stress-energy tensor of matter and all possible non-gravitational fields. Assuming the matter source purely made of perfect fluid we have
\be
T^{\mu\nu}=(\rho+p) u^\mu u^\nu+ p \;g^{\mu\nu},
\label{T}
\ee
where $p$ and $\rho$ are the locally measured pressure and energy density, respectively, and $u^\mu$ is the four-vector of velocity of an element of fluid. The gravitational piece of the effective stress-energy pseudotensor, $\Lambda^{\mu\nu}$, is given by
\be
\Lambda^{\mu\nu} = 16 \pi (-g) t^{\mu\nu}_{LL}+h^{\mu\alpha}_{,\beta} h^{\nu\beta}_{,\alpha}-h^{\mu\nu}_{,\alpha\beta} h^{\alpha\beta},
\label{Lambda}
\ee
where $t^{\mu\nu}_{LL}$ is the {\em Landau-Lifshitz} pseudotensor which is given by
\bea\label{landaulifshitz}
16 \pi (-g) t^{\mu\nu}_{LL} &\define& g_{\lambda\alpha} g^{\beta\rho} h^{\mu\lambda}_\beta h^{\nu\alpha}_{,\rho}+\frac{1}{2} g_{\lambda\alpha} g^{\mu\nu} h^{\lambda\beta}_{,\rho} h^{\rho\alpha}_{,\beta}-2g_{\alpha\beta} g^{\lambda(\mu} h^{\nu)\beta}_{,\rho} h^{\rho\alpha}_{,\lambda}\nonumber\\
&&+ \frac{1}{8} (2 g^{\mu\lambda} g^{\nu\alpha}-g^{\mu\nu} g^{\lambda\alpha}) (2 g_{\beta\rho} g_{\sigma\tau}-g_{\rho\sigma} g_{\beta\tau}) h^{\beta\tau}_{,\lambda} h^{\rho\sigma}_{,\alpha}.
\label{tLL}
 \eea

To derive the relaxed form of Einstein's equations in Eq. (\ref{relaxed}) from their regular form in Eq. (\ref{einstein}) the following key identity is useful. This identity is valid in any coordinate system and for any spacetime metric:
\be
H^{\mu\alpha\nu\beta}_{,\alpha\beta} = (-g) (2 G^{\mu\nu}+16 \pi t_{LL}^{\mu\nu}),
\label{identity}
\ee
where $G^{\mu\nu}$ and $t_{LL}^{\mu\nu}$ are the Einstein tensor and the Landau-Lifshitz pseudotensor, respectively, and
\be
H^{\mu\alpha\nu\beta} \define \gothg^{\mu\nu}\gothg^{\alpha\beta}-\gothg^{\alpha\nu}\gothg^{\beta\mu}.
\ee
The tensor $H^{\mu\alpha\nu\beta}$ has the same symmetry properties as the Riemann tensor, and if we apply $\partial_{\alpha\beta}$ operator to it we immediately obtain
\be
H^{\mu\alpha\nu\beta}_{,\alpha\beta} = -\Box_\eta h^{\mu\nu}+h^{\alpha\beta} \partial_{\alpha\beta} h^{\mu\nu}-\partial_\beta h^{\mu\nu} \partial_\alpha h^{\nu\beta},
\ee
which together with identity Eq. (\ref{identity}) leads to Eq. (\ref{relaxed}).

Before proceeding, we shall discuss some important points about this relaxed form of the field equations compared to its regular form. Up to this point we have not applied any approximation, neither weak-field nor slow-motion approximation. The relaxed Einstein equations in Eq. (\ref{relaxed}) in harmonic coordinates are {\em as exact as} the standard Einstein equations in Eq. (\ref{einstein}).

Eqs. ( note that although Eq. (\ref{relaxed}) takes the form of a simple wave equation in harmonic coordinates and doesn't look as difficult as Eq. (\ref{einstein}), it is actually still very complicated to solve from many aspects. On the right-hand side of the relaxed equation,  $\tau^{\mu\nu}$ is a function of the field, $h^{\mu\nu}$ and the derivatives (see Eqs. (\ref{tau}, \ref{Lambda}, \ref{tLL})). In addition, there is a second derivative term, namely $h^{\mu\nu}_{,\alpha\beta} h^{\alpha\beta}$, which properly belongs on the left-hand side of the equation where the other second derivative terms in the d'Alembertian operator are. In another words, while we do know the formed Green function solutions for $\Box_\eta h^{\mu\nu}=\eta^{\mu\nu}\partial_\mu\partial_\nu$ we do not for $(\eta^{\mu\nu}-h^{\mu\nu})\partial_\mu\partial_\nu$, because we are solving for $h^{\mu\nu}$ and do not know it before solving the equation. This term causes a deviation from the flat null cones of the  background Minkowski spacetime and therefore a modification in the propagation of the field. Fortunately, it has been shown that DIRE recovers the leading manifestations of this effect. Notice that in the regular form of Einstein's equations we have all the geometrical properties of spacetime on the left-hand side and all the matter distribution information (energy-momentum tensor) on the right-hand side. This symmetry does not hold in the relaxed Einstein's equations any more. Generally speaking, by converting to the relaxed form we have not decreased the level of complexity of the equations, we have only changed from a complicated form which we don't know any formalism to solve analytically, to another complicated form for which at least we do have a well-known mechanism for obtaining analytic, if approximate, solutions.. 

Second, as we mentioned, the right-hand side of Eq. (\ref{relaxed}) depends on $h^{\mu\nu}$, and $h^{\mu\nu}$ is the same quantity for which we are trying to solve the equations. Comparing with the classic concept of wave equation, it means what is waving in the left-hand side of the wave equation also plays a role in the source term on the right-hand side of the field (wave) equations. This means that not only the localized matter source generates gravity but also gravity itself generates gravity which is basically everywhere. This is a consequence of non-linearity of the field equations in general relativity.

Third, since $\Lambda^{\mu\nu}$ is at least quadratic in $h$, the relaxed field equations in Eq. (\ref{relaxed}) are very naturally amenable to a perturbative non-linear expansion. If we assume that $h^{\mu\nu}$ is suitably small everywhere, then iteration methods can be applied to solve these equations with some hope that the solutions might converge (possibly asymptotically) at the higher orders. 

Fourth, as an immediate consequence of the harmonic gauge condition, the right-hand side of the relaxed equations Eq. (\ref{relaxed}) is conserved in the sense that $\tau^{\mu\nu}_\nu=0$. This can be shown to be equivalent to the covariant equations of motion of matter:
\be
\tau_{,\nu}^{\mu\nu}=0 \Leftrightarrow T_{;\nu}^{\mu\nu}=0,
\ee
where comma and semicolon represent normal partial derivative and covariant derivative operators, respectively.
\subsection{Source, Near Zone and Radiation Zone}

Consider two non-spinning compact objects, for example two black-holes, two neutron stars, or one black-hole and one neutron star, with masses $m_1$ and $m_2$, orbiting around each other and radiating gravitational waves. We assume that the size $\mathcal{S}$ of these compact bodies is very small compared to the separation distance $r$ between them ($\mathcal{S}\ll r $). According to an observer at the center of mass of the system, the companions are located at the positions ${\bf x}_1$ and ${\bf x}_2$, and rotate about the common center of mass (denoted by the small red cross in Fig.~\ref{fig:center}) in orbits with the larger mass having the smaller orbit, ${\bf x}_1$, and smaller linear velocity, ${\bf v}_1$. Here we choose the center of mass to be at the origin of our coordinate system, i.e. $x^{\mu}_{CM}=(t, 0, 0, 0)$. For simplicity. Fig. \ref{fig:center} shows the orbits to be circular.

We are mainly interested in solving the field equations for the field at a point close to the source objects, in order to compute the equations of motion of the system. The vector ${\bf x}$ shows the position of the field point relative to the origin. We define $R$ to be the distance between the field point and the center of mass of the binary-system. Since we chose the origin to be at the center of mass, $R$ is equal to $\mid{\bf x}\mid$ here. This situation is illustrated in Fig.~\ref{fig:center}. After this point we also assume slow-motion ($v \ll 1$) and weak-field ($u \ll 1$). 
\begin{figure}[t]
\centering
\includegraphics[width=0.6\textwidth]{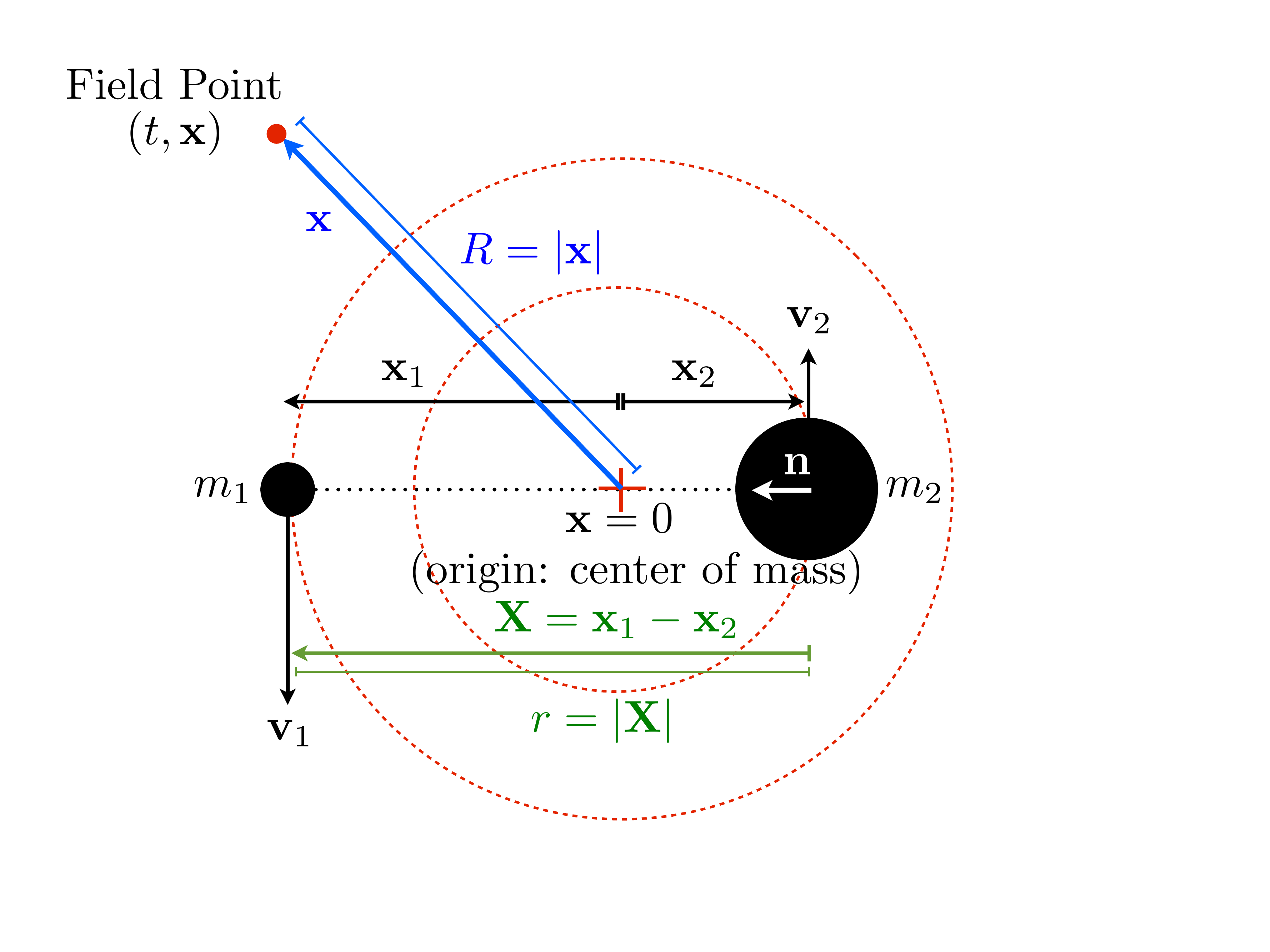}
\caption{Position of two compact objects ($m_1$, $m_2$) at ${\bf x}_1$ and ${\bf x}_2$ relative to the coordinate origin, orbiting in quasi-circular orbits around the center of mass of the binary system with velocities ${\bf v}_1$ and ${\bf v}_2$, respectively. The vector ${\bf x}$ indicates the position of field point relative to the origin. The origin is chosen to be at the center of mass; $r$ is the distance between the masses and $R$ is defined to be the distance between the field point and the center of mass.}
\label{fig:center}
\end{figure}
We define three spacetime zones around the center of mass of the binary system: (1) The {\em source zone}, which includes any point in the world tube $\mathcal{T}=\{x^{\mu}\mid  R<\mathcal{S}, -\infty<t<\infty\}$, where $\mathcal{S}$ is the radius of a sphere that contains all the matter. Any event that happens inside the source area at anytime belongs to this zone. (2) The {\em near zone}, which includes any point inside the world tube $\mathcal{D}=\{x^{\mu}\mid R<\mathcal{R}, -\infty<t<\infty\}$ where  $\mathcal{R}\sim \mathcal{S}/v\sim \lambda/2\pi$; $\lambda$ and $v$ are the wavelength of the radiated gravitational-wave, and the relative velocity of the source bodies, respectively. Note that the near zone includes the source zone. (3) The {\em far zone} (radiation zone), which includes all the spacetime outside the near zone, or equivalently $\mathcal{F}=\{x^{\mu}\mid R>\mathcal{R}, -\infty<t<\infty\}$. Fig.~\ref{fig:zones} shows these zones in the spacetime around a binary-system source. For most of the evolution, up to the point where the post-Newtonian approximation breaks down, $\mathcal{R}\gg \mathcal{S}$.\\
\begin{figure}[t]
\centering
\includegraphics[width=0.6\textwidth]{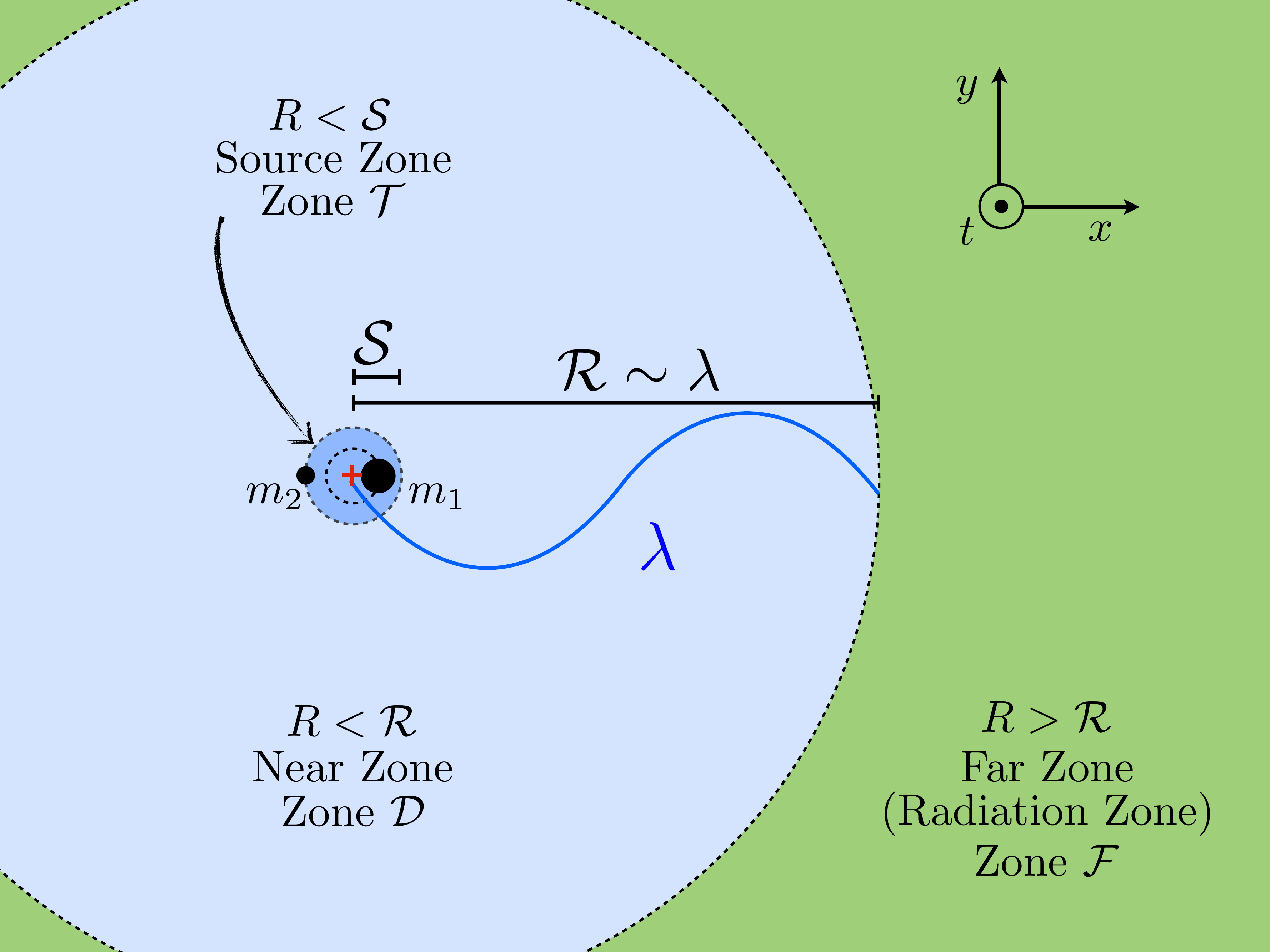}
\caption{Near-zone and far-zone. At one wavelength ($\lambda$) away from the source, $\mathcal{R}$ divides the spacetime around the binary system to two regions: near-zone ($R<\mathcal{R}$) and far-zone ($R>\mathcal{R}$). We treat the field points in different zones a bit differently but in the end, the final result must be independent of $\mathcal{R}$. The quantity $\mathcal{S}$ represents the radial size of the source.}
\label{fig:zones}
\end{figure}
After defining the near zone and far zone we are ready to go back and discuss the standard solutions of the relaxed Einstein equations in Eq.\ref{relaxed} which are retarded, flat-spacetime Green functions in their integral form:
\be
h^{\mu\nu}(t,{\bf x})= 4 \int \frac{\tau^{\mu\nu} (t', {\bf x}')\; \delta (t'-[t-\mid {\bf x}-{\bf x}'\mid ])}{\mid {\bf x}-{\bf x}'\mid }d^4x'.
\label{integral}
\ee
This integral is taken over all spacetime. But the delta function in the integrand reduces the integral to one over the past null cone $C$ emanating from the field point $(t,{\bf x})$. That is because the integrand is zero everywhere except when $t'=t-\mid {\bf x}-{\bf x}'\mid $. This is illustrated in Fig.~\ref{fig:retarded-cone}. 
\begin{figure}[t]
\centering
\includegraphics[width=0.4\textwidth]{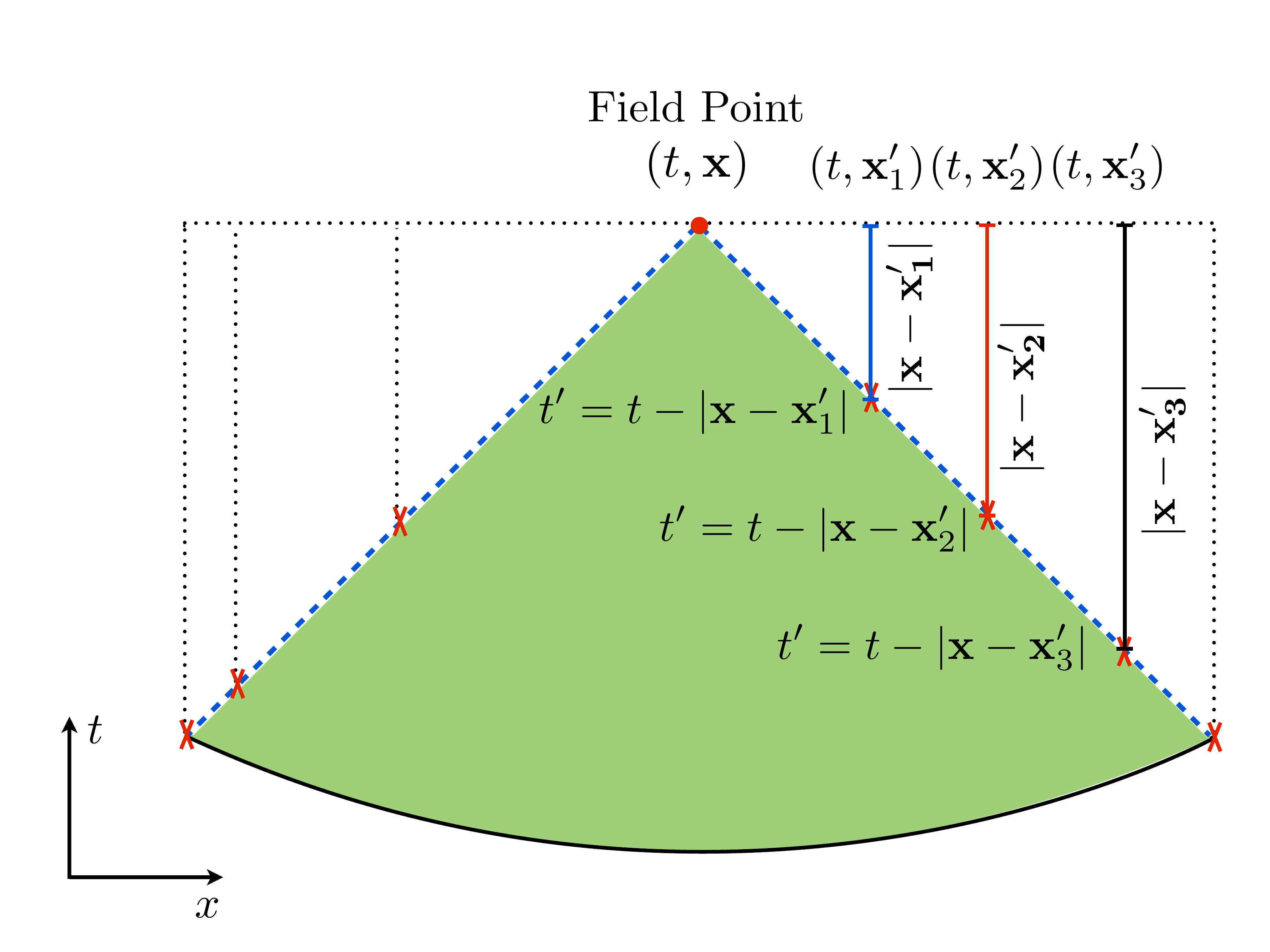}
\caption{Past null cone of a field point. This figure illustrates in 2D how the integration over the whole spacetime for a field point at $(t, {\bf x})$ (see Eq.\ref{integral}) reduces to integration only on the past null cone of that particular field point.}
\label{fig:retarded-cone}
\end{figure}

As long as the field point is inside the near zone we can approximately treat the gravitational fields as almost instantaneous functions of the source variables. We can also neglect the retarded solutions or treat them as a small perturbation of instantaneous solutions. However, in the far zone the fully retarded solutions should be evaluated. Anyhow, field point could be either in the near zone or in the far zone. Both of these situations are shown in Fig.~\ref{fig:cones}. The intersection of the near-zone world tube $\mathcal{D}$ and the hypersurface of the past null cone $\mathcal{C}$ is the region denoted by $\mathcal{N}$. We expect that the dominant contribution to the the integral will come from this region, because of the strong effect of the source in this area.

We break the integration of Eq. (\ref{integral}) over the whole past null cone into two pieces: (1) Integration over the hypersurface $\mathcal{N}$, where the points are close to the matter source and the most important effect comes from, (2) Over the rest of the past null cone i.e. $\mathcal{C}-\mathcal{N}$, where gravity alone contributes to the integral, so that
\be
h^{\mu\nu}= h_{\mathcal{N}}^{\mu\nu}+h_{\mathcal{C}-\mathcal{N}}^{\mu\nu}.
\ee

We treat these two pieces of the integral a bit differently. Fig.~\ref{fig:cones} shows the situation in two different cases. In the left we see the case in which the field point is inside the near zone. This is relevant to the case that we want to calculate the equations of motion of the compact objects in a binary-system. The right panel of Fig. \ref{fig:cones} shows the relevant case for evaluating the gravitational waveform and the energy flux in radiation-zone, when the field point is in the far-zone and very far away from the matter source. 
\begin{figure}[t]
\centering
\includegraphics[width=0.6\textwidth]{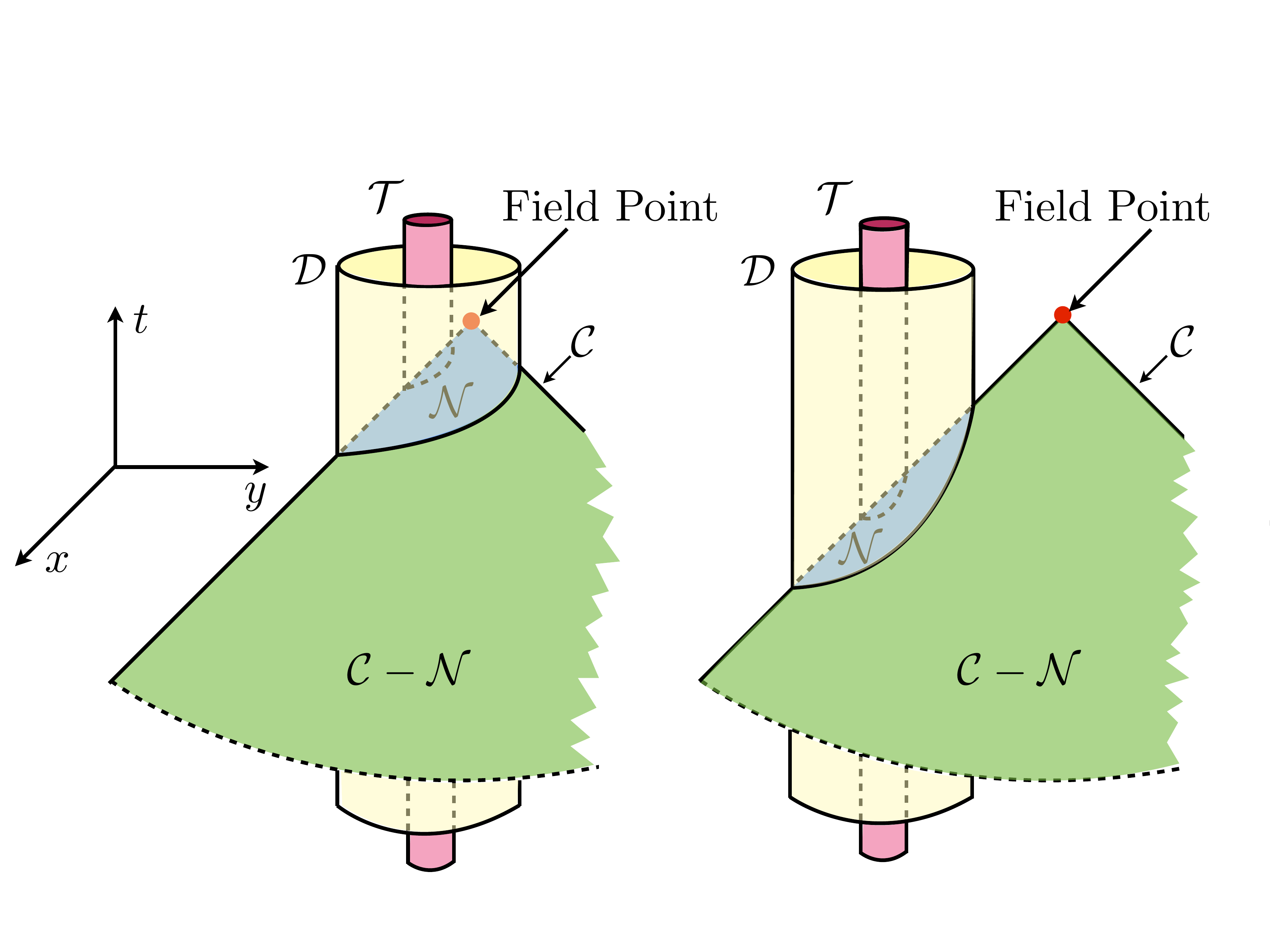}
\caption{Intersectional regions in 3D. Past harmonic null cone $\mathcal{C}$ of the field point $(t, {\bf x})$ intersects the near-zone world tube $\mathcal{D}$ in the hypersurface $\mathcal{N}$. In the left panel the field point is inside the near-zone while in the right panel the field point is in the far-zone. Region $\mathcal{C}-\mathcal{N}$ indicates a part of the past null cone that is in far-zone. World tube $\mathcal{T}$ presents the source-zone.}
\label{fig:cones}
\end{figure}
Depending on if the field point is in the near zone or in the far zone, and if the integral is taken over hypersurface $\mathcal{N}$ or hypersurface $\mathcal{C}-\mathcal{N}$, we have four possible situations: 1) near-zone field point, near-zone integration 2) near-zone field point, far-zone integration, 3) far-zone field point, near-zone integration, 4) far-zone field point, far-zone integration. All of them have been discussed in ~\cite{wil96} in detail. To obtain the equations of motion we focus on {\bf near-zone field points}.

In this case, both ${\bf x}$ and ${\bf x}'$ in Eq.~\ref{integral} are within the near-zone, therefore $\mid {\bf x}-{\bf x}'\mid  \leq 2\mathcal{R}$. The value of $\tau^{\mu\nu}$ varies on a time scale $\mathcal{S}/v \sim \mathcal{R}$. Thereafter we can do a Taylor expansion in powers of the small quantity $\mid {\bf x}-{\bf x}'\mid $. We obtain
\be
h_\mathcal{N}^{\mu\nu} (t,{\bf x}) = 4 \sum_{m=0}^{\infty} \frac{(-1)^m}{m!} \frac{\partial^m}{\partial t^m} \int_\mathcal{M} \tau^{\mu\nu}  (t,{\bf x}) \mid {\bf x}-{\bf x}'\mid ^{m-1} d^3x,
\label{nearzone-integral}
\ee

where $\mathcal{M}$ is shown in Fig.~\ref{fig:cross-section} which represents the intersection of the hypersurface $t=constant$ and the near-zone word tube $\mathcal{D}$.
\begin{figure}[t]
\centering
\includegraphics[width=0.4\textwidth]{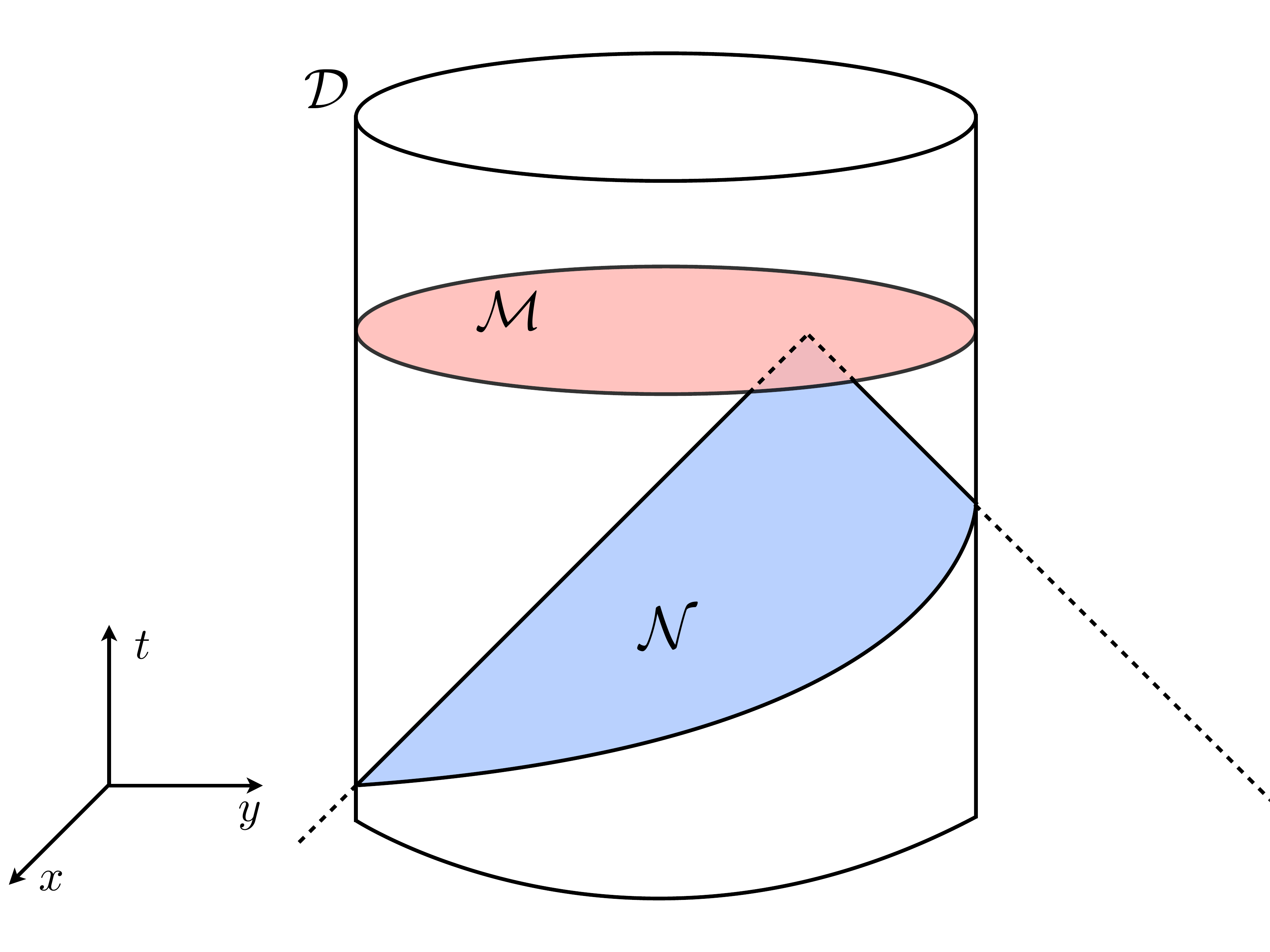}
\caption{Switching the region of integration in the near-zone from region $\mathcal{N}$ to time-independent region $\mathcal{M}$. This is because reactions in the near-zone are almost instantaneous.}
\label{fig:cross-section}
\end{figure}
We do not expect the integral in Eq. (\ref{integral}) to depend upon the arbitrary boundary $\mathcal{R}$. We integrate over the whole past null cone and the final answer of Eq. (\ref{integral}) must be independent of where the radial boundary between the near-zone and far-zone is located. But, each piece of this integral either $h_{\mathcal{N}}^{\mu\nu}$ or $h_{\mathcal{C}-\mathcal{N}}^{\mu\nu}$ individually depends upon $\mathcal{R}$. The only argument that one can make to avoid any inconsistency is that all $\mathcal{R}$-dependent terms must cancel between the inner and outer integrals. This cancellation of $\mathcal{R}$-dependent terms has been shown explicitly in~\cite{pat00}. 

Thus, to determine the field $h^{\mu\nu}$ we don't care about $\mathcal{R}$-dependent terms in $h_{\mathcal{N}}^{\mu\nu}$ and $h_{\mathcal{C}-\mathcal{N}}^{\mu\nu}$ because they all together will finally cancel out anyway. So, we just keep $\mathcal{R}$-independent terms in each expression, then add them up to obtain the overall $h^{\mu\nu}$.

It can be shown that for near-zone field points, the outer integral, i.e. $h_{\mathcal{C}-\mathcal{N}}^{\mu\nu}$, can be ignored until 3PN order. However, for far-zone field points the outer integrals begin to contribute at 2PN order. Will and Wiseman~\cite{wil96} have calculated the contribution of these terms to the gravitational waveform and energy flux up to 2PN order.


\subsection{Iteration of the Relaxed Einstein Equations}
Figure~\ref{fig:iteration} schematically shows the algorithm for solving the relaxed Einstein equations by iteration. Iteration is a useful tool here because the field itself $h^{\mu\nu}$ appears quadratically in the source of the field equation and is assumed to be small. The starting point is $h_0^{\mu\nu}=0$, then construct $\tau_0^{\mu\nu}(h_0)$ and find $h_1^{\mu\nu}$. In another words, starting from $N=1$ and knowing $h_0^{\mu\nu}$ based on our knowledge about $\tau^{00}$ up to Newtonian order (the only survived component of $\tau^{\mu\nu}$ at this order), in principle we are able to solve the field equation in the next order: $\Box h_1^{\mu\nu}=-16 \pi \tau^{\mu\nu} ({\scriptstyle	 h_0^{\mu\nu}})$. This gives us $h_1^{\mu\nu}$. We substitute this recent obtained solution, $h_1^{\mu\nu}$, to the next-order field equation to get $h_2^{\mu\nu}$. In principle, this iterative procedure can be continued until the order, needed to achieve a desired accuracy.
\begin{figure}[h]
\centering
\includegraphics[width=0.6\textwidth]{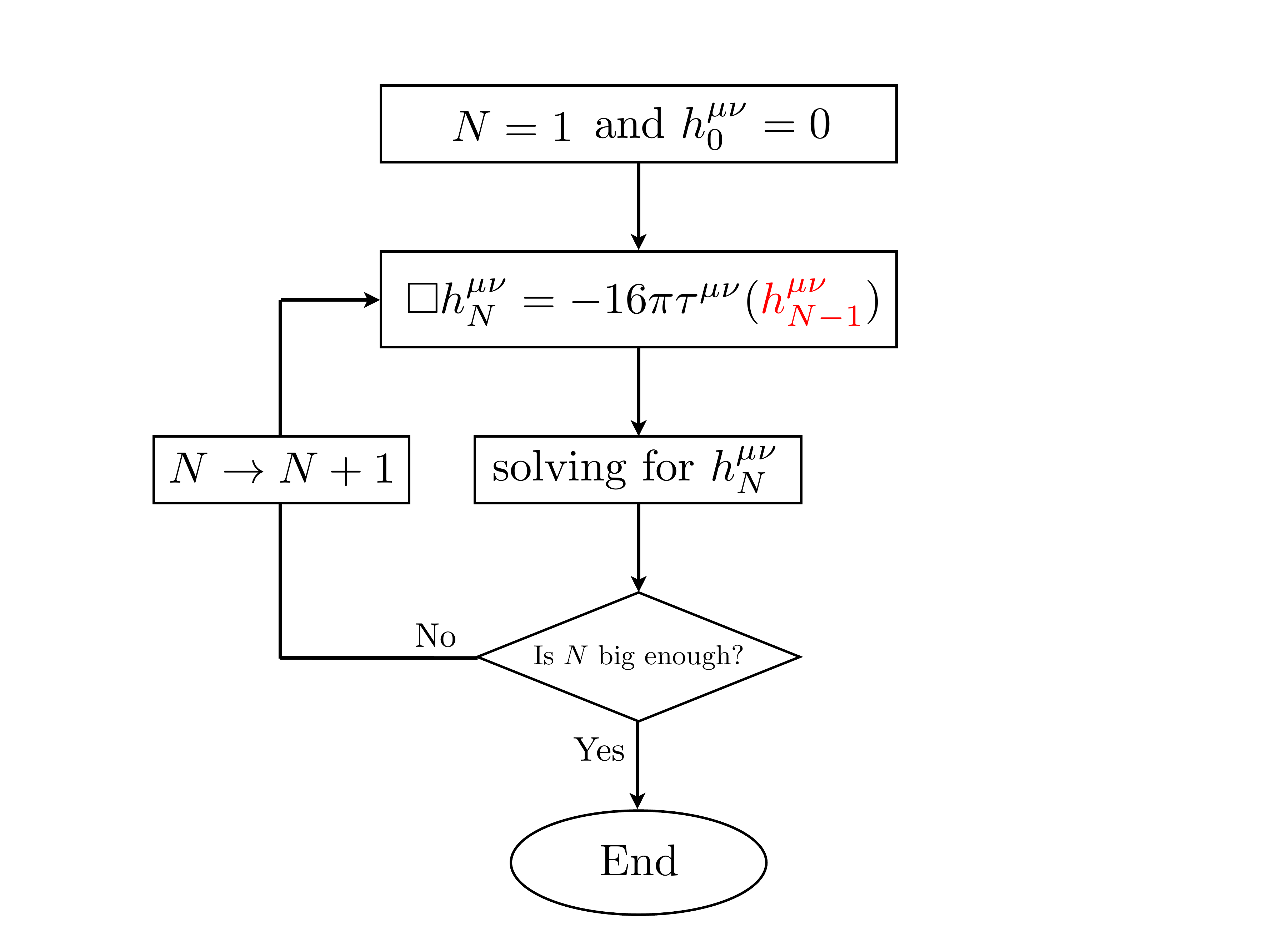}
\caption{Iteration Procedure. A simple, algorithmic illustration to show how the method of iteration works to solve the relaxed Einstein field equations in higher post-Newtonian orders by using the lower order solutions.}
\label{fig:iteration}
\end{figure}

To derive the equation of motion of the source from the field $h_N^{\mu\nu}$, first we have to construct the stress-energy tensor $T_N^{\mu\nu}$ up to the proper order from the field and then solve ${T_N^{\mu\nu}}_{;\mu}=0$ at its $N$-th order. The field $h_N^{\mu\nu}$ obtained from the $N$-th iteration is a functional of the matter variables. To compute the gravitational field as a function of spacetime one needs to solve the equations of motion ${T^{\mu\nu}}_{;\mu}=0$ to the $(N-1)$-th order to obtain the matter variables as functions of spacetime.

We have given a rough picture of the iteration procedure required to obtain the equations of motion of the source and to determine the gravitational waveform and energy flux via the DIRE approach. In the remaining sections of this chapter we will present some of the details. We will rederive the equations of motion of non-spinning compact binaries in general relativity up to 1PN order via the DIRE approach (This has been fully done up to 2.5PN order plus the 3.5PN order contributions by Pati and Will~\cite{pat00, pat02}). Where needed for future reference, we will quote the complete 2PN expressions. We will refer to them in the next part where we generalize the DIRE approach to calculate the equations of motion of non-spinning compact binaries in scalar-tensor theories of gravity up to 2.5PN order.

\section{Formal Structure of Near Zone Fields and Expansion to Higher PN Orders}

We introduce a simplified notation for the components of the gravitational field $h^{\mu\nu}$ and stress-energy tensor $T^{\mu\nu}$, to make the coming expressions a bit easier to work with:
\begin{eqnarray}\label{Ncomponents}
N &\equiv& h^{00} \sim \mathcal{O}(\epsilon) \,, \nonumber \\
K^i &\equiv& h^{0i} \sim \mathcal{O}(\epsilon^{3/2}) \,, \nonumber \\
B^{ij} &\equiv& h^{ij} \sim \mathcal{O}(\epsilon^2) \,, \nonumber \\
B &\equiv& h^{ii} \sim \mathcal{O}(\epsilon^2) \,, 
\end{eqnarray}
and
\begin{eqnarray}\label{sigma:components}
\sigma & \equiv& T^{00} \biggl( \sim \mathcal{O}(\rho)\biggr)+ T^{ii}  \biggl( \sim \mathcal{O}(\rho \epsilon)\biggr) \,, \nonumber \\
\sigma^i & \equiv& T^{0i} \sim \mathcal{O}(\rho \epsilon^{1/2}) \,, \nonumber \\
\sigma^{ij} & \equiv& T^{ij} \sim \mathcal{O}(\rho \epsilon) \,.
\label{sigmas}
\end{eqnarray}

where we show the leading order dependence on $\epsilon$ in the near zone.  Recall that $\epsilon\sim v^2\sim u\sim \rho/p \ll 1$.

From the definition Eq. (\ref{h}), one can invert the tensor $\gothg^{\mu\nu}$ to find
$g_{\alpha\beta}$ in terms of $h^{\mu\nu}$.  Expanding to the
required order, we find,
\begin{subequations}
\label{gr-metric}
\begin{eqnarray}
g_{00} &=& -1 +  \frac{1}{2} N {\color{red}\epsilon}
+  ( \frac{1}{2} B - \frac{3}{8} N^2 ) {\color{red}\epsilon^2}
+ (\frac{5}{16} N^3 - \frac{1}{4} NB + \frac{1}{2} K^j K^j   ) {\color{red}\epsilon^3}
+\mathcal{O}(\epsilon^4) \,, 
\label{metric00}
\\
g_{0i} &=& -  K^i  {\color{red}\epsilon^{3/2}}+   \frac{1}{2} N K^i  {\color{red}\epsilon^{5/2}} +\mathcal{O}(\epsilon^{7/2}) \,, 
\label{metric0i}
\\
g_{ij} &=& \delta^{ij} [ 1+ \frac{1}{2} N {\color{red}\epsilon}- ( \frac{1}{8} N^2 + \frac{1}{2} B  ){\color{red}\epsilon^2} ] +  B^{ij} {\color{red}\epsilon^2}
+\mathcal{O}(\epsilon^3) \,, 
\label{metricij}
\\
(-g) &=& 1+  N {\color{red}\epsilon} -  B {\color{red}\epsilon^2} + \mathcal{O}(\epsilon^3) \,,
\label{metricdet}
\end{eqnarray}
\end{subequations}
where $\epsilon$ helps us to keep track of different orders of magnitude for different terms. Note that in Eq. (\ref{gr-metric}) we have shown the full metric required for the 2.5PN equations of motion i.e. 
$g_{00}$ to $\mathcal{O}(\epsilon^{7/2})$, 
$g_{0i}$ to $\mathcal{O}(\epsilon^{3})$ , and
$g_{ij}$ to $\mathcal{O}(\epsilon^{5/2})$.
However, to obtain the equations of motion to 1.5PN order, determining
the components of the metric up to one order less than what is shown above for each component would be enough.

From above equations in Eq. (\ref{gr-metric}), also notice that in order to find the metric $g_{\alpha\beta}$ to the desired order. For the 1PN equations of motion we must obtain $N$ and $B$ to $\mathcal{O}(\epsilon^{7/2})$, $K^i$ to $\mathcal{O}(\epsilon^{5/2})$, and $B^{ij}$ to $\mathcal{O}(\epsilon^{3/2})$. Note that we treat $B^{ij}$ and its trace $B$ differently simply because $B$ appears in $g_{00}$ linearly. 

The next variable that must be evaluated to solve the relaxed Einstein equations is $\tau^{\mu\nu}$ which is made of two pieces: $T^{\mu\nu}$ and $\Lambda^{\mu\nu}$ (see Eq. (\ref{tau})). We leave the components of $T^{\mu\nu}$ in the form introduced in Eq. (\ref{sigmas}) until the final steps of the calculation. To evaluate the components of $\Lambda^{\mu\nu}$ in terms of the field components required for calculating equations of motion up to 2.5PN order, we use Eqs. (\ref{Lambda},\ref{tLL}) and obtain:
\begin{subequations}
\label{Lambdacomponents}
\begin{eqnarray}
\Lambda^{00} &=& - \frac{7}{8} (\nabla N)^2 + \left [ \frac{5}{8} {\dot N}^2 - \ddot N N -2 {\dot N}^{,k} K^k
+ \frac{1}{2} K^{i,j} (3 K^{j,i}+ K^{i,j}) 
\right . 
\nonumber \\
&& \left . 
+ {\dot K}^j N^{,j} 
- B^{ij} N^{,ij} 
+ \frac{1}{4} \nabla N \cdot
\nabla B + \frac{7}{8} N (\nabla N)^2 \right] + \mathcal{O}(\rho \epsilon^3) \,,
\\
\Lambda^{0i} &=& \biggl [ N^{,k}( K^{k,i}- K^{i,k}) + \frac{3}{4}\dot
N N^{,i} \biggr ] + \mathcal{O}(\rho \epsilon^{5/2}) \,, \\
\Lambda^{i j} &=& \frac{1}{4} {\Big [} N^{,i}N^{,j} - \frac{1}{2}
\delta^{ij} (\nabla N)^2 {\Big ]} 
+ {\biggl \{} 2 K^{k,(i}K^{j),k}- K^{k,i}K^{k,j}
-K^{i,k}K^{j,k}   \nonumber \\
&& 
+2N^{,(i} {\dot K}^{j)} 
+ \frac{1}{2}N^{,(i} B^{,j)} - \frac{1}{2}N \Big [ N^{,i}N^{,j} - \frac{1}{2}
\delta^{ij}  (\nabla N)^2 \Big ] 
 \nonumber \\
&& 
-\delta^{ij} \Big (K^{l,k}K^{[k,l]} 
+N^{,k}{\dot K}^{k}
+ \frac{3}{8}{\dot N}^2 + \frac{1}{4} \nabla N \cdot
\nabla B \Big ) \biggr \}
 + \mathcal{O}(\rho \epsilon^3) \,,
\\
\Lambda^{ii} &=& - \frac{1}{8} (\nabla N)^2 
+ \left [ K^{l,k}K^{[k,l]}-N^{,k}{ \dot K}^{k} - \frac{1}{4}\nabla N \cdot \nabla B - \frac{9}{8} {\dot N}^2+ \frac{1}{4}N ({\nabla N})^2 \right ] 
\nonumber \\
&&
+  \mathcal{O}(\rho \epsilon^3) \,.
\end{eqnarray}
\end{subequations}

As long as the field point is in the near-zone, we can use the Taylor expansion of the gravitational field introduced in Eq. (\ref{nearzone-integral}) and write the components of $h^{\mu\nu}$ as integrals over the time-constant region $\mathcal{M}$ (see Fig.~\ref{fig:cross-section}) and their time derivatives. The near-zone expansions of the field components i.e. $N$, $K^i$, and $B^{ij}$ are then given by
\begin{subequations}
\label{bigexpansion}
\begin{eqnarray}
%
%
N_{\cal N} &=& 4 {\color{red}\epsilon} \int_{\cal M} \frac{\tau^{00}(t,{\bf x}^\prime)}{\mid {\bf x}-{\bf
x}^\prime \mid } d^3x^\prime 
+2 {\color{red}\epsilon^2} \partial^2_t \int_{\cal M} \tau^{00}(t,{\bf x}^\prime) \mid {\bf x}-{\bf x}^\prime \mid  d^3x^\prime
-{2 \over 3} {\color{red}\epsilon^{5/2}} \stackrel{(3)\qquad}{{\cal I}^{kk}(t)} 
\nonumber \\
&&
+ {1 \over 6} {\color{red}\epsilon^3} \partial^4_t \int_{\cal M}
\tau^{00}(t,{\bf x}^\prime) \mid {\bf x}-{\bf x}^\prime \mid ^3 d^3x^\prime 
\nonumber \\
&&
- {1 \over 30} {\color{red}\epsilon^{7/2}} \biggl [ (4x^{kl}+2r^2\delta^{kl})
\stackrel{(5)\quad}{{\cal I}^{kl}(t)}
- 4 x^k \stackrel{(5)\qquad}{{\cal I}^{kll}(t)}
+ \stackrel{(5)\qquad}{{\cal I}^{kkll}(t)} \biggr
] 
+ N_{\partial {\cal M}} + \mathcal{O}(\epsilon^4),
\label{bigexpansiona} \\
%
%
K^i_{\cal N} &=& 4 {\color{red}\epsilon^{3/2}} \int_{\cal M} \frac{\tau^{0i}(t,{\bf x}^\prime)}{\mid {\bf x}-{\bf
x}^\prime \mid } d^3x^\prime +2 {\color{red}\epsilon^{5/2}} \partial^2_t \int_{\cal M}
\tau^{0i}(t,{\bf x}^\prime) \mid {\bf x}-{\bf x}^\prime \mid  d^3x^\prime
 \nonumber \\
&&
 +{2 \over 9} {\color{red}\epsilon^3} \biggl [ 3 x^k \stackrel{(4)\quad}{{\cal I}^{ik}(t)}
- \stackrel{(4)\qquad}{{\cal I}^{ikk}(t)}
+2 \epsilon^{mik}  \stackrel{(3)\qquad}{{\cal J}^{mk}(t)}
\biggr ] + K^i_{\partial {\cal M}} + \mathcal{O}(\epsilon^{7/2}) \,,
\label{bigexpansionb} \\
%
%
B^{ij}_{\cal N} &=& 4 {\color{red}\epsilon^2} \int_{\cal M} \frac{\tau^{ij}(t,{\bf x}^\prime)}{\mid {\bf x}-{\bf x}^\prime \mid } d^3x^\prime 
- 2 {\color{red}\epsilon^{5/2}} \stackrel{(3)\quad}{{\cal I}^{ij}(t)}
+2 {\color{red}\epsilon^3} \partial^2_t \int_{\cal M}
\tau^{ij}(t,{\bf x}^\prime) \mid {\bf x}-{\bf x}^\prime \mid  d^3x^\prime
\nonumber \\
&& - {1 \over 9} {\color{red}\epsilon^{7/2}} \biggl [
3 r^2 \stackrel{(5)\quad}{{\cal I}^{ij}(t)}
-2x^k \stackrel{(5)\qquad}{{\cal I}^{ijk}(t)}
- 8 x^k \epsilon^{mk(i} \stackrel{(4)\qquad}{{\cal J}^{m\mid j)}(t)}
+ 6 \stackrel{(3)\qquad}{M^{ijkk}(t)} \biggr ]
\nonumber \\
&& + B^{ij}_{\partial {\cal M}} + \mathcal{O}(\epsilon^4) \,,
\label{bigexpansionc}
\end{eqnarray}
\end{subequations}
where we have define the {\em moments} of the system by
\begin{subequations}
\label{momenas}
\begin{eqnarray}
{\cal I}^Q &\equiv&  \int_{\cal M} \tau^{00} x^Q d^3x \,,
\label{IQ}
\\
{\cal J}^{iQ} &\equiv&  \upepsilon^{iab}\int_{\cal M} \tau^{0b} x^{aQ} d^3x \,,
\label{JiQ}
\\
M^{ijQ} &\equiv& \int_{\cal M} \tau^{ij} x^Q d^3x \,,
\label{MijQ}
\end{eqnarray}
\label{genmoment}
\end{subequations}
The index $Q$ is a multi-index, such that $x^Q$ denotes $x^{i_1} \dots x^{i_q}$. The boundary terms $N_{\partial {\cal M}}$, $K^i_{\partial {\cal M}}$ and $B^{ij}_{\partial {\cal M}}$ can be found in Appendix C of~\cite{pat00}, but they will play no role in our analysis because they contribute at higher PN orders than we care about. Looking at ~\ref{bigexpansion}, all integrals are well-behaved such that all integrands are constructed from (1) a specific component of the stress-energy pseudo-tensor $\tau^{\mu\nu}$, (2) either a power of $\mid {\bf x}-{\bf x}'\mid $ (Poisson-like potentials and their generalizations) or a multiple combination of spatial coordinates i.e. $x^i$ (multipole moments), and (3) are integrated over a finite domain. Here we re-emphasize that all near-zone integrals are taken over time-constant region of $\mathcal{M}$ and we discard any possible $\mathcal{R}$-dependent term because it must cancel with a corresponding term from the far-zone integral. 

In the near zone, the potentials are either Poisson-like potentials $P$ (the most frequent kind of potential), super-potentials $S$, or super-duper-potentials $SD$. For a source $f$, they are given by the following definitions and satisfy the relevant Poisson equations,
\begin{subequations}
\begin{eqnarray}\label{definesuper-1}
P(f) \equiv {1 \over {4\pi}} \int_{\cal M} {{f(t,{\bf x}^\prime)}
\over {\mid {\bf x}-{\bf x}^\prime \mid  }} d^3x^\prime \,, &\quad \nabla^2
P(f) = -f \,,\\
S(f) \equiv {1 \over {4\pi}} \int_{\cal M} {f(t,{\bf x}^\prime)}
{\mid {\bf x}-{\bf x}^\prime \mid  } d^3x^\prime \,, &\quad \nabla^2
S(f) = 2 P(f) \,,\\
SD(f) \equiv {1 \over {4\pi}} \int_{\cal M} {f(t,{\bf x}^\prime)}
{\mid {\bf x}-{\bf x}^\prime \mid  ^3} d^3x^\prime \,, &\quad \nabla^2
SD(f) = 12 S(f) \,.
\end{eqnarray}
\end{subequations}
We also define potentials based on the {\em densities} $\sigma$,
$\sigma^i$ and $\sigma^{ij}$ 
\begin{subequations}
\label{definesuper0}
\begin{eqnarray}
\Sigma (f) &\equiv& \int_{\cal M} {{\sigma(t,{\bf x}^\prime)f(t,{\bf x}^\prime)}
\over {\mid {\bf x}-{\bf x}^\prime \mid  }} d^3x^\prime = P(4\pi\sigma f) \,,
\\
\Sigma^i (f) &\equiv& \int_{\cal M} {{\sigma^i(t,{\bf x}^\prime)f(t,{\bf
x}^\prime)}
\over {\mid {\bf x}-{\bf x}^\prime \mid  }} d^3x^\prime = P(4\pi\sigma^i f) \,,
\\
\Sigma^{ij} (f) &\equiv& \int_{\cal M} {{\sigma^{ij}(t,{\bf x}^\prime)f(t,{\bf
x}^\prime)}
\over {\mid {\bf x}-{\bf x}^\prime \mid  }} d^3x^\prime = P(4\pi\sigma^{ij} f) \,,
\end{eqnarray}
\end{subequations}
along with the {\em super-potentials}
\begin{subequations}
\label{definesuper1}
\begin{eqnarray}
X(f)  &\equiv& \int_{\cal M} {\sigma(t,{\bf x}^\prime)f(t,{\bf
x}^\prime)}
{\mid {\bf x}-{\bf x}^\prime \mid  } d^3x^\prime   = S(4\pi\sigma f) \,,\\
X^i(f)  &\equiv& \int_{\cal M} {\sigma^i(t,{\bf x}^\prime)f(t,{\bf
x}^\prime)}
{\mid {\bf x}-{\bf x}^\prime \mid  } d^3x^\prime  = S(4\pi\sigma^i f) \,,\\
X^{ij}(f)  &\equiv& \int_{\cal M} {\sigma^{ij}(t,{\bf x}^\prime)f(t,{\bf
x}^\prime)}
{\mid {\bf x}-{\bf x}^\prime \mid  } d^3x^\prime   = S(4\pi\sigma^{ij} f) \,,
\end{eqnarray}
\end{subequations}
and {\em super-duper-potensials}
\begin{subequations}
\label{definesuper2}
\begin{eqnarray}
Y(f)  &\equiv& \int_{\cal M} {\sigma(t,{\bf x}^\prime)f(t,{\bf
x}^\prime)}
{\mid {\bf x}-{\bf x}^\prime \mid ^3} d^3x^\prime   = SD(4\pi\sigma f) \,.
\end{eqnarray}
\end{subequations}
Super-duper-potentials begin to show up at 2PN order (only $Y$ at 2PN order and $Y^{i}$ and $Y^{ij}$ at higher orders) while super-potentials begin to contribute at 1PN order. However, Poisson potentials are everywhere; including Newtonian, 1PN, and 2PN terms.


A number of potentials occur sufficiently frequently in the PN expansion that it is easier to redefine them specifically, just to make the calculations easier to follow.  At Newtonian order there is the Newtonian potential,
\begin{eqnarray}\label{U}
U \equiv \int_{\cal M} {{\sigma(t,{\bf x}^\prime)}
\over {\mid {\bf x}-{\bf x}^\prime \mid  }} d^3x^\prime = P(4\pi\sigma) =
\Sigma(1) \,,
\end{eqnarray}
At 1PN order, frequent potentials are:
\begin{eqnarray}\label{1PNpotentials}
V^i \equiv \Sigma^i(1)  \,, \qquad & \Phi_1^{ij} \equiv \Sigma^{ij}(1) \,, \qquad & \Phi_1 \equiv \Sigma^{ii}(1) \,,\nonumber \\
\Phi_2 \equiv \Sigma(U) \,, \qquad & X \equiv X(1)  \,. \qquad &
\end{eqnarray}
In \eref{bigexpansion} we have the {\em implicit} integral form of the components of gravitational field $h^{\mu\nu}$ in terms of stress-energy pseudo-tensor components. Armed with \eref{Lambdacomponents} and starting from \eref{tau}, we can evaluate the {\em explicit} form of the near-zone field components in terms of Poisson-like potentials (see \eref{U}, \eref{1PNpotentials}) and multiple-moments (see \eref{genmoment}). To do that we need to evaluate the contribution at each order and be very careful about it. The {\em leading} order of magnitude of each field component is shown in \eref{Ncomponents} but here we need to keep track of the contribution in each PN order separately. So we use the following useful notation. Notice that in this chapter we will do the calculation up to 1.5PN order but here we show the expansion through 2.5PN order, one PN order beyond what we need for the 1.5PN equations of motion:
\begin{subequations}
\label{expandNKB}
\begin{eqnarray}
N &=& {\color{red}\epsilon} (N_0 + {\color{red}\epsilon} N_1+ {\color{red}\epsilon^{3/2}} N_{1.5}+ {\color{red}\epsilon^2} N_2+
{\color{red}\epsilon^{5/2}} N_{2.5}
)
+\mathcal{O}(\epsilon^4) \,, \\
K^i &=& {\color{red}\epsilon^{3/2}} (K_1^i + {\color{red}\epsilon} K_2^i +{\color{red}\epsilon^{3/2}} K_{2.5}^i
) 
+\mathcal{O}(\epsilon^{7/2}) \,, \\
B &=& {\color{red}\epsilon^2} (B_1 + {\color{red}\epsilon^{1/2}} B_{1.5} 
+{\color{red}\epsilon} B_2 + {\color{red}\epsilon^{3/2}} B_{2.5} 
)
+\mathcal{O}(\epsilon^4) \,, \\
B^{ij} &=& {\color{red}\epsilon^2} (B_2^{ij} + {\color{red}\epsilon^{1/2}} B_{2.5}^{ij} +
) +\mathcal{O}(\epsilon^3) \,,
\end{eqnarray}
\end{subequations}
where the subscript on each term indicates the relevant level of PN order in which that particular term leads. For example, $N_0$ is the leading Newtonian order of the field component $N$, while $N_1$ is its leading 1PN contribution, and so on. In other words, in 1PN calculations we do {\em not} expect any terms except those with the subscript of 1. Consequently, the subscript of the first term in each line shows the PN order in which the relevant field component begins to contribute. For instance, one can read from \eref{expandNKB} that $B$ and $B^{ij}$ show up at 1PN and 2PN for the first time, respectively. From \eref{expandNKB} we expect a specific order of magnitude for each subscripted term in these relations, for example $N_0\sim \mathcal{O}(\epsilon)$ and $N_1\sim\mathcal{O}(\epsilon^2)$. In fact, one can check this after evaluating the explicit values of the terms later.
Notice that our separate treatment of $B$ and $B^{ij}$ leads to the slightly awkward notational circumstance that, for example, $B_2^{ii} = B_1$.

At this point we are ready to deal with the relaxed field equations \eref{relaxed} at the first level of iteration (Newtonian order). At lowest order in the PN expansion (shown as subscript $0$ in \eref{expandNKB}), we only need to evaluate $\tau^{00}$ with $h_0^{\mu\nu}=0$, $g_{\mu\nu}=\eta_{\mu\nu}$, so that (see \eref{sigmas})
\be
\tau^{00}=(-g) T^{00}+\mathcal{O}(\rho \epsilon)=\sigma+\mathcal{O}(\rho \epsilon).
\label{Ntau}
\ee
Other components of $\tau^{\mu\nu}$ are of higher orders. 

As a result, at the Newtonian order the tensorial relaxed Einstein equations reduce to a single equation 
\be
\Box N_0 = -16 \pi \sigma,
\ee
which, with the definition of the Newtonian potential $U$ in \eref{U}, has the solution in near-zone
\be
N_0 = 4\int_\mathcal{M} \frac{\sigma\, d^3 x' }{\mid {\bf x}-{\bf x}'\mid }= 4 U.
\label{N0}
\ee
This result reproduces Newtonian gravity and confirms the fact that general relativity contains Newtonian gravity at its lowest order when post-Newtonian theory is used.

In the next step, using~\ref{metricdet}, \eref{tau}, \eref{Lambda}, and \eref{N0} and keeping only the next generation of higher order terms compared to the first survived terms in the first generation i.e. \ref{Ntau}, we have
\begin{eqnarray}
\tau^{00} &=& \sigma - \sigma^{ii} + 4\, \sigma U - {7 \over {8\pi}} (\nabla U)^2 + \mathcal{O}(\rho\epsilon^2) \,, \nonumber \\
\tau^{0i} &=& \sigma^i + \mathcal{O}(\rho\epsilon^{3/2}) \,, \nonumber \\
\tau^{ii} &=& \sigma^{ii} -{1 \over {8\pi}} (\nabla U)^2 + \mathcal{O}(\rho\epsilon^2) \,, \nonumber \\
\tau^{ij} &=&  \mathcal{O}(\rho\epsilon) \,. 
\label{tauPN}
\end{eqnarray}
Substituting into Eqs. (\ref{bigexpansion}), and calculating terms
through 1.5PN order (e.g. $\mathcal{O}(\epsilon^{5/2})$ in $N$), we obtain
\begin{subequations}
\begin{eqnarray}
N_1 &=& 7U^2-4\Phi_1+2\Phi_2+2{\ddot X} \,, \\
K_{1}^i &=& 4V^i \,,\\
B_1 &=& U^2+4\Phi_1-2\Phi_2 \,, \\
N_{1.5} &=& -{2 \over 3} \stackrel{(3)\quad}{{\cal I}^{kk}(t)}
\,,\\
B_{1.5} &=& -{2} \stackrel{(3)\quad}{{\cal I}^{kk}(t)} \,.
\end{eqnarray}
\label{1PNNKB}
\end{subequations}
To rederive above equations one needs to use the identities introduced in appendix D of \cite{pat00}, specially the following identity:
\be
P(\mid \nabla U\mid ^2) = -\frac{1}{2} U^2 +\Phi_2\,.
\ee

Using \eref{1PNNKB} in \eref{gr-metric} to the appropriate order, the physical metric to 1.5PN order is obtained as
\begin{subequations}
\label{1.5pnmetric}
\begin{eqnarray}
g_{00} &=& -1 + 2U - 2U^2 + \ddot X 
- {4 \over 3}\stackrel{(3)\quad}{{\cal I}^{kk}(t)} 
+ \mathcal{O}(\epsilon^3) \,,\\
g_{0i} &=& -4V^i + \mathcal{O}(\epsilon^{5/2}) \,,\\
g_{ij} &=& \delta_{ij} (1+2U) + \mathcal{O}(\epsilon^2)\,.
\end{eqnarray}
\end{subequations}
and will be needed in deriving the equations of motion later on.
\section{Conversion to the Baryon Density $\rho^*$ and Equations of Motion in Terms of Potentials}
\label{sec:conserved-density}
We treat the source bodies as pressure-free balls of baryons characterized by the ``conserved'' baryon mass density $\rho^*$, given by
\be
\rho^*= m n \sqrt{-g} u^0,
\label{rhostar}
\ee
where $m$ is the rest mass per baryon and $n$ is the baryon number density. From the conservation of baryon density, expressed in covariant terms by $(n u^{\mu})_{;\mu}=0=((\sqrt{-g})^{-1}(\sqrt{-g} n u^\mu)_{,\mu}$, we see that $\rho^*$ obeys the non-covariant, but exact, continuity equation (see Fig.~\ref{fig:current})
\be
\frac{\partial\rho^*}{\partial t}+\nabla\cdot j=0,
\label{continuity}
\ee
where $j=\rho^* v$, $v^i=u^i/u^0$, and spatial gradients and dot products use the Cartesian metric. In terms of $\rho^*$, the stress-energy tensor is given by
\be
T^{\mu\nu}=\rho^* \frac{1}{\sqrt{-g}} u^0 v^\mu v^\nu,
\label{Tstar}
\ee
where $v^\mu=(1, v^i)$. We define the baryon rest mass as
\be
m_A\equiv \int_A \rho^* d^3x,
\ee
such that
\be
{\bf x}_A\equiv \frac{1}{m_A} \int_A \rho^* {\bf x}\; d^3x,
\ee
indicates the baryonic center-of-mass. Therefore, the velocity and acceleration of each body are defined by
\bea
{\bf v}_A \equiv \frac{d{\bf x}_A}{dt}=\frac{1}{m_A} \int_A \rho^* {\bf v}\; d^3x,\\
{\bf a}_A \equiv \frac{d{\bf v}_A}{dt}=\frac{1}{m_A} \int_A \rho^* {\bf a}\; d^3x.
\label{acceleration}
\eea
\begin{figure}[htb]
\centering
\includegraphics[width=0.6\textwidth]{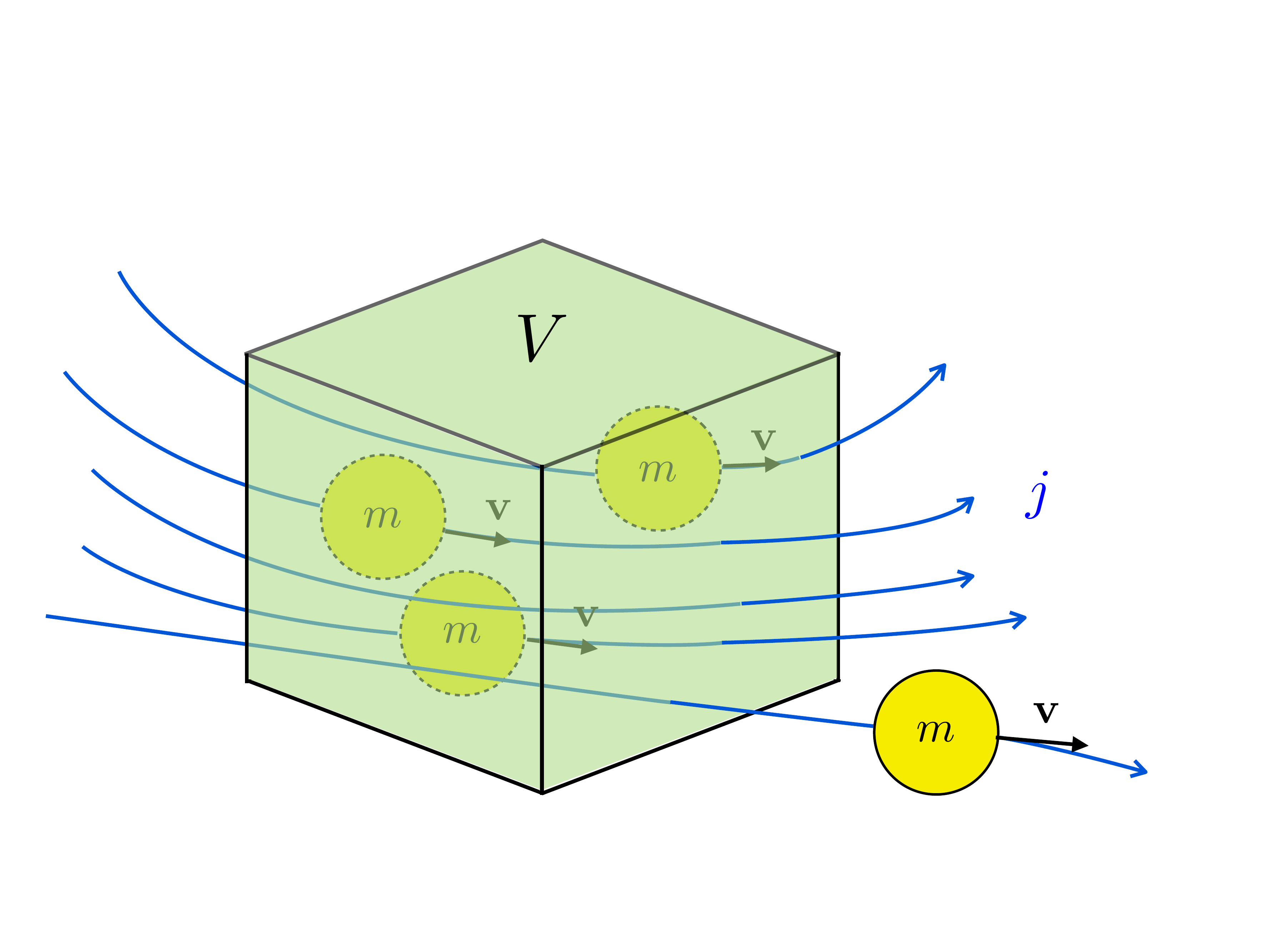}
\caption{Illustration of $m$, $\rho^*$, and $j$. $\rho^*=dm/dV$ is the amount of $m$ per unit volume (in the box), $j=\rho^* v$ represents the flux and $m$ is the mass carried by a baryonic particle.}
\label{fig:current}
\end{figure}

Using the equations of motion, $T^{\mu\nu}_{;\mu}=0$ for each fluid element it is not difficult to show that
\be
a^i\equiv \frac{dv^i}{dt}=-\Gamma^i_{\mu\nu} v^\mu v^\nu +\Gamma^0_{\mu\nu} v^\mu v^\nu v^i,
\label{ai}
\ee
where $\Gamma^\gamma_{\mu\nu}$ are the components of the Christoffel symbols computed from the metric via
\be
\Gamma^\alpha_{\mu\nu}=\frac{1}{2} g^{\alpha\beta} (g_{\beta\mu,\nu}+g_{\beta\nu,\mu}-g_{\mu\nu,\beta}).
\label{christoffel}
\ee

Our task therefore, is to determine the Christoffel symbols through a PN order sufficient for equations of motion valid through 1.5PN order using the 1.5PN accurate expressions of the metric in Eq. (\ref{1.5pnmetric}) (different components of $\Gamma^\alpha_{\mu\nu}$ are needed to different accuracy, depending on the number of factors of velocity which multiply them); re-express the Poisson potentials contained in the metric in terms of $\rho^*$, rather than in terms of the ``densities'' $\sigma$, $\sigma^i$ and $\sigma^{ij}$, substitute into Eq. (\ref{acceleration}), and integrate over the $A$-th body, keeping only terms that do not depend on the bodies' finite size.

We must now convert all potentials from integrals over $\sigma$, $\sigma^i$
and $\sigma^{ij}$ to integrals over the conserved baryon density $\rho^*$,
defined by Eq. (\ref{rhostar}).  
From Eqs. (\ref{sigmas}, \ref{Tstar}), we find
\begin{eqnarray}
\sigma &=&  \frac{\rho^* u^0}{\sqrt{-g}} \, (1+v^2)\,,
\nonumber \\
\sigma^i &=& \frac{\rho^* u^0}{\sqrt{-g}} \,  v^i\,,
\nonumber \\
\sigma^{ij} &=& \frac{ \rho^* u^0}{\sqrt{-g}} \, v^i v^j\,,
\label{sigmatorho}
\end{eqnarray}
where $u^0 = (-g_{00} -2 g_{0i}v^i - g_{ij}v^iv^j)^{-1/2}$.  Substituting the expansions for the metric, Eqs. (\ref{gr-metric}), and for the field components Eqs. (\ref{expandNKB}) from \eref{N0} and \eref{1PNNKB}, we obtain, to the order required for the 1.5PN equations of motion, 
\begin{subequations}
\begin{eqnarray}
\sigma &=&  \rho^* \biggl [ 1 + 
	\epsilon \left ({3 \over 2} v^2 - U_\sigma \right )
	+  O(\epsilon^2) \biggr] \,, \\
\sigma^i &=& \rho^*  v^i \biggl [ 1 + O(\epsilon) \biggr ] \,, \\
\sigma^{ij} &=&  O(\epsilon) \,, \\
\sigma^{ii} &=& \rho^* v^2 \biggl [ 1 + O(\epsilon) \biggr ] \,. 
\end{eqnarray}
\label{sigmatorhoPN}
\end{subequations}
Substituting these formulae into the definitions for $U_\sigma$ and the other potentials defined in Eqs.~(\ref{1PNpotentials}), and iterating successively, we convert all such potentials into  new potentials defined using $\rho^*$, plus PN corrections.  For example, we find that
\begin{eqnarray}
U_\sigma &=&  U + {\color{red}\epsilon} \left ({3 \over 2} \Phi_1 - \Phi_2 \right )  + O(\epsilon^2) \,,\\
V_\sigma^i&=& V^i +  {\color{red}\epsilon} \left ( \half \Sigma (v^i v^2) -V_2^i \right ) + O(\epsilon^2) \,,
\label{Usigmaexpand}
\end{eqnarray}
where henceforth, $U$, $V^i$, $V_2^i$, $\Phi_1$, $\Phi_2$, and $\Sigma$ are 
defined in terms of $\rho^*$ (see Appendix A of \cite{pat00}).

At this point everything depends on the conserved baryonic density $\rho^*$, and we are ready to calculate the acceleration of body-$A$ from \eref{acceleration} and \eref{ai} as
\be
a^i_A=\frac{1}{m_A} \int_A \rho^* (-\Gamma^i_{\mu\nu} v^\mu v^\nu+\Gamma^0_{\mu\nu} v^\mu v^\nu v^i) d^3x.
\ee

To do the above integration, first we have to calculate the integrand, which is equal to $\rho^*$ times $a^i$. The Christoffel symbols are given in terms of the metric components and their derivatives in \eref{christoffel}. Metric components are functions of the field components (see \eref{gr-metric}), which we already derived as explicit functions of the potentials defined in \eref{1PNpotentials}, up to 1.5PN order in Eqs.~(\ref{N0}, \ref{1PNNKB}) (also see \eref{expandNKB}). Applying all these and inserting the iterated forms of all potentials, we obtain the acceleration of a given element of matter through 1.5PN order in the general form of
\be
a^i=\frac{dv^i}{dt}=a_N^i+ a_{1PN}^i+ a^i_{1.5PN},
\label{fluidEOM}
\ee
where $a^i_{1.5PN}=0$ because the 1.5PN contributions to the metric are all functions of time, which do not survive the gradient used to calculate the Christoffel symbols.
\bea
a_N^i&=&U^{,i}\,,\\\nonumber
a_{1PN}^i &=&
	v^2 U^{,i} -4 v^i v^j U^{,j} - 3v^i \dot U - 4 U U^{,i} 
	+ 8 v^j V^{[i,j]} \\
&&	+ 4 \dot V^i + \half \ddot X^{,i} + {3 \over 2} \Phi_1^{,i}
	-\Phi_2^{,i} \,.
	\label{1.5PNEOM}
\eea


\section{Two-Body Equations of Motion}
\label{sec:Two-Body}

We must now integrate all potentials that appear in the equation of motion, as well as the equation of motion itself given in \eref{fluidEOM} over the bodies in the binary system.  We treat each body as a non-rotating, spherically symmetric fluid ball (as seen in its momentary rest frame), whose characteristic size $\mathcal{S}$ is much smaller than the orbital separation ($\mathcal{S}\ll r$).  We shall discard all terms in the resulting equations that are proportional to positive powers of $\mathcal{S}$: these correspond to multipolar interactions and their relativistic corrections. We also discard all terms that are proportional to negative powers of $\mathcal{S}$: these correspond to self-energy corrections of PN and higher order. We retain only terms that are proportional to $\mathcal{S}^0$. Such terms will generally depend only on the mass of each body, but it is conceivable that terms could arise that are proportional to $\mathcal{S}^0$, but that still depend on the internal structure of each body.  It can be shown~\cite{pat02} that such terms cannot appear at 1PN order by a simple symmetry argument.  At 2PN order, terms of this kind {\it could} appear in certain non-linear potentials, but in fact vanish identically by a subtler symmetry.  At 3PN order, such $\mathcal{S}^0$ structure-dependent terms definitely appear, but whether they survive in the final equations of motion is an open question at present.

Our assumption that the bodies are non-rotating will imply simply that every element of fluid in the body has the same coordinate velocity, so that $v^i$ can be pulled outside any integral. This assumption can be easily modified in order to deal, for example, with rotating bodies. We also assume that each body is suitably spherical.  By this we mean that, in a local inertial frame co-moving with the body and centered at its baryonic center of mass, the baryon density distribution is static and spherically symmetric in the coordinates of that frame.

We shall evaluate the acceleration consistently for body-1;  the corresponding equation for body-2 can be obtained by interchange. At the end, we shall find the centre-of-mass and relative equations of motion.

The Newtonian acceleration is straightforward: 
\begin{eqnarray}\label{newtonterm} 
{(a^i_1)}_N &=& -(1/m_1) \int_1  \rho^* \int {\rho^*}^\prime {{(x^i -
{x^i}^\prime)} \over {\mid {\bf x} -
{\bf x}^\prime \mid ^3}} d^3x^\prime  d^3x 
\\\nonumber
&=&  
-(1/m_1) \int_1 \int_1  \rho^* {\rho^*}^\prime {{(x^i -
{x^i}^\prime)} \over {\mid {\bf x} -
{\bf x}^\prime \mid ^3}} d^3x d^3x^\prime  -
 (1/m_1) \int_1  \rho^* d^3x \int_2 {\rho^*}^\prime {{(x^i -
{x^i}^\prime)} \over {\mid {\bf x} -
{\bf x}^\prime \mid ^3}} d^3x^\prime  \,,
\end{eqnarray}
where $\rho^*$ and ${\rho^*}^\prime$ are conserved densities at spatial points ${\bf x}$ and ${\bf x'}$. The denisty $\rho^*$ and ${\rho^*}^\prime$ vanish anywhere outside the bodies. The first term in the last line of \eref{newtonterm} in which both integral points ${\bf x}$ and ${\bf x'}$ are in the same body vanishes by symmetry, irrespective of any relativistic flattening or any other effect (Newton's third law). In the second term in which ${\bf x}$ is in body-1 and ${\bf x'}$ is in body-2, we find that all contributions apart from the leading term are of positive powers in $\mathcal{S}$, and thus are dropped. This is equivalent to fixing ${\bf x}$ at ${\bf x}_1$ and ${\bf x'}$ at ${\bf x}_2$. The integral result is as easy as
\bea\nonumber
{(a^i_1)}_N  = - m_2 \frac{n^i}{r^2}\,,\\
{(a^i_2)}_N  = + m_1 \frac{n^i}{r^2}\,,
\label{NEOM}
\eea 
with the second equation obtained from the first by the interchange $1\rightleftharpoons 2$. These are the well-known Newtonian equations of motion for body-1 and body-2, re-derived via the post-Newtonian DIRE approach at its lowest order.

The 1PN terms are similarly straightforward. A term such as $v^2 U^{,i}$ is  integrated over body-1 by setting $v=v_1$ and writing $U=U_1 + U_2$.  With $v^2$ pulled outside the integral, the integration is equivalent to that of the Newtonian term in \eref{newtonterm}, with the result $v^2 U^{,i} \to -m_2 v_1^2 n^i /r^2$.  Other 1PN terms involving quadratic powers of velocity ($v^i \dot U$, $v^j V^{[i,j]}$, $\Phi_1^{,i}$ and the velocity-dependent parts of ${\dot V}^i$ and ${\ddot X}^{,i}$ ) are treated similarly.

In the non-linear term $UU^{,i}$, the term involving $U_1 U_1^{,i}$ is of order $\mathcal{S}^i/\mathcal{S}^4$, where $\mathcal{S}^i$ represents a vector, like $(x-x^\prime)^i$ that resides entirely within the body. In the two cross terms $U_1 U_2^{,i}$ and $U_2 U_1^{,i}$,  $U_1$ and $U_1^{,i}$ are of order $1/\mathcal{S}$ and $\mathcal{S}^i/\mathcal{S}^3$ respectively. It can be shown (see \cite{pat02} for details) that the only terms in the product that vary overall  as $\mathcal{S}^0$ will have odd numbers of vectors $\mathcal{S}^i$, whose integral over body-1 vanishes by spherical symmetry. Only the term from $U_2 U_2^{,i}$ contributes. The result is $UU^{,i} \to -m_2^2 n^i/r^3$. 

In the terms ${\dot V}^i$ and ${\ddot X}^{,i}$, the acceleration $dv^i/dt$ appears.  Working to 1PN order, we must insert the Newtonian equation of motion; but working to 2PN order (or higher), we must insert the 1PN (or higher) equations of motion. For ${\dot V}^i$, the result using the Newtonian equation of motion is
\begin{equation}
{\dot V}^i = - \int \int {\rho^{*\prime} \over 
	{\mid {\bf x} - {\bf x}^\prime\mid }} {{\rho^{*\prime\prime} 
	(x^\prime - x^{\prime\prime} )^i } \over 
  	{\mid {\bf x}^\prime - {\bf x}^{\prime\prime}\mid ^3}} 
	d^3x^\prime d^3x^{\prime\prime}
	+ \int {{\rho^{*\prime} v^{i \prime} {\bf v}^\prime 
	\cdot ({\bf x} - {\bf x}^{\prime} )} \over 
	{\mid {\bf x} - {\bf x}^\prime\mid ^3}} d^3x^\prime \,.
\end{equation}
The double integral is integrated over body-1 similarly to the term $UU^{,i}$, and the velocity-dependent term is integrated similarly to the term $v^2U^{,i}$.  The general result of these considerations is that, at 1PN order, only terms are kept in which, in the quantity ${\bf x}-{\bf x}^\prime$, the two vectors are evaluated at the baryonic  center of mass of the two different bodies, respectively, and never within the same body.

The resulting 1PN equation of motion is 
\begin{eqnarray}
\label{1PNEOM}
a_{1\, (1PN)}^{i} &=&  {m_2 \over r^2} {\color {red}{\epsilon}} \biggl\{ n^i \biggl [ 
	4{m_2 \over r} + 5{m_1 \over r} - v_1^2 +4\vonevtwo 
	- 2v_2^2 + {3 \over 2}(\vtwon)^2 \biggr ] 
\\
&& + (v_1 - v_2)^i (4\vonen -3\vtwon) \biggr\} \,, 
\nonumber 
\\\nonumber
a_{2\, (1PN)}^{i} &=&  a_{1\, (1PN)}^{i} \text{ with } 1 \rightleftharpoons 2 \,.
\end{eqnarray}
Note that as a natural consequence of the interchange $1 \rightleftharpoons 2$, we have to also convert $n^i \to -n^i$, because the vector ${\bf n}$ is a unit vector from body-2 toward the direction of body-1. 


\section{Relative Equations of Motion}
\label{sec:relativecoords}

In the previous section we derived the equation of motion up to 1PN order for each star of a compact binary system. In this section we convert the already obtained equations of motion in \sref{sec:Two-Body} to their equivalent equations in the center of mass frame. It is useful to note that the Newtonian equations, given in \eref{NEOM}, admit a first integral that corresponds to uniform motion of a ``center of mass'' quantity, namely
\begin{equation}
m_1 v_1^i  + m_2 v_2^i  = C^i \,,
\label{cmintegral}
\end{equation}
where $C^i$ is a constant.  Choosing the coordinates  so that $C^i =0$,  we obtain the transformation from individual to relative velocities,  to Newtonian order, 
\begin{eqnarray}
v_1^i &=& +{m_2 \over m} v^i \,,
\nonumber \\
v_2^i &=& -{m_1 \over m} v^i\,.
\label{1PNtransform}
\end{eqnarray}
These expressions can be used in 1PN terms in the equation of motion. Calculating $a_1^i - a_2^i$, using Eqs.~(\ref{NEOM}, \ref{1PNEOM}), and substituting Eqs.~(\ref{1PNtransform}), we obtain the final relative equation of motion through 1PN order as 
\be
{{d^2 {\bf X}} \over {dt^2}} = -{m \over r^2} {\bf n}
+ {m \over r^2} \bigl[ \, {\bf n} A_{PN} + {\dot r}{\bf v} B_{PN} \bigr]  {\color {red}{\epsilon} }\label{1PNeomfinal}
\ee
where ${\bf X} \equiv {\bf x}_1 - {\bf x}_2$,\footnote{Notice that ${\bf X}$ is {\em not} the coordinate of the center of mass.} ${\bf v} \equiv {\bf v}_1 - {\bf v}_2$, $r \equiv \mid {\bf X}\mid $, ${\bf n}
\equiv {\bf X}/r$, $m \equiv m_1 + m_2$, $\eta \equiv m_1m_2/m^2$, 
and $\dot
r = dr/dt$. The coefficients $A$ and $B$ are given by
\begin{subequations}\label{1PNcoeffs}
\bea
A_{PN} &=& -(1+3\eta)v^2 + {3 \over 2}\eta {\dot r}^2 +2(2+\eta)m/r \,,
 \\
B_{PN} &=&  2(2-\eta) \,. 
\eea
\end{subequations}
Equation~(\ref{1PNeomfinal}) with \eref{1PNcoeffs} describes the relative motion of the companions in a compact binary system in general relativity with the accuracy of one order of magnitude in $\epsilon$ beyond the Newtonian limit, where the components are non-spinning, spherical, very small compared to the separation distance, slowly moving compared to the speed of light, and far away enough from each other such that the tidal gravitational field of each body at the other body can be neglected. We showed, in this chapter, how the DIRE method works to order 1PN in GR. To learn how DIRE is applied at 2PN order in general relativity see~\cite{pat00, pat02} and at 2PN order in scalar-tensor theories of gravity see part~\ref{part:3} of this dissertation. 

In the following we quote the 2PN and 2.5PN coefficients in the relative equations of motion for an inspiralling compact binary system in general relativity~\cite{pat02}. The reader might compare Eqs. (\ref{1PNcoeffs}, \ref{2PNcoeffs}) in general relativity with the corresponding expressions for scalar-tensor theories given by Eqs. (\ref{EOMcoeffs}).
\begin{subequations}\label{2PNcoeffs}
\begin{eqnarray}
A_{2PN} &=& -\eta(3-4\eta)v^4 + {1 \over 2}\eta(13-4\eta)v^2m/r
	+{3 \over 2}\eta(3-4\eta)v^2{\dot r}^2 
\nonumber \\
&&
	+(2 +25\eta+2\eta^2){\dot r}^2m/r
	-{15 \over 8}\eta(1-3\eta){\dot r}^4
	-{3 \over 4}(12+29\eta)(m/r)^2 \,,
 \\
B_{2PN} &=& {1 \over 2}\eta(15+4\eta)v^2
	-{3 \over 2}\eta(3+2\eta){\dot r}^2
	-{1 \over 2}(4 +41\eta+8\eta^2)m/r \,,
\label{eomfinalcoeffs2PN}
\\
A_{2.5PN} &=& 3v^2+{17 \over 3}m/r\,,
 \\
B_{2.5PN} &=& v^2 + 3m/r\,,
\label{eomfinalcoeffs2.5PN}
\end{eqnarray}
\end{subequations}

\begin{savequote}[0.55\linewidth]
{\scriptsize ``The scientist is not a person who gives the right answers, he's one who asks the right questions.''}
\qauthor{\scriptsize---Claude L\'{e}vi-Strauss}
\end{savequote}
\chapter{Parameter Estimation} 
\label{chapter7} 
We begin this chapter with a general discussion of data analysis methods in gravitational-wave astronomy. We then focus on the matched filtering technique and introduce the basics of this method. We end this chapter with an example to show how matched filtering method can be applied to do parameter estimation for a compact binary source of gravitational-waves. We will use these same methods in Part~\ref{part:4} where we apply Fisher matrix analyses to bound the graviton mass and to constrain the deviation from Lorentz symmetry in quantum-mechanical inspired, Lorentz-violating theories of gravity.
\thispagestyle{myplain}
\lhead[\thepage]{Chapter 6. \emph{Parameter Estimation}}      
\rhead[Chapter 6. \emph{Parameter Estimation}]{\thepage}

\section{Gravitational-Wave Data Analysis}
As we discussed earlier in \cref{chapter4}, the observation of gravitational waves requires a very precise data analysis strategy, which is different from conventional astronomical data analysis in many ways. There are several reasons why this is so. Sathyaprakash and Schutz~\cite{sat09} have listed some of them as:
\begin{itemize}
\item Data analysis systems have to carry out {\em all-sky} searches, because gravitational wave detectors are essentially omni-directional, with their response better than $50\%$ of the root-mean-square over $75\%$ of the sky.
\item Interferometer detectors are typically broadband, covering three to four orders of magnitude in frequency. This allows searches to be carried out over a wide range of frequencies, and helps to track sources whose frequency changes rapidly.
\item Measuring the polarization of gravitational waves is possible only via data analysis of multiple detectors. Using multiple detectors also helps coincidence analysis and the efficiency of event recognition. Polarization measurement is of fundamental importance and has astrophysical implications too. 
\item Unlike typical detection techniques for electromagnetic radiation from astronomical sources, most astrophysical gravitational waves are detected coherently, by following the phase of the radiation, rather than just the energy. The phase evolution contains more information than the amplitude does and the signal structure is a rich diagnostic of the underlying physics.
\item Detection of gravitational wave is computationally very expensive. Gravitational wave detectors acquire data continuously for many years at the rate of several megabytes per second.  
\end{itemize}

In this chapter we consider the problem of detection of gravitational-wave signals embedded in a background of noise of a detector, and the question of estimation of their parameters. This led data analysts to develop a useful set of tools to search for gravitational-wave signals. A very powerful method to detect a signal in noise that is optimal by several criteria consists of correlating the data with a template that is matched to the expected signal. This {\em matched-filtering} technique is a special case of the {\em maximum likelihood} detection method. In this chapter we review the theoretical foundation of the method and we show how it can be applied to the case of a very general deterministic gravitational-wave signal buried in a {\em stationary} and {\em Gaussian} noise. Among all the potential candidates of gravitational-wave sources, inspiralling compact binaries are amongst the most promising. This is a result of the ability to model the phase and amplitude of the signals quite accurately and consequently to achieve maximum signal-to-noise ratio (SNR) by using matched filtering techniques.

Even though gravitational-wave signals have not been detected yet, we can already investigate the performance of the detectors from a parameter estimation point of view. The relevant information is the distribution of the measured values (e.g., component masses, time of coalescence) and the error bounds on their variances. The Fisher information matrix is a convenient tool to obtain these error bounds. More details on the Fisher matrix analysis and the matched filtering technique will be given in \sref{sec:matched}.  Indeed, in the cases that will interest us, the Fisher information matrix can easily be computed because inspiralling compact binaries can be modeled analytically. The covariance matrix was derived in~\cite{cut94, fin93, jar94} using Newtonian waveforms, extended to second post-Newtonian order (2PN) ~\cite{poi95, kro95}, and revisited up to 3.5 PN order~\cite{aru05a, aru05b}. The main advantage of the covariance matrix is that once analytical expressions are available, expected error bounds can be calculated quickly for any type of component masses. Moreover, the errors are expected to fall off as the inverse of the SNR. However, the analytical expressions are valid in the strong-signal approximation case only. Since the first detection of gravitational-wave signals is expected to be in a low-SNR regime (below 20), the Fisher information matrix may not be the best tool to estimate error bounds in practice.

There are other methods for estimating errors bounds that are based on simulations, and they should be able to correctly estimate error bounds even at low SNRs. However, these methods are computationally much more intensive compared to the Fisher information matrix formalism. For instance, in~\cite{rov06}, the authors use a Bayesian analysis framework (for binary neutron star signals) so as to estimate the signal's parameters and their errors. The posterior integration is carried out using Markov Chain Monte Carlo (MCMC) methods. In~\cite{bal96a, bal96b}, the authors compared the error bounds given by the Fisher information matrix with those from Monte Carlo simulations. They found that in the case of black-hole neutron-star binaries ($(1.4, 10)\;M_\odot$), the covariance matrix underestimates the error bounds by a factor of 2 at a SNR of 10 (chirp mass errors). This discrepancy vanishes when the SNR is approximately 15 for a Newtonian waveform and 25 for a 1PN waveform. It was also stated that the inclusion of higher order terms would be computationally quite intensive~\cite{cok08}.

A very important development was the work by Cutler et al.~\cite{cut92} where it was realized that for the case of coalescing binaries matched filtering was sensitive to very small post-Newtonian effects of the waveform. Thus these effects can be detected. This leads to a much better verification of Einstein's theory of relativity and provides a wealth of astrophysical information that would make a laser interferometric gravitational-wave detector a true astronomical observatory complementary to those utilizing the electromagnetic spectrum.

Figure~\ref{fig:known} shows a schematic outline of the way in which LIGO and Virgo searches can be broken down. As one moves from left to right on the diagram, waveforms increase in duration, while as one moves from top to bottom, {\em a priori} waveform definition decreases. Populating the upper left corner is the extreme of an inspiraling compact binary system of two neutron stars in the regime where corrections to Newtonian orbits can be calculated with great confidence. Populating the upper right corner are isolated, known, non-glitching spinning neutron stars with smooth rotational spindown and measured orientation parameters. Populating the lower left corner of the diagram are supernovae, rapid bursts of gravitational radiation for which phase evolution cannot be confidently predicted, and for which it is challenging to make even coarse spectral predictions. At the bottom right one finds a stochastic, cosmological background of radiation for which phase evolution is random, but with a spectrum stationary in time. Between these extremes can live sources on the left such as the merger phases of a BH-BH coalescence. On the right one  finds, for example, an accreting neutron star in a low-mass X-ray binary system where  fuctuations in the accretion process lead to unpredictable wandering phase. 

\begin{figure}[h!]
\centering
\includegraphics[width=0.6\textwidth]{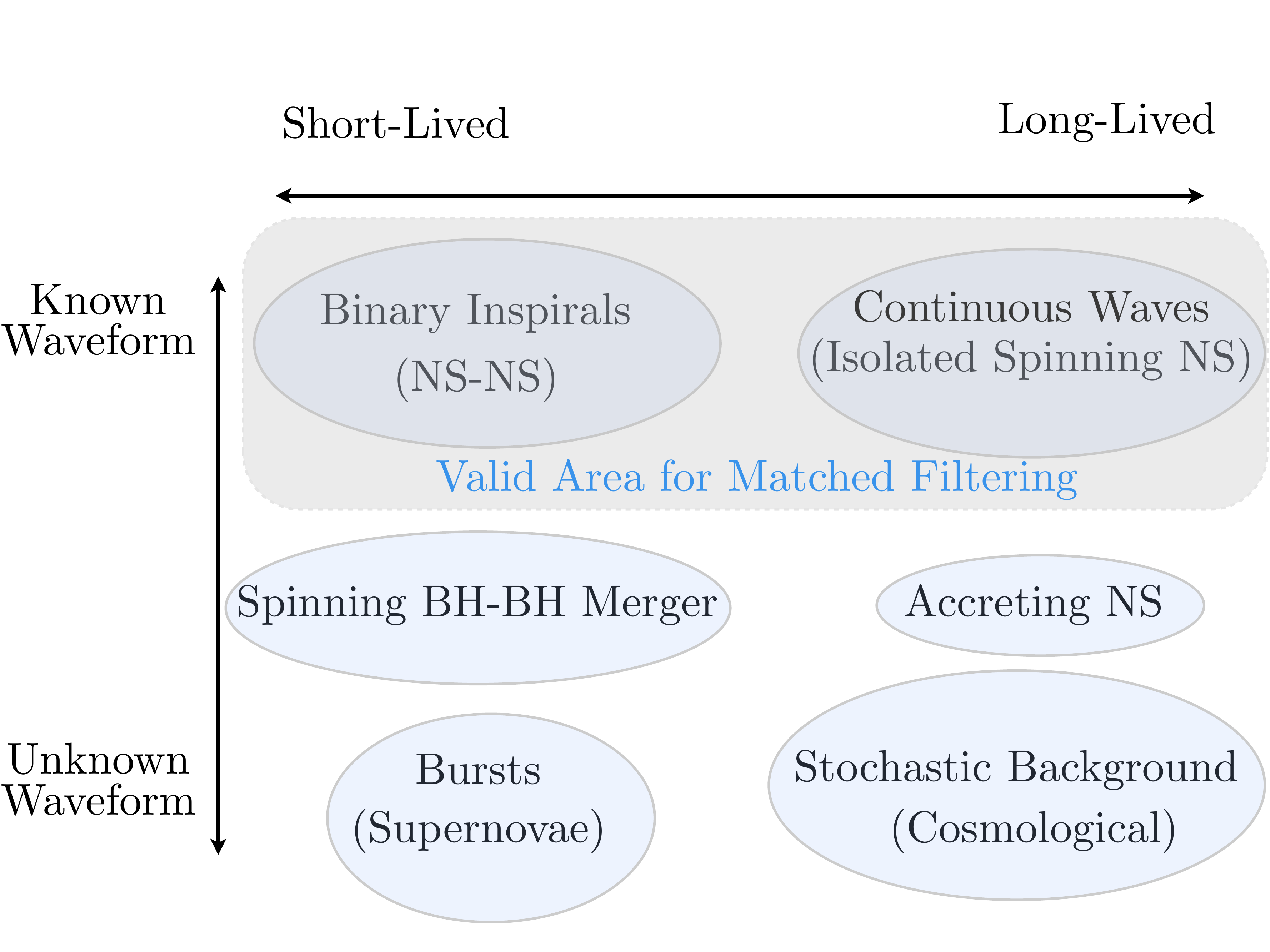}
\caption{A schematic illustration of different gravitational-wave sources in terms of duration and our knowledge about the waveform~\cite{Riles:2012fk}. Gray area shows where we can use matched filtering technique with no problem.}
\label{fig:known}
\end{figure}

The matched filtering technique can be applied as long as the waveform is known (gray area in \fref{fig:known}). Solving the field equations and obtaining the gravitational waveform as a known expression, one can use it as a template to do matched filtering and hence measure the properties of gravitational source. For instance, in \sref{sec:matched-example} we show how we can use the matched filtering method to estimate the parameters of a compact binary system, such as the masses and spins of the companions. In Part~\ref{part:4} we use the same method to bound the graviton's mass~\cite{poi95, mir12} and to constrain the deviation from Lorentz symmetry~\cite{mir12} in alternative theories of gravity.

\section{Matched Filtering: Theory}
\label{sec:matched}

Various work by various authors, including Finn \cite{fin92} and Cutler and Flanagan \cite{cut94} have put the theory and measurement of gravitational-wave signals on a firm statistical foundation, rather similar to that underlying the theory of radar detection \cite{wai62, hel68}. Here in this section we introduce the theory of {\em matched filtering}  and parameter estimation for our future purposes in Part~\ref{part:4}. 

To extract the gravitational-wave signal $h(t;\boldsymbol\theta)$ from noisy detector data, we need to be armed with some standard mechanism. When a signal of the form $h(t;\boldsymbol\theta)$ has passed through the detectors (a network of detectors), this data analysis mechanism should allow us to determine the value of the source parameters $\boldsymbol\theta$ and the measurement error $\Delta \boldsymbol\theta = \boldsymbol\theta - \vec{\tilde{\theta}}$, where $\vec{\tilde{\theta}}$ denotes the true value.

It is useful to define $p(\boldsymbol\theta\mid s)$ as the probability that the gravitational-wave signal is characterized by the parameters $\boldsymbol\theta$, where the detector output is $s(t)$ and a signal $h(t;\boldsymbol\theta)$ ---for {\it any} value of the parameters $\boldsymbol\theta$--- is present. Finn in \cite{fin92} has derived an expression for $p(\boldsymbol\theta\mid s)$. The detector output signal is composed of gravitational-wave signal $h(t;\boldsymbol\theta)$ and the {\em stationary random (Gaussian)} function of detector noise $n(t)$ such that
\begin{equation} 
s(t) = h(t;\boldsymbol\theta) + n(t).
\label{2.1}
\end{equation}
Note that being a stationary and Gaussian random process for the detector noise, $n(t)$, is a crucial assumption. Finn shows that
\begin{equation}
p(\boldsymbol\theta\mid s) \propto p^{(0)}(\boldsymbol\theta)
\exp \Bigl[ -{\textstyle \frac{1}{2}} \Bigl( h(\boldsymbol\theta) - s\mid  h(\boldsymbol\theta) - s \Bigr) \Bigr],
\label{2.2}
\end{equation}
where $p^{(0)}(\boldsymbol\theta)$ is the a priori probability that the signal is characterized by $\boldsymbol\theta$ (this represents our prior information regarding the possible value of the parameters) and where the constant of proportionality is independent of $\boldsymbol\theta$. In a given measurement, characterized by the particular detector output $s(t)$, the true values of the source parameters can be estimated by maximizing the value of probability distribution function and locating the parameter $\boldsymbol\theta$ at this maximized $p(\boldsymbol\theta\mid s)$ which in this case $\boldsymbol\theta=\vec{\hat\theta}$. This is the so-called {\em maximum-likelihood} estimator~\cite{wai62}.

The inner product operator $(\cdot \mid  \cdot)$ is defined such that  \cite{cut94} 
\begin{equation}
(g \mid  h) = 2 \int_0^\infty
\frac{\tilde{g}^*(f) \tilde{h}(f) +
\tilde{g}(f) \tilde{h}^*(f)}{S_n(f)} df.
\label{2.4}
\end{equation}
The inner product in \eref{2.4} is defined so that the probability for the noise $n(t)$ to have a particular realization $n_0(t)$ is given by $p(n=n_0) \propto \exp[-(n_0\mid n_0)/2]$. The noise spectral density $S_n(f)$  in \eref{2.4} is twice the Fourier transform of the autocorrelation function of the noise detector
\begin{equation}
S_n(f) = 2 \int_{-\infty}^\infty
C_n(\tau) e^{2\pi i f \tau} d\tau,
\label{2.3}
\end{equation}
which is defined for $f>0$ only, and $C_n(\tau)$ is the {\em autocorrelation function} of the noise detector 
\be
C_n(\tau) = \langle n(t) n(t+\tau) \rangle, 
\ee
where $\langle \cdot \rangle$ denotes a time average (It is assumed that the noise has zero mean). All of the statistical properties of the detector noise can be summarized by its autocorrelation function. Notice that in \eref{2.4} ``$\;{}^*\;$'' denotes complex conjugation and ``$\; \tilde{} \;$'' shows the Fourier transformation e.g.
\begin{equation}
\tilde{g}(f) = \int_{-\infty}^\infty
g(t) e^{2\pi i f t} dt.
\label{2.5}
\end{equation}

We define $\rho$, the signal-to-noise ratio (SNR) associated with the measurement, to be the norm of the signal $h(t;\boldsymbol\theta)$,
\begin{equation}
\rho^2 = (h\mid h) = 4 \int_0^\infty
\frac{ \mid \tilde{h}(f)\mid ^2 }{S_n(f)} df,
\label{2.6}
\end{equation}
evaluated at $\boldsymbol\theta = \vec{\hat{\theta}}$, where $p(\boldsymbol\theta\mid s)$ is maximum and therefore $\boldsymbol\theta=\vec{\hat\theta}$ is the estimated value of the source parameters. In the limit of large values of SNR, to which we henceforth specialize,  $p(\boldsymbol\theta\mid s)$ will be strongly peaked about this value. We now derive a simplified expression for $p(\boldsymbol\theta\mid s)$ appropriate for this limiting case of high SNR values.

First of all, we assume that $p^{(0)}(\boldsymbol\theta)$ is nearly uniform near $\boldsymbol\theta = \vec{\hat{\theta}}$. This indicates that the prior information is practically irrelevant to the determination of the source parameters; we shall relax this assumption below. Then, denoting 
\be
\xi(\boldsymbol\theta) \define (h(\boldsymbol\theta) -s\mid h(\boldsymbol\theta)-s), 
\ee
we have that $\xi$ is minimum at $\boldsymbol\theta = \vec{\hat{\theta}}$. It follows that this can be expanded as
\begin{equation}
\xi(\boldsymbol\theta) = \xi(\vec{\hat{\theta}}) +
{\displaystyle \frac{1}{2}}\;
\xi_{,a b} (\vec{\hat{\theta}})
\Delta \theta^a \Delta \theta^b + \cdots,
\label{2.7}
\end{equation}
where $\Delta \theta^a = \theta^a - \hat{\theta}^a$, comma represents partial derivative with respect to the parameter $\theta$ (for example $\xi_{,a}={\partial\xi}/{\partial\theta^a}$), and summation over repeated indices is understood. We assume that $\rho$ is sufficiently large that the higher-order terms can be neglected. Calculation yields 
\be
\xi_{,ab} = (h_{,ab}\mid h-s) + (h_{,a}\mid h_{,b}), 
\ee
and we again assume that $\rho$ is large enough that the first term can be neglected (see Cutler and Flanagan \cite{cut94} for details). Therefore, in the limit of high SNR values, \eref{2.2} can be well approximated by a Gaussian form distribution as
\begin{equation}
p(\boldsymbol\theta\mid s) \propto p^{(0)}(\boldsymbol\theta)
\exp \Bigl[ - {\textstyle \frac{1}{2}}
\Gamma_{ab} \Delta \theta^a \Delta \theta^b \Bigr],
\label{2.8}
\end{equation}
where 
\begin{equation}
\Gamma_{ab} = \bigl( h_{,a} \mid  h_{,b} \bigr),
\label{2.9}
\end{equation}
evaluated at $\boldsymbol\theta = \vec{\hat{\theta}}$, is the {\em Fisher information matrix} \cite{hel68} that is the most crucial quantity that has to be evaluated in the matched filtering technique. From Eq.~(\ref{2.8}) it can be established that the variance-covariance matrix $\Sigma^{ab}$ is given by
\begin{equation}
\Sigma^{ab} \equiv \langle \Delta \theta^a
\Delta \theta^b \rangle = (\vec{\Gamma}^{-1})^{ab}.
\label{2.10}
\end{equation}
Here, $\langle \cdot \rangle$ denotes an average over the probability distribution function Eq.~(\ref{2.8}), and $\vec{\Gamma}^{-1}$ represents the inverse of the Fisher matrix. We define the {\em measurement error} in the parameter $\theta^a$ to be
\begin{equation}
\sigma_a =
\bigl\langle (\Delta \theta^a)^2 \bigr\rangle^{1/2}
= \sqrt{\Sigma^{aa}}
\label{2.11}
\end{equation}
(no summation over repeated indices), and ---based on the above defined $\sigma_a$ and $\sigma_b$--- the {\em correlation coefficient} between parameters $\theta^a$ and $\theta^b$ as
\begin{equation}
c^{ab} = \frac{\langle \Delta \theta^a
\Delta \theta^b \rangle}{\sigma_a \sigma_b} =
\frac{\Sigma^{ab}}{
\sqrt{\Sigma^{aa} \Sigma^{bb}}};
\label{2.12}
\end{equation}
by definition each $c^{ab}$ must lie in the range $(-1,1)$. In the next section, we clarify how to use the method of matched filtering by giving an example. For a specific SNR value, with knowing $(a)$ the anticipated noise spectral density of a gravitational-wave detector and $(b)$ the waveform template accurate to the appropriate post-Newtonian order, in \sref{sec:matched-example} we describe how one can use the Fisher matrix approach to calculate $\sigma_a$ and $c^{ab}$.



\section{Matched Filtering to Parameter Estimation: An Example}
\label{sec:matched-example}

In this section we apply the matched filtering Fisher matrix analysis to a specific example. This example has been studied by Poisson and Will~\cite{poi95} and we review it here. The techniques and methods that we show in this example are same as those that we will apply in Part~\ref{part:4} of this dissertation. 

The detailed expression for the post-Newtonian waveform is complicated: the dependence on the various angles (position of the source in the sky, orientation of the detector, orientation of the polarization axes) is not simple, and the waves have several frequency components given by the harmonics of the orbital frequency (assuming that the orbit is circular~\cite{pet64, lin90}). A Fourier domain waveform (the so-called as TaylorF2 template), which is the most often employed PN approximant, is given by
\begin{equation}
\tilde{h}(f) = {\cal A} f^{-7/6} e^{i \psi(f)},
\label{3.5}
\end{equation}
where the amplitude ${\cal A} \propto {\cal M}^{5/6} Q(\mbox{angles})/r$, ($r$ is the distance to the source, $Q$ is a function of the various angles mentioned above) and the phase is
\begin{eqnarray}\label{7.17}
\psi(f) &=& 2\pi f t_c - \phi_c - \frac{\pi}{4} +
\frac{3}{128} (\pi {\cal M} f)^{-5/3}
\biggl[1 + \frac{20}{9} \biggl( \frac{743}{336} +
\frac{11}{4} \eta \biggr) (\pi M f)^{2/3}
\nonumber \\ & & \mbox{}
- 4(4\pi - \beta) (\pi M f) +
10 \biggl( \frac{3058673}{1016064} + \frac{5429}{1008}\eta
+ \frac{617}{144} \eta^2 - \sigma \biggr) (\pi M f)^{4/3}
\biggr].
\label{3.6}
\end{eqnarray}
Here we introduce all the variables in \eref{3.6}: 
\begin{itemize}
\item $f$ is the Fourier transform variable. Notice the difference between this variable, $f$, and the gravitational-wave frequency $F$ in this section. 
\item $t_c$ and $\phi_c$ are constants of the problem and represent the time and phase at the time of coalescence, respectively. The explicit functionality of $t$ and $\Phi$ (the phase $\Phi(t)=\int2\pi F(t) dt$) in terms of wave frequency $F$ is given by
\begin{eqnarray}
t(F) &=& t_c - \frac{5}{256} {\cal M} (\pi {\cal M} F)^{-8/3}
\biggl[1 + \frac{4}{3} \biggl( \frac{743}{336} +
\frac{11}{4} \eta \biggr) (\pi M F)^{2/3} -
\frac{8}{5} (4\pi - \beta) (\pi M F)
\nonumber \\ & & \mbox{} +
2 \biggl( \frac{3058673}{1016064} + \frac{5429}{1008}\eta
+ \frac{617}{144} \eta^2 - \sigma \biggr) (\pi M F)^{4/3}
\biggr],
\label{3.3}\\
\Phi(F) &=& \phi_c - \frac{1}{16} (\pi {\cal M} F)^{-5/3}
\biggl[1 + \frac{5}{3} \biggl( \frac{743}{336} +
\frac{11}{4} \eta \biggr) (\pi M F)^{2/3}
- \frac{5}{2} (4\pi - \beta) (\pi M F)
\nonumber \\ & & \mbox{} +
5 \biggl( \frac{3058673}{1016064} + \frac{5429}{1008}\eta
+ \frac{617}{144} \eta^2 - \sigma \biggr) (\pi M F)^{4/3}
\biggr],
\label{3.2}
\end{eqnarray}
where $\phi_c$ and $t_c$ are (formally) the values of $\Phi$ and $t$ at $F=\infty$. Of course, the signal can not be allowed to reach arbitrarily high frequencies; it must be cut off at a frequency $F=F_i$ corresponding to the end of the inspiral. We put $\pi M F_i = (M/r_i)^{3/2} = 6^{-3/2}$; $r_i = 6M$ is the Schwarzschild radius of the innermost circular orbit for a test mass moving in the gravitational field of a mass $M$ \footnote{Strictly speaking, our expression for $F_i$  is only valid in the limit $\eta \to 0$. For simplicity, and because this will not affect our results significantly, we shall ignore the corrections to the innermost circular orbit which are due to the finite value of the mass ratio. These are computed by L.E. Kidder, C.M. Will, and A.G. Wiseman, Phys. Rev. D {\bf 47}, 3281 (1993) using post-Newtonian theory, and by S. Detweiler and J.K. Blackburn, Phys. Rev. D {\bf 46}, 2318 (1992) and G.B. Cook, Phys. Rev. D {\bf 50}, 5025 (1994) using numerical relativity.}.
\item $\beta$ and $\sigma$ represent respectively {\em spin-orbit} and {\em spin-spin} effects such that
\bea
\beta = \frac{1}{12} \sum_{i=1}^2
\bigl[ 113 (m_i/M)^2 + 75\eta \bigr]
\vec{\hat{L}} \cdot \vec{\chi}_i,
\label{1.4}\\
\sigma = \frac{\eta}{48} \bigl(
-247\; \boldsymbol{\chi}_1 \cdot \boldsymbol{ \chi}_2 +
721 \;\vec{\hat{L}} \cdot \boldsymbol{\chi}_1
\vec{\hat{L}} \cdot \boldsymbol{\chi}_2 \bigr).
\label{1.5}
\eea
where $\boldsymbol{\chi}_i = \vec{S}_i/{m_i}^2$; $\vec{S}_1$, $\vec{S}_2$ are the spin angular momentum of each companion, and $\hat{\vec{L}}$ is the unit vector in the direction of total orbital angular momentum.
\item $M$ (total mass), $\mu$ (reduced mass), and $\mathcal{M}$ ({\em chirp} mass) are three different characteristic masses of the system (all have dimension of mass) in terms of each companion mass i.e. $m_1$ and $m_2$ as
\be\label{massdefinitions}
M\define m_1+m_2, \qquad \mu \define \frac{m_1 m_2}{m_1+m_2}, \qquad \mathcal{M}\define \frac{(m_1 m_2)^{3/5}} {(m_1+m_2)^{1/5}}.
\ee
Defining dimensionless, symmetric, mass-ratio parameter $\eta$ as 
\be\label{massdefinitions2}
\eta\define\frac{\mu}{M},
\ee
we can rewrite the last equation in \eref{massdefinitions} as $\mathcal{M}=\eta^{3/5} M$.
\end{itemize}
The main purpose of this section is to estimate the {\em anticipated} accuracy with which the various parameters such as $\mathcal{M}, \eta, \beta$, and $\sigma$ can be determined during a gravitational-wave measurement.

At this point we have to specify the anticipated noise spectral density of the detector. In this example we just follow Poisson and Will~\cite{poi95} and use the following analytic expression for LIGO-VIRO-type detectors. 
\begin{equation}
S_n(f) = {\textstyle \frac{1}{5}} S_0
\bigl[ (f_0/f)^4 + 2 + 2(f/f_0)^2 \bigr],
\label{3.1}
\end{equation}
where $S_0$ is a normalization constant irrelevant for our purposes, and $f_0$ the frequency at which $S_n(f)$ is minimum; we set $f_0 = 70$Hz, which is appropriate for advanced LIGO sensitivity \cite{ligo}. To mimic seismic noise we assume that Eq.~(\ref{3.1}) is valid for $f > 10 {\rm Hz}$ only, and that $S_n(f) = \infty$ for $f < 10 {\rm Hz}$. Although Eq.~(\ref{3.1}) is not the most updated analytic expression, it is ideal for our purposes to show the application of the Fisher matrix analysis method. We will use the most updated version of the noise spectral density for different detectors in Part~\ref{part:4}.

We now substitute Eq.~(\ref{3.5}) into Eq.~(\ref{2.6}) and calculate the signal-to-noise ratio. We readily obtain 
\begin{equation}
\rho^2 = 20 {\cal A}^2 {S_0}^{-1} {f_0}^{-4/3} I(7),
\label{3.7}
\end{equation}
where the integrals $I(q)$ represent various moments of the noise spectral density:
\begin{equation}
I(q) \equiv \int_{1/7}^{(6^{3/2} \pi M f_0)^{-1}}
\frac{x^{-q/3}}{x^{-4} + 2 + 2x^2} dx,
\label{3.8}
\end{equation}
where $x =f/f_0$ and the minimum and maximum of $x$ in this case is put equal to $1/7$ (corresponding to $f_{\text min}=10$ and $f_0=70$ for LIGO) and $(6^{3/2} \pi M f_0)^{-1}$ (corresponding to $f_{max}=f_{\text ISCO}$), respectively.

As the next step toward the computation of the Fisher matrix, we calculate the derivatives of $\tilde{h}(f)$ with respect to the following seven parameters
\begin{equation}
\boldsymbol{\theta} = (\ln{\cal A},\; f_0 t_c,\; \phi_c,\; \ln {\cal M},\;
\ln \eta,\; \beta,\; \sigma).
\label{3.9}
\end{equation}
By taking derivatives of the Fourier domain waveform in Eqs.~(\ref{7.17}, \ref{3.5}) with respect to all parameters in Eq.~(\ref{3.9}) we obtain
\begin{subequations}
\begin{eqnarray}
\tilde{h}_{,1} &=& \tilde{h},  \\
\tilde{h}_{,2} &=& 2\pi i (f/f_0) \tilde{h},  \\
\tilde{h}_{,3} &=& -i \tilde{h},  \\
\tilde{h}_{,4} &=& - \frac{5 i}{128} (\pi {\cal M} f)^{-5/3}
(1 + A_4 v^2 - B_4 v^3 + C_4 v^4) \tilde{h}, \label{3.10} \\
\tilde{h}_{,5} &=& -\frac{i}{96} (\pi {\cal M} f)^{-5/3}
(A_5 v^2 - B_5 v^3 + C_5 v^4) \tilde{h},  \\
\tilde{h}_{,6} &=& \frac{3i}{32} \eta^{-3/5}
(\pi {\cal M} f)^{-2/3} \tilde{h},  \\
\tilde{h}_{,7} &=& -\frac{15 i}{64} \eta^{-4/5}
(\pi {\cal M} f)^{-1/3} \tilde{h}, 
\end{eqnarray}\label{3.10}
\end{subequations}
where $v\equiv(\pi M f)^{1/3}$ and the index numbers in the left hand sides of the above equations correspond to different components of $\boldsymbol\theta$ in Eq.~(\ref{3.9}), respectively. Notice that $\tilde{h}_{,1}$ which corresponds to $\ln \mathcal{A}$ is the only one among the above expressions which does not have an imaginary part \footnote{The parameter $\theta^1 = \ln \cal A$ is therefore entirely uncorrelated with the other parameters, and we find $\sigma_1 = \Delta {\cal A} / {\cal A} = 1/\rho$, $c^{1a}=0$, in all cases. We shall no longer be concerned with this parameter.}. In Eq.~(\ref{3.10}), we also have defined
\begin{subequations}
\begin{eqnarray}
A_4 &=& \frac{4}{3} \biggl( \frac{743}{336} +
\frac{11}{4} \eta \biggr),  \\
B_4 &=& \frac{8}{5} (4\pi - \beta),\\
C_4 &=& 2 \epsilon \biggl(
\frac{3058673}{1016064} + \frac{5429}{1008} \eta
+ \frac{617}{144} \eta^2 - \sigma \biggr), 
\end{eqnarray}\label{3.11}
\end{subequations}
and
\begin{subequations}
\begin{eqnarray}
A_5 &=& \frac{743}{168} - \frac{33}{4} \eta,  \\
B_5 &=& \frac{27}{5} (4\pi - \beta), \\
C_5 &=& 18 \epsilon \biggl(
\frac{3058673}{1016064} - \frac{5429}{4032} \eta
- \frac{617}{96} \eta^2 - \sigma \biggr). 
\end{eqnarray}\label{3.12}
\end{subequations}

Finally, the components of $\vec{\Gamma}$ can be obtained by evaluating the inner products $(h_{,a}\mid h_{,b})$ using Eq.~(\ref{2.4}) as 
\be
\Gamma_{ab}=(h_{,a}\mid h_{,b})=2\int_{f_{min}}^{f_{max}} \displaystyle\frac{\tilde h^*_{,a}(f) \tilde h_{,b}(f) + \tilde h_{,a}(f) \tilde h^*_{,b}(f)} {S_n(f)}
\ee
where different components of $\tilde h_{,a}$ are given by Eqs.~(\ref{3.10}, \ref{3.11}, \ref{3.12}) and $S_n(f)$ in this specific example is given by Eq.~(\ref{3.1}). The $\Gamma_{ab}$'s can all be expressed in terms of the parameters $\boldsymbol\theta$, the signal-to-noise ratio $\rho$, and the integrals $I(q)$. The resulting expressions are too numerous and lengthy to be displayed here. To double check this example and for our future use, we developed a computer code \footnote{Written in Mathematica$\copyright$}. Starting with the same initial conditions, we re-produced exactly the results shown in tables II\&III of \cite{poi95}.

The variance-covariance matrix $\Sigma^{ab}$ can now be obtained from Eqs.~(\ref{2.10}), and the measurement errors and correlation coefficients computed from Eqs.~(\ref{2.11}, \ref{2.12}). Before doing so, however, we must first state our assumptions regarding the prior information available on the source parameters. We assume the SNR value to be $\rho=10$ everywhere and that the companions are spin-less so that $\beta=\sigma=0$.

\mypart{Motion and Gravitational Radiation in Scalar-Tensor Gravity}{{\vfill \small{\em This part is based on a published paper in Physical Review D.~\cite{mir13} in which we adapt the Newtonian method of DIRE to scalar-tensor theory, coupled with an approach pioneered by Eardley for incorporating the internal gravity of compact, self-gravitating bodies. Explicit equations of motion for non-spinning binary systems (including neutron stars and black holes) are derived to $2.5$ post-Newtonian order or ${O} (v/c)^5$ beyond Newtonian gravity.}}}
{\begin{framed}
\begin{itemize}
\item \cref{chapter8}--- \nameref{chapter8}
\item \cref{chapter9}--- \nameref{chapter9}
\item \cref{chapter10}--- \nameref{chapter10}
\item \cref{chapter11}--- \nameref{chapter11}
\item \cref{chapter12}--- \nameref{chapter12}
\end{itemize}
\end{framed}
} {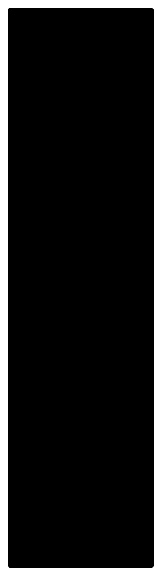}\label{part:3}
\begin{savequote}[0.65\linewidth]
{\scriptsize ``A scientific truth does not triumph by convincing its opponents and making them see the light, but rather because its opponents eventually die and a new generation grows up that is familiar with it.''}
\qauthor{\scriptsize---Max Planck}
\end{savequote}
\chapter{Introduction and Basics} 

\label{chapter8} 
\thispagestyle{myplain}
\lhead[\thepage]{Chapter 7. \emph{Introduction and Basics}}      
\rhead[Chapter 7. \emph{Introduction and Basics}]{\thepage}
\section{Introduction} 
\label{sec:intro}\ClearWallPaper

The anticipated detection of gravitational waves by a network of ground-based laser-interferometric observatories promises a new way of ``listening'' to the universe in the high-frequency band.  A future space-borne interferometer would open the low-frequency band and pulsar timing arrays may soon begin exploring the nano-Hertz region of the gravitational-wave spectrum.  In addition to providing a wealth of astrophysical information, these observations also hold the promise of providing tests of Einstein's theory of general relativity in the strong-field, dynamical regime.

The ``inspiralling compact binary''-- a binary system of neutron stars or black holes (or one of each) in
the late stages of inspiral and coalescence -- is a leading potential source for detection. Given the expected sensitivity of the ground-based interferometers, stellar-mass compact binaries could be detected out to hundreds of megaparsecs, while for a space interferometer, inspirals involving supermassive black holes could be heard to cosmological distances.

In order to maximize the detection capability and the science return of these observatories, extremely accurate, theoretically generated ``templates'' for the gravitational waveform emitted during the inspiral phase must be available.   This means that correction terms in the equations of motion and gravitational-wave signal must be calculated to high orders in the post-Newtonian (PN) approximation to general relativity, which, roughly speaking, is an expansion in powers of $v/c \sim (Gm/rc^2)^{1/2} $ (for a review and references see~\cite{sat09}).   Contributions to the waveform from the merger phase of the two objects and from the ``ringdown'' phase of the final vibrating black hole also play an important role.

The detected gravitational-wave signals can also be used to test Einstein's theory in the radiative regime, particularly for waves emitted by sources characterized by strong-field gravity, such as inspiraling compact binaries. One way to study the potential for this is to check the consistency of a hypothetical observed waveform with the predicted higher-order terms in the general relativistic PN sequence, which depend on very few parameters (only the two masses, for non-spinning, quasi-circular inspirals).  Another is to examine the constraints that could be placed on specific alternative theories using gravitational-wave observations~\cite{willST,Will:1997bb,scharrewill,Will:2004xi,BBW,BBW2,stavridis,arunwill09,sopuertayunes}.  Most of these analyses have incorporated only the dominant effect that distinguishes the chosen theory from general relativity, such as dipole radiation or the wavelength-dependent propagation of a massive graviton (see, however~\cite{yunespanicardoso}).   

Some authors have taken a different approach by proposing parametrized versions of the gravitational waveform model~\cite{Yunes:2009ke,mishra:2010tp,mir12}, inspired by the parametrized post-Newtonian (PPN) formalism used for solar-system experiments, and analysing the bounds that could be placed on those theory-dependent parameters by various gravitational-wave observations.   Yet the authors of these frameworks were limited by the fact that for many alternative theories of gravity, only the leading terms in the waveform model have been derived.  

In addition, the existing parametrizations of the gravitational waveform make the implicit assumption that the gravitational wave signal during the inspiral depends only on the masses of the orbiting compact bodies (in the spinless case), and not on their internal structure.  This is true in general relativity, which satisfies the Strong Equivalence Principle, but is known to be violated by almost every alternative theory that has ever been studied.   In scalar-tensor theory, for example, the internal gravitational binding energy of neutron stars has a definite effect on the motion and gravitational-wave emission, and since the binding energy can amount to as much as 20 percent of the total mass-energy of the body, the effects can be significant.  In order to determine the full nature of the gravitational-wave signal in an alternative theory of gravity, the strong internal gravity of each body must be accounted for somehow, even in a PN expansion.

To make the situation even more interesting, binary black holes play a special role within the scalar-tensor class of alternative theories.   Based on evidence from a 1972 theorem by Hawking~\cite{hawking}, together with known results from first-post-Newtonian theory, it is likely that in a broad class of scalar-tensor theories,  {\em binary black hole motion and gravitational radiation emission are observationally indistinguishable from their GR counterparts}. This conjecture will be discussed in more detail later in \cref{chapter12}.   

Scalar-tensor gravity is the most popular and well-motivated class of alternative theories to general relativity.   Apart from the long history of such theories, dating back more than 50 years to Jordan, Fierz, Brans and Dicke~\cite{brans}, scalar-tensor gravity has been postulated as a possible low-energy limit of string theory.  In addition, a wide class of so-called $f(R)$ theories, designed to provide an alternative explanation for the acceleration of the universe to the conventional dark-energy model, can be recast into the form of a scalar-tensor theory (for reviews, see~\cite{fujiimaeda,tsujikawa}).  

Measurements in the solar system and in binary pulsar systems already place strong constraints on key parameters of such theories, notably the coupling parameter $\omega_0$.   Yet these tests probe only the lowest-order, first post-Newtonian limit of these theories, some aspects of their strong-field regimes (related to the strong internal gravity of the neutron stars in binary pulsars) and the lowest-order, dipolar aspects of gravitational radiation damping.    

These considerations have motivated us to develop the full equations of motion and gravitational waveform for compact bodies in a class of scalar-tensor theories to a high order in the PN sequence. 

It should be acknowledged that we do not expect any big surprises.   Damour and Esposito-Far\`ese~\cite{DamourEsposito96}  have shown on general grounds that the available constraints on the scalar-tensor coupling constant $\omega_0$ derived from solar-system experiments  imply that scalar-tensor differences from GR will be small to essentially all PN orders, except for certain regions of scalar-tensor theory space where non-linear effects inside neutron stars, called ``spontaneous scalarization'',  can occur.   It is therefore unlikely that we will be able to point to a qualitatively new test of scalar-tensor gravity to be performed with gravitational waves.

Nevertheless we expect to provide a complete and consistent waveform model to an order in the PN approximation comparable to the best models from GR.  With this model it will be possible to carry out parameter estimation analyses for gravitational waves from binary inspiral, and to compare the bounds with those from earlier work that either confined attention to the leading dipole term, such as~\cite{BBW}, or assumed extreme mass ratios, such as~\cite{yunespanicardoso}.

\section{An Overview}
\label{sec:intro}

We will use the approach known as Direct Integration of the Relaxed Einstein Equations (DIRE) that we described in \cref{chapter6}. DIRE is based on a framework originally developed by Epstein and Wagoner~\cite{wag76, ew2b, eps75}, extended by Will, Wiseman and Pati~\cite{wis92,wil96,pat00,pat02}, and applied to numerous problems in post-Newtonian gravity~\cite{kww,kidder,willspinorbit,wangwill,zengwill,mitchellwill}. As we discussed earlier in \cref{chapter6}, DIRE is a self-contained approach in which the Einstein equations are cast into their ``relaxed'' form of a flat-spacetime wave equation together with a harmonic gauge condition, and are solved  formally as a retarded integral over the past null cone of the field point.  The ``inner'', or near-zone part of this integral within a sphere of radius $\lambda$,  a gravitational wavelength, is approximated in a slow-motion expansion using standard techniques; the ``outer'' part, extending over the radiation zone, is evaluated using a null integration variable.

DIRE is rather easily adapted to scalar-tensor theories, so that the same methods that have been worked out for GR can be applied here.  It is possible that many other theories that generalize the standard action of general relativity in four spacetime dimensions by adding various fields could be cast in a similar form, permitting a systematic study of their predictions for compact binary inspiral beyond the lowest order in the PN approximation.   Indeed another motivation for this work is to lay out a template for possible extensions to other theories of gravity, such as the Einstein-Aether theory~\cite{aether} or TeVeS~\cite{teves}.

Specifically, the theories we address here are described by the action given by Eq.~(\ref{STaction1}) that we recall here as
\begin{equation}
S = (16\pi)^{-1} \int \left [ \phi R - \phi^{-1} \omega(\phi) g^{\alpha\beta} \partial_\alpha \phi \partial_\beta \phi \right ] \sqrt{-g} d^4x + S_m ( \mathfrak{m}, g_{\alpha\beta}) \,,
\label{STaction}
\end{equation}
where $R$ is the Ricci scalar of the spacetime metric $g_{\alpha\beta}$, $\phi$ is the scalar field, of which $\omega$ is a function.  Throughout, we use the so-called ``metric'' or ``Jordan'' representation, in which the matter action $S_m$ involves the matter fields $\mathfrak{m}$ and the metric only; $\phi$ does not couple directly to the matter (see~\cite{DamourEsposito92} for example, for a representation of this class of theories in the so-called ``Einstein'' representation).  We exclude the possibility of a potential or mass for the scalar field.

In order to incorporate the internal gravity of compact, self-gravitating bodies, we adopt an  approach pioneered by Eardley~\cite{eardley}, based in part on general arguments dating back to Robert Dicke,  in which one treats the matter energy-momentum tensor as a sum of delta functions located at the position of each body,  but assumes that the mass of each body is a function $M_A(\phi)$ of the scalar field.  This reflects the fact that the gravitational binding energy of the body is controlled by the value of the gravitational constant, which is directly related to the value of the background scalar field in which the body finds itself.    Consequently, the matter action will have an {\em effective} dependence on $\phi$, and as a result the field equations will depend on the ``sensitivity'' of the mass of each body to variations in the scalar field, holding the total number of baryons fixed.  The sensitivity of body $A$ is defined by
\begin{equation}
s_A \equiv \left ( \frac{d \ln M_A(\phi)}{d \ln \phi} \right ) \,.
\end{equation}
For neutron stars, the sensitivity depends on the mass and equation of state of the star and is typically of order $0.2$; in the weak-field limit, $s_A$ is proportional to the Newtonian self-gravitational energy per unit mass of the body.  From the theorem of Hawking, for stationary black holes, it is known that $s_{\rm BH} = 1/2$.   

This part of the dissertation (Chapters~\ref{chapter8}-\ref{chapter12}) reports the results of a calculation of the explicit equations of motion for binary systems of non-spinning compact bodies, through $2.5$PN order, that is, to order $(v/c)^5$ beyond Newtonian theory. The post-Newtonian corrections at 1PN and 2PN orders are conservative; we obtain from them expressions for the conserved total energy and linear momentum, and obtain the 2-body Lagrangian from which they can be derived.  There are also terms in the equations of motion at $1.5$PN and $2.5$PN orders.  These are gravitational-radiation reaction terms.  Terms at $1.5$PN order do not occur in general relativity (see \sref{sec:Two-Body}), but in scalar-tensor theories with compact bodies, they are the result of the emission of {\em dipole} gravitational radiation.  At $2.5$PN order, one finds the analogue of the general relativistic quadrupole radiation, together with PN correction effects related to monopole and dipole radiation.   

Not surprisingly the expressions for these quantities are complicated, much more so than their counterparts in general relativity. On the other hand, they depend on a relatively small number of parameters, related to the value of $\omega(\phi)$ far from the system, where $\phi = \phi_0$, along with its derivatives with respect to $\varphi \equiv \phi/\phi_0$, and the sensitivities $s_1$ and $s_2$ of the two bodies, and their derivatives with respect to $\phi$. The parameters and their definitions are shown in Table \ref{tab:params}.

At Newtonian order, the ``bare'' gravitational coupling constant $G$ is related to the asymptotic value of the scalar field, but for two-body systems of compact objects, the coupling is given by the combination $G\alpha$, where
\begin{equation}
\alpha = \frac{3 + 2\omega_0}{4 + 2\omega_0} + \frac{(1-2s_1)(1-2s_2)}{4 + 2\omega_0} \,,
\end{equation}
where $\omega_0 = \omega(\phi_0)$.  At $1$PN order there are two body-dependent parameters, $\bar{\gamma}$ and $\bar{\beta}_A$, $A = 1,2$ (see Table \ref{tab:params} for definitions of the parameters).  For non-compact objects, where $s_A \ll 1$,  $\bar{\gamma}= \gamma -1$ and $\bar{\beta}_A = \beta -1$, where $\gamma$ and $\beta$ are precisely the PPN parameters for scalar-tensor theory, as listed in~\ref{ST-PPN-paras}.  At $2$PN order, there are two additional parameters $\delta_A$ and $\chi_A$.  Most of the parameters in Table \ref{tab:params} can be related directly to parameters defined in~\cite{DamourEsposito96,DamourEsposito92}.

\begin{table}
\centering
\begin{tabular}{c l}
\hline
Parameter&Definition\\
\hline\hline\\
$G$&$\phi_0^{-1} (4+2\omega_0)/(3+2\omega_0)$\\
$\zeta$&$1/(4+2\omega_0)$
\\
$\lambda_1$&$(d\omega/d\varphi)_0 \zeta^2/(1-\zeta)$\\
$\lambda_2$&$(d^2\omega/d\varphi^2)_0 \zeta^3/(1-\zeta)$
\\
\\
{\bf Sensitivities}&
\\
$s_A$&$[d \ln M_A(\phi)/d \ln \phi]_0$
\\
$s'_A$&$[d^2 \ln M_A(\phi)/d \ln \phi^2]_0$
\\
$s''_A$&$[d^3 \ln M_A(\phi)/d \ln \phi^3]_0$\\
\\
\hline\\
\multicolumn{2}{l}{\bf Newtonian}\\
$\alpha $&$1 - \zeta + \zeta (1-2s_1)(1- 2s_2) $
\\
\\
\multicolumn{2}{l}{\bf post-Newtonian}\\
$\bar{\gamma}$ & $-2 \alpha^{-1}\zeta (1-2s_1)(1-2s_2)$
\\
$\bar{\beta}_1 $&$\alpha^{-2} \zeta (1-2s_2)^2 \left ( \lambda_1 (1-2s_1) + 2 \zeta s'_1 \right )$
\\
$\bar{\beta}_2 $&$\alpha^{-2} \zeta (1-2s_1)^2 \left ( \lambda_1 (1-2s_2) + 2 \zeta s'_2 \right )$
\\
\\
\multicolumn{2}{l}{\bf 2nd post-Newtonian}\\
$\bar{\delta}_1$ &$ \alpha^{-2} \zeta (1-\zeta) (1-2s_1)^2$
\\
$\bar{\delta}_2$ &$\alpha^{-2} \zeta (1-\zeta) (1-2s_2)^2 $
\\
$\bar{\chi}_1 $&$ \alpha^{-3} \zeta (1-2s_2)^3  \left [ (\lambda_2 -4\lambda_1^2 + \zeta \lambda_1 ) (1-2s_1) -6 \zeta \lambda_1 s'_1 + 2 \zeta^2 s''_1 \right ] 
$
\\ 
$\bar{\chi}_2 $&$ \alpha^{-3} \zeta (1-2s_1)^3  \left [ (\lambda_2 -4\lambda_1^2 + \zeta \lambda_1 ) (1-2s_2) -6 \zeta \lambda_1 s'_2 + 2 \zeta^2 s''_2 \right ] $\\
\\
\hline
\end{tabular}
\caption{\label{tab:params} Parameters used in the equations of motion at Newtonian, 1PN, and 2PN orders.}
\end{table}

Here we will quote the bottom-line result: the two-body equation of motion, expressed in relative coordinates, ${\bf X} \equiv {\bf x}_1 - {\bf x}_2$, through $2$PN order.  This equation is ready-to-use, for example in calculating time derivatives of radiative multipole moments in determining the gravitational-wave signal. The equation has the form
\begin{eqnarray}\label{finaleom}
{{d^2 {\bf X}} \over {dt^2}} &=& -{G\alpha m \over r^2} {\bf n} 
+ {G\alpha m \over r^2} \bigl[ \, {\bf n} (A_{1PN} + A_{2PN}) 
	+ {\dot r}{\bf v} (B_{1PN} + B_{2PN} ) \bigr]  
\nonumber \\
&&
+ {8 \over 5} \eta {(G\alpha m)^2 \over r^3} 
	\bigl[\dot r {\bf n} (A_{1.5PN}+A_{2.5PN})
	- {\bf v}(B_{1.5PN}+B_{2.5PN})\bigr] \,,
\label{eomfinal}
\end{eqnarray}
where  $r \equiv \mid {\bf X}\mid $, ${\bf n} \equiv {\bf X}/r$, $m \equiv m_1 + m_2$, $\eta \equiv m_1m_2/m^2$,  ${\bf v} \equiv {\bf v}_1 - {\bf v}_2$, and $\dot r = dr/dt$.  We use units in which $c=1$ but for reasons to be discussed later, we will not set $G=1$, in contrast to the notation used earlier in this dissertation. The leading term is Newtonian gravity. The next group of terms are the conservative terms, of integer PN order, while the final group are dissipative radiation-reaction terms, of half-odd-integer PN order.  The coefficients $A$ and $B$ are given explicitly in Eqs.~(\ref{EOMcoeffs}).

Several things are worth noting about these equations (and indeed about all the two-body equations shown later in the next chapters).  In the general relativistic limit $\omega_0 \to \infty$, or $\zeta \to 0$, the equations (including the $2.5$PN terms) reduce to those of general relativity, as determined by many authors~\cite{damourderuelle,damour300,kopeikin85,GK86,bfp98,futamase01,pat02}. Considering scalar-tensor theories, one might compare the values of coefficients  $A$ and $B$ in Eq.~(\ref{finaleom}) given in Eqs.~(\ref{EOMcoeffs}) with their corresponding values in general relativity given in Eqs.~(\ref{1PNcoeffs}, \ref{2PNcoeffs}). At $1$PN order, the equations agree with the standard scalar-tensor equations, both for weakly self-gravitating bodies in the general class of theories~\cite{nor70b} (shown within the PPN framework in Sec.\ 6.2 and 7.3 of~\cite{wil74a}), and for arbitrarily compact bodies in pure Brans-Dicke theory (as displayed in Sec.\ 11.2 of~\cite{wil74a}).

Although a number of authors have obtained partial results in scalar-tensor theory at $2$PN order, notably the metric sufficient to study light deflection at $2$PN order~\cite{niST2pn,deng2pn}, and the generic structure of the $2$PN Lagrangian for $N$ compact bodies~\cite{DamourEsposito96}, our explicit formulae for the $2$PN and $2.5$PN contributions to the two-compact-body equations of motion are new.   

The energy loss that results from the $1.5$ PN and $2.5$ PN terms in the equations of motion is in complete agreement with the energy flux calculated to the corresponding order by Damour and Esposito-Far\`ese~\cite{DamourEsposito92}.

The other interesting limit is that in which both bodies are black holes.  Assuming that Hawking's result that $s_{\rm BH} = 1/2$ applies equally for binary black holes as for isolated black holes, we find that the parameters $\bar{\gamma}$, $\bar{\beta}_A$, $\bar{\delta}_A$ and $\bar{\chi}_A$ all vanish, and $\alpha = 1-\zeta = (3+2\omega_0)/(4+2\omega_0)$.  In this case the equations reduce {\em identically} to those of general relativity through $2.5$PN order, with $G\alpha m_A $ replacing of $Gm_A$ for each body.  In other words, if each mass is rescaled by $(4+2\omega_0)/(3+2\omega_0)$, the scalar-tensor equations of motion for binary black holes, including the 2.5PN terms, become {\em identical} to those in general relativity.   Again this applies to all the equations of motion and related quantites (total energy, Lagrangian), whether for the individual bodies or for the relative motion.  Since the masses of bodies in binary systems are measured purely via the Keplerian dynamics of the system, the rescaling is unmeasurable, and therefore, the dynamics of binary black holes in this class of theories is observationally indistinguishable from the dynamics in general relativity.   Assuming, as we believe will be the case, that this is also true for the gravitational wave emission, the conclusion is that gravitational-wave observations of binary black hole systems will be unable to distinguish between these two theories.   

If only one member of the binary system is a black hole, then $\alpha = 1-\zeta$, and $\bar{\gamma} = \bar{\beta}_A = 0$, so that even at $1$PN order, the equations of motion are {\em identical} to those of general relativity, after rescaling each mass.  Only at $1.5$PN order and above do differences between the two theories occur for the mixed binary system, because of the non-vanishing of ${\cal S}_{-}$ in the dipole radiation reaction term, and the non-vanishing of $\bar{\delta}_1$ (if body 1 is the neutron star) in the $2$PN terms.  However, in this case {\em all} the deviations from general relativity depend on a single parameter $Q$, given by
\begin{equation}
Q \equiv \zeta (1-\zeta)^{-1} (1-2s_1)^2 \,,
\end{equation}
where $s_1$ is the sensitivity of the neutron star.  In particular, all reference to the parameters $\lambda_1$ and $\lambda_2$ disappears, and the motion through $2.5$PN order is identical to that predicted by pure Brans-Dicke theory.   If this conclusion holds true for the gravitational-wave emission, then gravitational-wave observations of mixed black-hole neutron-star binaries will be unable to distinguish between Brans-Dicke theory and its generalizations. The only caveat is that, for a given neutron star, generalized scalar-tensor theories can predict very different values of its un-rescaled mass and its sensitivity from those predicted by pure Brans-Dicke. Now we turn to the detailed calculations.

\section{Foundations: Relaxed Field Equations}
\label{sec:relaxed}

\subsection{Field equations and equations of motion}
\label{sec:fieldequations}

We begin by recasting the field equations of scalar-tensor theory into a form that parallels as closely as possible the ``relaxed Einstein equations'' used to develop post-Minkowskian and post-Newtonian theory in general relativity. Referring back to \sref{sec:relaxedeq} will be useful. The original field equations of scalar-tensor theory as derived from the action of Eq.\ (\ref{STaction}) take the form of Eqs.~(\ref{STfieldEqs1}) that we recall here as
\begin{subequations}\label{STfieldEqs}
\begin{eqnarray}
G_{\mu\nu} &=& \frac{8\pi}{\phi} T_{\mu\nu} + \frac{\omega(\phi)}{\phi^2} \left ( \phi_{,\mu} \phi_{,\nu} - \frac{1}{2} g_{\mu\nu} \phi_{,\lambda} \phi^{,\lambda} \right ) + \frac{1}{\phi} \left ( \phi_{;\mu\nu} - g_{\mu\nu} \Box_g \phi \right ) \,, 
\label{fieldeq1}\\
\Box_g \phi  &=& \frac{1}{3 + 2\omega(\phi)} \left ( 8\pi T - 16\pi \phi \frac{\partial T}{\partial \phi} - \frac{d\omega}{d\phi} \phi_{,\lambda} \phi^{,\lambda} \right ) \,,
\label{fieldeq2}
\end{eqnarray}
\end{subequations}
where $T_{\mu\nu}$ is the stress-energy tensor of matter and non-gravitational fields, $G_{\mu\nu}$ is the Einstein tensor constructed from the physical metric $g_{\mu\nu}$, $\phi$ is the scalar field, $\omega(\phi)$ is a coupling function, $\Box_g$ denotes the scalar d'Alembertian with respect to  the metric, and commas and semicolons denote ordinary and covariant derivatives, respectively.  We work throughout in the metric or ``Jordan'' representation of the theory, in contrast to the ``Einstein'' representation used, for example in \sref{sec:ST} and ~\cite{DamourEsposito92}.

Normally, such as for a perfect-fluid source, the matter stress-energy tensor depends only on the matter field variables and the physical metric $g_{\mu\nu}$, not on the scalar field, and accordingly the term $\partial T/\partial \phi$ does not appear in the field equations.  But in dealing with a system of self-gravitating bodies, we will adopt an approach pioneered by Eardley~\cite{eardley}.  
Because $\phi$ controls the local value of the gravitational constant in and near each body in this class of theories, the total mass of each body, including its self-gravitational binding energy, may depend on the scalar field.  Thus, as long as each body can be regarded as being in stationary equilibrium during its motion, Eardley proposed letting each mass be a function of $\phi$, namely $M_A(\phi)$.     With this assumption, $T^{\mu\nu}$ takes the form
\begin{eqnarray}
T^{\mu\nu} (x^\alpha) &=& (-g)^{-1/2} \sum_A \int d\tau M_A (\phi) u_A^\mu u_A^\nu \delta^4 (x_A^\alpha (\tau) - x^\alpha)  
\nonumber \\
&=& (-g)^{-1/2} \sum_A M_A (\phi) u_A^\mu u_A^\nu (u_A^0)^{-1} 
\delta^3 ({\bf x} - {\bf x}_A) \,,
\label{Tmunu}
\end{eqnarray}
where $\tau$ is proper time measured along the world line of body $A$ and  $u_A^\mu$ is its four-velocity.  The {\em indirect} coupling of $\phi$ to matter via the binding energy is responsible for the term $\partial T/\partial \phi$ in the field equations.

From the Bianchi identity applied to Eq.\ (\ref{fieldeq1}), the equation of motion is
\begin{equation}
{T^{\mu\nu}}_{;\nu} = \frac{\partial T}{\partial \phi} \phi^{,\mu} \,,
\end{equation}
with the right-hand-side vanishing in the perfect-fluid case. From the compact body form of $T^{\mu\nu}$ in Eq.\ (\ref{Tmunu}), it can then be shown that the equation of motion for each compact body takes the modified geodesic form
\begin{equation}
u^\nu \nabla_\nu (M_A(\phi) u^\mu ) = - \frac{dM_A}{d\phi} \phi^{,\mu} \,,
\end{equation}
or in terms of coordinate time and ordinary velocities $v^\alpha$,
\begin{equation}
\frac{dv^j}{dt} + \Gamma^j_{\alpha\beta} v^\alpha v^\beta
-\Gamma^0_{\alpha\beta} v^\alpha v^\beta v^j 
= - \frac{1}{M_A(u^0)^2} \frac{dM_A}{d\phi} ( \phi^{,j} - \phi^{,0} v^j ) \,.
\label{geodesiceq}
\end{equation}
These equations of motion could also be derived directly from the effective matter action, $S_m = \sum_A \int_A M_A (\phi) d\tau$.  Equation (\ref{Tmunu}) can equally well be taken to describe a pressureless perfect fluid (dust), simply by letting the mass of each particle be a constant, independent of $\phi$.

\subsection{Relaxed field equations in scalar-tensor gravity}
\label{sec:relaxed2}
 
To recast Eq.~(\ref{fieldeq1}) into the form of a ``relaxed'' Einstein equation, we recall the discussion of \sref{sec:relaxedeq}. Defining the 
quantities 
\begin{subequations}
\begin{eqnarray}
\gothg^{\mu\nu} &\equiv& \sqrt{-g} g^{\mu\nu} \,, 
\label{gothg}\\
H^{\mu\alpha\nu\beta} &\equiv& \gothg^{\mu\nu}  \gothg^{\alpha\beta} -  \gothg^{\alpha\nu} \gothg^{\beta\mu} \,,
\label{Hdef}
\end{eqnarray}
\end{subequations}
we show that the following is an identity, valid for any spacetime,
\begin{equation}
{H^{\mu\alpha\nu\beta}}_{,\alpha\beta} = (-g) (2G^{\mu\nu} + 16\pi t_{LL}^{\mu\nu} ) \,,
\label{Hidentity}
\end{equation}
where $t_{LL}^{\mu\nu}$ is the Landau-Lifshitz pseudotensor [see Eqs.~(\ref{landaulifshitz}) for an explicit formula].

To incorporate scalar-tensor theory into this framework, we assume that, far from any isolated source, the metric takes its Minkowski form $\eta_{\mu\nu}$, and that the scalar field $\phi$ tends to a constant value $\phi_0$.   We define the rescaled scalar field $\varphi \equiv \phi/\phi_0$.  We next define the conformally transformed metric $\tilde{g}_{\mu\nu}$ by 
\begin{equation}
\tilde{g}_{\mu\nu} \equiv \varphi {g}_{\mu\nu} \,,
\label{gtilde}
\end{equation}
and the gravitational field $\tilde{h}^{\mu\nu}$ by the equation
\begin{equation}
\tilde{\gothg}^{\mu\nu} \equiv \sqrt{-\tilde{g}} \tilde{g}^{\mu\nu} \equiv \eta^{\mu\nu} - \tilde{h}^{\mu\nu} \,.
\label{hdef}
\end{equation}
From Eq.\ (\ref{gtilde}) it can be shown that this is equivalent to
\begin{equation}
\gothg^{\mu\nu} \equiv \varphi^{-1} (\eta^{\mu\nu} - \tilde{h}^{\mu\nu}) \,.
\end{equation}
We now impose the ``Lorentz'' gauge condition
\begin{equation}
{\tilde{h}^{\mu\nu}}_{,\nu} = 0 \,,
\label{gauge}
\end{equation}
which is equivalent to
\begin{equation}
{\gothg^{\mu\nu}}_{,\nu} =- \varphi^{-2} \varphi_{,\nu} (\eta^{\mu\nu}- \tilde{h}^{\mu\nu} ) \,.
\end{equation}
Substituting Eqs.\ (\ref{fieldeq1}), (\ref{fieldeq2}), (\ref{hdef}) and (\ref{gauge}) into (\ref{Hidentity}), we can recast the field equation (\ref{fieldeq1}) into the form
\begin{equation}
\Box_\eta \tilde{h}^{\mu\nu} =  -16\pi \tau^{\mu\nu} \,,
\label{heq0}
\end{equation}
where $\Box_\eta$ is the flat spacetime d'Alembertian with respect to $\eta_{\mu\nu}$, and where
\begin{equation}
16\pi \tau^{\mu\nu}= 16\pi (-g) \frac{\varphi}{\phi_0} T^{\mu\nu}  + \Lambda^{\mu\nu} + \Lambda_S^{\mu\nu} \,,
\label{heq}
\end{equation}
where
\begin{subequations}
\begin{eqnarray}
\Lambda^{\mu\nu} &\equiv& 16\pi \left [ (-g) t_{LL}^{\mu\nu} \right ](\tilde{\gothg}^{\mu\nu})
+ {\tilde{h}^{\mu\alpha}}_{,\beta} {\tilde{h}^{\nu\beta}}_{,\alpha}
-\tilde{h}^{\alpha\beta} {\tilde{h}^{\mu\nu}}_{,\alpha\beta} \,,
\\
\Lambda_S^{\mu\nu} &\equiv&  \frac{(3+ 2\omega)}{\varphi^2} \varphi_{,\alpha} \varphi_{,\beta} \left ( \tilde{\gothg}^{\mu\alpha}\tilde{\gothg}^{\nu\beta} - \frac{1}{2} \tilde{\gothg}^{\mu\nu}\tilde{\gothg}^{\alpha\beta} \right ) \,,
\end{eqnarray}
\label{Lambdadef}
\end{subequations}
where the notation $[(-g) t_{LL}^{\mu\nu}](\tilde{\gothg}^{\mu\nu})
$ denotes that the Landau-Lifshitz piece should be calculated using only $\tilde{\gothg}$, in other words, exactly as in general relativity, except using the conformal metric, rather than the physical metric.   
The scalar field equation can also be rewritten in terms of a flat-spacetime wave equation, of the form
\begin{equation}
\Box_\eta \varphi = -8\pi \tau_s \,,
\label{phieq}
\end{equation}
where
\begin{eqnarray}
 \tau_s &=& -\frac{1}{3+2\omega} \sqrt{-g} \frac{\varphi}{\phi_0} \left ( T -2 \varphi \frac{\partial T}{\partial \varphi} \right )
- \frac{1}{8\pi}  \tilde{h}^{\alpha\beta} \varphi_{,\alpha\beta}
\nonumber \\
&&
+\frac{1}{16\pi} \frac{d}{d\varphi} \left [ \ln \left ( \frac{3+2\omega}{\varphi^2} \right ) \right ] \varphi_{,\alpha} \varphi_{,\beta} \tilde{\gothg}^{\alpha\beta} \,.
\label{taustar}
\end{eqnarray}
In principle, Eqs.\ (\ref{gothg}) and (\ref{hdef}) can be combined to give $g_{\mu\nu}$ in terms of $\varphi$ and $\tilde{h}^{\mu\nu}$, although in practice, we will express it as a PN expansion.  The final result will be the relaxed field equations (\ref{heq0}) - (\ref{taustar}) expressed entirely in terms of $\tilde{h}^{\mu\nu}$, $\varphi$, and the matter variables.   The next task will be to solve these equations iteratively in a post-Newtonian expansion in the near-zone. Formally the solutions of these wave equations can be expressed using the standard retarded Green function, in the form
\begin{eqnarray}
\tilde{h}^{\mu\nu} (t,{\bf x}) &=& 4 \int \frac{\tau^{\mu\nu} (t-\mid {\bf x}-{\bf x}'\mid ,{\bf x}')}{\mid {\bf x}-{\bf x}'\mid } d^3x' \,,
\nonumber \\
\varphi (t,{\bf x}) &=& 2 \int \frac{\tau_s (t-\mid {\bf x}-{\bf x}'\mid ,{\bf x}')}{\mid {\bf x}-{\bf x}'\mid } d^3x' \,,
\end{eqnarray}
where the integration is over the past flat spacetime null cone of the field point $(t,{\bf x})$.   We will expand these integrals in the near-zone, and incorporate a slow-motion, weak-field expansion in terms of a small parameter $\epsilon \sim v^2 \sim m/r$; the strong-field internal gravity effects will be encoded in the functions $M_A(\phi)$.

\begin{savequote}[0.55\linewidth]
{\scriptsize ``Science never solves a problem without creating ten more.''}
\qauthor{\scriptsize---George Bernard Shaw}
\end{savequote}
\chapter{Formal Structure and Expansion of The Near-Zone Fields} 

\label{chapter9} 
\thispagestyle{myplain}
\lhead[\thepage]{Chapter 8. \emph{Formal Structure and Expansion of The Near-Zone Fields}}      
\rhead[Chapter 8. \emph{Formal Structure and Expansion of The Near-Zone Fields}]{\thepage}

\section{Formal Structure of The Near-Zone Fields}
\label{sec:nearzonefields}

Following Eq.~(\ref{Ncomponents}), we reintroduce a simplified notation for the field
$\tilde{h}^{\mu\nu}$ and the scalar field $\varphi$:
\begin{eqnarray}
N &\equiv& \tilde{h}^{00} \sim O(\epsilon) \,, \nonumber \\
K^i &\equiv& \tilde{h}^{0i} \sim O(\epsilon^{3/2}) \,, \nonumber \\
B^{ij} &\equiv& \tilde{h}^{ij} \sim O(\epsilon^2) \,, \nonumber \\
B &\equiv& \tilde{h}^{ii} \equiv \sum_i h^{ii} \sim O(\epsilon^2) \,,
\nonumber \\
\Psi &\equiv& \varphi - 1 \sim O(\epsilon) \,,
\end{eqnarray}
where we show the leading order dependence on $\epsilon$ in the near
zone.  To obtain
the equations of motion to 2.5PN order, we need to determine
the components of the physical metric and $\varphi$ to the following orders:
$g_{00}$ to $O(\epsilon^{7/2})$,
$g_{0i}$ to $O(\epsilon^{3})$ ,
$g_{ij}$ to $O(\epsilon^{5/2})$, and 
$\varphi$ to $O(\epsilon^{7/2})$.
From the Eqs.~(\ref{gothg}, \ref{hdef}), one can invert to find
$g_{\mu\nu}$ in terms of $\tilde{h}^{\mu\nu}$ and $\varphi$ to the appropriate order in $\epsilon$, as in PWI, Eq.\ (4.2).  Expanding to the required order, we find (compare to \eref{gr-metric})
\begin{subequations}
\label{STmetricexpand}
\begin{eqnarray}
g_{00} &=& -1 +  \left ( \frac{1}{2} N + \Psi \right) {\color {red} \epsilon}
+  \left ( \frac{1}{2} B - \frac{3}{8} N^2 - \frac{1}{2} N \Psi
- \Psi^2 \right ) {\color {red} \epsilon^{2}}\\
&& +  \left (\frac{5}{16} N^3 - \frac{1}{4} NB + \frac{1}{2} K^j K^j +\frac{3}{8} N^2 \Psi - \frac{1}{2} B \Psi +\frac{1}{2} N \Psi^2 + \Psi^3 \right ) {\color {red} \epsilon^{3}}
\nonumber +O(\epsilon^4) \,, 
\label{metric00}
\\
g_{0i} &=& -  K^i  {\color {red} \epsilon^{3/2}}+  \left ( \frac{1}{2} N + \Psi \right ) K^i  {\color {red} \epsilon^{5/2}}+O(\epsilon^{7/2}) \,, 
\label{metric0i}
\\
g_{ij} &=& \delta^{ij} \left \{ 1+  \left (\frac{1}{2} N - \Psi \right ) {\color {red} \epsilon}-  \left ( \frac{1}{8} N^2 + \frac{1}{2} B + \frac{1}{2} N \Psi - \Psi^2 \right ) {\color {red} \epsilon^{2}} \right \} +  B^{ij} {\color {red} \epsilon^{2}}
+O(\epsilon^3) \,, 
\label{metricij}
\\
(-g) &=& 1+  (N - 4\Psi)\; {\color {red} \epsilon}-  (B +4N \Psi - 10 \Psi^2 )\; {\color {red} \epsilon^{2}}+ O(\epsilon^3) \,.
\label{metricdet}
\end{eqnarray}
\label{metric}
\end{subequations}

In Eqs.~Eqs.~((\ref{STmetricexpand}) we do not distinguish between covariant and contravariant components of quantities such as $K^i$ or $B^{ij}$, since their indices are assumed to be raised or lowered using the Minkowski metric, whose spatial components are $\delta_{ij}$.

We now define a set of provisional ``densities'' following the convention of Blanchet and Damour~\cite{bla89} (given in Eqs.~(\ref{sigma:components})), but adding a separate density for the scalar field equation:
\be
\sigma_s \equiv - T + 2\varphi \partial T/\partial \varphi \,.
\ee
The second contribution to $\sigma_s$ will be non-zero only in the case where our system consists of gravitationally bound bodies, whose internal structure could depend on the environmental value of $\varphi$.

Because of the way we have formulated the relaxed scalar-tensor equations, the quantity $\Lambda^{\mu\nu}$ has {\em exactly} the same form as in Eqs.~(\ref{Lambdacomponents}) to the 2PN order needed for our work. The additional scalar stress-energy pseudotensor is new and given by
\begin{subequations}
\label{Lambdascalar}
\begin{eqnarray}
\Lambda_S^{00} &=& \frac{3+2\omega_0}{2} (\nabla \Psi )^2 {\color {red} \epsilon}
+\frac{3+2\omega_0}{2} \left \{ N (\nabla \Psi )^2 - 2 \left ( 1- \frac{\omega_0'}{3+2\omega_0} \right ) \Psi  (\nabla \Psi )^2 + {\dot \Psi}^2 \right \} {\color {red} \epsilon^{2}}
\nonumber \\
&& + O(\rho \epsilon^3) \,,
\\
\Lambda_S^{0i} &=& -(3+2\omega_0) {\dot \Psi} \Psi^{,i} {\color {red} \epsilon^{3/2}} +  O(\rho \epsilon^{5/2}) \,, 
\\
\Lambda_S^{ij} &=&  (3+2\omega_0) \left \{ \Psi^{,i}\Psi^{,j} - {1 \over 2}
\delta^{ij} (\nabla \Psi)^2 \right \} {\color {red} \epsilon} \nonumber \\
&&-(3+2\omega_0) \left \{  2\left ( 1- \frac{\omega_0'}{3+2\omega_0} \right ) \Psi 
\left [\Psi^{,i}\Psi^{,j} - {1 \over 2}\delta^{ij} (\nabla \Psi)^2 \right ] -\frac{1}{2} \delta^{ij} {\dot \Psi}^2
\right \} {\color {red} \epsilon^{2}}\nonumber\\
&& + O(\rho \epsilon^3) \,,
\\\nonumber
\Lambda_S^{ii} &=& -\frac{3+2\omega_0}{2}(\nabla \Psi)^2  {\color {red} \epsilon}
+ (3+2\omega_0) \left \{ \left ( 1- \frac{\omega_0'}{3+2\omega_0} \right ) \Psi (\nabla \Psi)^2+ \frac{3}{2} {\dot \Psi}^2 \right \} {\color {red} \epsilon^{2}}\\
&& +  O(\rho \epsilon^3) \,,
\end{eqnarray}
\end{subequations}
where $\omega'_0 \equiv (d\omega/d\varphi)_0$.

The near-zone expansions of the fields $N$, $K^i$, $B^{ij}$ and $\Psi$ are then given by \eref{bigexpansion} and 
\begin{eqnarray}
%
%
\Psi_{\cal N} &=& 2 {\color {red} \epsilon} \int_{\cal M} \frac{\tau_s(t,{\bf x}^\prime)}{\mid {\bf x}-{\bf x}^\prime \mid } d^3x^\prime 
- 2 {\color {red} \epsilon^{3/2}} \dot{M_s}  
+ {\color {red} \epsilon^2} \partial^2_t \int_{\cal M} \tau_s(t,{\bf x}^\prime) \mid {\bf x}-{\bf x}^\prime \mid  d^3x^\prime
\nonumber 
\\
&&
-{1 \over 3} {\color {red} \epsilon^{5/2}} \left ( r^2 \stackrel{(3)\quad}{M_s(t)}
-2x^j \stackrel{(3)\quad}{{\cal I}^{j}_s(t)}
+ \stackrel{(3)\quad}{{\cal I}^{kk}_s(t)} \right )
+ {1 \over 12} {\color {red} \epsilon^3} \partial^4_t \int_{\cal M}
\tau_s(t,{\bf x}^\prime) \mid {\bf x}-{\bf x}^\prime \mid ^3 d^3x^\prime 
\nonumber \\
&&- {1 \over 60} {\color {red} \epsilon^{7/2}} \left \{ r^4 \stackrel{(5)\quad}{M_s(t)} 
- 4r^2 x^j \stackrel{(5)\quad}{{\cal I}^{j}_s(t)}
+(4x^{kl}+2r^2\delta^{kl})
\stackrel{(5)\quad}{{\cal I}^{kl}_s(t)}
- 4 x^k \stackrel{(5)\qquad}{{\cal I}^{kll}_s(t)}
+ \stackrel{(5)\qquad}{{\cal I}^{kkll}_s(t)} \right
\} \nonumber \\
&&
+ \Psi_{\partial {\cal M}}+ O(\epsilon^4) \,,
 \label{bigexpansiondPsi}
\end{eqnarray}
where the 
scalar moments ${\cal I}^{Q}_s$ and $M_s$ are defined by
\begin{subequations}
\begin{eqnarray}
{\cal I}^{Q}_s &\equiv& \int_{\cal M} \tau_s x^Q d^3x \,,
\label{IsQ}
\\
M_s &\equiv& \int_{\cal M} \tau_s d^3x \,.
\label{Ms}
\end{eqnarray}
\label{genmomentPsi}
\end{subequations}
Again, the index $Q$ is a multi-index, such that $x^Q$ denotes $x^{i_1} \dots x^{i_q}$. The integrals are taken over a constant time hypersurface $\cal M$ at time $t$ out to a radius $\cal R$, which represents the boundary between the near zone and the far zone. The structure of the expansions for ${N}_{\cal N}$, ${K}^i_{\cal N}$ and ${B}^{ij}_{\cal N}$ is identical to the structure in Chapter~\ref{chapter6} because the source $\tau^{\mu\nu}$ satisfies the conservation law ${\tau^{\mu\nu}}_{,\nu}=0$, a consequence of the Lorentz gauge condition.    However, no such explicit conservation law applies to $\tau_s$; nevertheless, in a post-Newtonian expansion, we will be able to show, for example, that the term $\epsilon^{3/2} \dot{M}_s$ actually vanishes to lowest PN order, and thus contributes only beginning at $\epsilon^{5/2}$ order; the other terms involving time derivatives of $M_s$ will also be boosted to one higher PN order.  The time derivatives of the dipole moments ${\cal I}^{j}_s$ do {\em not} vanish in general; this is related to the well-known phenomenon of dipole gravitational radiation that can occur in scalar-tensor theories. The boundary terms $N_{\partial {\cal M}}$, $K^i_{\partial {\cal M}}$ and $B^{ij}_{\partial {\cal M}}$ can be found in Appendix C of PWI, but they will play no role in our analysis.  As in Chapter~\ref{chapter6}, we will discard all terms that depend on the radius $\cal R$ of the near-zone; these necessarily cancel against terms that arise from integrating over the remainder of the past null cone; those ``outer'' integrals can be shown to make no contribution to the near zone metric to the PN order at which we are working.

In the near zone, the potentials are Poisson-like potentials and their generalizations.  Most were defined in \cite{pat00}, but we will need to define additional potentials associated with the scalar field.  For a source $f$, we use the definition of the Poisson potential $P(f)$ in \eref{definesuper-1}. We also use the definition of potentials based on the ``densities'' $\sigma$, $\sigma^i$ and $\sigma^{ij}$ and $\sigma_s$ constructed from $T^{\alpha\beta}$ and from $T-2\varphi \partial T/\partial \varphi$ in Eqs.~(\ref{definesuper0}) plus a new potential
\begin{eqnarray}
\Sigma_s (f) &\equiv& \int_{\cal M} {{\sigma_s(t,{\bf x}^\prime)f(t,{\bf
x}^\prime)}
\over {\mid {\bf x}-{\bf x}^\prime \mid  }} d^3x^\prime = P(4\pi\sigma_s f) \,,
\end{eqnarray}
along with the super- and superduper-potentials defined in Eq.~(\ref{definesuper1},~\ref{definesuper2}) and their obvious counterparts $X^i$,  $X_s$,  and so on.
A number of potentials occur sufficiently frequently in the PN
expansion that is it useful to define them specifically.  There are the the ``Newtonian'' potentials, 
\begin{subequations}
\begin{eqnarray}
U \equiv \int_{\cal M} {{\sigma(t,{\bf x}^\prime)}
\over {\mid {\bf x}-{\bf x}^\prime \mid  }} d^3x^\prime = P(4\pi\sigma) =
\Sigma(1) \,,
\\
U_s\equiv \int_{\cal M} {{\sigma_s (t,{\bf x}^\prime)}
\over {\mid {\bf x}-{\bf x}^\prime \mid  }} d^3x^\prime = P(4\pi\sigma_s) =
\Sigma_s(1) \,.
\end{eqnarray}
\end{subequations}
The potentials needed for the post-Newtonian limit are (compare to \eref{1PNpotentials}):
\begin{eqnarray}
V^i \equiv \Sigma^i(1) \,,& \qquad &\Phi_1^{ij} \equiv \Sigma^{ij}(1) \,, \nonumber \\
\Phi_1 \equiv \Sigma^{ii}(1) \,, &\qquad &\Phi_1^s \equiv \Sigma_s (v^2) \,,\nonumber \\
\Phi_2 \equiv \Sigma(U) \,, &\qquad &\Phi^s_2 \equiv \Sigma_s(U) \,,\nonumber \\
\Phi_{2s} \equiv \Sigma (U_s) \,, &\qquad &\Phi^s_{2s} \equiv \Sigma_s (U_s) \,,
\nonumber \\
X \equiv X(1)\,, &\qquad & X_s \equiv X_s(1) \,. \quad
\end{eqnarray}
Useful 2PN potentials include:
\begin{eqnarray}
V_2^i \equiv \Sigma^i(U) \,, &\qquad & V_{2s}^i \equiv \Sigma^i(U_s) \,,
\nonumber \\
\Phi_2^i \equiv \Sigma(V^i) \,, &\qquad & Y \equiv Y(1) \,,
\nonumber \\
X^i \equiv X^i(1) \,, &\qquad & X_1 \equiv X^{ii}(1) \,, \nonumber
\\
X_2 \equiv  X(U) \,, &\qquad & X_{2s} \equiv X(U_s) \,,\nonumber \\
X_2^s \equiv X_s(U) \,,&\qquad & X_{2s}^s \equiv X_s(U_s) \,,\nonumber \\
P_2^{ij} \equiv P(U^{,i}U^{,j}) \,,& \qquad & P_2 \equiv P_2^{ii}=\Phi_2
-{1 \over 2}U^2 \,,\nonumber \\
P_{2s}^{ij} \equiv P(U^{,i}_sU^{,j}_s) \,, &\qquad & P_{2s} \equiv P_{2s}^{ii}=\Phi_{2s}^s
-{1 \over 2}U^{2}_s \,,\nonumber \\
G_1 \equiv P({\dot U}^2)  \,, &\qquad & G_{1s} \equiv P({\dot U}_s^{2}) \,,
\nonumber \\
G_2 \equiv P(U {\ddot U})  \,, &\qquad & G_{2s} \equiv P(U {\ddot U}_s) \,,
\nonumber \\
G_3 \equiv -P({\dot U}^{,k} V^k) \,, &\qquad & G_{3s} \equiv
 -P({\dot U}_s^{,k} V^k) \,,\nonumber \\
G_4 \equiv P(V^{i,j}V^{j,i}) \,, &\qquad & G_5 \equiv -P({\dot V}^k U^{,k}) \,,\nonumber \\
G_6 \equiv P(U^{,ij} \Phi_1^{ij}) \,, &\qquad & G_{6s} \equiv P(U_s^{,ij} \Phi_1^{ij}) \,,\nonumber \\
G_7^i \equiv P(U^{,k}V^{k,i}) + {3 \over 4} P(U^{,i}\dot U ) \,,& \qquad & H \equiv P(U^{,ij} P_2^{ij}) \,,  \nonumber \\
H_s \equiv P(U^{,ij} P_{2s}^{ij}) \,,  &\qquad & H^s \equiv P(U_s^{,ij} P_2^{ij}) \,, \nonumber \\
H_s^s \equiv P(U_s^{,ij} P_{2s}^{ij}) \,.&\qquad &  
\label{potentiallist}
\end{eqnarray}

\section{Expansion of Near-Zone Fields to 2.5PN Order}
\label{sec:2.5expansion}

In evaluating the contributions at each order, we shall use the notation defined in ~\ref{expandNKB} plus a similar notation for the scalar sector as
\label{expandNKBPsi}
\begin{eqnarray}
\Psi &=& {\color {red} \epsilon} (\Psi_0 + {\color {red} \epsilon^{1/2}} \Psi_{0.5} +{\color {red} \epsilon} \Psi_1+ {\color {red} \epsilon^{3/2}} \Psi_{1.5}+ {\color {red} \epsilon^2} \Psi_2+
{\color {red} \epsilon^{5/2}} \Psi_{2.5})
+O(\epsilon^4) \,,
\end{eqnarray}
where the subscript on each term indicates the level (1PN, 2PN, 2.5PN,
etc.) of its leading contribution to the equations of motion.

\subsection{Newtonian, 1PN and 1.5PN solutions}
\label{sec:N1.5PNsolution}

At lowest order in the PN expansion, we only need to evaluate
$\tau^{00} = (-g)T^{00}/\phi_0 + O(\rho\epsilon) = \sigma/\phi_0  +
O(\rho\epsilon)$ (recall that $\sigma^{ii} \sim \epsilon \sigma$), and $\tau_s = \sigma_s/[\phi_0(3+2\omega_0)]$, where $\omega_0 \equiv \omega(\phi_0)$. 
Since both densities have compact support, the outer integrals vanish, and we
find
\begin{eqnarray}
N_0 &=& \frac{4U}{\phi_0} \,,
\\
\Psi_0 &=& \frac{2U_s}{\phi_0 (3+2\omega_0)}  \,.
\label{newtonian}
\end{eqnarray}
Consider the case where we are dealing with pure perfect fluids, with no compact bodies having sensitivity factors $s_A$. Then to Newtonian order, $\sigma = \sigma_s$, $U=U_s$, and the metric to Newtonian order is given by the leading term in Eq.\ (\ref{metric00}), 
\begin{eqnarray}
g_{00} &=& -1 + \left ( \frac{1}{2} N + \Psi \right) 
\\
&=& -1 + 2 \frac{4+2\omega_0}{\phi_0 (3+2\omega_0)} U  \,.
\end{eqnarray}
We therefore identify the coefficient of $U$ in $g_{00}$ as the effective Newtonian gravitational coupling constant, $G$, given by  
\begin{equation}
G \equiv \frac{1}{\phi_0} \frac{4+2\omega_0}{3+2\omega_0} \,,
\label{Gdefinition}
\end{equation}
in agreement with our earlier definition of $G_{today}$ in Eq.~(\ref{today}). However, we will not set $G=1$ as is conventional in general relativity, in order to highlight the fact that it is an effective gravitational constant linked to the asymptotic value of $\phi$, which could, for example, vary with time as the universe evolves. For future use, we also recall $\zeta$ and $\lambda_1$ from Eq.~(\ref{st-para}) and define a new parameter $\lambda_2$ as
\begin{eqnarray}
\zeta &\equiv&  \frac{1}{4+2\omega_0}
\,, \nonumber 
\\
\lambda_1 &\equiv&  \frac{(d\omega/d\varphi)_0 \zeta}{3+2\omega_0} \,,
\nonumber 
\\
\lambda_2 &\equiv& \frac{(d^2\omega/d\varphi^2)_0 \zeta^2}{3+2\omega_0} \,.
\end{eqnarray}
A consequence of these definitions is that 
\begin{eqnarray}
\frac{1}{\phi_0} &=& G (1-\zeta) \,,
\nonumber \\
\frac{1}{\phi_0 (3+2\omega_0)} &=& G \zeta \,.
\end{eqnarray}
It is worth pointing out that $\omega_0$ enters at Newtonian order, via the modified coupling constant $G$ of Eq.\ (\ref{Gdefinition}).  It is then clear, by virtue of the expansion $\omega(\phi) = \omega_0 + (d\omega/d\varphi)_0 \Psi +
 (d^2\omega/d\varphi^2)_0 \Psi^2/2 + \dots$, that the parameter $\lambda_1$ will first contribute at $1$PN order, $\lambda_2$ will first contribute at $2$PN order, and so on.
 
To this order, $(-g)= 1+ 4GU(1-\zeta)-8GU_s \zeta + O(\epsilon^2)$.
Then, through PN order, the required forms for $\tau^{\mu\nu}$ and $\tau_s$ are given by
\begin{subequations}
\begin{eqnarray}\nonumber
\tau^{00} &=&G(1-\zeta) \biggl \{  \sigma - \sigma^{ii} +G(1-\zeta) \bigl ( 4\sigma U
- \frac{7}{8\pi} (\nabla U)^2 \bigr ) - G\zeta \bigl ( 6\sigma U_s - \frac{1}{8\pi}  ({\nabla U_s})^{2} \bigr ) \biggr \}\\
&&
+ O(\rho\epsilon^2) \,, 
\label{tau00PN}
 \\
\tau^{0i} &=& G(1-\zeta) \sigma^i + O(\rho\epsilon^{3/2}) \,,
\label{tau0jPN}
 \\
\tau^{ii} &=& G(1-\zeta) \biggl \{  \sigma^{ii} - \frac{1}{8\pi} G(1-\zeta) (\nabla U)^2 - \frac{1}{8\pi} G\zeta (\nabla U_s)^{2} \biggr \}
+ O(\rho\epsilon^2) \,, 
\label{tauPN}
 \\
\tau^{ij} &=&  O(\rho\epsilon) \,, 
\label{tauijPN}
 \\ \nonumber
\tau_s &=& G \zeta \biggl \{ \sigma_s + 2G (1-\zeta) \sigma_s U
 - 2G ( 2\lambda_1 +\zeta ) \sigma_s U_s
 + \frac{1}{2\pi} G( \lambda_1 -\zeta) (\nabla U_s)^{2}   \biggr \} \\
 &&
 + O(\rho \epsilon^2) 
 \,. 
\label{tausPN}
\end{eqnarray}
\label{tauPN}
\end{subequations}
Substituting into Eqs. (\ref{bigexpansion}), and calculating terms through 1.5PN order (e.g. $O(\epsilon^{5/2})$ in $N$), we obtain
\begin{subequations}
\label{postnewtonian}
\begin{eqnarray}
N_1 &=& G(1-\zeta) \biggl \{ 7G (1-\zeta) U^2 -4 \Phi_1
+2G (1-\zeta) \Phi_2+2 {\ddot X} 
\nonumber \\
&& - G \zeta  U_s^{2} -24G  \zeta \Phi_{2s}
+ 2 G \zeta  \Phi^s_{2s} \biggr \}
 \,, \\
K_{1}^i &=& 4G(1-\zeta) V^i \,,\\
B_1 &=& G(1-\zeta) \biggl \{ G(1-\zeta)U^2+4\Phi_1-2G(1-\zeta)\Phi_2 + G\zeta U_s^{2} - 2G\zeta \Phi^s_{2s} \biggr \} \,, \\
\Psi_1 &=& G\zeta \biggl \{ -2G ( \lambda_1-\zeta ) U_s^{2} 
+ 4G (1-\zeta)\Phi^s_2 -4 G( \lambda_1+ 2\zeta ) \Phi^s_{2s}
+  \ddot{X}_s \biggr \} \,,
\\
N_{1.5} &=& -{2 \over 3} \stackrel{(3)\qquad}{{\cal I}^{kk}(t)}
\,,\\
B_{1.5} &=& -{2} \stackrel{(3)\qquad}{{\cal I}^{kk}(t)} \,,
\\
\Psi_{1.5} &=& -2 \dot{M}_s(t) 
+ \frac{2}{3} x^j \stackrel{(3)\quad}{{\cal I}_s^{j}(t)}
- \frac{1}{3} \stackrel{(3)\qquad}{{\cal I}_s^{kk}(t)} \,.
\label{psi15}
\end{eqnarray}
\end{subequations}
In  Eq.\ (\ref{psi15}), we have used the fact (to be verified later) that, because of the conservation of baryon number, and assuming that our compact bodies have  stationary internal structure, $M_s(t)$ is constant to the lowest PN order.  Thus, rather than contributing to $\Psi_{0.5}$ as shown in Eq.\ (\ref{bigexpansiondPsi}), the term $-2\dot{M}_s$ contributes to $\Psi_{1.5}$; similarly the term in $\Psi_{1.5}$ involving three time derivatives of $M_s$ actually contributes to $\Psi_{2.5}$.

The physical metric to 1.5PN order is then given by
\begin{subequations}
\label{1.5pnmetric}
\begin{eqnarray}
g_{00} &=& -1 + 2G(1-\zeta)U + 2G\zeta U_s  - 2G^2 (1-\zeta)^2 U^2 
- 2G^2 \zeta ( \zeta + \lambda_1) U_s^{2}
\nonumber \\
&& 
-4G^2 \zeta (1-\zeta) U\,U_s
+4G^2 \zeta (1-\zeta) \Phi^s_2
-12 G^2 \zeta (1-\zeta) \Phi_{2s}
\nonumber \\
&&-4G^2 \zeta (2 \zeta + \lambda_1) \Phi^s_{2s}
+ G(1-\zeta) \ddot X 
+G\zeta {\ddot X}_s
\nonumber \\
&&
- {4 \over 3}\stackrel{(3)\qquad}{{\cal I}^{kk}(t)} 
-2 \dot{M}_s(t) 
+ \frac{2}{3} x^j \stackrel{(3)\quad}{{\cal I}_s^{j}(t)}
- \frac{1}{3} \stackrel{(3)\qquad}{{\cal I}_s^{kk}(t)}
+ O(\epsilon^3) \,,\\
g_{0i} &=& -4G(1-\zeta)V^i + O(\epsilon^{5/2}) \,,\\
g_{ij} &=& \delta_{ij} \biggl [1+2G(1-\zeta)U -2G\zeta U_s \biggr ] + O(\epsilon^2) \,.
\end{eqnarray}
\end{subequations}

\subsection{2PN and 2.5PN solutions}
\label{sec:22.5PNsolutions}

At 2PN and 2.5PN order, we obtain, from Eqs.~ (\ref{heq}), (\ref{Lambda}) and (\ref{Lambdascalar}),
\begin{subequations}
\begin{eqnarray}
\tau^{ij} &=& G(1-\zeta) \sigma^{ij} 
+ \frac{1}{4\pi} G^2 (1-\zeta)^2 \biggl [ U^{,i}U^{,j} - \frac{1}{2}
\delta^{ij} (\nabla U)^2 \biggr ] 
\nonumber \\
&& + \frac{1}{4\pi} G^2 \zeta (1-\zeta) \biggl [U_s^{,i}U_s^{,j} - \frac{1}{2}
\delta^{ij} (\nabla U_s)^2 \biggr ] 
+ O(\rho\epsilon^2) \,, \\
\tau^{0i} &=& G(1-\zeta) \sigma^i + G^2 (1-\zeta)^2 \biggl ( 4\sigma^i U
+ \frac{2}{\pi} U^{,j}V^{[j,i]}
+ \frac{3}{4\pi} \dot U U^{,i} \biggr )
\nonumber \\
&& 
- G^2 \zeta (1-\zeta) \biggl ( 6\sigma^i U_s
+ \frac{1}{4 \pi}  \dot{U}_s U_s^{,i} \biggr )
+ O(\rho\epsilon^{5/2}) \,.
\end{eqnarray}
\end{subequations}
Outer integrals and boundary terms contribute nothing, so we obtain
\begin{subequations}
\begin{eqnarray}\nonumber
B_2^{ij} &=& 4G(1-\zeta) \Phi_1^{ij} 
+ G^2 (1-\zeta)^2 \biggl [ 4P_2^{ij}-\delta^{ij}(2\Phi_2-U^2) \biggr ]\\
&&
+G^2 \zeta (1-\zeta) \biggl [ 4{P}_{2s}^{ij}-\delta^{ij}(2\Phi^s_{2s}-U_s^{2}) \biggr ] \,,
\\\nonumber
K_2^i &=& G^2 (1-\zeta)^2 \biggl ( 8V_2^i -8\Phi_2^i + 8UV^i + 16G_7^i \biggr )\\
&& + 2G(1-\zeta){\ddot X}^i -G^2 \zeta (1-\zeta) \biggl [ 24 V^i_{2s} +4 P(\dot{U}_s U_s^{,i} ) \biggr ] \,,
\\
B_{2.5}^{ij} &=&
-2 \stackrel{(3)\qquad}{{\cal I}^{ij}(t)}  \,, \\
K_{2.5}^i &=& \frac{2}{3} x^k \stackrel{(4)\qquad}{{\cal
I}^{ik}(t)}
- \frac{2}{9} \stackrel{(4)\qquad}{{\cal I}^{ikk}(t)}
+ \frac{4}{9} \epsilon^{mik}  \stackrel{(3)\qquad}{{\cal J}^{mk}(t)}
\,.
\end{eqnarray}
\end{subequations}
All solutions obtained so far must be substituted into Eqs.\ (\ref{heq}), (\ref{taustar}), (\ref{Lambda}) and (\ref{Lambdascalar}) to obtain $\tau^{00}$, $\tau^{ii}$ and $\tau_s$ to the required order,
\begin{subequations}
\begin{eqnarray}
\tau^{00} &=& G(1-\zeta) \biggl \{ \sigma - \sigma^{ii} 
+G(1-\zeta) \bigl ( 4\sigma U- \frac{7}{8\pi} (\nabla U)^2 \bigr ) 
- G\zeta \bigl ( 6\sigma U_s - \frac{1}{8\pi}  (\nabla U_s)^{2} \bigr ) \biggr \} 
\nonumber \\
&&
+ G^2 (1-\zeta)^2 \biggl \{ \sigma \biggl [ 7G(1-\zeta) U^2 - 8\Phi_1 + 2G (1-\zeta) \Phi_2 + 2\ddot{X} \biggr ]  - 4\sigma^{ii} U 
\nonumber \\
&& \quad + \frac{1}{4\pi} \biggl [ \frac{5}{2} \dot{U}^2 - 4U \ddot{U}
- 8 \dot{U}^{,k} V^k  +2V^{i,j}(3V^{j,i}+V^{i,j}) +4{\dot V}^jU^{,j}
-4U^{,ij}\Phi_1^{ij}
+8\nabla U \cdot \nabla \Phi_1
\nonumber \\
&& \quad -{7 \over 2}  \nabla U \cdot \nabla \ddot X
- G(1-\zeta) \left ( 4 \nabla U \cdot \nabla \Phi_2 
+10U(\nabla U)^2 
+4U^{,ij} P_2^{ij} \right ) \biggr ] \biggr \}
\nonumber \\
&&
+ G^2 \zeta (1-\zeta) \Biggl \{ \sigma \biggl [
G(6\lambda_1 - 1 + 19\zeta) U_s^{2}
-G(1-\zeta) \bigl ( 24 UU_s
+24 \Phi_{2s}
+12  \Phi_2^s \bigr )
\nonumber \\
&& \quad
+2 G( 6\lambda_1+1+ 11\zeta) \Phi_{2s}^s
-3 \ddot{X}_s \biggr ] + 6\sigma^{ii} U_s
\nonumber \\
&& \quad
+\frac{1}{4\pi} \biggl [ 
 G(1-\zeta) \biggl ( 2U (\nabla U_s)^2 
+4 U_s \nabla U \cdot \nabla U_s
+42  \nabla U \cdot \nabla \Phi_{2s}
+2 \nabla U_s \cdot \nabla \Phi_{2}^s
\nonumber \\
&& \quad
-4 \nabla U \cdot \nabla \Phi_{2s}^s
-4 U^{,ij} P_{2s}^{ij}   \biggr )
+\frac{1}{2} \dot{U}_s^{2} 
-2 G(\lambda_1 + 2\zeta)  \nabla U_s \cdot \nabla \Phi_{2s}^s
+\frac{1}{2} \nabla U_s \cdot \nabla \ddot{X}_s
 \biggr ] \Biggr \}
 \nonumber \\
&& 
+ G(1-\zeta) \biggl \{ \sigma \biggl [
 \frac{4}{3}   \stackrel{(3)\qquad}{{\cal I}^{kk}(t)} + 6\dot{M}_s(t)
 -2 x^j  \stackrel{(3)\quad}{{\cal I}_s^{j}(t)}
 +  \stackrel{(3)\quad}{{\cal I}_s^{kk}(t)} \biggr ]
+ \frac{1}{2\pi} U^{,ij}   \stackrel{(3)\quad}{{\cal I}^{ij}(t)}
+ \frac{1}{12\pi} U_s^{,j}  \stackrel{(3)\quad}{{\cal I}_s^{j}(t)} \biggr \} 
 \nonumber \\
&& +O(\rho \epsilon^3) \,,
\\
\tau^{ii} &=& G(1-\zeta) \biggl \{ \sigma^{ii}  
- \frac{1}{8\pi} G(1-\zeta)(\nabla U)^2 
-\frac{1}{8\pi} G\zeta (\nabla U_s)^{2} \biggr \}
 \nonumber \\
&&
+ G^2 (1-\zeta)^2 \biggl \{ 4 \sigma^{ii} U 
- \frac{1}{4\pi} \biggl [ {9 \over 2}{\dot U}^2 
+4V^{i,j}V^{[i,j]} +4{\dot V}^jU^{,j} 
+{1 \over 2} \nabla U \cdot \nabla \ddot X \biggr ] \biggr \}
 \nonumber \\
&&
- G^2 \zeta(1-\zeta) \biggl \{ 6 \sigma^{ii} U_s 
- \frac{1}{4\pi} \biggl [ \frac{3}{2} \dot{U}_s^{2} 
-G(1-\zeta) \bigl ( 2\nabla U_s \cdot \nabla \Phi_2^s 
-6 \nabla U \cdot \nabla \Phi_{2s} \bigr )
 \nonumber \\
&& \quad
+2G (\lambda_1 + 2\zeta) \nabla U_s \cdot \nabla \Phi_{2s}^s 
-\frac{1}{2}  \nabla U_s \cdot \nabla \ddot{X}_s \biggr ] \biggr \}-\frac{1}{12\pi} G(1-\zeta)U_s^{,j}  \stackrel{(3)\quad}{{\cal I}_s^{j}(t)}  +O(\rho \epsilon^3)\,, \\
\tau_s &=& G\zeta \biggl \{ \sigma_s + 2G(1-\zeta) \sigma_s U
-2G (2\lambda_1 + \zeta) \sigma_s U_s 
+ \frac{1}{2\pi} G (\lambda_1 - \zeta) (\nabla U_s)^{2} \biggr \}
\nonumber \\
&&
+ G^2 \zeta \sigma_s \biggl \{ 
G(1-\zeta) \biggl [ 2(1-\zeta) U^2 
- 4(2\lambda_1+\zeta) \bigl (UU_s + \Phi_2^s \bigr )
-12 \zeta \Phi_{2s}  \biggr ] 
-(1-\zeta) \bigl (  4\Phi_1 - \ddot{X} \bigr )
\nonumber \\
&& \quad
+ G (20 \lambda_1^2 - 4 \lambda_2 + 6 \zeta \lambda_1 +2\zeta^2 ) U_s^{2}
+4G (2\lambda_1 + \zeta)(\lambda_1 + 2\zeta) \Phi_{2s}^s
- (2\lambda_1 + \zeta) \ddot{X}_s \biggr \}
\nonumber \\
&&
- \frac{1}{8\pi} G^2 \zeta \biggl \{
(1-\zeta) \bigl ( 8U \ddot{U}_s + 16 V^j \dot{U}_s^{,j} + 8 \Phi_1^{ij} U_s^{,ij} \bigr ) 
+ 4(\lambda_1 - \zeta) \bigl ( \dot{U}_s^{2} -  \nabla U_s \cdot \nabla \ddot{X}_s \bigr )
\nonumber \\
&& \quad
- G(1-\zeta) \biggl [ 16 (\lambda_1-\zeta) \nabla U_s \cdot \nabla \Phi_{2}^s
- 8(1-\zeta) U_s^{,ij} P_2^{ij} - 8\zeta U_s^{,ij} P_{2s}^{ij} \biggr ]
\nonumber \\
&& \quad
+ 16 G (\lambda_1 +2\zeta)(\lambda_1 - \zeta)\nabla U_s \cdot \nabla \Phi_{2s}^s 
-8 G (\lambda_2 -4\lambda_1^2 + 4\zeta \lambda_1 - \zeta^2 ) U_s (\nabla U_s)^{2} \biggr \} 
\nonumber \\
&&
+ G\biggl \{ \sigma_s \left [
 \frac{2}{3}  \zeta \stackrel{(3)\qquad}{{\cal I}^{kk}(t)} + \frac{1}{3} (2\lambda_1 + \zeta) \left (6\dot{M}_s(t) 
 -2 x^j  \stackrel{(3)\quad}{{\cal I}_s^{j}(t)}
 +  \stackrel{(3)\qquad}{{\cal I}_s^{kk}(t)} \right ) \right ]
 \nonumber \\
&& \quad
+ \frac{1}{2\pi} \zeta U_s^{,ij}   \stackrel{(3)\quad}{{\cal I}^{ij}(t)}
+ \frac{1}{3\pi} (\lambda_1 - \zeta) U_s^{,j}  \stackrel{(3)\quad}{{\cal I}_s^{j}(t)} \biggr \} 
+ O(\rho \epsilon^3) \,.
\end{eqnarray}
\end{subequations}
Substituting into Eqs.\ (\ref{bigexpansiona}), (\ref{bigexpansionc}) and (\ref{bigexpansiondPsi}) 
and evaluating terms
through $O(\epsilon^{7/2})$, and verifying that the outer integrals and
surface terms make no ${\cal R}$-independent contributions, we obtain,
\begin{subequations}
\begin{eqnarray}
N_2 &=& G(1-\zeta) \biggl \{ \frac{1}{6} \stackrel{(4)}{Y} 
- 2 \ddot{X}_1
+ G(1-\zeta) \biggl [ 7U\ddot{X} -16 U\Phi_1 -4 V^i V^i 
-16 \Sigma(\Phi_1) + \Sigma(\ddot{X}) + 8 \Sigma^i (V^i) 
+ \ddot{X}_2
 \nonumber \\
&& \quad
 - 4G_1 - 16G_2 + 32 G_3 + 24 G_4 
- 16 G_5 - 16 G_6 \biggr ]
+G^2(1-\zeta)^2 \biggl [  8U\Phi_2 + \frac{20}{3} U^3 -16 H \biggr ]
\biggr \}
 \nonumber \\
&&
+ G^2 \zeta (1-\zeta) \biggl \{
24 \Sigma^{ii} (U_s) -U_s \ddot{X}_s -12 \Sigma(\ddot{X}_s)
+ \Sigma_s (\ddot{X}_s) -12 \ddot{X}_{2s} + \ddot{X}_{2s}^s
+4 G_{1s}  
 \nonumber \\
&& \quad
+ G(1-\zeta) \biggl [ 8U \Phi_{2s}^s - 4 U_s \Phi_2^s - 84 U \Phi_{2s} - 
4 UU_s^{2} -12\Sigma(\Phi_{2s}) -48 \Sigma(\Phi_2^s)
 \nonumber \\
&& \quad
+4 \Sigma_s(\Phi_2^s) +4\Sigma_s(UU_s) -12 \Sigma(UU_s) -16H_s
\biggr ] + 24 G(\lambda_1 + 3\zeta) \Sigma(U_s^{2}) \nonumber \\
&& \quad
+ 4G(\lambda_1 + 2\zeta)\biggl [ 12 \Sigma(\Phi_{2s}^s)
- \Sigma_s(\Phi_{2s}^s) - \Sigma_s(U_s^{2}) +  U_s \Phi_{2s}^s
 \biggr ]
\biggr \}  \,,
\\
B_2 &=& G(1-\zeta) \biggl \{ 2\ddot{X}_1 
+ G(1-\zeta) \biggl [ U\ddot{X} + 4 V^iV^i - \Sigma(\ddot{X})
-8 \Sigma^i(V^i) +16 \Sigma^{ii}(U) 
 \nonumber \\
&& \quad
 - \ddot{X}_2 - 20G_1
+ 8G_4
+16 G_5 \biggr ] \biggr \}
+ G^2 \zeta(1-\zeta) \biggl \{ U_s \ddot{X}_s -24 \Sigma^{ii} (U_s) 
-\Sigma_s (\ddot{X}_s)  -\ddot{X}_{2s}^s
+4 G_{1s}
 \nonumber \\
&& \quad
+ G(1-\zeta) \biggl [ 4 U_s \Phi_2^s - 12 U \Phi_{2s} 
 +12\Sigma(\Phi_{2s}) 
-4 \Sigma_s(\Phi_2^s) -4\Sigma_s(UU_s) +12 \Sigma(UU_s) 
\biggr ]
 \nonumber \\
&& \quad
+ 4G (\lambda_1 + 2\zeta)\biggl [ 
 \Sigma_s(\Phi_{2s}^s) + \Sigma_s(U_s^{2}) -  U_s \Phi_{2s}^s
 \biggr ] \biggr \} \,,
 \\
\Psi_2 &=&
G \zeta \biggl \{ \frac{1}{12} \stackrel{(4)}{Y_s} + G(1-\zeta) \biggl [ 2 \Sigma_s(\ddot{X}) 
-8 \Sigma_s(\Phi_1)
-8 G_{2s} + 16 G_{3s} - 8G_{6s} + 2\ddot{X}_2^s \biggr ]
  \nonumber \\
&& \quad
 -2 G(\lambda_1 +2\zeta) \bigl ( \Sigma_s(\ddot{X}_s) 
+ \ddot{X}_{2s}^s \bigr )
-2G  (\lambda_1-\zeta) U_s\ddot{X}_s
- 8G^2 (1-\zeta)(\lambda_1 +2\zeta) \biggl [\Sigma_s (\Phi_2^s)
+ \Sigma_s(UU_s) \biggr ]
 \nonumber \\
&& \quad
+ 8 G^2 (\lambda_1 +2\zeta) \biggl [
 (\lambda_1-\zeta) U_s \Phi_{2s}^s
 +(\lambda_1 +2\zeta) \Sigma_s(\Phi_{2s}^s)
  \biggr ] -8 G^2  (1-\zeta)(\lambda_1-\zeta) U_s \Phi_2^s 
  \nonumber \\
&& \quad
+G^2 (1-\zeta)^2 \bigl ( 4\Sigma_s(U^2) -8H^s \bigr )
- G^2 \zeta (1-\zeta)  \bigl ( 24 \Sigma_s(\Phi_{2s}) + 8H_s^s \bigr )
 \nonumber \\
&& \quad
- \frac{4}{3} G^2 ( \lambda_2 -4\lambda_1^2 + 4\zeta \lambda_1 - \zeta^2 ) U_s^{3}
-4 G^2 (\lambda_2-4\lambda_1^2 -5\zeta\lambda_1-4\zeta^2) \Sigma_s(U_s^{2})  \biggr \} \,,
\\
N_{2.5} &=& 
 -\frac{1}{15}(2x^{kl}+r^2\delta^{kl})\stackrel{(5)\quad}{{\cal
I}^{kl}(t)} + \frac{2}{15} x^k\stackrel{(5)\qquad}{{\cal I}^{kll}(t)}
- \frac{1}{30} \stackrel{(5)\qquad}{{\cal I}^{kkll}(t)}
+ G(1-\zeta) \biggl [ \frac{16}{3} U\stackrel{(3)\qquad}{{\cal I}^{kk}(t)}
-4X^{,kl}\stackrel{(3)\qquad}{{\cal I}^{kl}(t)} 
\nonumber \\
&&
\qquad +24 U \dot{M}_s(t)
-8(x^k U - X^{,k})\stackrel{(3)\quad}{{\cal I}_s^{k}(t)}
+4U \stackrel{(3)\qquad}{{\cal I}_s^{kk}(t)}
-\frac{2}{3} X_s^{,k}\stackrel{(3)\quad}{{\cal I}_s^{k}(t)} \biggr ]
\,, \\
B_{2.5} &=&  
-\frac{1}{3}  r^2 \stackrel{(5)\quad}{{\cal I}^{kk}(t)}
+\frac{2}{9} x^k \stackrel{(5)\qquad}{{\cal I}^{kll}(t)}
+ \frac{8}{9}  x^k \epsilon^{mkj} \stackrel{(4)\qquad}{{\cal J}^{mj}(t)}
- \frac{2}{3}  \stackrel{(3)\qquad}{M^{kkll}(t)}
+\frac{2}{3}G(1-\zeta) X_s^{,k}\stackrel{(3)\quad}{{\cal I}_s^{k}(t)} \,,
\\
\Psi_{2.5} &=&
-\frac{1}{30}(2x^{kl}+r^2\delta^{kl})\stackrel{(5)\quad}{{\cal
I}_s^{kl}(t)} 
+ \frac{1}{15} x^k\stackrel{(5)\qquad}{{\cal I}_s^{kll}(t)}
- \frac{1}{60} \stackrel{(5)\qquad}{{\cal I}_s^{kkll}(t)}
+ \frac{1}{15} r^2 x^k\stackrel{(5)\quad}{{\cal I}_s^{k}(t)}
-\frac{1}{3} r^2 \stackrel{(3)\quad}{M_s(t)}
\nonumber \\
&&
+ G\zeta \biggl [ \frac{4}{3} U_s \stackrel{(3)\qquad}{{\cal I}^{kk}(t)}
-2 X_s^{,kl} \stackrel{(3)\quad}{{\cal I}^{kl}(t)} \biggr ]
+ \frac{4}{3} G(\lambda_1+2\zeta) X_s^{,k} \stackrel{(3)\quad}{{\cal I}_s^{k}(t)}
\nonumber \\
&&
+\frac{2}{3}G(2\lambda_1 +\zeta) U_s \biggl [ 6\dot{M}_s(t) 
- 2x^k\stackrel{(5)\quad}{{\cal I}_s^{k}(t)}
+  \stackrel{(3)\quad}{{\cal I}_s^{kk}(t)} \biggr ] \,.
\end{eqnarray}
\end{subequations}
 
\begin{savequote}[0.55\linewidth]
{\scriptsize ``In questions of science, the authority of a thousand is not worth the humble reasoning of a single individual.''}
\qauthor{\scriptsize---Galileo Galilei}
\end{savequote}
\chapter{Matter Source and Equations of motion} 
\label{chapter10} 
\thispagestyle{myplain}
\lhead[\thepage]{Chapter 9. \emph{Matter Source and Equations of motion}}      
\rhead[Chapter 9. \emph{Matter Source and Equations of motion}]{\thepage}

\section{Energy-Momentum Tensor and The Conserved Density}
\label{sec:conserveddensity}

We now must expand the effective energy-momentum tensor, Eq.\ (\ref{Tmunu}) in a PN expansion to the required order, including the $\phi$ dependence of the masses $M_A$.  
We first expand $M_A (\phi)$ about the asymptotic value $\phi_0$:  
\begin{equation}
M_A(\phi) = M_{A0} + \delta\phi \left (\frac{dM_A}{d\phi} \right )_0
+ \frac{1}{2} \delta \phi^2  \left (\frac{d^2M_A}{d\phi^2} \right )_0
+ \frac{1}{6} \delta \phi^3  \left (\frac{d^3M_A}{d\phi^3} \right )_0
+ \dots \,.
\end{equation}  
We then define the dimensionless ``sensitivities''
\begin{eqnarray}
s_A &\equiv& \left ( \frac{d \ln M_A(\phi)}{d \ln \phi} \right )_0 \,,
\nonumber \\
s'_A &\equiv& \left ( \frac{d^2 \ln M_A(\phi)}{d (\ln \phi)^2} \right )_0 \,,
\nonumber \\
s''_A &\equiv& \left ( \frac{d^3 \ln M_A(\phi)}{d (\ln \phi)^3} \right )_0 \,.
\end{eqnarray}
Note that the definition of $s'_A$ used in \cite{tegp} and \cite{alsing} has the opposite sign from our definition.
Recalling that $\phi = \phi_0 (1 + \Psi)$ we can write
\begin{eqnarray}
M_A(\phi) &=&
 m_{A}\left [1  +  s_A \Psi + \frac{1}{2}  (s_A^2 +s'_A   -s_A) \Psi^2 
 \right .
 \nonumber \\
&&\left .+ \frac{1}{6}  (s''_A + 3 s'_A s_A - 3s'_A + s_A^3 - 3s_A^2 + 2s_A)\Psi^3 + O(\Psi^4)  \right ] 
\nonumber \\
&\equiv& m_{A} \left [1 +  {\cal S}(s_A; \Psi) \right ] \,,
\end{eqnarray}
where we define the constant mass for each body $m_A \equiv M_{A0}$ and the definition of ${\cal S}(s_A; \Psi)$ is clearly given above in terms of the sensitivities.

In general relativity, neglecting pressure, the stress energy tensor can be written as (see Eqs.~(\ref{Tstar}))
\begin{equation}
T^{\mu\nu} = \rho^* (-g)^{-1/2} u^\mu u^\nu / u^0 \,,
\end{equation}
where $\rho^*$ is identified as the ``baryonic'', or ``conserved'' mass density, $\rho^* = mn\sqrt{-g} \,u^0$, where $n$ is the number density of baryons, and $m$ is the rest mass per baryon.  It satisfies an exact continuity equation $\partial \rho^*/\partial t + \nabla \cdot (\rho^* {\bf v} ) = 0$, and implies that the baryonic mass of any isolated body is constant.   Here we identify the ``baryons'' as our compact point masses with constant mass $m_{A}$, so that 
\begin{equation}
\rho^* = \sum_A m_{A} \delta^3 ({\bf x} - {\bf x}_A) \,,
\label{rhostarsum}
\end{equation}
Thus, we can rewrite Eq.\ (\ref{Tmunu}) in the form
\begin{equation}
T^{\mu\nu} = \rho^* (-g)^{-1/2} u^0 v^\mu v^\nu  \left [1 +  {\cal S}(s; \Psi) \right ] \,,
\end{equation}
where $\rho^*$ is given by Eq.\ (\ref{rhostar}), and where we have substituted $u^\mu = u^0 v^\mu$, with $v^\mu = dx^\mu/dt =  (1, {\bf v})$ being the ordinary velocity.  We have dropped the subscript from the variable $s$ in $\cal S$ because it will be assigned a label $A$ wherever the delta function that is implicit in $\rho^*$ corresponds to body $A$.   Thus, we arrive at a conversion from the $\sigma$-densities of Eq.\ (\ref{sigmas}) to $\rho^*$, given by
\begin{eqnarray}
\sigma &=& \rho^*  (-g)^{-1/2} u^0 (1+v^2)  \left [1 +  {\cal S}(s; \Psi) \right ]
 \,,
\nonumber \\
\sigma^i &=&  \rho^*  (-g)^{-1/2} u^0 v^i  \left [1 +  {\cal S}(s; \Psi) \right ]
 \,,
\nonumber \\
\sigma^{ij} &=&  \rho^*  (-g)^{-1/2} u^0 v^i v^j  \left [1 +  {\cal S}(s; \Psi) \right ]
 \,.
\end{eqnarray} 
To convert $\sigma_s$, recall that 
\begin{eqnarray}
T &=& g_{\mu\nu} T^{\mu\nu}  
\nonumber \\
&=&
-\rho^* (-g)^{-1/2} (u^0)^{-1}  \left [1 +  {\cal S}(s; \Psi) \right ] \,,
\end{eqnarray}
and that $\varphi = 1 +  \Psi$, $\partial/\partial \varphi =  \partial/\partial \Psi$.  Consequently
\begin{eqnarray}
\sigma_s &=& - T + 2\varphi \frac{\partial T}{\partial \varphi} 
\nonumber \\
&=& \rho^* (-g)^{-1/2} (u^0)^{-1} 
\left [ 1 +  {\cal S} - 2(1+ \Psi) \frac{\partial {\cal S}}{\partial \Psi} \right ]  
\nonumber \\
&=& \rho^* (-g)^{-1/2} (u^0)^{-1} \bigl [ (1-2s) + {\cal S}_s (s; \Psi) \bigr ] \,.
\end{eqnarray}
Defining
\begin{eqnarray}
a_s &\equiv& s^2 +s'   - \frac{1}{2} s  \,,
\nonumber \\
{a_s}' &\equiv& s'' + 2ss' - \frac{1}{2} s'  \,,
\nonumber \\
b_s &\equiv& {a_s}' - a_s + sa_s \,,
\end{eqnarray}
we can write
\begin{eqnarray}
{\cal S} (s; \Psi) &=& s \Psi + \frac{1}{4} (2a_s -s) \Psi^2 
+ O(\Psi^3) \,,
\nonumber \\
{\cal S}_s (s; \Psi) &=&  - 2a_s \Psi - b_s \Psi^2 
+ O(\Psi^3) \,.
\end{eqnarray}

Substituting the expansion for the metric, Eq.\ (\ref{metric}), and for the metric potentials, Eq.\ (\ref{expandNKB}), we obtain to the 2.5PN order required for the equations of motion,
\begin{subequations}
\begin{eqnarray}
\sigma &=& \rho^*\left[1+\epsilon \left(\frac{3}{2} v^2 -G(1-\zeta) U_\sigma + G\zeta (5+2s) U_{s \sigma} \right) 
+\epsilon^2  \left(  \frac{7}{8} v^4 
+\frac{5}{2} G^2 (1-\zeta)^2 U_\sigma^2  \right. \right.
\nonumber
\\
&& 
\left. \left .+ \frac{1}{2} G(1-\zeta) v^2 U_\sigma -4 G (1-\zeta) v^i V_\sigma^i +\frac{3}{2} (5+2s) G\zeta v^2 U_{s \sigma}
-  (5 + 2s) G^2 \zeta (1-\zeta)U_\sigma
U_{s \sigma} 
\right. \right.
\nonumber
\\
&& \left. \left . 
+\frac{1}{2} (15 + 18s + 4 a_s) G^2 \zeta^2 U_{s \sigma}^2
+\frac{3}{4} B_1-\frac{1}{4} N_1 
+\frac{1}{2} (5+ 2s)  \Psi_1 \right) 
\right .
\nonumber
\\
&&
\left. + \epsilon^{5/2}\left(2 N_{1.5}+\frac{1}{2} (5+ 2s)  \Psi_{1.5}\right) 
+{O}(\epsilon^3)  \right]  \,,
\label{sigmaPN}
\\
\sigma^i&=&\rho^*v^i\left[1+\epsilon \biggl( \frac{1}{2} v^2 
- G(1-\zeta)U_\sigma
+G \zeta (5+2s) U_{s \sigma} \biggr) 
+{O}(\epsilon^2) \right]  \,,
\label{sigmaiPN}
\\
\sigma^{ij}&=&\rho^* v^i v^j \biggl [1 +{O}(\epsilon) \biggr ] \,,
\label{sigmaijPN}
\\
\sigma^{ii}&=&\rho^* v^2 \left[1+\epsilon\biggl( \frac{1}{2}v^2
-G(1-\zeta)U_\sigma
+G \zeta (5+2s) U_{s \sigma}\biggr)+{O}(\epsilon^2) \right] \,,
\label{sigmaiiPN}
\\
\sigma_s &=& \rho^*\left [ (1-2s) - \epsilon  \left \{ \frac{1}{2}  (1-2s) v^2 + 3 G (1-\zeta)  (1-2s) U_\sigma 
- 3 G \zeta \left ( 1-2s - \frac{4}{3} a_s \right ) U_{s \sigma} \right \}
\right .
\nonumber
\\
&& \left .+\epsilon^2\biggr\{ 
-\frac{1}{8} (1-2s)v^4
+ \frac{21}{2} G^2(1-\zeta)^2 (1-2s) U_\sigma^2
-\frac{1}{2}G(1-\zeta)(1-2s)v^2 U_\sigma
\right .
\nonumber
\\
&&
\left . 
+4G(1-\zeta)(1-2s)v^iV_\sigma^i 
-\frac{3}{2} G \zeta \left((1-2s) -\frac{4}{3}a_s\right)v^2 U_{s\sigma}
\right .
\nonumber
\\
&&
\left .
-9 G^2 \zeta (1-\zeta) \left ( 1-2s - \frac{4}{3}a_s \right ) U_\sigma U_{s\sigma} + \frac{3}{2} G^2 \zeta^2 \left ( 1-2s- 8a_s- \frac{8}{3} b_s\right )
U_{s\sigma}^2
\right .
\nonumber
\\
&&
\left.
+\frac{1}{4} (1-2s)B_1
-\frac{3}{4}(1-2s)N_1 
+\frac{3}{2}\left(1-2s-\frac{4}{3}a_s\right) \Psi_1\biggr\}
\right.
\nonumber
\\
&&
\left .
+\epsilon^{5/2} \biggr\{ \frac{3}{2} \left(1-2s-\frac{4}{3} a_s\right) \Psi_{1.5}\biggr\}+{O}(\epsilon^3)  \right] \,,
\label{sigmasPN}
\end{eqnarray}
\label{sigmaPN}
\end{subequations}
where $U_\sigma$, $U_{s\sigma}$ and $V^i_{\sigma}$ are defined in terms of the $\sigma$-densities.  

Substituting these formulas into the definitions of $U_\sigma$, $U_{s\sigma}$ and the other potentials defined in terms of $\sigma$, we can convert all potentials into new versions defined in terms of $\rho^*$, plus PN corrections.  For example, we find that the ``Newtonian'' potentials $U_\sigma$ and $U_{s\sigma}$ become
\begin{eqnarray}
U_\sigma&=&
U+
\epsilon \left\{
\frac{3}{2}\Phi_1-G(1-\zeta)\Phi_2
+ 6G\zeta\Phi_{2s}- G\zeta\Phi_{2s}^s\right\}
\nonumber
\\
&&
+\epsilon^2 \left\{
\frac{7}{8}\Sigma(v^4)
+ \frac{5}{2}G(1-\zeta)\Sigma(\Phi_1)
+\frac{1}{2}G(1-\zeta)\Sigma(v^2 U)
-4G(1-\zeta)\Sigma(v^i V^i)
\right.
\nonumber
\\
&&
\quad-\frac{1}{2}G(1-\zeta)\,\Sigma(\ddot{X})-G^2(1-\zeta)^2\Sigma(\Phi_2) 
+\frac{3}{2}G^2(1-\zeta)^2\Sigma(U^2)
+9G\zeta\Sigma(v^2 U_s)
\nonumber
\\
&&
 \quad-\frac{3}{2}G\zeta \Sigma_s(v^2 U_s)
+\frac{1}{2}G\zeta\,\Sigma_s(\Phi_1^s)-3G\zeta\,\Sigma(\Phi_1^s) 
+3G\zeta\,\Sigma(\ddot{X}_s)
-\frac{1}{2}G\zeta\,\Sigma_s(\ddot{X}_s)
\nonumber
\\
&&
\quad-G^2\zeta(1+12\lambda_1+5\zeta)\Sigma(\Phi_{2s}^s)
+G^2\zeta(2\lambda_1+\zeta)\Sigma_s(\Phi_{2s}^s)
+G^2\zeta(1+17\zeta-6\lambda_1)\Sigma(U_s^2)
\nonumber
\\
&&
\quad- \frac{1}{2}G^2\zeta(11\zeta-2\lambda_1)\Sigma_s(U_s^2)
-6G^2\zeta(1-\zeta)\Sigma(U U_s)
+G^2\zeta(1-\zeta)\Sigma_s(U U_s)
\nonumber
\\
&&
\quad+2G^2\zeta^2\Sigma(a_s U_s^2)-6G^2\zeta(1-\zeta)\Sigma(\Phi_2^s) 
+G^2\zeta(1-\zeta)\Sigma_s( \Phi_2^s) 
 -24 G^2 \zeta^2\,\Sigma(\Sigma(a_s U_s)) 
 \nonumber
\\
&&\quad+4G^2\zeta^2\,\Sigma_s(\Sigma(a_s U_s)) 
  \biggr \}
 \nonumber
\\
&&
+\epsilon^{5/2}\biggl\{ 
-{4\over 3}\stackrel{(3)\quad}{{\cal I}^{kk}(t)} U 
-\frac{1}{6}\stackrel{(3)\quad}{{\cal I}_s^{kk}(t)} (6U - U_s )
+ \frac{1}{3} \stackrel{(3)\quad}{{\cal I}_s^{j}(t)} 
\left ( 6x^j U - x^j U_s - 6X^{,j} + X_s^{,j} \right )
\nonumber
\\
&&
-\dot{M}_s(t) (6U - U_s) \biggr\}+{O}(\epsilon^3) \,,\\
U_{s\sigma}&=&U_s
+\epsilon \biggr\{
-\frac{1}{2}\Phi_1^s
-3G(1-\zeta)\Phi_2^s
+3G\zeta\Phi_{2s}^s
-4G\zeta \Sigma(a_s U_s)
 \biggr\}
\nonumber
\\
&&+\epsilon^2 \biggl \{
-\frac{1}{8}\Sigma_s(v^4)
-\frac{1}{2}G(1-\zeta)\Sigma_s(\Phi_1)
-\frac{1}{2}G(1-\zeta)\Sigma_s(v^2 U)
+4G(1-\zeta)\Sigma_s(v^i V^i)
\nonumber
\\
&&
\quad-\frac{3}{2}G(1-\zeta)\Sigma_s(\ddot{X})
+G^2(1-\zeta)^2\Sigma_s(\Phi_2)
+\frac{11}{2}G^2(1-\zeta)^2\Sigma_s(U^2)
-\frac{3}{2}G\zeta\Sigma_s(v^2 U_s)
\nonumber
\\
&&
\quad+2G\zeta\Sigma(a_s v^2 U_s) 
-\frac{3}{2}G\zeta\Sigma_s(\Phi_1^s)
+2G\zeta\Sigma(a_s \Phi_1^s)-2G\zeta\Sigma(a_s\ddot{X}_s)
+\frac{3}{2}G\zeta\Sigma_s(\ddot{X}_s)
\nonumber
\\
&&
\quad+4G^2\zeta(2\lambda_1+\zeta)\Sigma(a_s \Phi_{2s}^s)
+G^2\zeta(1-4\zeta-6\lambda_1)\Sigma_s(\Phi_{2s}^s)
-4G^2\zeta(4\zeta-\lambda)\Sigma(a_s U_s^2) 
\nonumber
\\
&&
\quad+ \frac{1}{2}G^2\zeta(2+7\zeta-6\lambda_1)\Sigma_s(U_s^2)
-4G^2\zeta^2\Sigma(b_s U_s^2) 
-9G^2\zeta(1-\zeta)\Sigma_s(U U_s)
\nonumber
\\
&&
\quad+12G^2\zeta(1-\zeta)\Sigma(a_s U U_s)
-12G^2\zeta^2\Sigma_s(\Sigma(a_s U_s))
-3G^2\zeta(1-\zeta)\Sigma_s(\Phi_2^s)
\nonumber
\\
&&
\quad+4G^2\zeta(1-\zeta)\Sigma(a_s\Phi_2^s)
+16 G^2\zeta^2\,\Sigma(a_s \Sigma(a_s U_s)) 
 \biggr \}
\nonumber
\\
&&
+\epsilon^{5/2}\biggl \{  
-\frac{1}{6} \stackrel{(3)\quad}{{\cal I}_s^{kk}(t)} \left (3U_s - 4  \Sigma(a_s) \right )
+ \frac{1}{3} \stackrel{(3)\quad}{{\cal I}_s^{j}(t)}  \left ( 3 x^jU_s - 4x^j \Sigma (a_s) - 3 X_s^{,j} + 4 X (a_s)^{,j} \right ) 
\nonumber
\\
&&
\quad-\dot{M}_s(t)  \left ( 3U_s - 4 \Sigma(a_s) \right ) \biggr \}+{O}(\epsilon^3) \,,
\end{eqnarray}
while the relevant PN potentials become
\begin{eqnarray}
\Phi_{1\sigma} &=& \Phi_1 
+\epsilon \left\{
\frac{1}{2} \Sigma(v^4) - G(1- \zeta) \Sigma(v^2 U) 
+ 6G\zeta \Sigma(v^2 U_s) - G \zeta \Sigma_s (v^2 U_s)
\right\}
\nonumber
\\
&&
+{O}(\epsilon^2)\,,
\\
\Phi_{2\sigma} &=& \Phi_2
+\epsilon \biggl \{
\frac{3}{2} \Sigma(v^2 U) + \frac{3}{2} \Sigma(\Phi_1)
- G(1- \zeta) \Sigma(U^2)
- G(1- \zeta) \Sigma(\Phi_2)
\nonumber 
\\
&& 
+ 6G\zeta \Sigma(UU_s) - G\zeta \Sigma_s(UU_s) 
+6G\zeta \Sigma(\Phi_{2s}) - G\zeta \Sigma(\Phi_{2s}^s)
\biggr \}
+{O}(\epsilon^2)\,,
\\
\Phi^s_{2\sigma} &=& \Phi^s_2
+\epsilon \biggl \{
-\frac{1}{2} \Sigma_s(v^2 U) + \frac{3}{2} \Sigma_s(\Phi_1)
-3 G(1- \zeta) \Sigma_s(U^2)
- G(1- \zeta) \Sigma_s(\Phi_2)
\nonumber 
\\
&& 
+ 3G\zeta \Sigma_s (UU_s)
-4 G\zeta \Sigma(a_s UU_s)
+6G\zeta \Sigma_s(\Phi_{2s}) 
- G\zeta \Sigma_s(\Phi_{2s}^s)
\biggr \}
+{O}(\epsilon^2)\,,
\\
\Phi_{2s\sigma} &=& \Phi_{2s}
+\epsilon \biggl \{
\frac{3}{2} \Sigma(v^2 U_s) - \frac{1}{2} \Sigma(\Phi_1^s)
- G(1- \zeta) \Sigma(UU_s)
- 3G(1- \zeta) \Sigma(\Phi_2^s)
\nonumber 
\\
&& 
+ 6G\zeta \Sigma(U_s^2)
- G\zeta \Sigma_s(U_s^2)
+3 G\zeta \Sigma(\Phi_{2s}^s)
-4 G\zeta \Sigma(\Sigma(a_s U_s))
\biggr \}
+{O}(\epsilon^2)\,,
\\
\Phi_{2s\sigma}^s &=& \Phi_{2s}^s
+\epsilon \biggl \{
-\frac{1}{2} \Sigma_s(v^2 U_s) - \frac{1}{2} \Sigma_s(\Phi_1^s)
- 3G(1- \zeta) \Sigma_s(UU_s)
- 3G(1- \zeta) \Sigma_s(\Phi_2^s)
\nonumber 
\\
&& 
+ 3G\zeta \Sigma_s(U_s^2)
- 4 G\zeta \Sigma(a_s U_s^2)
+3 G\zeta \Sigma_s(\Phi_{2s}^s)
-4 G\zeta \Sigma_s(\Sigma(a_s U_s))
\biggr \}
\nonumber
\\
&&
+{O}(\epsilon^2)\,,
\\
\ddot{X}_\sigma &=& \ddot{X} 
+\epsilon \biggl \{\frac{3}{2} \ddot{X}(v^2) - G(1-\zeta) \ddot{X}(U)
+ 6G \zeta \ddot{X}(U_s) - G\zeta \ddot{X}_s (U_s) 
\biggr \}
+{O}(\epsilon^2)\,,
\\
\ddot{X}_{s\sigma} &=& \ddot{X}_s 
+\epsilon \biggl \{- \frac{1}{2} \ddot{X}_s(v^2) - 3G(1-\zeta) \ddot{X}_s(U)
+ 3G \zeta \ddot{X}_s(U_s) - 4G\zeta \ddot{X}(a_s U_s) 
\biggr \}
\nonumber
\\
&&
+{O}(\epsilon^2)\,,
\\
V_\sigma^i&=&V^i+\epsilon \left\{\frac{1}{2}\Sigma(v^i v^2)-G(1-\zeta) V_2^i+6G\zeta V_{2s}^i-G\zeta \Sigma_s(v^i U_s)  \right\}
+{O}(\epsilon^2)\,,
\end{eqnarray}
where all potentials are now defined
in terms of the density $\rho^*$, and including, where needed, the sensitivity factors $s$, $a_s$ and $b_s$.  In manipulating these expressions, we have made use of the identities, valid for any function $f$,  $\Sigma(sf) = [\Sigma(f) - \Sigma_s(f)]/2$ and $\Sigma(x^i \, f) = x^i \Sigma(f) - X^{,i}(f)$.  The potentials $U$ and $U_s$ will henceforth be given by 
\begin{eqnarray}
U &=& \int_{\cal M} {{\rho^*(t,{\bf x}^\prime)}
\over {\mid {\bf x}-{\bf x}^\prime \mid  }} d^3x^\prime \,,
\nonumber \\
U_s &=&  \int_{\cal M} {\bigl (1-2s({\bf x}^\prime) \bigr ){\rho^* (t,{\bf x}^\prime)}
\over {\mid {\bf x}-{\bf x}^\prime \mid  }} d^3x^\prime \,.
\end{eqnarray}
In some cases we will use the same notation as before, to avoid a proliferation of hats, tildes or subscripts.
We redefine the $\Sigma$, $X$ and $Y$ potentials by
\begin{subequations}
\begin{eqnarray}
\Sigma (f) &\equiv& \int_{\cal M} {{\rho^*(t,{\bf x}^\prime)f(t,{\bf x}^\prime)}
\over {\mid {\bf x}-{\bf x}^\prime \mid  }} d^3x^\prime = P(4\pi\rho^* f) \,,
\\
\Sigma^i (f) &\equiv& \int_{\cal M} {{\rho^* (t,{\bf x}^\prime) v'^if(t,{\bf
x}^\prime)}
\over {\mid {\bf x}-{\bf x}^\prime \mid  }} d^3x^\prime = P(4\pi\rho^* v^i f) \,,
\\
\Sigma^{ij} (f) &\equiv& \int_{\cal M} {{\rho^* (t,{\bf x}^\prime)
v'^iv'^j f(t,{\bf
x}^\prime)}
\over {\mid {\bf x}-{\bf x}^\prime \mid  }} d^3x^\prime = P(4\pi\rho^* v^iv^j f) \,,
\\
\Sigma_s (f) &\equiv& \int_{\cal M} {\bigl (1-2s({\bf x}^\prime) \bigr ){\rho^* (t,{\bf x}^\prime)f(t,{\bf
x}^\prime)}
\over {\mid {\bf x}-{\bf x}^\prime \mid  }} d^3x^\prime = P(4\pi (1-2s) \rho^* f) \,,
\\
X(f)  &\equiv& \int_{\cal M} {\rho^* (t,{\bf x}^\prime)f(t,{\bf
x}^\prime)}
{\mid {\bf x}-{\bf x}^\prime \mid  } d^3x^\prime  \,,
\\
Y(f) &\equiv& \int_{\cal M} {\rho^* (t,{\bf x}^\prime)f(t,{\bf x}^\prime)}
{\mid {\bf x}-{\bf x}^\prime \mid ^3 } d^3x^\prime  \,,
\end{eqnarray}
\end{subequations}
and their obvious counterparts $X^i$, $X^{ij}$, $X_s$, $Y^i$, $Y^{ij}$, $Y_s$, and so on.   With this new convention, all the potentials defined in Eqs.\ (\ref{potentiallist}) can be redefined appropriately.  

\section{Equations of Motion in Terms of  Potentials}

Pulling together all the potentials expressed in terms of $\rho^*$, inserting into the metric, Eq.\ (\ref{metric}), calculating the Christoffel symbols, we obtain from Eq.\ (\ref{geodesiceq}) the equation of motion
\begin{equation}
dv^i/dt = a_N^i + \epsilon a_{PN}^i + \epsilon^{3/2} a_{1.5PN}^i
+ \epsilon^{2} a_{2PN}^i + \epsilon^{5/2} a_{2.5PN}^i + O(\epsilon^3) \,,
\label{dvdt25pn}
\end{equation}
where
\begin{eqnarray}
a_N^i &=& G(1-\zeta) U^{,i} + G\zeta (1-2s) U_s^{,i} \,,
\label{afluidN}
\end{eqnarray}
\begin{eqnarray}\label{aPN}
a_{PN}^i &=& 
v^2 \left [ G(1-\zeta) U^{,i} - G\zeta (1-2s) U_s^{,i} \right ] 
- 4G(1-\zeta) v^i v^j U^{,j}
- v^i \left [ 3G(1-\zeta) \dot{U} - G\zeta (1-2s) \dot{U}_s \right ]
\nonumber \\
&&
- 4G^2 (1-\zeta )^2 UU^{,i}
- 4G^2  \zeta (1-\zeta ) (1-2s) UU_s^{,i}
- 2G^2 \zeta \left [ \lambda_1 (1-2s)  + 2\zeta s' \right ] U_s U_s^{,i}
\nonumber \\
&&
+8G(1-\zeta) v^j V^{[i,j]} + 4G(1-\zeta) \dot{V}^i
+\frac{1}{2} G(1-\zeta) \ddot{X}^{,i}
+ \frac{1}{2} G\zeta (1-2s) \ddot{X}_s^{,i}
\nonumber \\
&&
+ \frac{3}{2} G(1-\zeta) \Phi_1^{,i}  - \frac{1}{2} G\zeta (1-2s)  {\Phi^s_1}^{,i} 
- G^2(1-\zeta)^2 \Phi_2^{,i} 
- G^2  \zeta (1-\zeta ) (1-2s) {\Phi^s_2}^{,i}
\nonumber \\
&&
- G^2 \zeta \left [ 1-\zeta + (2 \lambda_1 + \zeta)(1-2s) \right ] 
{\Phi^s_{2s}}^{,i}
- 4G^2 \zeta^2 (1-2s) \Sigma^{,i}(a_s U_s) \,,
\label{afluidPN}
\end{eqnarray}
\begin{eqnarray}
a_{1.5PN}^i &=& \frac{1}{3} (1-2s) \stackrel{(3)}{{\cal I}_s^{i}} \,,
\label{afluid15PN}
\end{eqnarray}
\begin{eqnarray}\label{a2PN}
a_{2PN}^i &=&
4 G(1-\zeta)  v^i v^j v^k V^{j,k} + v^2 v^i \left [ G(1-\zeta) \dot{U} - G\zeta (1-2s) \dot{U}_s \right ]
\nonumber \\
&&
+v^i v^j  \left [ 4G^2(1-\zeta)^2 \Phi_2^{,j} + 4G^2 \zeta (1-\zeta) {\Phi^s_{2s}}^{,j}-2 G(1-\zeta) \Phi_1^{,j} -2 G(1-\zeta) \ddot{X}^{,j} \right ]
\nonumber \\
&&
+ v^j v^k \left [ 2 G(1-\zeta) \Phi_1^{jk,i} - 4 G(1-\zeta) \Phi_1^{ij,k}
+2 G^2(1-\zeta)^2 P_2^{jk,i} - 4 G^2(1-\zeta)^2 P_2^{ij,k}
\right .
\nonumber \\
&& \qquad
\left .
+ 2 G^2 \zeta (1-\zeta) P_{2s}^{jk,i} - 4 G^2 \zeta(1-\zeta) P_{2s}^{ij,k} \right ]
\nonumber \\
&&
+ v^2 \left [ -\frac{1}{2} G(1-\zeta) \Phi_1^{,i} 
+\frac{1}{2} G \zeta (1-2s) {\Phi^s_1}^{,i} 
- G^2 (1-\zeta)^2 \Phi_2^{,i}
+ G^2 \zeta (1-\zeta)(1-2s) {\Phi^s_2}^{,i}
\right .
\nonumber \\
&& \qquad
\left .
- G^2 \zeta \left [ 1-\zeta - (2 \lambda_1 + \zeta)(1-2s) \right ] 
{\Phi^s_{2s}}^{,i}
+ 2G^2 \zeta \left [ \lambda_1 (1-2s) + 2\zeta s' \right ]
U_s U_s^{,i}
 \right .
\nonumber \\
&& \qquad
\left .
+ 4 G^2 \zeta^2 (1-2s) \Sigma^{,i}(a_s U_s)
+ \frac{1}{2} G(1-\zeta) \ddot{X}^{,i}
- \frac{1}{2} G\zeta (1-2s) \ddot{X}_s^{,i}
\right ]
\nonumber \\
&&
+ v^i \left [ 
3  G^2 (1-\zeta)^2 \dot{\Phi}_2
- G^2 \zeta (1-\zeta) (1-2s) \dot{\Phi}^s_2
+ G^2 \zeta \left [ 3(1- \zeta) - (2\lambda_1 + \zeta) (1-2s) \right ]
\dot{\Phi}^s_{2s}
 \right .
\nonumber \\
&& \qquad
\left .
- 4G^2 \zeta^2 (1-2s) \dot{\Sigma}(a_s U_s)
- 2 G^2 \zeta \left [\lambda_1 (1-2s) + 2\zeta s' \right ]
U_s \dot{U}_s
-\frac{1}{2} G(1-\zeta) \dot{\Phi}_1
 \right .
\nonumber \\
&& \qquad
\left .
-\frac{1}{2} G \zeta (1-2s) \dot{\Phi}^s_1
- \frac{3}{2} G(1-\zeta) \stackrel{(3)}{X}
+\frac{1}{2} G \zeta (1-2s) \stackrel{(3)}{X}_s
 \right .
\nonumber \\
&& \qquad
\left .
+ 4 G^2 (1-\zeta)^2 V^k U^{,k}
+ 4 G^2 \zeta (1-\zeta) (1-2s)V^k U_s^{,k}
\right ]
\nonumber \\
&&
+ v^j \left [
8G^2 (1-\zeta)^2 V_2^{[i,j]}
+ 8 G^2  \zeta (1-\zeta) \Sigma_s^{,[i}(v^{j]} U_s) 
-16 G^2 (1-\zeta)^2 \Phi_2^{[i,j]}
+4 G (1-\zeta) \ddot{X}^{[i,j]}
 \right .
\nonumber \\
&& \qquad
\left .
+ 32 G^2 (1-\zeta)^2 G_7^{[i,j]}
- 8 G^2 \zeta (1-\zeta) P(\dot{U}_s U_s^{,[i})^{,j]}
- 16  G^2 (1-\zeta)^2 U V^{[i,j]}
 \right .
\nonumber \\
&& \qquad
\left .
-4 G (1-\zeta) \Sigma^{,[i}(v^{j]} v^2)
+ 8 G^2 (1-\zeta)^2 V^i U^{,j}
+ 8 G^2 \zeta (1-\zeta) (1-2s) V^j U_s^{,i}
\right .
\nonumber \\
&& \qquad
\left . 
-4 G(1-\zeta) \dot{\Phi}_1^{ij}
-4 G^2 (1-\zeta)^2 \dot{P}_2^{ij}
-4 G^2 \zeta (1-\zeta) \dot{P}_{2s}^{ij}
\right ]
\nonumber \\
&&
+ \frac{1}{24} G(1-\zeta) \stackrel{(4)}{Y^{,i}}
+ \frac{1}{24} G\zeta (1-2s) \stackrel{(4)}{Y_s^{,i}}
+2 G(1-\zeta) \stackrel{(3)}{X^{i}}
+ \frac{3}{4} G(1-\zeta) \ddot{X}_1^{,i}
\nonumber \\
&&
- \frac{1}{4} G \zeta (1-2s) \ddot{X}_s^{,i}(v^2)
+ 2 G(1-\zeta) \dot{\Sigma}(v^iv^2)
+ \frac{7}{8} G(1-\zeta) \Sigma^{,i}(v^4)
- \frac{1}{8} G \zeta (1-2s)\Sigma_s^{,i}(v^4)
\nonumber \\
&&
+ \frac{9}{2} G^2 (1-\zeta)^2 \Sigma^{,i}(v^2 U)
- \frac{1}{2} G^2 \zeta \left [ 3(1-\zeta) -(2\lambda_1 + \zeta)(1-2s)
\right ] \Sigma_s^{,i}(v^2 U_s)
\nonumber \\
&&
- \frac{3}{2} G^2 \zeta (1-\zeta) (1-2s) \Sigma_s^{,i}(v^2 U)
+ 2  G^2 \zeta^2 (1-2s)  \Sigma^{,i}(v^2 a_s U_s)
-4 G^2 (1-\zeta)^2 \Sigma^{,i}(v^j V^j)
\nonumber \\
&&
+ 4 G^2 \zeta (1-\zeta) (1-2s) \Sigma_s^{,i}(v^j V^j)
- \frac{3}{2} G^2 (1-\zeta)^2 \Sigma^{,i}(\Phi_1)
- \frac{3}{2} G^2 \zeta (1-\zeta) (1-2s) \Sigma_s^{,i}(\Phi_1)
\nonumber \\
&&
+ 2 G^2 \zeta^2 (1-2s) \Sigma^{,i}(a_s \Phi^s_1)
+ \frac{1}{2} G^2 \zeta \left [ 1- \zeta + (2 \lambda_1 + \zeta)(1-2s) \right ]  \Sigma_s^{,i}(\Phi^s_1)
- 6 G^2 (1-\zeta)^2 U \Phi_1^{,i}
\nonumber \\
&&
+ 2 G^2 \zeta (1-\zeta) (1-2s) U {\Phi^s_1}^{,i}
+ G^2 \zeta \left [ \lambda_1 (1-2s) + 2 \zeta s' \right ]
U_s {\Phi^s_1}^{,i}
-2 G^2 (1-\zeta)^2 \Phi_1U^{,i} 
\nonumber \\
&&
- 2 G^2 \zeta (1-\zeta) (1-2s) \Phi_1 U_s^{,i}
+ G^2 \zeta \left [ \lambda_1 (1-2s) + 2 \zeta s' \right ]
\Phi^s_1 U_s^{,i}
- 4 G^2 (1-\zeta)^2 \Phi_1^{ij}U^{,j} 
\nonumber \\
&&
- 4 G^2 \zeta (1-\zeta) (1-2s)\Phi_1^{ij}U_s^{,j}
+ 8G^2 (1-\zeta)^2  V^j V^{j,i}
+ 4 G^2 (1-\zeta)^2  V^i \dot{U}
- 4 G^2 \zeta (1-\zeta) (1-2s)V^i \dot{U}_s
\nonumber \\
&&
-2 G^2 (1-\zeta)^2 U \ddot{X}^{,i}
-2 G^2 \zeta (1-\zeta) (1-2s) U \ddot{X}_s^{,i}
- G^2 \zeta \left [ \lambda_1 (1-2s) + 2 \zeta s' \right ]
U_s \ddot{X}_s^{,i}
\nonumber \\
&&
-2 G^2 (1-\zeta)^2 \ddot{X} U^{,i}
-2 G^2 \zeta (1-\zeta) (1-2s)  \ddot{X} U_s^{,i}
- G^2 \zeta \left [ \lambda_1 (1-2s) + 2\zeta s' \right ]
\ddot{X}_s U_s^{,i}
\nonumber \\
&&
-8 G^2 (1-\zeta)^2 U \dot{V}^{i}
-\frac{1}{2} G^2 (1-\zeta)^2 \Sigma^{,i}(\ddot{X})
-\frac{1}{2} G^2 \zeta (1-\zeta) (1-2s) \Sigma_s^{,i}(\ddot{X})
\nonumber \\
&&
- \frac{1}{2} G^2 \zeta \left [ 1- \zeta + (2 \lambda_1 + \zeta)(1-2s) \right ]  \Sigma_s^{,i}(\ddot{X}_s)
-2 G^2 \zeta^2 (1-2s) \Sigma^{,i}(a_s \ddot{X}_s)
\nonumber \\
&&
-\frac{1}{2} G^2 (1-\zeta)^2 \ddot{X}_2^{,i}
- 2 G^2 \zeta^2 (1-2s) \ddot{X}^{,i}(a_s U_s)
-\frac{1}{2} G^2 \zeta (1-\zeta) (1-2s)  \ddot{X}_s^{,i}(U)
\nonumber \\
&&
- \frac{1}{2} G^2 \zeta \left [ 1- \zeta + (2 \lambda_1 + \zeta)(1-2s) \right ]  \ddot{X}_s^{,i}(U_s)
+ 4 G^2 (1-\zeta)^2  \dot{V}_2^i
- 4 G^2 \zeta (1-\zeta) \dot{\Sigma}_{s}(v^i\, U_s)
\nonumber \\
&&
- 8G^2 (1-\zeta)^2  \dot{\Phi}_2^i
- 6 G^2 (1-\zeta)^2 G_1^{,i}
+2 G^2 \zeta (1-\zeta) G_{1s}^{,i}
-4 G^2 (1-\zeta)^2 G_2^{,i}
\nonumber \\
&&
-4 G^2 \zeta (1-\zeta) (1-2s) G_{2s}^{,i}
+8 G^2 (1-\zeta)^2 G_3^{,i}
+8 G^2 \zeta (1-\zeta) (1-2s) G_{3s}^{,i}
+8 G^2 (1-\zeta)^2 G_4^{,i}
\nonumber \\
&&
-4 G^2 (1-\zeta)^2 G_6^{,i}
-4 G^2 \zeta (1-\zeta) (1-2s) G_{6s}^{,i}
+16 G^2 (1-\zeta)^2 \dot{G}_7^{i}
- 4 G^2 \zeta (1-\zeta) \dot{P}(\dot{U}_s U_s^{,i})
\nonumber \\
&&
+ 4 G^3 (1-\zeta)^3 U \Phi_2^{,i}
+ 4G^3 \zeta (1-\zeta) \left [ 1-\zeta + (2\lambda_1+\zeta)(1-2s) \right ] U {\Phi^s_{2s}}^{,i}
\nonumber \\
&& 
+ 4G^3 \zeta (1-\zeta)^2 (1-2s) U {\Phi^s_{2}}^{,i}
+ 2 G^3 \zeta (1-\zeta) \left [ \lambda_1 (1-2s) + 2\zeta s' \right ] U_s {\Phi^s_{2}}^{,i}
\nonumber \\
&&
+ 16 G^3 \zeta^2 (1-\zeta) (1-2s) U \Sigma^{,i}(a_s U_s)
+ 8 G^3 \zeta^2  \left [ \lambda_1 (1-2s) + 2\zeta s' \right ] U_s \Sigma^{,i}(a_s U_s)
\nonumber \\
&&
+ 2 G^3 \zeta (2\lambda_1+\zeta)\left [ \lambda_1 (1-2s) + 2\zeta s' \right ] U_s {\Phi^s_{2s}}^{,i}
+ 4 G^3 (1-\zeta)^3 \Phi_2 U^{,i}
+ 4 G^3 \zeta (1-\zeta)^2 \Phi^s_{2s} U^{,i}
\nonumber \\
&&
+2 G^3 \zeta (1-\zeta) \left [ \lambda_1 (1-2s) + 2\zeta s' \right ] \Phi^s_2 U_s^{,i}
+ 8 G^3 \zeta^2  \left [ \lambda_1 (1-2s) + 2\zeta s' \right ] \Sigma(a_s U_s) U_s^{,i}
\nonumber \\
&&
+ 2 G^3 \zeta \left \{ 2\zeta (1-\zeta)(1-2s) + (2\lambda_1 + \zeta)[\lambda_1  (1-2s) + 2\zeta s'] \right \} \Phi^s_{2s} U_s^{,i}
\nonumber \\
&&
+ 4 G^3 \zeta (1-\zeta)^2 (1-2s)  \Phi_{2} U_s^{,i}
+ 8 G^3 (1-\zeta)^3 U^2 U^{,i}
+ 8 G^3 \zeta (1-\zeta)^2 (1-2s) U^2 U_s^{,i}
\nonumber \\
&&
+ G^3 \zeta \left [ (8 \lambda_1^2 -2\zeta \lambda_1 - 2\lambda_2)(1-2s) +12 \lambda_1 \zeta s' -4 \zeta^2 s'' \right ] U_s^2 U_s^{,i}
\nonumber \\
&&
+ 8 G^3 \zeta (1-\zeta) \left [ \lambda_1 (1-2s) + 2 \zeta s' \right ] U U_s U_s^{,i}
- G^3 (1-\zeta)^3 \Sigma^{,i}(\Phi_2) 
- G^3  \zeta(1-\zeta)^2 \Sigma^{,i}(\Phi^s_{2s})
\nonumber \\
&&
+ G^3 \zeta \left \{ (2\lambda_1 + \zeta) \left [1-\zeta+(2\lambda_1+\zeta)(1-2s) \right] - \zeta (1-\zeta)(1-2s) \right \} \Sigma_s^{,i}(\Phi^s_{2s})
\nonumber \\
&&
- G^3 \zeta(1-\zeta)^2 (1-2s)  \Sigma_s^{,i}(\Phi_2)
+  G^3 \zeta (1-\zeta) \left [ 1-\zeta + (2\lambda_1+\zeta)(1-2s) \right ] \Sigma_s^{,i}(\Phi^s_{2})
\nonumber \\
&&
+ 4 G^3 \zeta^2 (1-\zeta)(1-2s)  \Sigma^{,i}(a_s \Phi^s_{2})
+4 G^3 \zeta^2 (2\lambda_1 + \zeta)(1-2s) \Sigma^{,i}(a_s \Phi^s_{2s})
\nonumber \\
&&
+ 16 G^3 \zeta^3 (1-2s) \Sigma^{,i}(a_s \Sigma(a_s U_s))
+ 4 G^3 \zeta^2 \left [ 1-\zeta + (2\lambda_1+\zeta)(1-2s) \right ] \Sigma_s^{,i}(\Sigma(a_s U_s))
\nonumber \\
&&
+\frac{3}{2} G^3 (1-\zeta)^3 \Sigma^{,i}(U^2)
+ \frac{3}{2} G^3 \zeta (1-\zeta)^2 (1-2s) \Sigma_s^{,i}(U^2)
+ G^3 \zeta (1-\zeta)^2 \Sigma^{,i}(U_s^2)
\nonumber \\
&&
+ \frac{1}{2} G^3 \zeta \left \{ (2\lambda_1 + \zeta)(1-\zeta)+ (1-2s)\left [\zeta(1-\zeta) + \zeta(2\lambda_1+1)+16 \lambda_1^2 - 4\lambda_2 \right ]\right \} \Sigma_s^{,i}(U_s^2)
\nonumber \\
&&
+ G^3 \zeta (1-\zeta) \left [ 1-\zeta + (2\lambda_1+\zeta)(1-2s) \right ] \Sigma_s^{,i}(U_s U)
+ 2 G^3 \zeta^2 \left [1-\zeta + 6\lambda_1 (1-2s) \right ]\Sigma^{,i}(a_s U_s^2)
\nonumber \\
&&
-4 G^3 \zeta^3 (1-2s) \Sigma^{,i}(b_s U_s^2)
+4 G^3 \zeta^2 (1-\zeta) (1-2s) \Sigma^{,i}(a_s U_s U)
\nonumber \\
&&
- 4 G^3 (1-\zeta)^3 P_2^{ij} U^{,j}
-4 G^3 \zeta (1-\zeta)^2 P_{2s}^{ij} U^{,j}
-4 G^3 \zeta (1-\zeta)^2 (1-2s) P_{2}^{ij} U_s^{,j}
\nonumber \\
&&
-4 G^3 \zeta^2 (1-\zeta) (1-2s) P_{2s}^{ij} U_s^{,j} 
- 4 G^3 (1-\zeta)^3 H^{,i}
-4 G^3 \zeta (1-\zeta)^2 H_s^{,i}
\nonumber\\
&&
-4 G^3 \zeta (1-\zeta)^2 (1-2s) H^{s\,,i}
-4 G^3 \zeta^2 (1-\zeta) (1-2s) {H^s_s}^{,i} \,,
\label{afluid2PN}
\end{eqnarray}
\begin{eqnarray}
a^i_{2.5PN} &=&
\frac{3}{5} x^j \left ( \stackrel{(5)\,}{{\cal I}^{ij}} - \frac{1}{3} \delta^{ij}  \stackrel{(5)\,}{{\cal I}^{kk}} \right )
+ 2 v^j \stackrel{(4)}{{\cal I}^{ij}}
+2 \left [  G(1-\zeta) U^{,j} + G\zeta (1-2s) U_s^{,j} \right ] \stackrel{(3)}{{\cal I}^{ij}}
\nonumber \\
&&
+ \frac{4}{3} \left [ G(1-\zeta) U^{,i} + G\zeta (1-2s) U_s^{,i} \right ] 
\stackrel{(3)\,}{{\cal I}^{kk}}
- \left [ G(1-\zeta) X^{,ijk} + G  \zeta  (1-2s)X_s^{,ijk} \right ]
\stackrel{(3)\,}{{\cal I}^{jk}}
\nonumber \\
&&
- \frac{2}{15} \stackrel{(5)\,}{{\cal I}^{ijj}}
+ \frac{2}{3} \epsilon^{qij} \stackrel{(4)\,}{{\cal J}^{qj}} 
- \frac{1}{15} (1-2s) x^j \left ( \stackrel{(5)\,}{{\cal I}_s^{ij}} + \frac{1}{2} \delta^{ij}  \stackrel{(5)\,}{{\cal I}_s^{kk}} \right )
+ \frac{1}{15} (1-2s) \left ( x^i x^j + \frac{1}{2} r^2 \delta^{ij} \right )
 \stackrel{(5)}{{\cal I}_s^{j}}
 \nonumber \\
&&
 + \frac{1}{30} (1-2s) \stackrel{(5)\quad}{{\cal I}_s^{ijj}} 
-\frac{1}{3}v^2 (1-2s) \stackrel{(3)}{{\cal I}_s^{i}}
-\frac{4}{3} G (1-\zeta) (1-2s) U \stackrel{(3)}{{\cal I}_s^{i}}
 \nonumber \\
&&
+ \frac{1}{6} v^i (1-2s) \left ( 2x^j \stackrel{(4)}{{\cal I}_s^{j}}
- \stackrel{(4)\,}{{\cal I}_s^{kk}}
- 6 \ddot{M}_s \right )
-\frac{1}{3} (1-2s) x^i \stackrel{(3)}{M_s}
\nonumber \\
&&
- \frac{1}{6} G \biggl \{ \left [ 1-\zeta + (4 \lambda_1 + \zeta)(1-2s) + 4\zeta s' \right ] U_s^{,i} 
+4  \zeta (1-2s) \Sigma^{,i}(a_s) \biggr \}  \left ( 2x^j \stackrel{(3)}{{\cal I}_s^{j}}
- \stackrel{(3)\,}{{\cal I}_s^{kk}}
- 6 \dot{M}_s \right )
\nonumber \\
&&
- \frac{1}{3} G \biggl \{ \left [ 1-\zeta + (4 \lambda_1 + \zeta)(1-2s) +  4\zeta s' \right ]   U_s +4  \zeta (1-2s)  \Sigma(a_s) \biggr \} \stackrel{(3)}{{\cal I}_s^{i}}
\nonumber \\
&&
+ \frac{1}{3} G \biggl \{ \left [ 1-\zeta + (2 \lambda_1+ \zeta)(1-2s) \right ]
X_s^{,ij} +4\zeta (1-2s) X_s^{,ij}(a_s) \biggr \} \stackrel{(3)}{{\cal I}_s^{j}}
 \,.
\label{afluid25PN}
\end{eqnarray}
We next turn to the problem of expressing these equations explicitly in terms of positions and velocities of each body in a two-body system.
 
\begin{savequote}[0.60\linewidth]
{\scriptsize ``Science is a way of thinking much more than it is a body of knowledge.''}
\qauthor{\scriptsize---Carl Sagan}
\end{savequote}
\chapter{Equations of Motion for Two Compact Objects} 
\label{chapter11} 
\thispagestyle{myplain}
\lhead[\thepage]{Chapter 10. \emph{Equations of Motion for Two Compact Objects}}      
\rhead[Chapter 10. \emph{Equations of Motion for Two Compact Objects}]{\thepage}

We now wish to calculate the equation of motion for a member of a compact binary system.   To do this, we integrate $\rho^* dv^i/dt$ over body 1, and substitute Eq.\ (\ref{dvdt25pn}) and then Eqs.\ (\ref{afluidN}) -- (\ref{afluid25PN}).  We follow closely the methods already detailed in Chapter~\ref{chapter6} based on ~\cite{pat02} (hereafter referred to as PWII) for evaluating the integrals of the various potentials, and so we will not repeat those details here. Readers should consult Sec.\ III and Appendices B, C, and D of PWII for details.  In  structural terms almost all of the potentials that appear in the 2PN terms in scalar-tensor theory also appear in general relativity, apart from the differences in the types of densities that generate the potentials, for example $U_s$ vs. $U$, $X_s$ vs. $X$, $\Phi^s_{2s}$ vs. $\Phi_2$, and so on.   The only 2PN term that does not appear in GR involves the potential $P(\dot{U}_s U_s^{,i})$, but this can be evaluated using the methods described in Chapter~\ref{chapter6}.  

Similarly, at 2.5PN order most of the moments that appear here also appear in GR, only a few, notably the scalar monopole and dipole moments $M_s$ and ${\cal I}_s^i$ are new.  Particularly new is the appearance of a $1.5$PN order term generated by the scalar dipole moment; this, of course, is the radiation-reaction counterpart of the well-known dipole gravitational radiation prediction of scalar-tensor theories.

\section{Conservative $1$PN and $2$PN Terms}

We begin with the conservative Newtonian, 1PN and 2PN terms.  The results are, at Newtonian and 1PN orders.
\begin{eqnarray}\label{1pnEOM}
a_{1\, (PN)}^{i} &=& -  \frac{G\alpha m_2}{r^2} n^i +  \frac{G\alpha m_2}{r^2} n^i \biggl \{ 
 -  (1+ \bar{\gamma}) v_1^2  - (2+ \bar{\gamma})(v_2^2 - 2{\bf v}_1 \cdot {\bf v}_2)
      + \frac{3}{2}(\nvb)^2
\nonumber \\
  && \qquad
	+ \left [4+ 2\bar{\gamma} + 2 \bar{\beta}_1 \right ] \frac{G\alpha m_2}{r} 
	+ \left[5+2\bar{\gamma} + 2 \bar{\beta}_2 \right ] \frac{G\alpha m_1}{r} 
    \biggr \}
\nonumber 
\\
&& \quad + \frac{G \alpha m_2}{r^2} (v_1 - v_2)^i \left [(4 +2\bar{\gamma})\nva -(3 + 2\bar{\gamma}) \nvb \right ] \,, 
\nonumber 
\\
a_{2\, (PN)}^{i} &=& \{1 \rightleftharpoons 2,\; {\bf n} \to - {\bf n}\} \,,
\label{1PNeom}
\end{eqnarray}
where $r \equiv \mid {\bf x}_1 - {\bf x}_2 \mid $, ${\bf n} \equiv ({\bf x}_1 - {\bf x}_2)/r$,
and where the parameters $\alpha$, $\bar{\gamma}$, and $\bar{\beta}_A$ are defined in Table \ref{tab:params}.
Note that under the interchange $(1 \rightleftharpoons 2)$, ${\bf n} \to - {\bf n}$.  At 2PN order, we find
\begin{eqnarray}\label{2pnEOM}
a_{1\,(2PN)}^i &=& \frac{G \alpha m_2}{r^2}n^i\bigg \{
-(2 + \bar{\gamma}) \bigl  [v_2^4
-2 v_2^2(\vonevtwo)
+ (\vonevtwo)^2
+3(\nvb)^2 (\vonevtwo) \bigr]
\nonumber\\
&&
\quad
+\frac{3}{2} (1+\bar{\gamma})v_1^2(\nvb)^2
+\frac{3}{2} \left (3+ \bar{\gamma} \right )v_2^2(\nvb)^2
-\frac{15}{8}(\nvb)^4
\nonumber\\
&&
\quad 
+ \frac{G \alpha m_2}{r}\biggl (
2 (2 +\bar{\gamma}) \bigl [ v_2^2-2\vonevtwo \bigr ] - 2 \bar{\beta}_1 v_1^2
\nonumber\\
&&
\qquad
+\frac{1}{2} \left ( (2+\bar{\gamma})^2 + 4\bar{\delta}_2 \right ) \bigl [(\nva)^2
-2(\nva)(\nvb) \bigr ]
\nonumber\\
&&
\qquad  - \frac{1}{2} \left ((6-\bar{\gamma})(2+\bar{\gamma}) +8\bar{\beta}_1 - 4\bar{\delta}_2 \right ) (\nvb)^2
\biggr )
\nonumber\\
&&
\quad +\frac{G \alpha m_1}{r} \biggl (
\frac{1}{4} \left ( 5 + 4 \bar{\beta}_2 \right )   \left [ v_2^2 - 2\vonevtwo \right ]
-\frac{1}{4} \left ( 15 + 8\bar{\gamma} + 4 \bar{\beta}_2 \right ) v_1^2
\nonumber\\
&&
\qquad \quad +\frac{1}{2} \left ( 17 + 18 \bar{\gamma}+\bar{\gamma}^2 -16 \bar{\beta}_2 + 4 \bar{\delta}_1 \right ) (\nvb)^2
\nonumber\\
&&
\qquad \quad
+\frac{1}{2}  \left ( 39 +26\bar{\gamma} + \bar{\gamma}^2 -8\bar{\beta}_2 + 4 \bar{\delta}_1 \right )
\left [(\nva)^2
-2(\nva)(\nvb) \right ] \biggr )
\nonumber \\
&&
\quad 
-\frac{1}{4} \frac{G^2 \alpha^2 m_1^2}{r^2} \left ( 57 +44\bar{\gamma}+ 9\bar{\gamma}^2 + 16(3+\bar{\gamma})\bar{\beta}_2 +4 \bar{\delta}_1 - 8 \bar{\chi}_2 \right )
\nonumber \\
&&
\quad 
-\frac{1}{2}\frac{G^2 \alpha^2 m_1m_2}{r^2} \left ( 69 + 48\bar{\gamma}+ 8\bar{\gamma}^2 + 8(3+\bar{\gamma})\bar{\beta}_2 +2(15+4\bar{\gamma}) \bar{\beta}_1  -48\bar{\gamma}^{-1}\bar{\beta}_1 \bar{\beta}_2 \right )
\nonumber \\
&&
\quad 
-\frac{1}{4} \frac{G^2 \alpha^2m_2^2}{r^2} \left ( 9(2+\bar{\gamma})^2 + 16 (2+\bar{\gamma}) \bar{\beta}_1 + 4\bar{\delta}_2
-8 \bar{\chi}_1 \right )
\bigg \}
\nonumber \\
&&
+\frac{G \alpha m_2}{r^2}(v_1^i - v_2^i)\bigg \{
2 (2+\bar{\gamma}) \left [v_2^2(\nva) + \vonevtwo (\nvb - \nva)
-\frac{3}{2}(\nva)(\nvb)^2 \right ]
\nonumber \\
&&
 \quad
+(1+\bar{\gamma})v_1^2(\nvb)
-(5+3\bar{\gamma}) v_2^2(\nvb)
+\frac{3}{2} (3+2\bar{\gamma}) (\nvb)^3
\nonumber \\
&&
 \quad
+\frac{G \alpha m_1}{4r}\bigg( \left ( 55 + 40\bar{\gamma} + 2\bar{\gamma}^2 - 16 \bar{\beta}_2 + 8 \bar{\delta} _1 \right )\nvb
- \left (63 + 40 \bar{\gamma}+ 2\bar{\gamma}^2 -8\bar{\beta}_2 
+ 8\bar{\delta} _1 \right ) \nva \bigg)
\nonumber \\
&&
 \quad
-\frac{1}{2} \frac{ G \alpha m_2}{r}\bigg( \left ((2+\bar{\gamma})^2 + 4\bar{\delta}_2 \right ) \nva
+ \left (4-\bar{\gamma}^2 + 4\bar{\beta}_1 - 4\bar{\delta}_2 \right ) \nvb\bigg)
\bigg \} \,,
\nonumber \\
a_{2\, (2PN)}^{i} &=& \{1 \rightleftharpoons 2,\; {\bf n} \to - {\bf n}\} \,,
\label{2PNeom}
\end{eqnarray}
where $\bar{\delta}_A$ and $\bar{\chi}_A$ are defined in Table \ref{tab:params}. 

It is straightforward to show that these equations of motion can be derived from a two-body Lagrangian, given by
\begin{eqnarray}
L &=& - m_1\left (1 - \frac{1}{2} v_1^2 - \frac{1}{8} v_1^4 - \frac{1}{16} v_1^6 \right )
+ \frac{1}{2} \frac{G \alpha m_1 m_2}{r} 
\nonumber \\
&&
+ \frac{G \alpha m_1 m_2}{r} \left \{ \frac{1}{2} (3+2\bar{\gamma}) v_1^2 - \frac{1}{4} (7+4\bar{\gamma}) \vonevtwo -\frac{1}{4} (\nva)( \nvb)  - \frac{1}{2} (1 + 2 \bar{\beta}_2) \frac{G \alpha m_1}{r} \right \} 
\nonumber \\
&&
+ \frac{G \alpha m_1 m_2}{r} \biggl \{ 
\frac{1}{8} (7+4\bar{\gamma})  \left [ v_1^4 - v_1^2 (\nvb)^2 \right ]
-(2+\bar{\gamma}) v_1^2 (\vonevtwo) + \frac{1}{8} (\vonevtwo)^2 
\nonumber \\
&&
\qquad + \frac{1}{16} (15 + 8\bar{\gamma}) v_1^2 v_2^2 + \frac{3}{16} (\nva)^2 (\nvb)^2
+ \frac{1}{4} (3+2\bar{\gamma}) \vonevtwo (\nva)(\nvb) 
\nonumber \\
&&
\qquad
+ \frac{G \alpha m_1}{r} \biggl [ \frac{1}{8} \left (2+12\bar{\gamma}+7\bar{\gamma}^2+8\bar{\beta}_2 - 4\bar{\delta}_1 \right ) v_1^2 
+ \frac{1}{8} \left ( 14 + 20\bar{\gamma} + 7\bar{\gamma}^2 +4 \bar{\beta}_2 - 4 \bar{\delta}_1 \right ) v_2^2
\nonumber \\
&&
\qquad \quad
-\frac{1}{4} \left ( 7 +16\bar{\gamma}+ 7\bar{\gamma}^2 + 4 \bar{\beta}_2 - 4 \bar{\delta}_1 \right ) \vonevtwo
- \frac{1}{4} \left ( 14+12\bar{\gamma} + \bar{\gamma}^2 - 8 \bar{\beta}_2 + 4 \bar{\delta}_1 \right ) (\nva)(\nvb)
\nonumber \\
&&
\qquad \quad
+ \frac{1}{8} \left (28+ 20\bar{\gamma}+\bar{\gamma}^2 - 8 \bar{\beta}_2 + 4 \bar{\delta}_1 \right ) (\nva)^2
+ \frac{1}{8} \left (4 +4\bar{\gamma} +\bar{\gamma}^2  + 4 \bar{\delta}_1 \right ) (\nvb)^2 \biggr ]
\nonumber \\
&&
\qquad
+ \frac{1}{2} \frac{G^2 \alpha^2 m_1^2}{r^2} \left [ 1+ \frac{2}{3} \bar{\gamma}+ \frac{1}{6} \bar{\gamma}^2 + 2 \bar{\beta}_2 + \frac{2}{3} \bar{\delta}_1 - \frac{4}{3} \bar{\chi}_2 \right ]
\nonumber \\
&&
\qquad
+ \frac{1}{8} \frac{G^2 \alpha^2 m_1 m_2}{r^2} \left [ 19 +8\bar{\gamma}  + 8 \bar{\beta}_1+ 8 \bar{\beta}_2 - 32 \bar{\gamma}^{-1} \bar{\beta}_1 \bar{\beta}_2 \right ] \biggr \}
\nonumber \\
&&
- \frac{1}{8} G \alpha m_1 m_2 \biggl [ 2(7+4\bar{\gamma}) {\bf a}_1 \cdot {\bf v}_2 (\nvb) + {\bf n} \cdot {\bf  a}_1 (\nvb)^2 - (7+4\bar{\gamma}) {\bf n} \cdot {\bf a}_1 v_2^2 \biggr ] 
\nonumber \\
&&
+ \{1 \rightleftharpoons 2,\; {\bf n} \to - {\bf n}\} \,.
\label{lagrangian}
\end{eqnarray}
As in general relativity, the Lagrangian contains acceleration-dependent terms at $2$PN order, and thus the Euler-Lagrange equations are $(d^2/dt^2)(\delta L/\delta a^i) - (d/dt)(\delta L/\delta v^i) + \delta L/\delta x^i = 0$.
The equations of motion (absent radiation-reaction terms) admit the usual conserved quantities. The energy is given to 2PN order by
\begin{eqnarray}
E &=& m_1\left (\frac{1}{2} v_1^2 + \frac{3}{8} v_1^4 + \frac{5}{16} v_1^6 \right )
- \frac{1}{2} \frac{G \alpha m_1 m_2}{r} 
\nonumber \\
&&
+ \frac{G \alpha m_1 m_2}{r} \left \{ \frac{1}{2} (3+2\bar{\gamma}) v_1^2 - \frac{1}{4} (7+4\bar{\gamma}) \vonevtwo  -\frac{1}{4} (\nva)( \nvb)  + \frac{1}{2} (1 + 2 \bar{\beta}_2) \frac{G \alpha m_1}{r} \right \} 
\nonumber \\
&&
+ \frac{G \alpha m_1 m_2}{r} \biggl \{ 
\frac{3}{8} (7+4\bar{\gamma})  v_1^4 - \frac{1}{8} (13+8\bar{\gamma}) v_1^2 (\nvb)^2 
\nonumber \\
&&
\qquad - \frac{1}{8} (55+28\bar{\gamma}) v_1^2 (\vonevtwo) + \frac{1}{8} (17+8\bar{\gamma}) (\vonevtwo)^2 
\nonumber \\
&&
\qquad + \frac{1}{16} (31+16\bar{\gamma}) v_1^2 v_2^2 + \frac{3}{16} (\nva)^2 (\nvb)^2
+ \frac{1}{4} (3+2\bar{\gamma}) \vonevtwo (\nva)(\nvb) 
\nonumber \\
&&
\qquad
+ \frac{1}{8} (13+8\bar{\gamma}) \vonevtwo (\nva)^2
- \frac{1}{8} (9+4\bar{\gamma}) v_1^2 (\nva)(\nvb) + \frac{3}{8} \nva (\nvb)^3
\nonumber \\
&&
\qquad
+ \frac{G \alpha m_1}{r} \biggl [ -\frac{1}{8} \left (12 -4\bar{\gamma} -7\bar{\gamma}^2 -8\bar{\beta}_2 +4\bar{\delta}_1 \right ) v_1^2 
+ \frac{1}{8} \left ( 14 + 20\bar{\gamma} + 7\bar{\gamma}^2 +4 \bar{\beta}_2 - 4 \bar{\delta}_1 \right ) v_2^2
\nonumber \\
&&
\qquad \quad
- \frac{1}{4}\left ( 12\bar{\gamma} + 7\bar{\gamma}^2 + 4 \bar{\beta}_2 -4 \bar{\delta}_1 \right ) \vonevtwo
- \frac{1}{4} \left ( 13 +12\bar{\gamma} + \bar{\gamma}^2 -8 \bar{\beta}_2 + 4 \bar{\delta}_1 \right ) (\nva)(\nvb)
\nonumber \\
&&
\qquad \quad
+ \frac{1}{8} \left (58+36\bar{\gamma}+\bar{\gamma}^2 - 8 \bar{\beta}_2 + 4\bar{\delta}_1 \right ) (\nva)^2
+ \frac{1}{8} \left (4+ 4\bar{\gamma} +\bar{\gamma}^2  + 4\bar{\delta}_1 \right ) (\nvb)^2 \biggr ]
\nonumber \\
&&
\qquad
- \frac{1}{2} \frac{G^2 \alpha^2 m_1^2}{r^2} \left [ 1 + \frac{2}{3} \bar{\gamma} + \frac{1}{6} \bar{\gamma}^2 + 2 \bar{\beta}_2 + \frac{2}{3} \bar{\delta}_1 - \frac{4}{3} \bar{\chi}_2 \right ]
\nonumber \\
&&
\qquad
- \frac{1}{8} \frac{G^2 \alpha^2 m_1 m_2}{r^2} \biggl [ 19 +8\bar{\gamma}  + 8 \bar{\beta}_1+ 8 \bar{\beta}_2 -32 \bar{\gamma}^{-1} \bar{\beta}_1 \bar{\beta}_2 \biggr ] \biggr \}
\nonumber \\
&&
+ \{1 \rightleftharpoons 2,\; {\bf n} \to - {\bf n}\} \,,
\label{Econserved}
\end{eqnarray}
while the total momentum is given by
\begin{eqnarray}
P^j &=& m_1 v_1^j 
\left ( 1 + \frac{1}{2} v_1^2 + \frac{3}{8} v_1^4  \right )
- \frac{1}{2} \frac{G \alpha m_1 m_2}{r} \left [ v_1^j + n^j (\nva) \right ]
\nonumber \\
&&
+ \frac{G \alpha m_1 m_2}{r} v_1^j 
\biggl \{ \frac{1}{8} (5+4\bar{\gamma}) v_1^2 
- \frac{1}{8} (7+4\bar{\gamma}) \left ( 2\vonevtwo - v_2^2 \right ) 
 -\frac{1}{4} (\nva)( \nvb)  
\nonumber \\
&&
\qquad
+ \frac{1}{8} (13+8\bar{\gamma}) \left ( (\nva)^2 - (\nvb)^2 \right ) 
-  (3 +2\bar{\gamma} - \bar{\beta}_2) \frac{G \alpha m_1}{r}
 + \frac{1}{2} (7+4\bar{\gamma}) \frac{G \alpha m_2}{r} \biggr \} 
\nonumber \\
&&
+  \frac{G \alpha m_1 m_2}{r} n^j (\nva)
\biggl \{- \frac{1}{8} (9+4\bar{\gamma}) v_1^2 
+ \frac{1}{8} (7+4\bar{\gamma}) \left ( 2\vonevtwo - v_2^2 \right ) 
\nonumber \\
&&
\qquad
+ \frac{3}{8} \left ( (\nva)^2 + (\nvb)^2 \right ) 
+  \frac{1}{4} (29 +16\bar{\gamma} ) \frac{G \alpha m_1}{r}
 - \frac{1}{4} (9 +8\bar{\gamma} - 8 \bar{\beta}_1) \frac{G \alpha m_2}{r} \biggr \}
 \nonumber \\
&&
+ \{1 \rightleftharpoons 2,\; {\bf n} \to - {\bf n}\} \,.
\label{Pconserved}
\end{eqnarray}

\section{Radiation-Reaction Terms}

At $1.5$PN order, the leading dipole radiation reaction term is given by
\begin{eqnarray}
a_{1\, (1.5PN)}^{i} &=& \frac{1}{3} (1-2s_1) \stackrel{(3)}{{\cal I}_s^{i}} \,, 
\nonumber 
\\
a_{2\, (1.5PN)}^{i} &=& \frac{1}{3} (1-2s_2) \stackrel{(3)}{{\cal I}_s^{i}} \,.
\label{15PNeom}
\end{eqnarray}
Because we will be working to $2.5$PN order, the scalar dipole moment ${\cal I}_s^i$ must be evaluated to post-Newtonian order, and when time derivatives of that moment generate an acceleration, the post-Newtonian equations of motion must be inserted.  Explicit two-body expressions for ${\cal I}_s^i$ and the other moments needed for the radiation-reaction terms are provided in an Appendix.  In addition to evaluating the direct $2.5$PN terms from Eq.\ (\ref{afluid25PN}) for two bodies, we must include the $1.5$PN contributions to the accelerations that occur in the $1$PN terms $\dot{V}^i$, $\ddot{X}^{,i}$ and $\ddot{X}_s^{,i}$ that appear in Eq.\ (\ref{afluidPN}).

At $2.5$PN order, the final two-body expressions take the form
\begin{eqnarray}
a^i_{1 \,(2.5PN)} &=&
\frac{3}{5} x_1^j \left ( \stackrel{(5)\,}{{\cal I}^{ij}} - \frac{1}{3} \delta^{ij}  \stackrel{(5)\,}{{\cal I}^{kk}} \right )
+ 2 v_1^j \stackrel{(4)\,}{{\cal I}^{ij}}
-\frac{1}{3}  \frac{G\alpha m_2}{r^2} n^i \stackrel{(3)\,}{{\cal I}^{kk}}
- 3  \frac{G\alpha m_2}{r^2} n^i n^j n^k
\stackrel{(3)\,}{{\cal I}^{jk}}
- \frac{2}{15} \stackrel{(5)\,}{{\cal I}^{ijj}}
\nonumber \\
&&
+ \frac{2}{3} \epsilon^{qij} \stackrel{(4)\,}{{\cal J}^{qj}} 
- \frac{1}{15} (1-2s_1) x_1^j \left ( \stackrel{(5)\,}{{\cal I}_s^{ij}} + \frac{1}{2} \delta^{ij}  \stackrel{(5)\,}{{\cal I}_s^{kk}} \right )
+ \frac{1}{15} (1-2s_1) \left ( x_1^i x_1^j + \frac{1}{2} r_1^2 \delta^{ij} \right )
 \stackrel{(5)}{{\cal I}_s^{j}}
 \nonumber \\
&&
  + \frac{1}{30} (1-2s_1) \stackrel{(5)\,}{{\cal I}_s^{ijj}} 
+ \frac{1}{6} v_1^i (1-2s_1) \left ( 2x_1^j \stackrel{(4)}{{\cal I}_s^{j}}
- \stackrel{(4)}{{\cal I}_s^{kk}}
- 6 \ddot{M}_s \right )
-\frac{1}{3} (1-2s_1) x_1^i \stackrel{(3)}{M_s}
\nonumber \\
&&
+ \frac{1}{6} \frac{G\alpha m_2}{r^2} n^i \biggl \{  1-2s_2 - 4 \bar{\gamma}^{-1} \left [ (1-2s_1)\bar{\beta}_1 + (1-2s_2)\bar{\beta}_2 \right ] \biggr \}  
\left ( 2x_1^j \stackrel{(3)}{{\cal I}_s^{j}}
- \stackrel{(3)}{{\cal I}_s^{kk}}
- 6 \dot{M}_s \right )
\nonumber \\
&&
- \frac{1}{6} \frac{G\alpha m_2}{r} n^i n^j (1-2s_2)(1-8\bar{\beta}_2/\bar{\gamma} ) \stackrel{(3)}{{\cal I}_s^{j}}
-\frac{1}{6}  \frac{G\alpha m_2}{r}  (1-2s_1) (1 - 8 \bar{\beta}_1/\bar{\gamma} ) \stackrel{(3)}{{\cal I}_s^{i}} 
\nonumber \\
&&
-\frac{1}{3}v_1^2 (1-2s_1) \stackrel{(3)}{{\cal I}_s^{i}}
+ \frac{1}{3} \frac{G\alpha m_2}{r} (s_1-s_2) ( 7 +4\bar{\gamma})\stackrel{(3)}{{\cal I}_s^{i}} \,,
\nonumber \\
a_{2\, (2.5PN)}^{i} &=& \{1 \rightleftharpoons 2,\; {\bf n} \to - {\bf n}\} \,.
 \label{25PNeom}
\end{eqnarray}
We shall defer calculating the moments and their time derivatives explicitly until the next subsection, where we obtain the relative equation of motion.

\section{Relative Equation of Motion}

We now wish to find the equation of motion for the relative separation
${\bf x} = {\bf x}_1 - {\bf x}_2$, through $2.5$PN order.   We take the PN contributions to the equation of motion for body 1 and body 2 and calculate
$d^2 {\bf x}/dt^2 = {\bf a}_1 - {\bf a}_2$.  We must then express the individual velocities ${\bf v}_1$ and ${\bf v}_2$ that appear in post-Newtonian terms in terms of ${\bf v} \equiv {\bf v}_1-{\bf v}_2$.  Since velocity-dependent terms show up at $1$PN order, we need to find the transformation from ${\bf v}_1$ and ${\bf v}_2$ to ${\bf v}$ to $1.5$PN order so as to keep all corrections through $2.5$PN order. To do this we make use of the momentum conservation law which the momentum is given in \eref{Pconserved}.  But because of the contributions of dipole radiation reaction at $1.5$PN order, the momentum is not strictly conserved because of the recoil of the system in response to the radiation of linear momentum at dipole order.  By combining Eqs.\ (\ref{Pconserved}) and (\ref{15PNeom}), it is straightforward to show that the following quantity is constant through $1.5$PN order:
\begin{eqnarray}
m_1 v_1^i
 ( 1 + \frac{ v_1^2}{2}   )
- \frac{G \alpha m_1 m_2}{2 r} \left [ v_1^i + n^i (\nva) \right ] + \frac{m_1}{3} (1-2s_1)  \ddot{\cal I}_s^i 
+ \{1 \rightleftharpoons 2, {\bf n}\to-{\bf n}\}   = C^i \,.
\end{eqnarray}
Setting $C^i = 0$ and combining this with the definition of $\bf v$, we find that 
\begin{eqnarray}
v_1^i &=& +\frac{m_2}{m} v^i + \delta^i \,,
\nonumber \\
v_2^i  &=& -\frac{m_1}{m} v^i + \delta^i \,,
\label{relative1}
\end{eqnarray}
where
\begin{equation}
\delta^i =   {1 \over 2} \eta \psi
\left [ \left (v^2 - {G \alpha m \over r}
 \right ) v^i - {G\alpha m \over r^2} {\dot r x^i} \right ]
- \frac{2}{3} \zeta \eta {\cal S}_{-} ({\cal S}_{+} + \psi {\cal S}_{-} )   \left (\frac{G \alpha m}{r} \right )^2 n^i + O(\epsilon^2) \,,
\label{relative2}
\end{equation}
where $m$ and $\eta$ are the total mass and reduced mass ratio, 
$\psi = \delta m /m = (m_1 - m_2)/m$, and 
\begin{eqnarray}
\label{cals}
{\cal S}_{-}  &\equiv& - \alpha^{-1/2} (s_1 - s_2) \,,
\nonumber \\
{\cal S}_{+} &\equiv&  \alpha^{-1/2} (1 -s_1 - s_2) \,.
\end{eqnarray}

We also need to evaluate the multipole moments that appear in the radiation-reaction terms to the appropriate order, and then calculate their time derivatives, inserting the equations of motion to the appropriate order as required.  Explicit formulae for the moments are displayed in Appendix \ref{sec:moments}.  Combining all the various PN contributions consistently, we arrive finally at the relative equation of motion through $2.5$PN order, as given in Eq.\ (\ref{eomfinal}) i.e.
\begin{eqnarray}
{{d^2 {\bf X}} \over {dt^2}} &=& -{G\alpha m \over r^2} {\bf n} 
+ {G\alpha m \over r^2} \bigl[ \, {\bf n} (A_{1PN} + A_{2PN}) 
	+ {\dot r}{\bf v} (B_{1PN} + B_{2PN} ) \bigr]  
\nonumber \\
&&
+ {8 \over 5} \eta {(G\alpha m)^2 \over r^3} 
	\bigl[\dot r {\bf n} (A_{1.5PN}+A_{2.5PN})
	- {\bf v}(B_{1.5PN}+B_{2.5PN})\bigr] \,,
\label{eomfinal}
\end{eqnarray}
where again $r \equiv \mid {\bf X}\mid $, ${\bf n}
\equiv {\bf X}/r$, $m \equiv m_1 + m_2$, $\eta \equiv m_1m_2/m^2$, 
${\bf v} \equiv {\bf v}_1 - {\bf v}_2$, and $\dot
r = dr/dt$. Here we display the coefficients $A$ and $B$ as:
\begin{subequations}
\label{EOMcoeffs}
\begin{eqnarray}
A_{1PN} &=& -(1+3\eta + \bar{\gamma})v^2 + {3 \over 2}\eta {\dot r}^2 +2(2+\eta + \bar{\gamma} + \bar{\beta}_{+} - \psi \bar{\beta}_{-} ) \frac{G\alpha m}{r} \,,
 \\
B_{1PN} &=&  2(2-\eta + \bar{\gamma}) \,, \label{eomfinalcoeffsPN}
 \\
A_{2PN} &=& -\eta(3-4\eta +\bar{\gamma})v^4 
	+ {1 \over 2}\left [\eta(13-4\eta + 4\bar{\gamma}) - 4 (1-4\eta) \bar{\beta}_+  
	+4 \psi (1-3\eta) \bar{\beta}_{-}  \right ]v^2\frac{G\alpha m}{r}
\nonumber \\
&&
	-{15 \over 8}\eta(1-3\eta){\dot r}^4 +{3 \over 2}\eta(3-4\eta + \bar{\gamma})v^2{\dot r}^2 
\nonumber \\
&&
	+\left [2 +25\eta+2\eta^2  +2(1+9\eta)\bar{\gamma}
	+ \frac{1}{2} \bar{\gamma}^2 -4\eta (3 \bar{\beta}_{+} - \psi \bar{\beta}_{-})
	+ 2\bar{\delta}_+ + 2\psi \bar{\delta}_{-} \right ]{\dot r}^2\frac{G\alpha m}{r}
\nonumber \\
&&	
	-\biggl [ 9+\frac{87}{4}\eta + (9+8\eta) \bar{\gamma}
	  + \frac{1}{4}(9 -2\eta) \bar{\gamma}^2 
	  + (8+15\eta+4\bar{\gamma}) \bar{\beta}_{+} 
	  - \psi (8+7\eta+4\bar{\gamma})\bar{\beta}_{-} 
	  \nonumber \\
&&	\qquad
        +(1-2\eta)(\bar{\delta}_+ - 2\bar{\chi}_+) 
        + \psi (\bar{\delta}_{-} + 2\bar{\chi}_{-})
	 - 24\eta \frac{\bar{\beta}_1 \bar{\beta}_2}{\bar{\gamma}}
	\biggr ] \left (\frac{G\alpha m}{r} \right )^2 \,,
 \\
B_{2PN} &=& {1 \over 2}\eta(15+4\eta + 8\bar{\gamma})v^2
	-{3 \over 2}\eta(3+2\eta +2\bar{\gamma}){\dot r}^2
 \\
&&	\nonumber
	-{1 \over 2}\left [ 4 +41\eta+8\eta^2  
	  +4(1+7\eta)\bar{\gamma} + \bar{\gamma}^2 
	  -8\eta (2\bar{\beta}_+ - \psi \bar{\beta}_{-} ) + 4\bar{\delta}_+
	  + 4 \psi \bar{\delta}_{-}
	\right ] \frac{G\alpha m}{r} \,,
\label{eomfinalcoeffs2PN}
\\ 
A_{1.5PN} &=& \frac{5}{2} \zeta {\cal S}_{-}^2\,,
 \\
B_{1.5PN} &=& \frac{5}{6} \zeta {\cal S}_{-}^2 \,.
\label{eomfinalcoeffs1.5PN}\\
A_{2.5PN} &=& a_1 v^2 + a_2 \frac{G\alpha m}{r} + a_3 \dot{r}^2 \,,
 \\
B_{2.5PN} &=& b_1 v^2 + b_2 \frac{G\alpha m}{r} + b_3 \dot{r}^2 \,,
\label{25pnAB}
\end{eqnarray}
\label{eomfinalcoeffs}
\end{subequations}
where in the two last equations
\begin{subequations}
\begin{eqnarray}
a_1 &=& 3 - \frac{5}{2} \bar{\gamma}  + \frac{15}{2} \bar{\beta}_+   +\frac{5}{8} \zeta {\cal S}_{-}^2 (9 + 4\bar{\gamma} -2\eta)
+ \frac{15}{8} \zeta \psi  {\cal S}_{-} {\cal S}_{+} \,,
\\
a_2 &=& \frac{17}{3}  + \frac{35}{6} \bar{\gamma} - \frac{95}{6} \bar{\beta}_+ 
- \frac{5}{24}\zeta {\cal S}_{-}^2  \left [ 135 + 56\bar{\gamma} + 8\eta + 32\bar{\beta}_+  \right ]
 +30 \zeta {\cal S}_{-} \left ( \frac{{\cal S}_{-} \bar{\beta}_+ + {\cal S}_{+} \bar{\beta}_{-}}{\bar{\gamma}} \right )
 \nonumber \\
&&  
-\frac{5}{8} \zeta \psi  {\cal S}_{-} \left ( {\cal S}_{+} - \frac{32}{3} {\cal S}_{-} \bar{\beta}_{-} +16 \frac{{\cal S}_{+} \bar{\beta}_+ + {\cal S}_{-} \bar{\beta}_{-}}{\bar{\gamma}} \right )  -40 \zeta \left (\frac{{\cal S}_{+} \bar{\beta}_+ + {\cal S}_{-} \bar{\beta}_{-}}{\bar{\gamma}} \right )^2 \,,
\\
a_3 &=& \frac{25}{8}  \left [ 2\bar{\gamma} - \zeta  {\cal S}_{-}^2 (1-2\eta)
  - 4\bar{\beta}_+ - \zeta \psi {\cal S}_{-} {\cal S}_{+}\right ]  \,, 
\\
b_1 &=& 1 - \frac{5}{6}\bar{\gamma} +\frac{5}{2} \bar{\beta}_+ 
-\frac{5}{24} \zeta {\cal S}_{-}^2 (7 + 4\bar{\gamma} -2\eta)
+ \frac{5}{8} \zeta \psi  {\cal S}_{-} {\cal S}_{+} \,,
\\
b_2 &=& 3 + \frac{5}{2} \bar{\gamma} -\frac{5}{2} \bar{\beta}_+ 
 -\frac{5}{24} \zeta   {\cal S}_{-}^2  \left [ 23 + 8\bar{\gamma} - 8\eta + 8\bar{\beta}_+  \right ]
 +\frac{10}{3} \zeta {\cal S}_{-}  \left ( \frac{{\cal S}_{-} \bar{\beta}_+ + {\cal S}_{+} \bar{\beta}_{-}}{\bar{\gamma}} \right )
  \nonumber \\
 &&
 -\frac{5}{8} \zeta \psi  {\cal S}_{-} \left ( {\cal S}_{+} - \frac{8}{3} {\cal S}_{-} \bar{\beta}_{-} + \frac{16}{3} \frac{{\cal S}_{+} \bar{\beta}_+ + {\cal S}_{-} \bar{\beta}_{-}}{\bar{\gamma}} \right ) \,,
 \\
b_3 &=& \frac{5}{8} \left [ 6\bar{\gamma} + \zeta  {\cal S}_{-}^2 (13 + 8\bar{\gamma}+2\eta)
  - 12\bar{\beta}_+  - 3 \zeta \psi {\cal S}_{-} {\cal S}_{+}  \right ]
  \,.
\end{eqnarray}
\label{25PNcoeffs}
\end{subequations}
Here the subscripts ``$+$'' and ``$-$'' on various parameters denote sums and differences, so that, for a chosen parameter  $\tau_i$ we define
\begin{eqnarray}
\tau_+ &\equiv& \frac{1}{2} ( \tau_1 + \tau_2) \,,
\nonumber \\
\tau_{-} &\equiv& \frac{1}{2} (\tau_1 - \tau_2) \,.
\end{eqnarray}
where $\tau$ can be either $\bar{\beta}$, $\bar{\delta}$, or $\bar{\chi}$. However, note that ${\cal S}_{+}$, ${\cal S}_{-}$ are already defined in Eqs.~(\ref{cals}) explicitly. 

Comparing relative equations of motion in scalar-tensor theories i.e. Eqs.~(\ref{EOMcoeffs}) with their correspondin expressions in general relativity i.e. Eqs.~(\ref{1PNcoeffs}, \ref{2PNcoeffs}), it clearly shows that scalar-tensor geavity gives general relativistic expressions plus some extra terms.
\section{Energy Loss Rate}

We now wish to evaluate the rate of energy loss that is induced by the radiation-reaction terms in the equations of motion.  Because those equations of motion contain both $1.5$PN  as well as $2.5$PN contributions, we will have not only the normal ``quadrupole'' order contributions to the energy loss rate analogous to those that appear in general relativity, but also dipole contributions that are in principle larger by a factor of $1/v^2$.  Since the conventional ``counter'' for keeping track of contributions to the waveform and energy flux in the wave-zone denotes the GR quadrupole terms as ``Newtonian'' or $0$PN order, the dipole terms will, by this reckoning, be of $-1$PN order.  

To evaluate the energy loss correctly through ``Newtonian'' order, we first express the conserved energy in relative coordinates to $1$PN order.  Using the transformations (\ref{relative1}) and (\ref{relative2}) to $1$PN  order, we obtain
\begin{eqnarray}
E &=& \frac{1}{2} \mu v^2 - \mu \frac{G\alpha m}{r} + \frac{3}{8} \mu (1-3\eta) v^4 
\nonumber \\
&&  + \frac{1}{2} \mu \frac{G\alpha m}{r} \left [ (3 + 2\bar{\gamma} +\eta)v^2
 + \eta \dot{r}^2 \right ] +  \frac{1}{2} \mu \left ( \frac{G\alpha m}{r} \right )^2 (1+2\bar{\beta}_+ - 2\psi \bar{\beta}_{-}) \,.
\end{eqnarray}
We then calculate $dE/dt$, inserting the $1.5$PN and $2.5$PN acceleration terms into the leading term ${\bf v} \cdot {\bf a}$, and inserting only the $1.5$PN terms wherever accelerations occur in the time derivative of the $1$PN terms.  

Beginning with the leading term, and expressing the $1.5$PN acceleration in the form ${\bf a}_{1.5PN}  = (D/r^3) (3\dot{r} {\bf n} - {\bf v} )$, where
$D = 4\eta \zeta (G\alpha m)^2 {\cal S}_{-}^2/3$, we find for the $-1$PN term
$(dE/dt)_{-1PN} = \mu (D/r^3)(3\dot{r}^2 - v^2)$.   This can be simplified by exploiting the identity
\begin{equation}
\frac{d}{dt} \left ( \frac{\dot{r}}{r^2} \right ) = \frac{v^2 - 3 \dot{r}^2 +  {\bf x} \cdot {\bf a}}{r^3} \,.
\label{edotidentity}
\end{equation}
Thus $(v^2 - 3\dot{r}^2)/r^3$ can be written as the total time derivative of a quantity that can be absorbed as a $1.5$PN correction to the definition of $E$, leaving $(dE/dt)_{-1PN} = \mu (D/r^3) ({\bf x} \cdot {\bf a})$.  Inserting the Newtonian acceleration for $\bf a$, we obtain
\begin{equation}
(dE/dt)_{-1PN} = -\frac{4}{3} \zeta  \frac{\mu \eta}{r}  \left ( \frac{G\alpha m}{r} \right )^3 {\cal S}_{-}^2 \,.
\label{eq:edotdipole}
\end{equation}
This is in agreement with earlier calculations of the energy flux due to dipole gravitational radiation~\cite{eardley,tegp}.

However, since we are working to Newtonian order in the energy loss, we also need to include the $1$PN contributions to the acceleration that appears in Eq.\ (\ref{edotidentity}), yielding a contribution given by $\mu D (G\alpha m/r^4) (A_{1PN} + \dot{r}^2 B_{1PN})$, where $A_{1PN}$ and $B_{1PN}$ are given by Eqs.\ (\ref{eomfinalcoeffsPN}). We then combine this with the other Newtonian order terms generated from $dE/dt$, leading to an expression of the general form
\begin{equation}
\frac{dE}{dt} = - \frac{8}{15} \frac{\mu \eta}{r} \left ( \frac{G \alpha m}{r} \right )^2 
\left [ p_1 \frac{G \alpha m}{r} v^2 + p_2  \frac{G \alpha m}{r} \dot{r}^2
+ p_3 v^2 \dot{r}^2
+ p_4 \left ( \frac{G \alpha m}{r} \right )^2 + p_5 v^4 
+ p_6 \dot{r}^4 \right ]
\label{edotgeneral}
\end{equation}
We now use an identity derived from the Newtonian equations of motion,
\begin{equation}
\frac{d}{dt} \left ( \frac{v^{2s}\dot{r}^p}{r^q} \right ) = \frac{v^{2s-2} \dot{r}^{p-1}}{r^{q+1}} \left( p v^4 - pv^2 \frac{G\alpha m}{r} - (p+q) v^2 \dot{r}^2 - 2s \frac{G\alpha m}{r} \dot{r}^2 \right ) \,.
\end{equation}
This is applicable at this PN order provided that the integers $s$ and $p$ are non-negative, $q \ge 2$ and $2s + p + 2q =7$.  Using the three possible cases 
$(s,\, p,\, q) = (1,\,1 ,\, 2), \, (0,\, 3,\, 2),\, (0,\,1,\,3)$,
we can freely manipulate the values of three of the six coefficients $p_i$ in Eq.\ (\ref{edotgeneral}).   The idea is to combine terms on the right-hand-side of Eq.\ (\ref{edotgeneral}) into a total time derivative, to move that to the left-hand-side and then to absorb it into a meaningless redefinition of $E$ (see for example, \cite{balacliff1,balacliff2} for discussion).  Thus one can easily arrange for $p_4$, $p_ 5$ and $p_6$ to vanish.   It then turns out that the coefficient $p_3$ of the term proportional to $v^2 \dot{r}^2$ is proportional to the combination of the $2.5PN$ equation-of-motion coefficients $5a_1 + 3a_3 - 15b_1 -5b_3$.  An inspection of Eqs.\ (\ref{25PNcoeffs}) reveals that this combination miraculously vanishes.   Pulling everything together, we obtain the final expression for the energy loss rate,
\begin{equation}
(dE/dt)_{0PN} = - \frac{8}{15} \frac{\mu \eta}{r} \left ( \frac{G \alpha m}{r} \right )^3 \left (\kappa_1 v^2  - \kappa_2 \dot{r}^2 \right ) \,,
\end{equation}
where
\begin{eqnarray}
\kappa_1 &=& 12 + 5\bar{\gamma} - 5\zeta {\cal S}_{-}^2 (3 + \bar{\gamma} + 2\bar{\beta}_+ )
+10 \zeta {\cal S}_{-} \left ( \frac{{\cal S}_{-} \bar{\beta}_+ + {\cal S}_{+} \bar{\beta}_{-}}{\bar{\gamma}} \right )
\nonumber \\
&& 
 +10 \zeta \psi   {\cal S}_{-}^2 \bar{\beta}_{-} - 10\zeta \psi  {\cal S}_{-} \left (  \frac{{\cal S}_{+} \bar{\beta}_+ + {\cal S}_{-} \bar{\beta}_{-}}{\bar{\gamma}} \right ) \,,
\nonumber \\
\kappa_2 &=& 11 + \frac{45}{4} \bar{\gamma} - 40 \bar{\beta}_+
- 5\zeta {\cal S}_{-}^2  \left [ 17+ 6\bar{\gamma} + \eta + 8\bar{\beta}_+  \right ]
 +90 \zeta {\cal S}_{-} \left ( \frac{{\cal S}_{-} \bar{\beta}_+ + {\cal S}_{+} \bar{\beta}_{-}}{\bar{\gamma}} \right )
 \nonumber \\
&&  
+ 40 \zeta \psi {\cal S}_{-}^2 \bar{\beta}_{-} - 30\zeta \psi  {\cal S}_{-} \left ( \frac{{\cal S}_{+} \bar{\beta}_+ + {\cal S}_{-} \bar{\beta}_{-}}{\bar{\gamma}} \right )  
-120 \zeta \left (\frac{{\cal S}_{+} \bar{\beta}_+ + {\cal S}_{-} \bar{\beta}_{-}}{\bar{\gamma}} \right )^2 \,.
\end{eqnarray}
These results are in complete agreement with the total energy flux to $-1PN$ and $0PN$ orders, as calculated by Damour and Esposito-Far\`ese~\cite{DamourEsposito92}. \footnote{Thanks to Michael Horbatsch for his invaluable help in verifying this agreement.}

\begin{savequote}[0.65\linewidth]
{\scriptsize ``Science, my lad, is made up of mistakes, but they are mistakes which it is useful to make, because they lead little by little to the truth.''}
\qauthor{\scriptsize---Jules Verne, Journey to the Center of the Earth}
\end{savequote}
\chapter{Discussion} 
\begin{quote}
\end{quote}

\label{chapter12} 
\thispagestyle{myplain}
\lhead[\thepage]{Chapter 11. \emph{Discussion}}      
\rhead[Chapter 11. \emph{Discussion}]{\thepage}
We have used the DIRE approach based on post-Minkowskian theory to derive the explicit equations of motion in a general class of massless scalar-tensor theories of gravity for compact binary systems through $2.5$PN order.   Here we discuss the results, and compare our work with related work on scalar-tensor gravity and equations of motion.   

\section{General Remarks and Comparison with Other Results}
\label{sec:genremarks}

We begin by noting that, not surprisingly, the expressions are considerably more complicated than the corresponding general relativistic expressions (compare Eqs.~(\ref{eomfinal}, \ref{EOMcoeffs}) with Eqs.~(\ref{1PNeomfinal}, \ref{eomfinalcoeffs1PN}) and Eqs.~(1.2, 1.3, 5.4) in PWII).  Given that the results depend on the function $\omega(\phi)$ and its first and second derivatives, on the masses of each body, and on the sensitivities of each body and their derivatives, it is somewhat remarkable that the final equations of motion depend on a rather small number of parameters, as shown in the right-hand column of Table \ref{tab:params}.   The parameter $\alpha$ combines with $G$ to yield an effective two-body Newtonian coupling constant. It is not a universal constant, as it depends symmetrically on the sensitivities of each body.   The parameter $\bar{\gamma}$ and the body-dependent parameter $\bar{\beta}_A$ govern the post-Newtonian corrections, while the body-dependent parameters $\bar{\delta}_A$ and $\bar{\chi}_A$ govern the $2$PN corrections.   In the radiation-reaction terms, the sensitivities $s_A$ occur explicitly along with  $\bar{\gamma}$ and $\bar{\beta}_A$.  

The relative simplicity of the parameters at $1$PN and $2$PN orders has been noted before.  Damour and Esposito-Far\`ese~\cite{DamourEsposito92,DamourEsposito96} (DEF hereafter) studied a class of multi-scalar-tensor theories, but worked in the Einstein representation, where the gravitational action was pure general relativity, augmented by a free action for the scalar fields.  This is a non-metric representation of the theory, since the scalar field(s) couple to normal matter via a function $A(\varphi)$ (here we will focus on a single scalar field).  For a compact body with mass $\tilde{m}(\varphi)$ (using the Eardley ansatz), the effective matter action depends on the product $A(\varphi)\tilde{m}(\varphi)$.  The scalar field $\phi$ of our Jordan representation is given by $\phi = A(\varphi)^{-2}$, and $3 + 2\omega(\phi) = (d\ln A/d\varphi)^{-2}$.  Using a diagrammatic approach, DEF showed that the important quantities involved derivatives of $A(\varphi)\tilde{m}(\varphi)$ with respect to $\varphi$, and consequently (in our language) $\omega$ and $s_A$ and their derivatives always combined in specific ways, leading to relatively few parameters.  Table \ref{tab:dictionary} gives a dictionary that translates from our parameters to those of DEF for the case of two bodies.  Interestingly, our parameters $\bar{\delta}_A$ do not appear in DEF's list, so far as we could tell.  

\begin{table}
\centering
\begin{tabular}{| c| ccc| }
\hline
This work&DEF&TEGP&PPN limit\\
\hline
$G\alpha$&$G_{12}$&${\cal G}_{12}$& 1\\
$\bar{\gamma}$&$\bar{\gamma}_{12}$&$\frac{3}{2} ({\cal B}_{12}/{\cal G}_{12} -1 )$&$ \gamma -1$\\
$\bar{\beta}_1$&$\beta^1_{22}$&$\frac{1}{2} ( {\cal D}_{122}/{\cal G}_{12}^2 -1)$&$ \beta -1 $\\
$\bar{\beta}_2$&$\beta^2_{11}$&$\frac{1}{2} ( {\cal D}_{211}/{\cal G}_{12}^2 -1)$&$ \beta -1 $\\
$\bar{\delta}_1$&$-$&$-$&$-$\\
$\bar{\delta}_2$&$-$&$-$&$-$\\
$\bar{\chi}_1$&$-\frac{1}{4}\epsilon^1_{222}$&$ -$&$- $\\
$\bar{\chi}_2$&$-\frac{1}{4}\epsilon^2_{111}$&$ -$&$- $\\
$\bar{\gamma}^{-1} \bar{\beta}_1 \bar{\beta}_2$&$-\frac{1}{2} \zeta_{1212}$&$-$&$ -$\\
\hline
\end{tabular}
\caption{\label{tab:dictionary} Dictionary of parameters used in the equations of motion.  DEF refers to Ref. \cite{DamourEsposito92,DamourEsposito96}; TEGP refers to Sec.\ 11.3 of Ref. \cite{tegp}; PPN refers to the parametrized post-Newtonian limit of weakly gravitating bodies}
\end{table}

In the $1$PN limit, Will~\cite{tegp} wrote down a general $N$-body Lagrangian for compact self-gravitating bodies that could span a wide class of metric theories of gravity that embody post-Galilean invariance (so-called ``semi-conservative'' theories of gravity), and that have no ``Whitehead'' potential in the post-Newtonian limit.   Comparing our Lagrangian of scalar-tensor theory with the 2-body limit of Eq.\ (11.62) of \cite{tegp}, we can translate between our parameters and the coefficients ${\cal G}_{ab}$, ${\cal B}_{ab}$, and ${\cal D}_{abc}$ of \cite{tegp}, as shown in Table \ref{tab:dictionary}.    

The factor $1-2s_A$ appears throughout these equations.   This quantity is often called the ``scalar charge'' of the object.  From the point of view of the Einstein representation of scalar-tensor theory, it is easy to see how this factor arises.  The scalar field appears in the gravitational part of the action only in a kinetic term $g^{\mu\nu} \varphi_{,\mu} \varphi_{,\nu}$ (we assume that there is no potential $V(\varphi)$).  It does not couple to gravity other than via the metric in the kinetic term.  The effective matter action for a compact body depends on the product $A(\varphi) M(\varphi)$.  Varying this product with respect to $\varphi$ yields the quantity
\begin{equation}
A(\varphi) M(\varphi) \left ( \frac{d \ln A}{d\varphi} + \frac{d \ln M}{d \ln \phi}
\frac{d \ln \phi}{d \varphi} \right )  \delta \varphi= A(\varphi) M(\varphi) \frac{d \ln A}{d\varphi} (1 - 2s)  \delta \varphi\,,
\end{equation}
where we used the fact that $\ln \phi = -2 \ln A(\varphi)$.  Thus the factor $1-2s$ and its derivatives naturally control the source of the scalar field, as can be seen clearly in Eq.\ (\ref{sigmasPN}).   Defining a scalar charge for body $A$ in a two-body system by 
\begin{equation}
q_A \equiv \alpha^{-1/2} (1 - 2s_A) \,,
\end{equation}
we see that the quantities ${\cal S}_{\pm}$ are given by 
\begin{eqnarray}
{\cal S}_+ &=& \frac{1}{2} (q_1 + q_2 ) \,,
\nonumber \\
{\cal S}_{-} &=& \frac{1}{2} (q_1 - q_2 ) \,.
\end{eqnarray}

The scalar charge, or sensitivity of a given body depends on its internal structure.  For weakly gravitating bodies, $s \approx -\Omega/M \ll 1$, where
$\Omega \equiv -(1/2)G\int\int \rho^* \rho'^* \mid {\bf x} - {\bf x}'\mid ^{-1} d^3x d^3x'$ is 
the Newtonian self-gravitational binding energy .  For neutron stars, values of the sensitivities range from $0.1$ to $0.3$, depending on the mass and equation of state of the body~\cite{willzaglauer,zaglauer} and can vary dramatically, depending on the specific form of $\omega(\phi)$~\cite{DamourEsposito92}.

\section{Weakly Self-Gravitating Systems}

In the post-Newtonian limit with weakly self-gravitating systems, the sensitivities $s_i$ are themselves of order $\epsilon$.  If one is working purely at $1$PN order, then the effects of sensitivities in the $1$PN terms of Eq.\ (\ref{1PNeom}) will be of $2$PN order.  So the only effect of the bodies' sensitivities in this case will come from the coefficient $\alpha$ in the Newtonian term.  Consider a specific example: body $1$ with sensitivity $s_1$ resides in the field of body $2$, with sensitivity zero.  The acceleration of body $1$ is then given by
\begin{equation}
{\bf a}_1 = - \frac{Gm_2}{r^2} n^i (1 - 2\zeta s_1) \,,
\end{equation}
and thus the body's Newtonian acceleration will depend on its internal structure, a violation of the Strong Equivalence Principle, commonly known as the Nordtvedt effect.  In the PPN framework~\cite{tegp}, the Nordtvedt effect is normally expressed in terms of $\Omega$. Alternatively, since $M \approx m_0 + \Omega$, we have that $\Omega/M = d \ln M/d\ln G$.   Taking into account Eq.\ (\ref{Gdefinition}), we can connect the sensitivity $s$ to $\Omega$ by
\begin{equation}
s = \left ( \frac{d \ln M}{d \ln G} \right )_0 \left ( \frac{d \ln G}{d \ln \phi} \right )_0  =  - \frac{\Omega}{M} \left [ 1 + 4\Lambda (2+\omega_0) \right ] \,,
\label{svsomega}
\end{equation}
where
\begin{equation}
\Lambda \equiv \frac{\phi_0 (d\omega/d\phi)_0}{(4+2\omega_0)^2(3+2\omega_0)} 
\end{equation}
is the parameter defined in TEGP (see Eqs.\ (5.36) and (5.38)) such that the PPN parameter $\beta = 1+\Lambda$ in scalar-tensor theory (note the relationship between $\phi_0$ and $G$, which is set equal to unity in TEGP).   We also have that $\gamma = 1 - 2\zeta$.  We can then express the acceleration of body $1$ as 
\begin{equation}
{\bf a}_1 = - \frac{Gm_2}{r^2} {\bf n} \left [1 + \left ( \frac{1}{2+\omega_0} + 4\Lambda \right ) \frac{\Omega_1}{m_1}  \right ] \,.
\end{equation}
The coefficient in front of $\Omega_1/m_1$ is precisely $4\beta - \gamma -3$, as in the standard PPN framework.  

In the $1$PN terms in Eq.\ (\ref{1PNeom}), for weakly self-gravitating systems, it is easy to see from Table \ref{tab:params} that in the limit $s_i \to 0$, $\alpha \to 1$, the parameters $\bar{\gamma}$ and $\bar{\beta}_i$ tend to the PPN parameters $\gamma -1$ and $\beta -1$, respectively, as shown in Table \ref{tab:dictionary}, and thus our equations of motion at $1$PN order agree with the standard ones for ``point'' masses in scalar-tensor theory.

The radiation-reaction results can also be compared with existing work.  The $-1PN$ energy loss due to dipole gravitational radiation reaction, Eq.\ (\ref{eq:edotdipole}) is in complete agreement with calculations of the dipole energy flux~\cite{eardley,willrad,tegp}.  In comparing Eq.\ (\ref{eq:edotdipole}) with Eqs.\ (10.84) and (10.136) of \cite{tegp}, the additional factor of $[1+4\Lambda (2+\omega_0)]^2$ arises from the relation (\ref{svsomega}) between $s$ and $\Omega/M$.  

For weakly self-gravitating bodies, the Newtonian-order energy loss simplifies by virtue of setting all sensitivities equal to zero.   In this case, with $\alpha =1$,
$\bar{\gamma} = -2\zeta$,  
$\bar{\beta}_+ = \beta - 1 = \Lambda$, $\bar{\beta}_{-} =0$, ${\cal S}_{-} =0$, and ${\cal S}_+ = 1$, we obtain
\begin{eqnarray}
\kappa_1 &=& 12 - \frac{5}{2 + \omega_0} \,,
\nonumber \\
\kappa_2 &=& 11 - \frac{45}{2}\zeta  - 40 \Lambda -30 \Lambda^2/\zeta
\nonumber \\
&=& 11 - \frac{45}{8 +4\omega_0} \left [ 1 + \frac{8}{9} \left (\frac{2\Lambda}{\zeta} \right ) + \frac{1}{3} \left (\frac{2\Lambda}{\zeta} \right )^2 \right ] \,.
\end{eqnarray}
These agree completely with Eq.\ (10.136) of \cite{tegp}.

\section{Binary Black Holes}

Roger Penrose was probably the first to conjecture, in a talk at the 1970 Fifth Texas Symposium, that black holes in Brans-Dicke theory are identical to their GR counterparts~\cite{thornedykla}.  Motivated by this remark, Thorne and Dykla showed that during gravitational collapse to form a black hole, the Brans-Dicke scalar field is radiated away, in accord with Price's theorem, leaving only its constant asymptotic value, and a GR black hole~\cite{thornedykla}.  Hawking~\cite{hawking} proved on general grounds that stationary, asymptotically flat black holes in vacuum in BD are the black holes of GR.  The basic idea is that black holes in vacuum with non-singular event horizons cannot support scalar ``hair''.   Hawking's theorem was extended to the class of $f(R)$ theories that can be transformed into generalized scalar-tensor theories by Sotiriou and Faraoni~\cite{sotirioufaraoni}.   
 
For a stationary single body, it is clear from Eq.\ (\ref{sigmasPN}) that, if $s = 1/2$ and all its derivatives vanish, the only solution for the scalar field is $\phi \equiv \phi_0$, and hence the equations reduce to those of general relativity.
In the Einstein representation, this corresponds to $A(\varphi) M(\varphi) =$ constant, so that the scalar field decouples from any source, and thus must be either constant or singular.    Consequently, stationary black holes are characterized by $s = 1/2$. 

Another way to see this is to note that, because all information about the matter that formed the black hole has vanished behind the event horizon, the only scale on which the mass of the hole can depend is the Planck scale, and thus $M \propto M_{Planck} \propto G^{-1/2} \propto \phi^{1/2}$.  Hence $s = 1/2$.

If $s_A=1/2$ for each black hole in a binary system, then, as we discussed in the introduction, all the parameters $\bar{\gamma}$, $\bar{\beta}_A$, $\bar{\delta}_A$, $\bar{\chi}_A$, and ${\cal S}_\pm$ vanish identically, and $\alpha = 1-\zeta$.  But since $\alpha$ appears only in the combination with $G\alpha m_A$, a simple rescaling of each mass puts all equations into complete agreement with those of general relativity, through $2.5$PN order.  

But is $s_A = 1/2$ really true for binary black holes?  If the orbital timescale is long compared to the dynamical (quasinormal mode) timescale of each black hole, then it is plausible to assume that Hawking's theorem holds for each black hole, at least up to some PN order.  On the other hand, one could imagine a situation where each hole is distorted by the tidal forces from the companion hole, or where gravitational radiation flowing across the event horizons disrupts the stationarity needed for Hawking's theorem.  In PN language, these kinds of effects are known to be of an order higher than the $2.5$PN order achieved in this paper, so perhaps some non-GR effects might emerge at sufficiently high PN order.   Can a perturbation of the scalar field be supported sufficiently by strong gravity or by time varying fields to make any difference?  Or, without matter to support it, does any scalar perturbation get radiated away on a quasinormal-mode timescale, which is short compared to the orbital timescale, except during the merger of the two black holes?  Preliminary evidence from numerical relativity supports the latter scenario: Healy {\em et al.}~\cite{healy} introduced a very large Brans-Dicke type scalar field into the initial data of a binary black hole merger and found that, while the field affected the inspiral while it lasted, it was radiated away rather quickly, although it was not possible from the numerical data to fully quantify this.

It should be pointed out that there {\em are} ways to induce scalar hair on a black hole.  One is to introduce a potential $V(\phi)$, which, depending on its form, can help to support a non-trivial scalar field outside a black hole.   Another is to introduce matter.  A companion neutron star is an obvious choice, and such a binary system  in scalar-tensor theory is clearly different from its general relativistic counterpart (see the next subsection).   Another possibility is a distribution of cosmological matter that can support a time-varying scalar field at infinity.  This possibility has been called ``Jacobson's miracle hair-growth formula'' for black holes, based on work by Jacobson~\cite{jacobsonhair,burgess}.  Whether it is possible to incorporate such ideas into our approach is a subject for future work.

These considerations motivate us to formulate a conjecture along the following lines:  Consider a scalar-tensor theory of gravity with no potential for the scalar field, and consider two black holes with non-singular event horizons in a vacuum (no normal matter), asymptotically flat spacetime with $\phi$ at spatial infinity constant in time.  Following an initial transient period short compared to the orbital period, the orbital evolution and gravitational radiation from the binary system are identical to those predicted by GR, after a mass rescaling, independent of the initial scalar field configuration.   Aspects of this conjecture could be addressed by numerical simulations that extend the work of~\cite{healy}.
It may also be possible to address it partially by generalizing Hawking's theorem to a situation that is not strictly stationary, but yet still retains some symmetry, such as a helical Killing vector.  This will be the subject of future work.  

\section{Black-Hole Neutron-Star Binary Systems}

Finally, we note the unusual circumstance that, if only one of the members of the binary system, say body 2, is a black hole, with $s_2 = 1/2$, then $\alpha = 1-\zeta$, $\bar{\gamma} = \bar{\beta}_A = 0$, and hence, through $1$PN order, the motion is again identical to that in general relativity.  This result is actually implicit in the post-Newtonian equations of motion for compact binaries in 
Brans-Dicke theory displayed in Eq.\ (11.91) of~\cite{tegp}, but was never stated explicitly there.  

At $1.5$PN order, dipole radiation reaction kicks in, since $s_1 < 1/2$.   In this case, ${\cal S}_{-} = {\cal S}_+ = \alpha^{-1/2} (1-2s_1)/2$, and thus the $1.5$PN coefficients in the relative equation of motion (\ref{eomfinal}) take the form
\begin{eqnarray}
A_{1.5PN} &=&  \frac{5}{8} Q \,,
\nonumber \\
B_{1.5PN} &=& \frac{5}{24} Q \,,
\end{eqnarray}
where
\begin{equation}
Q \equiv  \frac{\zeta}{1-\zeta} (1-2s_1)^2 = \frac{1}{3+2\omega_0} (1-2s_1)^2 \,.
\label{Qdefinition}
\end{equation}
At $2$PN order, $\bar{\chi}_A = \bar{\delta}_2 =0$, but $\bar{\delta}_1 = Q \ne 0$.   In this case, the $2$PN coefficients in (\ref{eomfinal}) take the form
\begin{eqnarray}
A_{2PN} &=& A_{2PN}^{GR} + Q \frac{G\alpha m_1}{r}
\left [ \dot{r}^2 -\frac{G\alpha m_1}{r} \right ] \,,
\nonumber \\
B_{2PN} &=& B_{2PN}^{GR} -2Q \frac{G\alpha m_1}{r} \,.
\end{eqnarray}
Finally, the $2.5$PN coefficients in Eq.\ (\ref{25pnAB}) have the form
\begin{eqnarray}
a_1 &=& 3 + \frac{5}{32} Q (9-2\eta+3\psi) \,,
\nonumber \\
a_2 &=& \frac{17}{3} - \frac{5}{96} Q (135+8\eta+ 3\psi) \,,
\nonumber \\
a_3 &=& - \frac{25}{32} Q (1-2\eta+\psi) \,,
\nonumber \\
b_1 &=& 1 - \frac{5}{96} Q (7-2\eta- 3\psi) \,,
\nonumber \\
b_2 &=& 3 - \frac{5}{96} Q (23 - 8 \eta + 3\psi) \,,
\nonumber \\
b_3 &=& \frac{5}{32}  Q (13 + 2\eta-3\psi) \,,
\end{eqnarray}
while the coefficients in the energy loss rate simplify to
\begin{eqnarray}
\kappa_1 &=& 12 - \frac{15}{4} Q \,,
\nonumber \\
\kappa_2 &=& 11 - \frac{5}{4} Q (17 + \eta) \,.
\end{eqnarray}
We  find, somewhat surprisingly, that the motion of a mixed compact binary system through $2.5$PN order differs from its general relativistic counterpart only by terms that depend on a single parameter $Q$, as defined by Eq.\ (\ref{Qdefinition}).  Furthermore, all reference to the parameters $\lambda_1$ and $\lambda_2$, related to derivatives of the coupling function $\omega(\phi)$, has disappeared, in other words, the motion of mixed compact binary systems in general scalar-tensor theories through $2.5$PN order is formally identical to that in standard Brans-Dicke theory.   The only way that a generalized scalar-tensor theory affects the motion differently than pure Brans-Dicke theory is through the value of the un-rescaled mass $m_1$ and the sensitivity $s_1$ for a neutron star of a given central density and total number of baryons.

The general conclusions reached in this work about binary black holes and mixed binaries in scalar-tensor gravity were obtained from the near-zone gravitational fields. If these conclusions continue to hold for the gravitational-wave signal, then gravitational-wave observations of binary black holes will be unable to distinguish between general relativity and scalar-tensor theories, and observations of mixed black-hole neutron-star binaries will be essentially unable to distinguish between general scalar-tensor theories and Brans-Dicke theory (Fig.~\ref{fig:binaries} illustrates this fact). The   radiative part of this problem, which will involve a derivation of the gravitational waveform to $2$PN order, together with the energy flux, will be the subject of future work.  

\begin{figure}[h]
\centering
\includegraphics[width=0.6\textwidth]{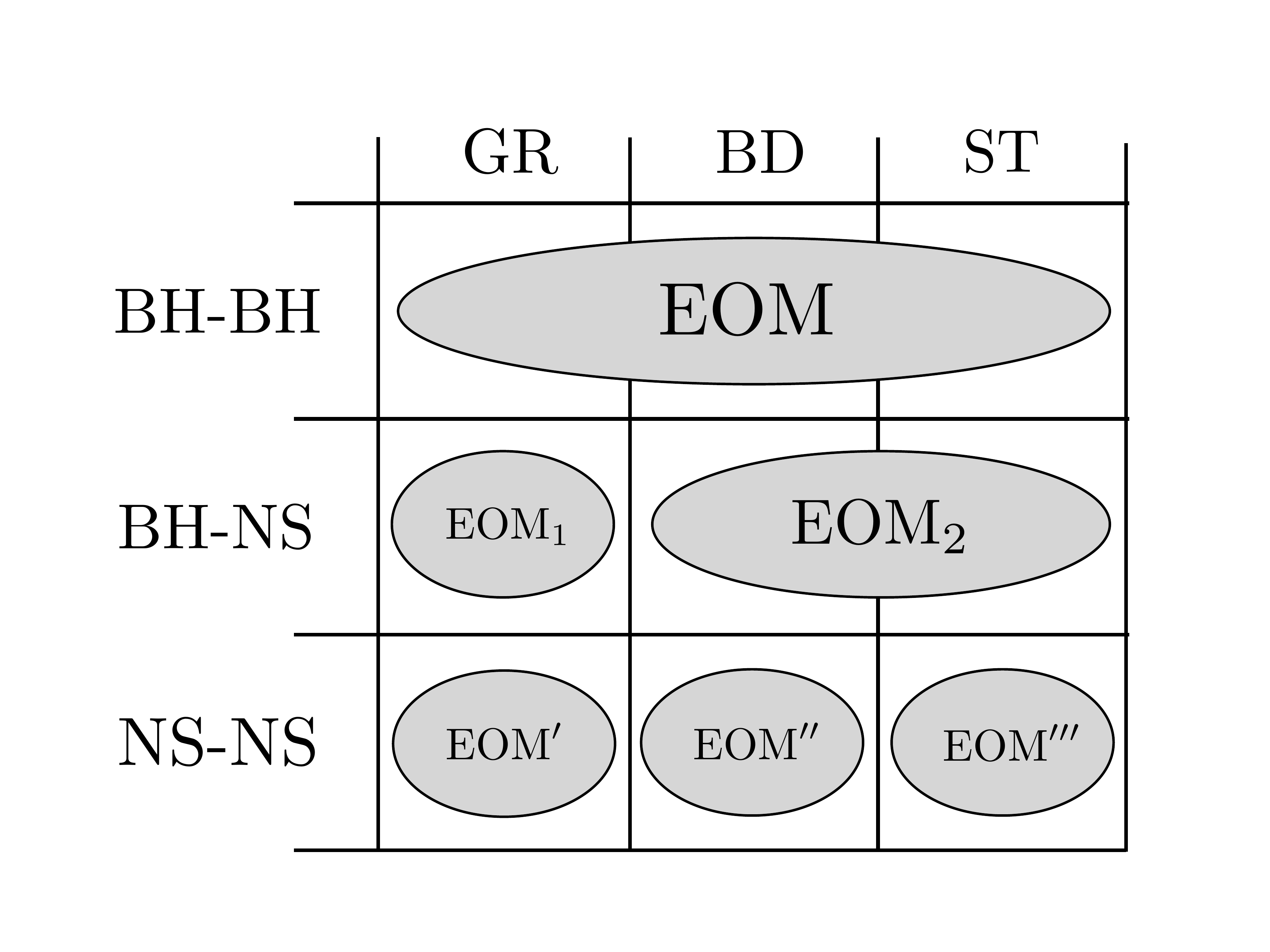}
\caption{For three different combinations of neutron-stars and black-holes in a binary system, this figure shows how the equations of motion can be able to distinguish between general relativity, Brans-Dicke theory, and general scalar-tensor theories of gravity. In the case of BH-BH all three theories are indistinguishable. For a BH-NS binary, the equations of motion in Brans-Dicke theory and general scalar-tensor theories are equivalent but both differ from GR. A binary system of two neutron-star is the only case where GR, BD, and ST each gives different equations of motion.}
\label{fig:binaries}
\end{figure}





\mypart{Constraining Lorentz-Violating, Modified Dispersion Relations with Gravitational Waves}{{\vfill \small{\em This part is based on a published paper in Physical Review D.~\cite{mir12} in which we construct a parametrized dispersion relation that can reproduce a range of known Lorentz-violating predictions and investigate their impact on the propagation of gravitational waves. We show how such corrections map to the waveform observable and to the parametrized post-Einsteinian framework, proposed to model a range of deviations from General Relativity. Given a gravitational-wave detection, the lack of evidence for such corrections could then be used to place a constraint on Lorentz violation. The constraints we obtain are tightest for dispersion relations that scale with small power of the graviton's momentum and deteriorate for a steeper scaling.}}}
{\begin{framed}
\begin{itemize}
\item  \cref{chapter13}--- \nameref{chapter13}
\item  \cref{chapter14}--- \nameref{chapter14}
\item  \cref{chapter15}--- \nameref{chapter15}
\end{itemize}
\end{framed}
}{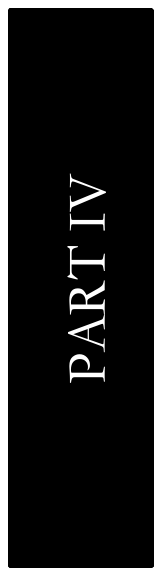}
\label{part:4}
\begin{savequote}[0.65\linewidth]
{\scriptsize ``Science may be described as the art of systematic oversimplification.''}
\qauthor{\scriptsize--- Karl R. Popper}
\end{savequote}
\chapter{Introduction and Foundations} 
\ClearWallPaper
In this chapter we first start with a brief introduction to declare the possibility of testing alternative theories of gravity by studying gravitational-wave signals emitted from inspiralling compact binary sources. Second, we propose a general, parametrized dispersion relation for Lorentz-violating theories which will be useful to do parameter estimation of the source and bounding the parameters of this modified dispersion relation, specially the parameter that presents the deviation from Lorentz symmetry. We also give an overview on the next following chapters of this part. The obtained bounds on the mass of graviton and on the deviation from Lorentz symmetry are also summarized.

\label{chapter13} 
\thispagestyle{myplain}
\lhead[\thepage]{Chapter 12. \emph{Introduction}}      
\rhead[Chapter 12. \emph{Introduction}]{\thepage}
\section{Introduction}

After a century of experimental success, Einstein's fundamental theories, ie.~the special theory of relativity and the General theory of Relativity (GR), are beginning to be questioned. As an example, consider the observation of ultra-high-energy cosmic rays. In relativity, there is a threshold of $\sim 5\times 10^{19} \; {\rm{eV}}$ (GZK limit) for the amount of energy that charged particles can carry, while cosmic rays have been detected with higher energies~\cite{Bird:1994uy}. On the theoretical front, theories of quantum gravity also generically predict a deviation from Einstein's theory at sufficiently large energies or small scales. In particular, Lorentz violation seems ubiquitous in such theories.  These considerations motivate us to study the effects of Lorentz violation on gravitational wave observables.

Einstein's theory will soon be put to the test through a new type of observation: gravitational-waves. Such waves are (far-field) oscillations of spacetime that encode invaluable and detailed information about the source that produced them. For example, the inspiral, merger and ringdown of compact objects (black holes or neutron stars) are expected to produce detectable waves that will access horizon-scale curvatures and energies. Gravitational waves may thus provide new hints as to whether Einstein's theory remains valid in this previously untested regime. For more details about gravitational-waves see \cref{chapter4}.

Gravitational-wave detectors are today a reality.  As we mentioned earlier in \cref{chapter4}, ground-based interferometers, such as the Advanced Laser Interferometer Gravitational Observatory (Ad.~LIGO)~\cite{ligo,Abbott:2007kv,2010CQGra..27h4006H} and Advanced~Virgo~\cite{virgo}, are currently being updated, and are scheduled to begin data acquisition by 2015. Second generation detectors, such as the Einstein Telescope (ET)~\cite{et,Punturo:2010zz} and the Laser Interferometer Space Antenna (LISA)~\cite{lisa,Prince:2003aa}, are also being planned for the next decade. Recent budgetary constraints in the United States have cast doubt on the status of LISA, but the European Space Agency is still considering a descoped, LISA-like mission (an NGO, or New Gravitational Observatory).  The detection of gravitational waves is, of course, not a certainty, as the astrophysical event rate is highly uncertain. However, there is consensus that advanced ground detectors should observe a few gravitational-wave events by the end of this decade.  

\begin{figure}[h]
\centering
\includegraphics[width=0.6\textwidth]{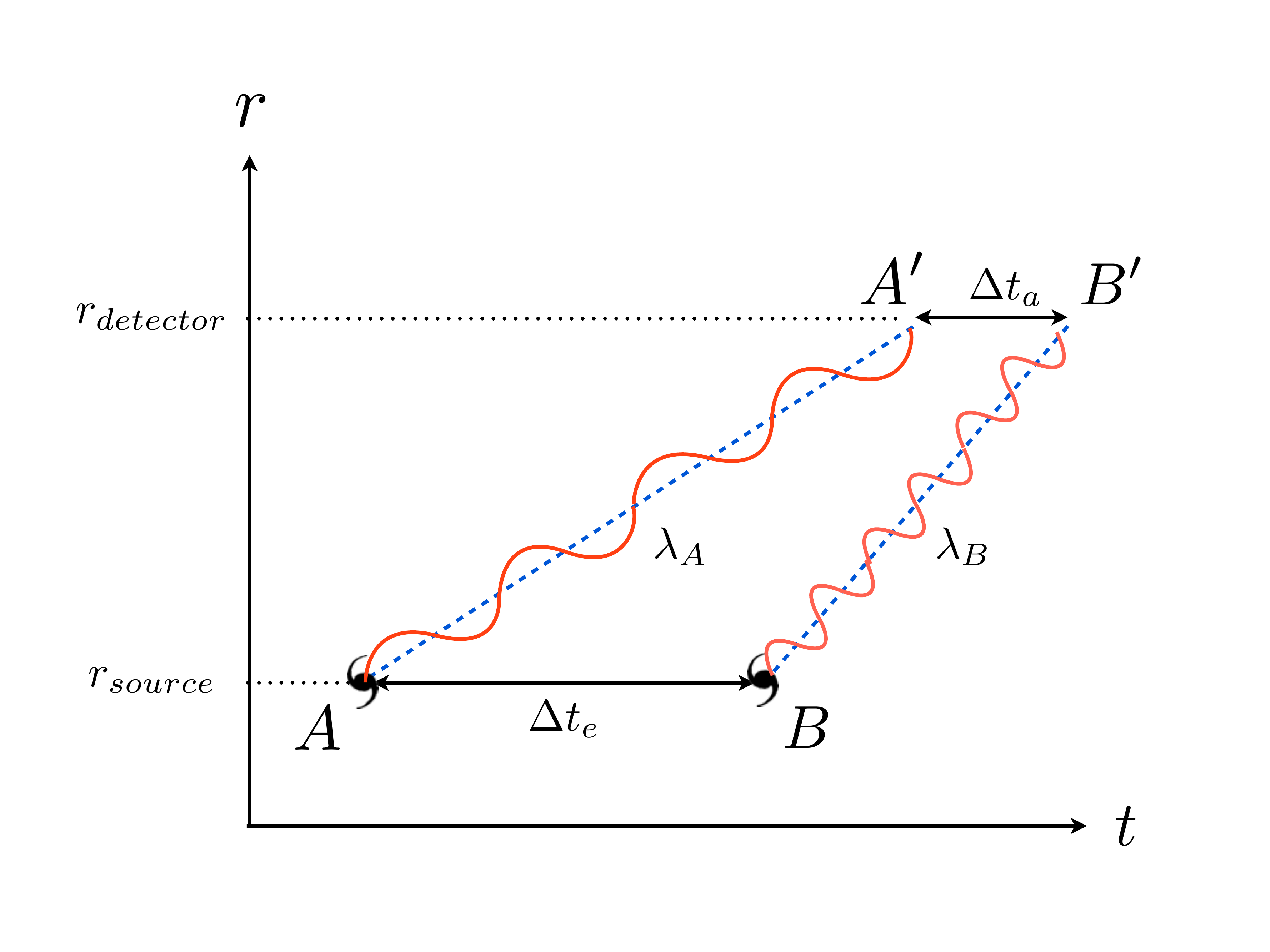}
\caption{Event $A$ is emission of a gravitational-wave signal from an inspiralling binary source with the wavelength of $\lambda_A$. We detect this signal at $A'$. After $\Delta t_e$ another signal is emitted by the same source at point $B$ in the spacetime, at a time closer to the merger. The wavelength of this signal, $\lambda_B$, is shorter therefore it travels faster than the first signal. We detect this second signal at  $B'$. Notice that $\Delta t_e \textgreater \Delta t_a$}
\label{fig:redshift}
\end{figure}

Some alternative gravity theories endow the graviton with a mass~\cite{tegp}.  Massive gravitons would travel slower than the speed of light, but most importantly, their speed would depend on their energy or wavelength. Since gravitational waves emitted by compact binary inspirals chirp in frequency, gravitons emitted in the early inspiral will travel more slowly than those emitted close to merger, leading to a frequency-dependent gravitational-wave dephasing, compared to the phasing of a massless general relativistic graviton. This fact is shown schematically in Fig.~\ref{fig:redshift}. If such a dephasing is not observed, then one could place a constraint on the graviton mass~\cite{Will:1997bb}.
A Lorentz-violating graviton dispersion relation leaves an additional
imprint on the propagation of gravitational waves, irrespective of the generation mechanism.  Thus a bound on the dephasing effect could also bound the degree of Lorentz violation.

Note that our use of the term ``graviton'' is not meant to imply that geavitational-wave detectors will observe individual gravitons. The detected waves are perfectly classical, i.e. they contain enourmous numbers of gravitons. In this work, we construct a framework to study the impact of a Lorentz-violating dispersion relation on the propagation of gravitational waves. We begin by proposing a generic, but quantum-gravitational inspired, modified dispersion relation, given by
\be
\label{dispersion}
E^2=p^2c^2 + m_{g}^2c^4 + \mathbb{A} p^\alpha c^{\alpha} ,
\ee
where $m_g$ is the mass of the graviton and $\mathbb{A}$ and $\alpha$ are two Lorentz-violating parameters that characterize the GR deviation ($\alpha$ is dimensionless while $\mathbb{A}$ has dimensions of $[{\rm{energy}}]^{2-\alpha}$).  We will assume that $\mathbb{A}/(cp)^{2-\alpha} \ll 1$. When either $\mathbb{A} = 0$ or $\alpha = 0$, the modification reduces to that of a massive graviton. When $\alpha=(3,4)$, one recovers predictions of certain quantum-gravitation inspired models. This modified dispersion relation introduces Lorentz-violating deviations in a continuous way, such that when the parameter $\mathbb{A}$ is taken to zero, the dispersion relation reduces to that of a simple massive graviton.

The dispersion relation of Eq.~(\ref{dispersion}) modifies the gravitational waveform observed at a detector by correcting the phase with certain frequency-dependent terms. In the {\em stationary-phase approximation} (SPA), the Fourier transform of the waveform is corrected by a term of the form $\zeta(\mathbb{A}) u^{\alpha-1}$, where $u = \pi {\cal{M}} f$ is a dimensionless measure of the gravitational-wave frequency with ${\cal{M}}$  being the ``chirp mass''. We show that such a modification can be easily mapped to the recently proposed {\em parametrized post-Einsteinian framework} (ppE)~\cite{Yunes:2009ke,Cornish:2011ys} for an appropriate choice of ppE parameters. 

In deriving the gravitational-wave Fourier transform we must assume a functional form for the waveform as emitted at the source so as to relate the time of arrival at the detector to the gravitational-wave frequency.  In principle, this would require a prediction for the equations of motion and gravitational-wave emission for each Lorentz violating theory under study.  However few such theories have reached a sufficient state of development to produce such predictions.  On the other hand, it is reasonable to assume that the predictions will be not too different from those of general relativity.  For example, Will argued~\cite{Will:1997bb} that for a theory with a massive graviton, the differences would be of order $(\lambda/\lambda_g)^2$, where $\lambda$ is the gravitational wavelength, and $\lambda_g$ is the graviton Compton wavelength, and $\lambda_g \gg \lambda$ for sources of interest.  Similar behavior might be expected in Lorentz violating theories.   The important phenomenon is the accumulation of dephasing over the enormous propagation distances from source to detector, not the small differences in the source behavior.   As a result, we will use the standard general relativistic wave generation framework for the source waveform.

\section{An Overview}
With this new waveform model described in the previous section, we then carry out a simplified (angle-averaged) Fisher-matrix analysis to estimate the accuracy to which the parameter $\zeta (\mathbb{A})$ could be constrained as a function of $\alpha$, given a gravitational-wave detection consistent with general relativity. We perform this study with a waveform model that represents a non-spinning, quasi-circular, compact binary inspiral, but that deviates from general relativity only through the effect of the modified dispersion relation on the propagation speed of the waves, via Eq.~(\ref{dispersion}).  

To illustrate our results, we show in Table~\ref{summary} the accuracy to which Lorentz-violation in the $\alpha=3$ case could be constrained, as a function of system masses and detectors for fixed {\em signal-to-noise ratio} (SNR). The case $\alpha=3$ is a prediction of ``doubly special relativity''. The bounds on the graviton mass are consistent with previous studies~\cite{Will:1997bb,Will:2004xi,BBW,arunwill09,keppel:2010qu,2010PhRvD..81f4008Y} (for a recent summary of current and proposed bounds on $m_g$ see~\cite{2011arXiv1107.3528B}).  The table here means that given a gravitational-wave detection consistent with GR, $m_{g}$ and ${\mathbb{A}}$ would have to be smaller than the numbers on the third and fourth columns respectively. \\

\setlength{\tabcolsep}{5pt} 
\setlength{\extrarowheight}{1.5pt}
\begin{table}[ht]
\centering
\begin{tabular}{| l| cc|  c c| }
\hline
Detector&$m_1$&$m_2$&					$m_g(eV)$		&${\mathbb A}(eV^{-1})$	\\
\hline\hline
Ad.~LIGO&			1.4&1.4&$			3.71\times 10^{-22}$&	$7.36\times 10^{-8}$\\
${\rm{SNR }} \;( \text{or } \rho)=10$&		1.4&10&$				3.56\times 10^{-22}$&	$3.54\times 10^{-7}$\\
&					10&10&$				3.51\times 10^{-22}$&	$6.83\times 10^{-7}$\\
\hline
ET&					10&10&$				2.99\times 10^{-23}$&	$2.32\times 10^{-8}$\\
${\rm{SNR }}\;( \text{or } \rho)=50$&		10&100&$			4.81\times 10^{-23}$&	$1.12\times 10^{-6}$\\
&					100&100&$			6.67\times 10^{-23}$&	$3.34\times 10^{-6}$\\
\hline
NGO&				$10^4$&$10^4$&$		3.05\times 10^{-25}$&	$2.16\times 10^{-2}$\\
${\rm{SNR }}\;( \text{or } \rho)=100$&		$10^4$&$10^5$&$		2.46\times 10^{-25}$&	$0.147$\\
&					$10^5$&$10^5$&$		2.03\times 10^{-25}$&	$0.189$\\
&					$10^5$&$10^6$&$		2.09\times 10^{-25}$&	$9.57$\\
&					$10^6$&$10^6$&$		1.49\times 10^{-25}$&	$23.2$\\
\hline
\end{tabular}

\caption{\label{summary} Accuracy to which graviton mass and the Lorentz-violating parameter ${\mathbb{A}}$ could be constrained for the $\alpha=3$ case, given a 
gravitational-wave detection consistent with GR. The first column lists the masses of the objects considered, the instrument analyzed and the signal-to-noise ratio (SNR).}
\end{table}

Let us now compare these bounds with current constraints. The mass of the graviton has been constrained dynamically to $m_{g} \leq 7.6 \times 10^{-20}$ eV through binary pulsar observations of the orbital period decay and statically to $4.4 \times 10^{-22}$ eV with Solar System constraints (see e.g.~\cite{2011arXiv1107.3528B}). We see then that even with the inclusion of an additional $\mathbb{A}$ parameter, the projected gravitational wave bounds on $m_{g}$ are still interesting. The quantity $\mathbb{A}$ has not been constrained in the gravitational sector. In the electromagnetic sector, the dispersion relation of the photon has been constrained: for example, for $\alpha=3$, $\mathbb{A} \lesssim 10^{-25} \; {\rm{eV}}^{-1}$ using TeV $\gamma$-ray observations~\cite{Biller:1998hg}. One should note, however, that such bounds on the photon dispersion relation are independent of those we study here, as in principle the photon and the graviton dispersion relations need not be tied together.

We must stress that, in this work, we only deal with Lorentz-violating corrections to the gravitational wave dispersion relation, and thus, we deal only with {\emph{propagation effects}} and not with {\emph{generation effects}}. Generation effects will in principle be very important, possible leading to the excitation of additional polarizations, as well as modifications to the quadrupole expressions. Such is the case in several modified gravity theories, such as Einstein-Aether theory and Ho\v{r}ava-Lifshitz theory~\cite{Blas:2011zd, 2004PhRvD..70b4003J, 2007PhRvD..76h4033F, 2010PhRvD..81f4031S, 2009PhRvD..80d4032S, 2011arXiv1105.2555H, 2010AIPC.1241.1128R, 2009PhRvD..79j4004B, 2011arXiv1103.3439M, 2011PhRvD..84j4035P, 2006JCAP...02..003D, 2011arXiv1110.5950Y, teves}. Generically studying the generation problem, however, is difficult, as there does not exist a general Lagrangian density that can capture all Lorentz-violating effects. Instead, one would have the gargantuan task of solving the generation problem within each specific theory. 

The goal of this piece of work, instead, is to consider generic Lorentz-violating effects in the dispersion relation and focus only on the propagation of gravitational waves. This will then allow us to find the corresponding ppE parameters that represent Lorentz-violating propagation. Thus, if future gravitational wave observations peak at these ppE parameters, then one could suspect that some sort of Lorentz-violation could be responsible for such deviations from General Relativistic. Future work will concentrate on the generation problem. 

The remainder of this part deals with the details of the calculations and is organized as follows. In \cref{chapter14}, we introduce and motivate the modified dispersion relation, given by \eref{dispersion}, and derive from it the gravitational-wave speed as a function of energy and the new Lorentz-violating parameters. In the same chapter, \sref{PropagationofGW}, we study the propagation of gravitons in a cosmological background as determined by the modified dispersion relation and graviton speed. We find the relation between emission and arrival times of the gravitational waves, which then allows us in \sref{modifiedWF} to construct a {\emph{restricted}} post-Newtonian gravitational waveform to $3.5$ PN order in the phase $[{\mathcal O}(v/c)^7]$.  We also discuss the connection to the ppE framework. In \cref{chapter15}, we calculate the Fisher information matrix for Ad.~LIGO, ET and a LISA-like mission and determine the accuracy to which the compact binary's parameters can be measured, including a bound on the graviton and Lorentz-violating Compton wavelengths. In secion~\ref{conclusions} we present some conclusions and discuss possible avenues for future research.



\begin{savequote}[0.55\linewidth]
{\scriptsize ``We know very little, and yet it is astonishing that we know so much, and still more astonishing that so little knowledge can give us so much power.''}
\qauthor{\scriptsize---Bertrand Russell}
\end{savequote}
\chapter{Gravitational Waves in Lorentz-Violating Gravity} 
\label{chapter14} 
\thispagestyle{myplain}
In this chapter we study how some specific properties of gravitational-waves change in Lorentz violating theories of gravity. We are specifically interested in how modifications in the speed of propagation affect the observed waveforms. Knowing about these modifications is required to do parameter estimation analyses for Lorentz violating theories in the next chapter. In this chapter we also show how one can map the calculations to the parametrized post-Einsteinian formalism.

\lhead[\thepage]{Chapter 13. \emph{Gravitational Waves in Lorentz-Violating Gravity}}      
\rhead[Chapter 13. \emph{Gravitational Waves in Lorentz-Violating Gravity}]{\thepage}


\section{The Speed of Gravitational Waves}
\label{speedofGW}

\subsection{Massive Graviton Theories}
In general relativity, gravitational waves travel at the speed of light $c$ because the gauge boson associated with gravity, the graviton, is massless. Modified gravity theories, however, predict modifications to the gravitational-wave dispersion relation, which would in turn force the waves to travel at speeds different than $c$. The most intuitive, yet purely phenomenological modification one might expect is to introduce a mass for the graviton, following the special relativistic relation 
\be
\label{SR}
E^2=p^2c^2+m_{g}^2c^4\,.
\ee
From this dispersion relation, together with the definition $v/c \equiv p/p^0$, or $v \equiv  c^2 p/E$ , one finds the graviton speed~\cite{Will:1997bb} 
\be
\label{SRvelocity}
\frac{v_g^2}{c^2}=1-\frac{m_g^2 c^4}{E^2},
\ee
where $m_g$, $v_{g}$ and $E$ are the graviton's rest mass, velocity and energy. 

\subsection{Lorentz-Violating Theories}

Different alternative gravity theories may predict different dispersion relations from Eq.~\eqref{SR}.  A few examples of such relations include the following:
\begin{itemize}
\item {\emph{Double Special Relativity Theory}}~\cite{2001PhLB..510..255A,2002PhRvL..88s0403M,AmelinoCamelia:2002wr,2010arXiv1003.3942A}: $E^2=p^2c^2+m_{g}^2c^4+\eta_{\rm dsrt} E^3 + \ldots$, where $\eta_{\rm dsrt}$ is a parameter of the order of the Planck length. 
\item {\emph{Extra-Dimensional Theories}}~\cite{2011PhLB..696..119S}: $E^2=p^2c^2+m_{g}^2c^4-\alpha_{\rm edt} E^4$, where $\alpha_{\rm edt}$ is a constant related to the square of the Planck length;
\item {\emph{Ho\v{r}ava-Lifshitz Theory}}~\cite{Horava:2008ih,Horava:2009uw,2010arXiv1010.5457V,Blas:2011zd}: $E^{2} = p^{2}c^{2} + (\kappa^{4}_{\rm hl} \mu^{2}_{\rm hl}/16) \; p^{4} + \ldots$, where $\kappa_{\rm hl}$ and $\mu_{\rm hl}$ are constants of the theory; 
\item {\emph{Theories with Non-Commutative Geometries}}~\cite{2011arXiv1102.0117G,Garattini:2011kp,Garattini:2011hy}: $\displaystyle{E^2 g_1^2(E)=m_{g}^2c^4+p^2c^2 g_2^2(E)}$ with $g_2=1$ and $\displaystyle{g_1=(1-\sqrt{\alpha_{\rm ncg}\pi}/2 )\exp({-\alpha_{\rm ncg} {E^2}/{E_p^2})}}$, with $\alpha_{\rm ncg}$ a constant.
\end{itemize} 
For more details about each of the alternative theories listed above, see \cref{chapter3}. Of course, the list above is just representative of a few models, but there are many other examples where the graviton dispersion relation is modified~\cite{Berezhiani:2007zf,Berezhiani:2008nr}. 
In general, a modification of the dispersion relation will be accompanied by a change in either the Lorentz group or its action in real or momentum space. Lorentz-violating effects of this type are commonly found in quantum gravitational theories, including loop quantum gravity~\cite{2008PhRvD..77b3508B} and string theory~\cite{2005hep.th....8119C,2010GReGr..42....1S}.  

Modifications to the standard dispersion relation are usually suppressed by the Planck scale, so one might wonder why one should study them. Recently, Collins, et~al.~\cite{2004PhRvL..93s1301C,2006hep.th....3002C} suggested that Lorentz violations in perturbative quantum field theories could be dramatically enhanced when one regularizes and renormalizes them. This is because terms that would vanish upon renormalization due to Lorentz invariance do not vanish in Lorentz-violating theories, leading to an enhancement after renormalization~\cite{2011arXiv1106.1417G}.

Although this is an appealing argument, we prefer here to adopt a more agnostic viewpoint and simply ask the following question: What type of modifications would enter gravitational-wave observables because of a modified dispersion relation and to what extent can these deviations be observed or constrained by current and future gravitational-wave detectors? In view of this, we postulate the parametrized dispersion relation
of Eq.\ (\ref{dispersion}).

One can see that this model-independent dispersion relation can be easily mapped to all the ones described above, in the limit where $E$ and $p$ are large compared to $m_g$, but small compared to the Planck energy $E_{p}$. More precisely, we have
\begin{itemize}
\item {\emph{Double Special Relativity}}: $ \mathbb{A}  = \eta_{\rm dsrt}$ and $\alpha = 3$. 
\item {\emph{Extra-Dim.~Theories}}: $ \mathbb{A}  = - \alpha_{\rm edt}$ and $\alpha = 4$. 
\item {\emph{Ho\v{r}ava-Lifshitz}}: $ \mathbb{A}  = \kappa^{4}_{\rm hl}  \mu^{2}_{\rm hl}/16$ 
and $\alpha = 4$, but with $m_{g} = 0$.
\item {\emph{Non-Commutative Geometries}}: $ \mathbb{A}  = 2 \alpha_{\rm ncg}/E_{p}^2 $ and $\alpha = 4$, after renormalizing $m_g$ and $c$. 
\end{itemize} 
Of course, for different values of $(\mathbb{A},\alpha)$ we can parameterize other Lorentz-violating corrections to the dispersion relation. One might be naively
tempted to think that a $p^{3}$ or $p^{4}$ correction to the above dispersion relation will induce a $1.5$ or $2$PN correction to the phase relative to the massive graviton term. This, however, would be clearly wrong, as $p$ is the graviton's momentum, not the momentum of the members of a binary system. 

With this modified dispersion relation the modified graviton speed takes the form
\be
\frac{v_{g}^{2}}{c^2}  = 1 - \frac{m_{g}^2 c^4}{E^{2}} - \mathbb{A} E^{\alpha-2} \left (\frac{v}{c} \right )^\alpha\,.
\ee
To first order in $\mathbb{A}$, this can be written as
\be
\frac{v_{g}^{2}}{c^2} = 1 - \frac{m_{g}^2 c^4}{E^{2}} - \mathbb{A} E^{\alpha-2} 
\left(1 - \frac{m_{g}^2 c^4}{E^{2}}\right)^{\alpha/2} \,,
\ee
and in the limit $E \gg m_{g}$ it takes the form
\be
\frac{v_{g}^{2}}{c^2}  = 1 - \frac{m_{g}^2 c^4}{E^{2}} - \mathbb{A} E^{\alpha-2}
\,.
\ee 
Notice that if $\mathbb{A} >0$ or if $m_{g}^{2}c^4/E^{2} > \mid \mathbb{A}\mid 
E^{\alpha-2}$, then the graviton travels slower than light speed. On
the other hand, if $\mathbb{A} < 0$ and $m_{g}^{2}c^4/E^{2} < \mid \mathbb{A}\mid 
E^{\alpha-2}$, then the graviton would propagate faster than light
speed.

\section{Propagation of Gravitational Waves}
\label{PropagationofGW}

We now consider the propagation of gravitational waves that satisfy the modified dispersion relation of Eq.~\eqref{dispersion}. Since we may consider sources at very great distances, we must consider the propagation in a cosmologycal background spacetime. Consider  the Friedman-Robertson-Walker background 
\be
ds^2=-dt^2+a^2(t)[d\chi^2+\Sigma^2(\chi)(d\theta^2+\sin^2\theta
\, d\phi^2)],
\ee
where $a(t)$ is the scale factor with units of length, and $\Sigma(\chi)$ is equal to $\chi$, $\sin\chi$ or $\sinh\chi$ if the universe is spatially flat, closed or open, respectively.   Here and henceforth,
we use units with $G=c=1$, where a useful conversion factor is $1 M_\odot=4.925 \times 10^{-6}$ s $= 1.4675$ km. 

In a cosmological background, we will assume that the modified dispersion relation takes the form
\begin{equation}
g_{\mu\nu} p^\mu p^\nu = - m_g^2 - \mathbb{A} \mid p\mid ^\alpha \,,
\label{dispersionRW}
\end{equation}
where $\mid p\mid  \equiv (g_{ij} p^i p^j)^{1/2}$.  
Consider a graviton emitted radially at $\chi=\chi_e$ and received at $\chi=0$.  By virtue of the $\chi$ independence of the $t - \chi$ part of the metric, the component $p_\chi$ of its 4-momentum is constant along its worldline. Using $E=p^0$, together with Eq.\ (\ref{dispersionRW}) and the relations
\be
\frac{p^\chi}{E}=\frac{d\chi}{dt},\;\; p^\chi=a^{-2}p_\chi,
\ee
we obtain
\be
\label{speed}
\frac{d\chi}{dt}=-\frac{1}{a}\left [1+\frac{m_g^2 a^2}{p_\chi^2}+\mathbb{A} \left (\frac{ a}{p_\chi}\right)^{2-\alpha} \right ]^{-\frac{1}{2}},
\ee
where $p_\chi^2=a^2(t_e)(E_e^2-m_g^2-\mathbb{A} \mid p_e\mid ^\alpha)$. The overall minus sign in the above equation is included because the graviton travels from the source to the observer.

Expanding to first order in $(m_g/E_e)\ll 1$, and $\mathbb{A}/p^{2-\alpha} \ll 1$ and integrating from emission time ($\chi=\chi_e$) to arrival time ($\chi = 0$), we find
\bea
\label{distance}
\chi_e&=&\int^{t_a}_{t_e}\frac{dt}{a(t)}-\frac{1}{2}\frac{m_g^2}{a^2(t_e) E^2_e}\int^{t_a}_{t_e}a(t)dt
\nonumber \\
&-&\frac{1}{2} \mathbb{A}\, \biggl(a(t_e) E_e\biggr)^{\alpha-2}\int^{t_a}_{t_e}a(t)^{1-\alpha} dt.
\eea

Consider gravitons emitted at two different times $t_e$ and $t'_e$, with energies $E_e$ and $E'_e$, and received at corresponding arrival times ($\chi_e$ is the same for both). Assuming $\Delta t_e\equiv t_e-t'_e\ll a/\dot{a}$, then
\bea
\label{deltat}
\Delta t_a &=&(1+Z)\left[ \Delta t_e +\frac{D_0}{2\lambda_g^2} \left(\frac{1}{f_e^2}-\frac{1}{{f}_e'{}^2}\right) 
\right.
\nonumber \\
&+& \left.
\frac{D_\alpha}{2 \lambda_{\mathbb{A}}^{2-\alpha}} \; \left(\frac{1}{f_e^{2-\alpha}}-\frac{1}{{f}_e'{}^{2-\alpha}}\right)  \right]\,,
\eea
where $Z\equiv a_0/a(t_e)-1$ is the cosmological redshift, and where we have defined 
\be
\lambda_{\mathbb{A}} \equiv h \; \mathbb{A}^{{1}/{(\alpha-2)}}\,,
\ee
and where $m_g/E_e=(\lambda_g f_e)^{-1}$, with $f_e$ the emitted gravitational-wave frequency, $E_{e} = h f_{e}$ and $\lambda_g=h/m_g\,$ the graviton Compton wavelength. Notice that when $\alpha=2$, then the $\mathbb{A}$ correction vanishes. Notice also that $\lambda_{\mathbb{A}}$ always has units of length, irrespective of the value of $\alpha$.  The distance measure $D_\alpha$ is defined by
\bea
D_\alpha\equiv \left(\frac{1+Z}{a_0}\right)^{1-\alpha}\int_{t_e}^{t_a}a(t)^{1-\alpha} dt
\eea
where $a_0=a(t_a)$ is the present value of the scale factor.
For a dark energy-matter dominated universe $D_\alpha$ and the luminosity distance $D_L$ have the form
\bea
D_\alpha &=& \frac{(1+Z)^{1-\alpha}}{H_0}\int_0^Z\frac{(1+z')^{\alpha-2}dz'}{\sqrt{\Omega_M(1+z')^{3}+\Omega_\Lambda}}\,,
\label{Dalpha-general}
\\
D_L&=&\frac{1+Z}{H_0}\int_0^Z\frac{dz'}{\sqrt{\Omega_M(1+z')^3+\Omega_\Lambda}},
\eea
where  $H_{0} \approx 72 \; {\rm{km}} \; {\rm{s}}^{-1} \; {\rm{Mpc}}^{-1}$ is the value of the Hubble parameter today and $\Omega_M=0.3$ and $\Omega_\Lambda=0.7$ are the matter and dark energy density parameters, respectively. 

Before proceeding, let us comment on the time shift found above in Eq.~\eqref{deltat}. First, notice that this equation agrees with the results of~\cite{Will:1997bb} in the limit $\mathbb{A} \to 0$. Moreover, in the limit $\alpha \to 0$, our results map to those of~\cite{Will:1997bb} with the relation $\lambda^{-2}_{g} \to \lambda_{g}^{-2} + \lambda_{\mathbb{A}}^{-2}$. Second, notice that in the limit $\alpha \to 2$, the $(a(t_{e}) E_{e})^{2-\alpha}$ in Eq.~\eqref{distance} goes to unity and the $\mathbb{A}$ correction becomes frequency independent. This makes sense, since in that case the Lorentz-violating correction we have introduced acts as a renormalization factor for the speed of light. 

\section{Modified Waveform in the Stationary Phase Approximation}\label{modifiedWF}

We consider the gravitational-wave signal generated by a non-spinning, quasi-circular inspiral in the post-Newtonian approximation. In this scheme, one assumes that orbital velocities are small compared to the speed of light ($v \ll 1$) and gravity is weak ($m/r \ll1$).  Neglecting any amplitude corrections (in the so-called {\emph{restricted}} PN approximation), the plus- and cross-polarizations of the metric perturbation can be represented as
\bea
\label{eq:h(t)}
h(t)&\equiv&A(t) e^{-i\Phi(t)},\\
\Phi(t)&\equiv&\Phi_c+2\pi\int^t_{t_c}f(t)dt,
\eea
where $A(t)$ is an amplitude that depends on the gravitational-wave polarization (see e.g.~Eq. $(3.2)$ in~\cite{Will:1997bb}), while $f(t)$ is the observed gravitational-wave frequency, and $\Phi_c$ and $t_c$ are a {\emph{fiducial}} phase and fiducial time, respectively, sometimes called the coalescence phase and time. 

The Fourier transform of Eq.~\eqref{eq:h(t)} can be obtained analytically in the stationary-phase approximation, where we assume that the phase is changing much more rapidly than the amplitude~\cite{Droz:1999qx,Yunes:2009yz}.  We then find
\be
\tilde{h}({f})=\frac{\tilde{A}({t})}{\sqrt{\dot{f}({t})}} e^{i\Psi({f})} \,,
\ee
where $f$ is the gravitational-wave frequency at the detector and
\bea
\tilde{A}({t})&=&\frac{4}{5}\frac{\mathcal{M}_e}{a_0 \Sigma(\kappa_e)} (\pi \mathcal{M}_e {f}_e)^{2/3},\\\label{phase}
\Psi({f})&=&2\pi{f}t_c-\Phi_c-\frac{\pi}{4}+2\pi\int_{f_c}^{{f}}(t-t_c)df.
\eea
In these equations, $\mathcal{M}_e=\eta^{3/5} m$ is the {\emph{chirp}} mass of the source, where $\eta = m_1m_2/(m_1+m_2)$ is the symmetric mass ratio. 

We can now substitute Eq.~(\ref{deltat}) into Eq.~(\ref{phase}) to relate the time at the detector to that at the emitter. Assuming that $\alpha \neq 1$, we find 
\bea
\Psi_{\alpha \neq 1}({f})&=& 2\pi{f} \bar{t}_c-\bar{\Phi}_c-\frac{\pi}{4}+2\pi \int_{f_{ec}}^{{f_e}}(t_e-t_{ec})df_e  -\frac{\pi D_0}{f_e \lambda_g^2} -\frac{1}{(1-\alpha)} \frac{\pi D_{\alpha}}{f_{e}^{1-\alpha}\lambda_{\mathbb{A}}^{2-\alpha} }\, ,
\eea
while for  $\alpha=1$, we find 
\bea
\label{alpha1-dephasing}
\Psi_{\alpha=1}({f})&=& 2\pi{f} \bar{\bar{t}}_c - \bar{\bar{\Phi}}_c-\frac{\pi}{4}+2\pi \int_{f_{ec}}^{{f_e}}(t_e-t_{ec})df_e  -\frac{\pi D_0}{f_e \lambda_g^2} +\frac{\pi D_{1}}{\lambda_{\mathbb{A}}} \ln\left(\frac{f_{e}}{f_{ec}}\right) \,.
\eea
The quantities $(\bar{t}_{c},\bar{\bar{t}}_{c})$ and $(\bar{\phi}_{c},\bar{\bar{\phi}}_{c})$ are new coalescence times and phases, into which constants of integration have been absorbed.  

We can relate $t_e - t_{ec}$ to $f_e$ by integrating the frequency chirp equation for non-spinning, quasi-circular inspirals from general relativity~\cite{Will:1997bb}:
\bea
\frac{df_e}{dt_e}&=&\frac{96}{5\pi\mathcal{M}_e^2} (\pi \mathcal{M}_e f_e)^{11/3}\biggl[ 1-\biggl(\frac{743}{336} +\frac{11}{4} \eta \biggr) (\pi M f_e)^{2/3} + 4\pi (\pi M f_e)\biggr],
\eea
where we have kept terms up to $1$PN order.  In the calculations that follow, we actually account for corrections up to $3.5$PN order, although we don't show these higher-order terms here (they can be found e.g.~in~\cite{2009PhRvD..80h4043B}). 

After absorbing further constants of integration into $(\bar{t}_c,\bar{\Phi}_c,\bar{\bar{t}}_c,\bar{\bar{\Phi}}_c)$, dropping the bars, and re-expressing everything in terms of the \emph{measured} frequency ${f}$ at the detector [note that $\dot{f}^{1/2}=(df_e/dt_e)^{1/2}/(1+Z)$], we obtain 
\bea\label{waveform}
\tilde{h}({f})=\left\{ \begin{array}{ll}
\tilde{A}({f}) e^{i\Psi({f})},& \mbox{for $0<{f}<{f}_{max}$}\\
0,&\mbox{for ${f}>{f}_{max}$} \,,
\end{array}
\right.
\eea
with the definitions
\begin{align}
\tilde{A}({f}) &\equiv \epsilon \; \mathcal{A} \; {u}^{-7/6}\,,
\qquad
{\mathcal{A}} =\sqrt{\frac{\pi}{30}}\frac{\mathcal{M}^2}{D_L}\,,
\\ \nonumber 
\label{psi}
\Psi({f})&= \Psi_{\GR}({f}) + \delta \Psi({f})\,,
\\
\Psi_{\GR}(f) &=2\pi{f}t_c-\Phi_c -\frac{\pi}{4}+\frac{3}{128}u^{-5/3} \sum_{n=0}^{\infty} \left[c_{n} + \ell_{n} \ln(u) \right] u^{n/3}\,,
\end{align}
where the numerical coefficient $\epsilon = 1$ for LIGO and ET, but $\epsilon = \sqrt{3}/2$ for a LISA-like mission (because when one angle-averages, the resulting geometric factors depend slightly on the geometry of the detector). The coefficients $(c_{n},\ell_{n})$ can be read up to $n=7$ in Appendix~\ref{sec:35PNwaveform}. In these equations, $u\equiv\pi\mathcal{M}{f}$ is a dimensionless frequency, while $\mathcal{M}$ is the measured chirp mass, related to the source chirp mass by $\mathcal{M}=(1+Z)\mathcal{M}_e$. The frequency ${f}_{max}$ represents an upper cut-off frequency where the PN approximation fails.

The dephasing caused by the propagation effects takes a slightly different form depending on whether  $\alpha \neq 1$  or  $\alpha=1$. In the general $\alpha \neq 1$ case, we find 
\be
\delta \Psi_{\alpha \neq 1}(f) = -\beta u^{-1} - \zeta u^{\alpha-1}\,,
\label{Dephasing-full}
\ee
where the parameters $\beta$ and $\zeta$ are given by
\bea\label{beta}
\beta&\equiv&\frac{\pi^2 D_0\mathcal{M}}{\lambda_g^2(1+Z)},
\\ \label{zeta}
\zeta_{\alpha \neq 1}&\equiv& \frac{\pi^{2-\alpha}}{(1-\alpha)} \frac{D_\alpha}{\lambda_{\mathbb{A}}^{2-\alpha}}  \frac{\mathcal{M}^{1-\alpha}}{(1+Z)^{1-\alpha}}\,. 
\\\nonumber&&
\eea
In the special $\alpha=1$ case, we find
\be
\delta \Psi_{\alpha = 1}(f) = -\beta u^{-1}  
+ \zeta_{\alpha=1} \ln\left(u\right)\,,
\label{newpsi}
\ee
where $\beta$ remains the same, while 
\be
\zeta_{\alpha=1} = \frac{\pi D_{1}}{\lambda_{\mathbb{A}}}\,,
\ee
and we have re-absorbed a factor into the phase of coalescence. 

As before, notice that in the limit $\mathbb{A} \to 0$, Eq.~\eqref{Dephasing-full} reduces to the results of~\cite{Will:1997bb} for a massive graviton. Also note that, as before, in the limit $\alpha \to 0$, we can map our results to those of~\cite{Will:1997bb} with $\lambda_{g}^{-2} \to \lambda_{g}^{-2} + \lambda_{\mathbb{A}}^{-2}$, i.e.~in this limit, the mass of the graviton and the Lorentz-violating $\mathbb{A}$ term become $100\%$ degenerate. In the limit $\alpha \to 2$, Eq.~\eqref{deltat} becomes frequency-independent, which then implies that its integral, Eq.~\eqref{phase}, becomes linear in frequency, which is consistent with the $\alpha \to 2$ limit of Eq.~\eqref{Dephasing-full}. Such a linear term in the gravitational-wave phase can be reabsorbed through a redefinition of the time of coalescence, and thus is not observable. This is consistent with the observation that the dispersion relation with $\alpha=2$ is equivalent to the standard massive graviton one with a renormalization of the speed of light. When $\alpha=1$, Eq.~\eqref{deltat} leads to a $1/f$ term, whose integral in Eq.~\eqref{phase} leads to a $\ln(f)$ term, as shown in Eq.~\eqref{alpha1-dephasing}. Finally, notice that, in comparision with the phasing terms that arise in the PN approximation to standard general relativity, these corrections are effectively of $(1+3\alpha/2)$PN order, which implies that the $\alpha=0$ term leads to a 1PN correction as in~\cite{Will:1997bb}, the $\alpha=1$ case leads to a $2.5$PN correction, the $\alpha=3$ case leads to a $5.5$PN correction and $\alpha=4$ leads to a $7$PN correction. This suggests that the accuracy to constrain $\lambda_{\mathbb{A}}$ will deteriorate very rapidly as $\alpha$ increases.

\section{Connection with the PPE Framework}

Recently, there has been an effort to develop a framework suitable for testing for deviations from general relativity in gravitational-wave data.  In analogy with the parametrized post-Newtonian (PPN) framework~\cite{wil71a,1972ApJ...177..775N,wil72,wil73, lrr-2006-3, tegp}, the parametrized post-Einsteinian (ppE) framework~\cite{Yunes:2009ke,Vigeland:2011ji,Cornish:2011ys} suggests that we deform the gravitational-wave observable away from our GR expectations in a well-motivated, parametrized fashion. In terms of the Fourier transform of the waveform observable in the SPA, the simplest ppE meta-waveform is
\be
\tilde{h}_{\ppE}(f) = \tilde{A}_{\GR} \left(1+\alpha_{\ppE} u^{a_{\ppE}} \right) e^{i \Psi_{\GR}(f) + i \beta_{\ppE}\;u^{b_{\ppE}}}\,,
\ee
where $(\alpha_{\ppE},a_{\ppE},\beta_{\ppE},b_{\ppE})$ are ppE, theory parameters. Notice that in the limit $\alpha_{\ppE} \to 0$ or $\beta_{\ppE} \to 0$, the ppE waveform reduces exactly to the SPA GR waveform. The proposal is then to match-filter with template families of this type and allow the data to select the best-fit ppE parameters to determine whether they are consistent with GR. 

We can now map the ppE parameters to those obtained from a generalized, Lorentz-violating dispersion relation:
\bea
\alpha_{\ppE} &=& 0\,
\qquad
\beta_{\ppE} = - \zeta
\qquad
b_{\ppE} = \alpha - 1\,.
\eea
Quantum-gravity inspired Lorentz-violating theories suggest modified dispersion exponents $\alpha=3$ or $4$, to leading order in $E/m_{g}$, which then implies ppE parameters $b_{\ppE} = 2$ and $3$. Therefore, if after a gravitational wave has been detected, a Bayesian analysis with ppE templates is performed that leads to values of $b_{\ppE}$ that peak around $2$ or $3$, this would indicate the possible presence of Lorentz violation~\cite{Cornish:2011ys}. Notice however that the $\alpha=1$ case cannot be recovered by the ppE formalism without generalizing it to include $\ln{u}$ terms. Such effects are analogous to memory corrections in PN theory.

At this point, we must spell out an important caveat.  The values of $\alpha$ that represent Lorentz violation for quantum-inspired theories ($\alpha=3,4$) correspond to very high PN order effects, i.e.~a relative $5.5$ or $7$ PN correction respectively. Any gravitational-wave test of Lorentz violation that wishes to constrain such steep momentum dependence would require a very accurate (high PN order) modeling of the general relativistic waveform itself. In the next chapter, we will employ $3.5$ PN accurate waveforms, which are the highest-order known, and then ask how well $\zeta$ and $\beta$ can be constrained. Since we are neglecting higher than $3.5$ PN order terms in the template waveforms, we are neglecting also any possible correlations or degeneracies between these terms and the Lorentz-violating terms. Therefore, any estimates made in the next section are at best optimistic bounds on how well gravitational-wave measurements could constrain Lorentz violation.


\begin{savequote}[0.55\linewidth]
{\scriptsize ``We are trying to prove ourselves wrong as quickly as possible, because only in that way can we find progress.''}
\qauthor{\scriptsize---Richard Feynman}
\end{savequote}
\chapter{Parameter Estimation in Lorentz-Violating Gravity} 
\label{chapter15} 
\thispagestyle{myplain}
In this chapter, we perform a simplified Fisher analysis, following the method outlined for compact binary inspiral in~\cite{cut94, fin93,poi95}, to get a sense of the bounds one could place on $\lambda_{g}$ and $\lambda_{\mathbb{A}}$ given a gravitational-wave detection that is consistent with general relativity. We begin by summarizing some of  the basic ideas behind a Fisher analysis, introducing some notation. We then apply this analysis to an Adv.~LIGO detector, an ET detector and a LISA-like mission.

\lhead[\thepage]{Chapter 14. \emph{Parameter Estimation in in Lorentz-Violating Gravity}}      
\rhead[Chapter 14. \emph{Parameter Estimation in in Lorentz-Violating Gravity}]{\thepage}

\section{Fisher-Matrix Parameter Estimation Method}

Based on the Fisher matrix method that we reviewed in \cref{chapter7}, we will work with an angle-averaged response function, so that the templates depend only on the following parameters: 
\be
\boldsymbol \theta= (\ln {\cal A}, \Phi_c, f_0
t_c, \ln {\cal M}, \ln \eta, \beta,\zeta)\,,
\ee 
where each component of the vector $\boldsymbol \theta$ is dimensionless. We recall that ${\cal{A}}$ is an overall amplitude that contains information about the gravitational-wave polarization and the beam-pattern function angles. The quantities $\Phi_{c}$ and $t_{c}$ are the phase and time of coalescence, where $f_0$ is a frequency characteristic of the detector, typically a ``knee'' frequency, or a frequency at which $S_n(f)$ is a minimum. The parameters ${\cal{M}}$ and $\eta$ are the chirp mass and symmetric mass ratio (see the definitions in Eq.~(\ref{massdefinitions})), which characterize the compact binary system under consideration. The parameters $\beta$ and $\zeta$ describe the  massive graviton and Lorentz-violating terms respectively.
   
Recalling Eq.~(\ref{2.6}), the SNR value for the templates in Eq.~\eqref{waveform} is simply 
\be
\rho= 2 \; \epsilon \; \mathcal{A} \; \left({\cal{M}} \pi \right)^{-7/6} \; f_0^{-2/3} \; I(7)^{1/2} S_0^{-1/2}\,,
\label{SNR}
\ee
where we have redefined the integrals $I(q)$ from Eqs.~(\ref{3.8}) (written specifically for Ad. LIGO) to a more general case as
\be
\label{moment}
I(q)\equiv \int_0^\infty \frac{x^{-q/3}}{g(x)} dx,
\ee
with $x \equiv f/f_{0}$. The quantity $g(x)$ is the rescaled power spectral density, defined via $g (x) \equiv S_h(f)/S_{0}$ for the detector in question, and $S_{0}$ is an overall constant. When computing the Fisher matrix, we will replace the amplitude ${\cal{A}}$ in favor of the SNR, using Eq.~\eqref{SNR}. This will then lead to bounds on $\beta$ and $\zeta$ that depend on the SNR and on a rescaled version of the moments $J(q)\equiv I(q)/I(7)$.

In the next sections, we will carry out the integrals in Eq.~\eqref{moment}, but we will approximate the limits of integration by certain $x_{\rm min}$ and $x_{\rm max}$~\cite{BBW}. The maximum frequency will be chosen to be the smaller of a certain instrumental maximum threshold frequency and that associated with a gravitational wave emitted by a particle in an innermost-stable circular orbit (ISCO) around a Schwarzschild black hole (BH): $f_{\rm max} = 6^{-3/2} \pi^{-1} \eta^{3/5} {\cal{M}}^{-1}$. The maximum instrumental frequency will be chosen to be $(10^{5},10^{3},1)$ Hz for Ad.~LIGO, ET and LISA-like, respectively. The minimum frequency will be chosen to be the larger of a certain instrumental minimum threshold frequency and, in the case of a space mission, the frequency associated with a gravitational wave emitted by a test-particle one year prior to reaching the ISCO. The minimum instrumental frequency will be chosen to be $(10,1,10^{-5})$ Hz for Ad.~LIGO, ET and a LISA-like mission, respectively.

Once the Fisher matrix has been calculated, we will invert it using a Cholesky decomposition to find the variance-covariance matrix, the diagonal components of which give us a measure of the accuracy to which parameters could be constrained. Let us then define the upper bound we could place on $\beta$ and $\zeta$ as $\Delta \beta \equiv \Delta^{1/2} /\rho$ and $\Delta \zeta \equiv \bar\Delta^{1/2} /\rho$, where $\Delta$ and $\bar\Delta$ are numbers. Combining these definitions with Eqs. (\ref{beta}) and (\ref{zeta}), we find, for $\alpha \ne 1$, the
bounds:
\bea
\lambda_g &>& \sqrt{ \frac{\rho \, D_0 \, {\cal M}}{(1+Z)}} \frac{\pi}{\Delta^{1/4}} \,,
\label{lambdabound}
\\
\lambda_{\mathbb{A}}^{\alpha-2} &<&\frac{\mid 1-\alpha\mid }{\pi^{2-\alpha}} \frac{\bar{\Delta}^{1/2}}{D_{\alpha} \rho} 
\frac{{\cal{M}}^{\alpha-1}}{(1 + Z)^{\alpha-1}}  \,,
\eea
Notice that the direction of the bound on $\lambda_{\mathbb{A}}$ itself depends on whether $\alpha > 2$ or $\alpha < 2$; but because $\mathbb{A} = (\lambda_{\mathbb{A}}/h)^{\alpha-2}$,  all cases yield an upper bound on $\mathbb{A}$. For the case $\alpha=1$
, we find
\bea
\lambda_{\mathbb{A}_{\alpha=1}} &>& \frac{\pi D_{1}}{\bar{\Delta}^{1/2}} \rho\,,
\eea
In the remaining sections, we set $\beta=0$ and $\zeta=0$ in all partial derivatives when computing the Fisher matrix, since we derive the error in estimating  $\beta$ and $\zeta$ about the nominal or {\it a priori} general relativity values, $(\beta,\zeta)=(0,0)$.

\section{Detector Spectral Noise Densities}

We model the Ad.~LIGO spectral noise density via~\cite{mishra:2010tp}
\bea
\frac{S_h(f)}{S_{0}} = \left\{ \begin{array}{ll}
\displaystyle10^ {\displaystyle16-4(x f_0-7.9)^2}+2.4\times 10^{-62} x^{-50}+0.08 x^{-4.69}&
\\  
+\displaystyle 123.35\biggl(\frac{1-0.23 x^2+0.0764 x^4}{1+0.17 x^2}\biggr), & f \geq f_s, \\
\infty, & f < f_s,
\end{array}
\right.
\label{advligo}
\eea
Here, $f_0=215$ Hz, $S_0=10^{-49}$ Hz$^{-1}$, and $f_s=10\; {\rm{Hz}}$ is a low-frequency cutoff below which $S_h(f)$ can be considered infinite for all practical purposes

The initial ET design postulated the spectral noise density~\cite{mishra:2010tp}
\bea
\frac{S_h(f)}{S_{0}}  =  \left\{ \begin{array}{ll}
\left[a_1 x^{b_1}+a_2 x^{b_2} +a_3 x^{b_3}+a_4 x^{b_4}\right]^{2}, & f \geq  f_s\, \\
\infty, & f < f_s,
\end{array}
\right.
\label{et}
\eea
where $f_0=100\,\mbox{Hz}$, $S_0 = 10^{-50}\,\mbox{Hz}^{-1}$, $f_s=1 \; {\rm{Hz}}$, and 
\begin{eqnarray}
a_1 =&2.39\times10^{-27},~\quad
b_1 &= -15.64,\nonumber\\
a_2 &= 0.349, \quad\quad\quad\quad
b_2 &= -2.145,\nonumber\\
a_3 &= 1.76, ~~\quad\quad\quad\quad
b_3 &= -0.12,\nonumber\\
a_4 &= 0.409, \quad\quad\quad\quad
b_4 &= 1.10.
\label{constants_sqrtpsd}
\end{eqnarray}
\begin{figure}[h!]
\centering
\includegraphics[width=12cm,clip=true]{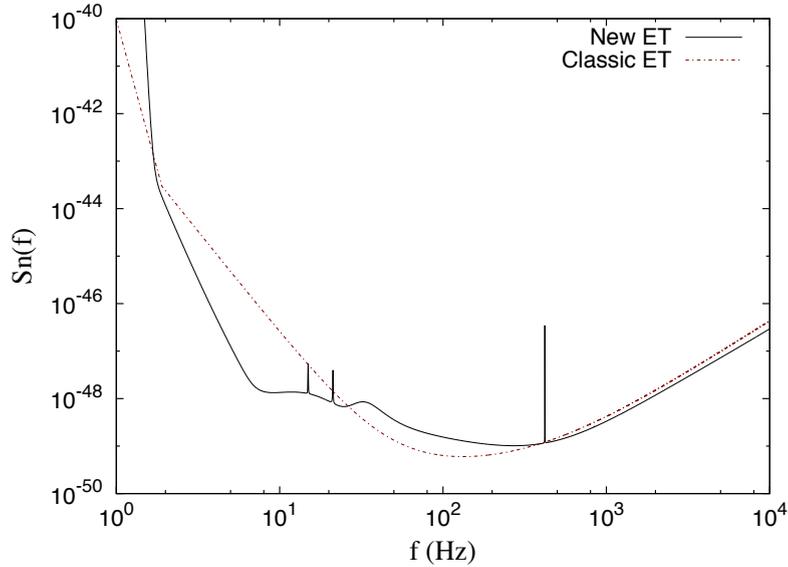} 
\caption{ET spectral noise density curves for the classic design (dotted) and the new design (solid).}
\label{ET-Noisecurve} 
\end{figure}

The classic LISA design had an approximate spectral noise density curve that could be
modeled via (see eg.~\cite{BBW, Barack:2003fp}):
\bea\nonumber\label{lisa}
S_h(f)&=&{\rm min}\biggl\{
\frac{S_h^{\rm NSA}(f)}{{\rm e}
\left(-\kappa T^{-1}_{\rm mission} dN/df\right)},\;S_h^{\rm NSA}(f)+S_h^{\rm gal}(f)
\biggr\}+S_h^{\rm ex-gal}(f)\,.
\label{Shtot}
\eea

where 
\bea
S_h^{\rm NSA}(f)&=& \biggl[9.18\times 10^{-52}\left(\f{f}{1~{\rm Hz}}\right)^{-4}\nonumber\\
&&+1.59\times 10^{-41}+9.18\times 10^{-38}\left(\f{f}{1~{\rm Hz}}\right)^2\biggr]~{\rm Hz}^{-1}\,.
\\
S_h^{\rm gal}(f) &=&
2.1\times 10^{-45}\left(\f{f}{1~{\rm Hz}}\right)^{-7/3}~{\rm Hz}^{-1}\,,
\\
S_h^{\rm ex-gal}(f) &=&
4.2\times 10^{-47}\left(\f{f}{1~{\rm Hz}}\right)^{-7/3}~{\rm Hz}^{-1}\,.
\eea
and 
\be
\f{dN}{df}=2\times 10^{-3}~{\rm Hz}^{-1}
\left(\f{1~{\rm Hz}}{f}\right)^{11/3}\,;
\ee
with $\Delta f=T^{-1}_{\rm mission}$ the bin size of the discretely
Fourier transformed data for a classic LISA mission lasting a time
$T_{\rm mission}$ and $\kappa\simeq 4.5$ the average number of
frequency bins that are lost when each galactic binary is fitted
out. 
\begin{figure}[h!]
\centering
\includegraphics[width=12cm,clip=true]{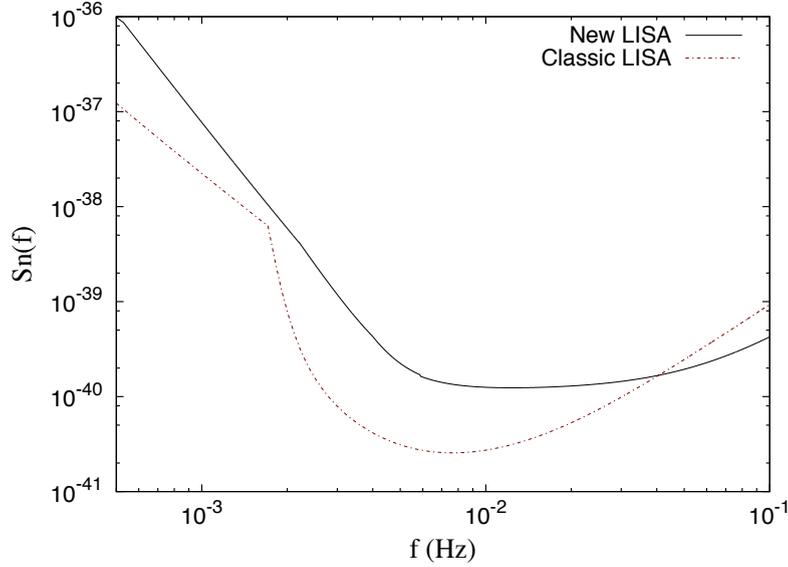}
\caption{LISA spectral noise density curves for the classic design (dotted) and the new NGO design (solid).}
\label{LISA-Noisecurve} 
\end{figure}
Recently, the designs of LISA and ET have changed somewhat. The new spectral noise density curves can be computed numerically~\cite{2011CQGra..28i4013H,Sathya,Berti} and are plotted in Fig.~\ref{ET-Noisecurve}, and Fig.~\ref{LISA-Noisecurve}. Notice that the bucket of the NGO noise curve has shifted to higher frequency, while the new ET noise curve is more optimistic than the classic one at lower frequencies. The spikes in the latter are due to physical resonances, but these will not affect the analysis. In the remainder of this chapter, we will use the {\em new} ET and NGO noise curves to estimate parameters.  

\section{Results}

We plot the bounds that can be placed on $\zeta$ by using different detectors in Fig.~\ref{Zeta-figs1}, Fig.~\ref{Zeta-figs2}, and Fig.~\ref{Zeta-figs3} as a function of the $\alpha$ parameter. Fig.~\ref{Zeta-figs1} corresponds to the bounds placed with Ad.~LIGO and $\rho = 10$ ($D_{L} \sim 160 \; {\rm{Mpc}}$, $Z \sim 0.036$ for a double neutron-star inspiral), Fig.~\ref{Zeta-figs2} corresponds to ET and $\rho = 50$ ($D_{L} \sim 2000 \; {\rm{Mpc}}$, $Z \sim 0.39$ for a double $10 M_{\odot}$ BH inspiral) and Fig.~\ref{Zeta-figs3} corresponds to NGO and $\rho = 100$ ($D_{L} \sim 20,000 \; {\rm{Mpc}}$, $Z \sim 2.5$ for a double $10^{5} M_{\odot}$ BH inspiral).  When $\alpha=0$ or $\alpha=2$, $\zeta$ cannot be measured at all, as it becomes $100\%$ correlated with either standard massive graviton parameters.  Thus we have drawn vertical lines in those cases.  As the figures clearly show, the accuracy to which $\zeta$ can be measured deteriorates rapidly as $\alpha$ becomes larger. In fact, once $\alpha > 4$, we find that $\zeta$ cannot be confidently constrained anymore because the Fisher matrix becomes non-invertible (its condition number exceeds $10^{16}$).

\begin{figure}[h!]
\centering
\hspace{-0.5cm} \includegraphics[width=11cm,clip=true]{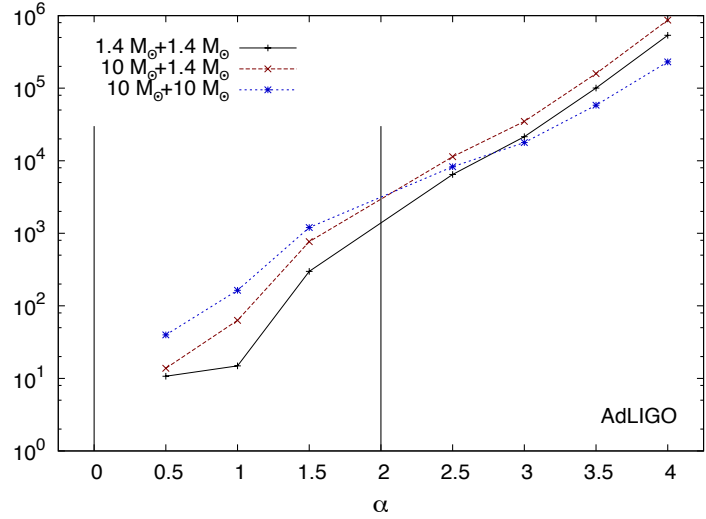} 
\caption{\label{Zeta-figs1} Bounds on the parameter $\zeta$ for different values of $\alpha$, using AdLIGO and $\rho=10$. Vertical lines at $\alpha=(0,2)$ show where the $\zeta$ correction becomes $100\%$ degenerate with other parameters. Figure contains several curves that show the bound for systems with different masses.}
\end{figure}

\begin{figure}[h!]
\centering
\hspace{-0.5cm} \includegraphics[width=11cm,clip=true]{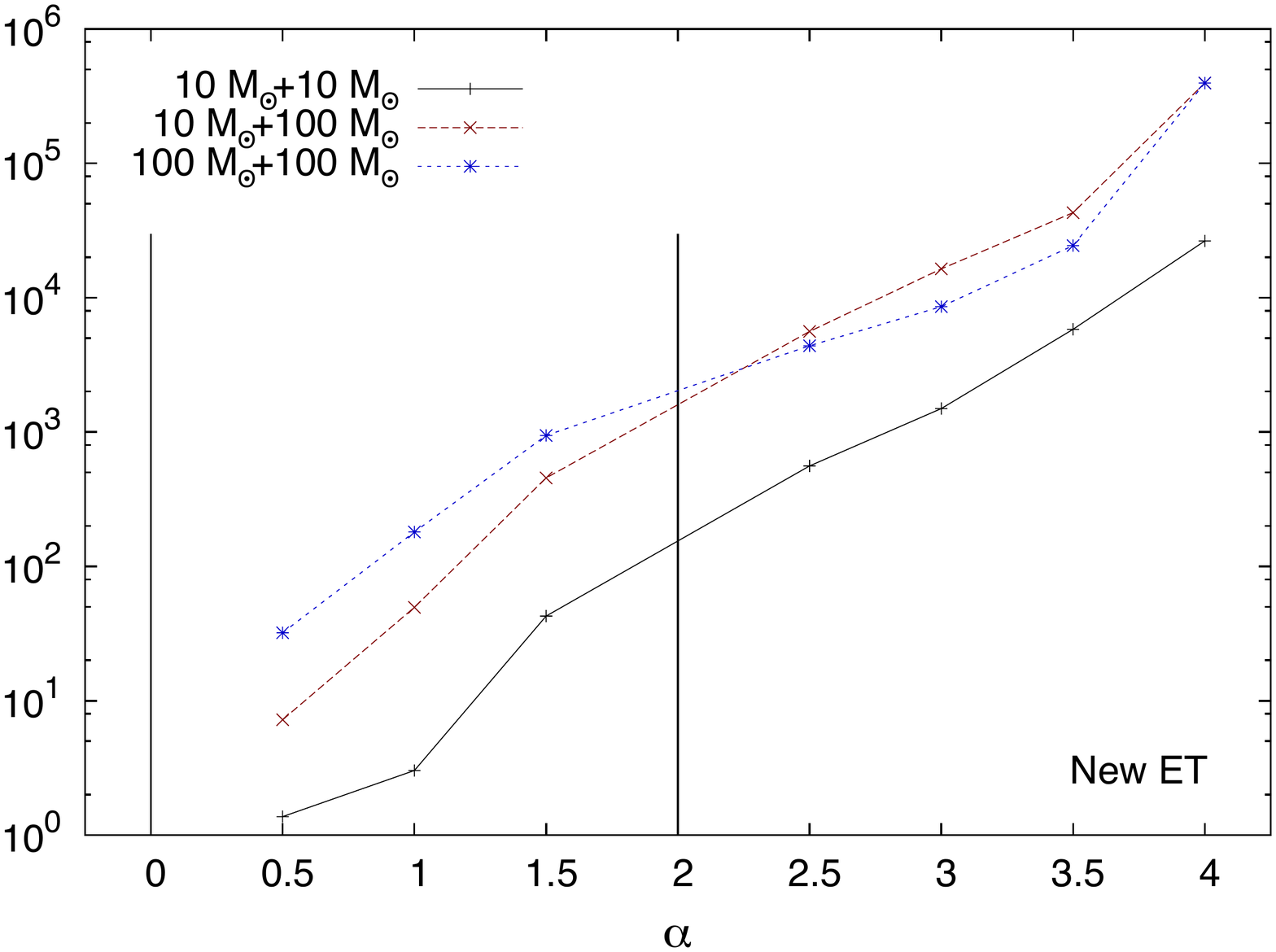} 
\caption{\label{Zeta-figs2} Bounds on the parameter $\zeta$ for different values of $\alpha$, using ET and $\rho=50$. Vertical lines at $\alpha=(0,2)$ show where the $\zeta$ correction becomes $100\%$ degenerate with other parameters. Figure contains several curves that show the bound for systems with different masses.}
\end{figure}

\begin{figure}[h!]
\centering
\hspace{-0.5cm} \includegraphics[width=11cm,clip=true]{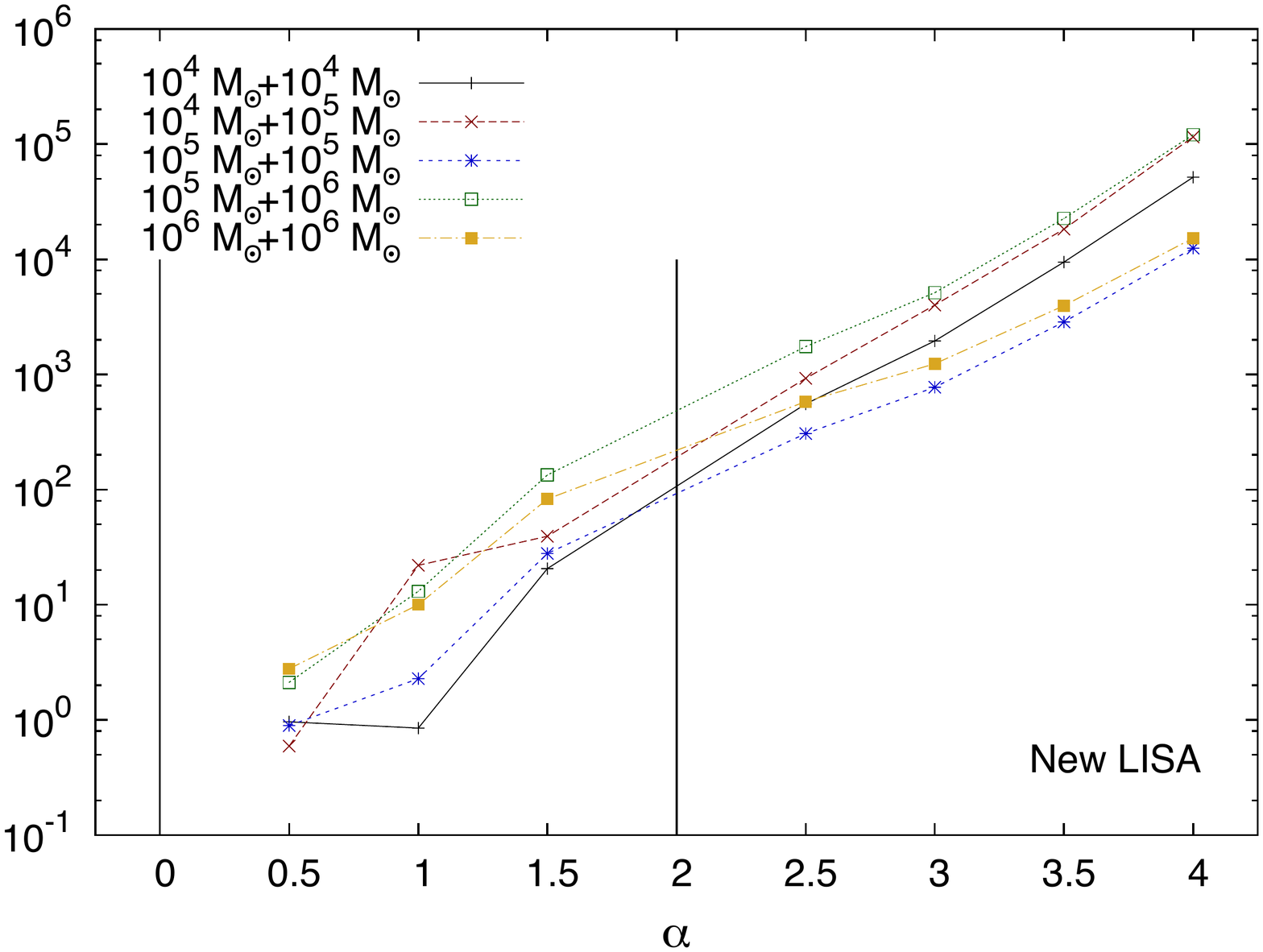}
\caption{\label{Zeta-figs3} Bounds on the parameter $\zeta$ for different values of $\alpha$, using NGO and $\rho = 100$. Vertical lines at $\alpha=(0,2)$ show where the $\zeta$ correction becomes $100\%$ degenerate with other parameters. Figure contains several curves that show the bound for systems with different masses.}
\end{figure}

Attempting to constrain values of $\alpha > 5/3$ becomes problematic not just from a data analysis point of view, but also from a fundamental one. The PN templates that we have constructed contain general relativity phase terms up to $3.5$ PN order. Such terms scale as $u^{2/3}$, which corresponds to $\alpha = 5/3$. Therefore, trying to measure values of $\alpha \geq 5/3$ without including the corresponding 4PN and higher-PN order terms is not well-justified. We have done so here, neglecting any correlations between these higher order PN terms and the Lorentz-violating terms, in order to get a rough sense of how well Lorentz-violating modifications could be constrained.

The bounds on $\beta$ and $\zeta$ are converted into  a lower bound on $\lambda_{g}$ and and upper bound on $\lambda_{\mathbb{A}}$ in Table~\ref{table-all} for $\alpha=3$ and binary systems with different component masses. Given a gravitational-wave detection consistent with general relativity, this table says that $\lambda_{g}$ and $\lambda_{\mathbb{A}}$ would have to be larger and smaller than the numbers in the seventh and eight columns of the table respectively. In addition, this table also shows the accuracy to which standard binary parameters could be measured, such as the time of coalescence, the chirp mass and the symmetric mass ratio, as well as the correlation coefficients between parameters. Different clusters of numbers correspond to constraints with Ad.~LIGO (top), New ET (middle) and NGO (bottom) (see caption for further details; specifically, notice the different units for the numbers in each section of the table)

\afterpage{%
    \thispagestyle{empty}
    \begin{landscape}
    \begin{center}
    \begin{table}[ht] 
\hsize\textwidth\columnwidth\hsize\csname
           @twocolumnfalse\endcsname
\centering
\begin{tabular}{ c c c | |  c c c c c c c c c c c c}
\hline\hline
&&&&&&\\
${\rm{Detector}}$ &$m_1$&$m_2$&$\Delta \phi_c$&$\Delta t_c$&$\Delta {\cal M}/{\cal M}$
&$\Delta \eta/\eta$&$\Delta\lambda_g$&$\Delta\lambda_\mathbb{A}$&$c_{{\cal M}\eta}$&$c_{{\cal M}
\beta}$&$c_{\eta \beta}$&$c_{{\cal M}\zeta}$&$c_{\eta
\zeta}$&$c_{\beta \zeta}$\\
&&&&&&\\

\hline\hline
&&&&&&\\
{\rm{Ad.~LIGO}} & 1.4&1.4&3.61&1.80&0.0374\%&6.80\%&3.34&0.911&-0.962&-0.991&0.989&-0.685&0.803&0.740\\
&1.4&10&3.34&9.99&0.267\%&12.8\%&3.48&4.36&-0.977&-0.993&0.917&-0.830&0.923&0.875\\
& 10&10&4.16&31.0&2.40\%&72.2\%&3.53&8.40&-0.978&-0.994&0.995&-0.874&0.947&0.915\\
&&&&&&\\
\hline\hline
&&&&&&\\
{\rm{ET}} & 10&10&       0.528&1.59&0.0174\%&1.70\%&4.15&0.0286&-0.952&-0.986&0.988&-0.742&0.875&0.813\\
& 10&100&    1.12&44.5&0.259\%&6.67\%&2.58&1.38&-0.974&-0.993&0.993&-0.872&0.951&0.915\\
& 100&100&  5.23&203&4.03\%&67.6\%&1.86&4.12&-0.983&-0.995&0.996&-0.914&0.969&0.947\\
&&&&&&\\

\hline\hline
&&&&&&\\
{\rm{NGO}} 
& $10^4$&$10^4$&  0.264&1.05&0.00124\%&0.368\% &4.06& 0.266 	&-0.957&-0.990&0.986&-0.636&0.761&0.687\\
& $10^4$&$10^5$&  0.264&5.42&0.00434\%& 0.383\%&5.04& 1.81 	&-0.955&-0.991&0.984&-0.757&0.884&0.809\\
& $10^5$&$10^5$&  0.295&9.54&0.0163\%& 1.33\%&6.12& 2.33 	&-0.944&-0.983&0.986&-0.749&0.891&0.823\\
& $10^5$&$10^6$&  0.351&142&0.0574\%& 2.03\%&5.93& 118 	&-0.961&-0.990&0.989&-0.938&0.942&0.891\\
& $10^6$&$10^6$&  0.415&228&0.138\%& 5.33\%&8.30& 286	&-0.956&-0.986&0.990&-0.820&0.935&0.885\\
&&&&&&\\
\hline \hline
\end{tabular}
\caption{Root-mean-squared errors for source parameters, the corresponding bounds on $\lambda_g$ and $\lambda_{\mathbb{A}}$, and the correlation coefficients, for the case $\alpha=3$ and for systems with different masses in units of $M_{\odot}$. 
The top cluster uses the Ad.~LIGO $S_{n}(f)$, $\rho =10$, $\lambda_{g}$ is in units of $10^{12} \; {\rm{km}}$, $\lambda_{\mathbb{A}}$ is in units of $10^{-16}\, \rm{km}$ and $\Delta t_c$ is in msecs.
The middle cluster uses the ET $S_{n}(f)$, $\rho = 50$, $\lambda_g$ is in units of $10^{13} \; {\rm{km}}$, $\lambda_{\mathbb{A}}$ is in units of $10^{-15}\, \rm{km}$ and $\Delta t_c$ is in msecs. 
The bottom cluster uses a NGO $S_{n}(f)$, $\rho = 100$, $\lambda_g$ is in units of $10^{15} \; {\rm{km}}$, $\lambda_{\mathbb{A}}$ is in units of $10^{-10}\, \rm{km}$ and $\Delta t_c$ is in secs.
} 
\label{table-all}
\end{table}
    \end{center}

\end{landscape}
    \clearpage
}


Although our results, presented in Fig.~\ref{Zeta-figs1}, Fig.~\ref{Zeta-figs2}, and Fig.~\ref{Zeta-figs3}, suggest bounds on $\zeta$ of ${\cal{O}}(10^{3}-10^{5})$ for the $\alpha=3$ case, the dimensional bounds in Table~\ref{table-all} suggest a strong constraint on $\lambda_{\mathbb{A}}$. This is because in converting from $\zeta$ to $\lambda_{\mathbb{A}}$ one must divide by the $D_{3}$ distance measure. This distance is comparable to (but smaller than) the luminosity distance, and thus, the longer the graviton propagates the more sensitive the constraints are to possible Lorentz violations.  Second, notice that the accuracy to which many parameters can be determined, e.g.~$t_c$, $\Delta{\mathcal M}$, and $\Delta\eta$, degrades with total mass because the number of observed gravitational-wave cycles decreases. Third, notice that the bound on the graviton Compton wavelength is not greatly affected by the inclusion of an additional parameter in the $\alpha=3$ case, and is comparable to the one obtained in~\cite{Will:1997bb} for LIGO. In fact, we have checked that in the absence of $\lambda_{\mathbb{A}}$ we recover Table~II in~\cite{Will:1997bb}.

We now consider how these bounds behave as a function of the mass ratio. Figure~\ref{Other-Bounds-figs1} plots the bound on the graviton Compton wavelength and Fig.~\ref{Other-Bounds-figs2} plots the Lorentz-violating Compton wavelength $\lambda_{\mathbb{A}}$ as a function of $\eta$ both for Ad.~LIGO and $\alpha =3$, with systems of different total mass. Notice that, in general, both bounds improve for comparable mass systems, even though the SNR is kept fixed. 

\begin{figure}[h!]
\centering
\includegraphics[width=12cm,clip=true]{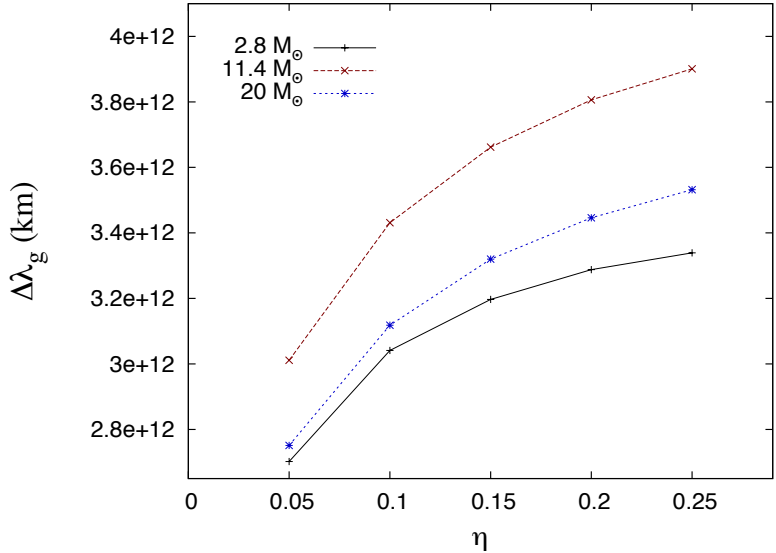}
\caption{\label{Other-Bounds-figs1} Bounds on $\lambda_g$ as a function of $\eta$ for different total masses. This is for Ad.~LIGO, with the SNR of $\rho = 10$ and assuming alternative theories in which $\alpha=3$. }
\end{figure}

\begin{figure}
\centering
\includegraphics[width=12cm,clip=true]{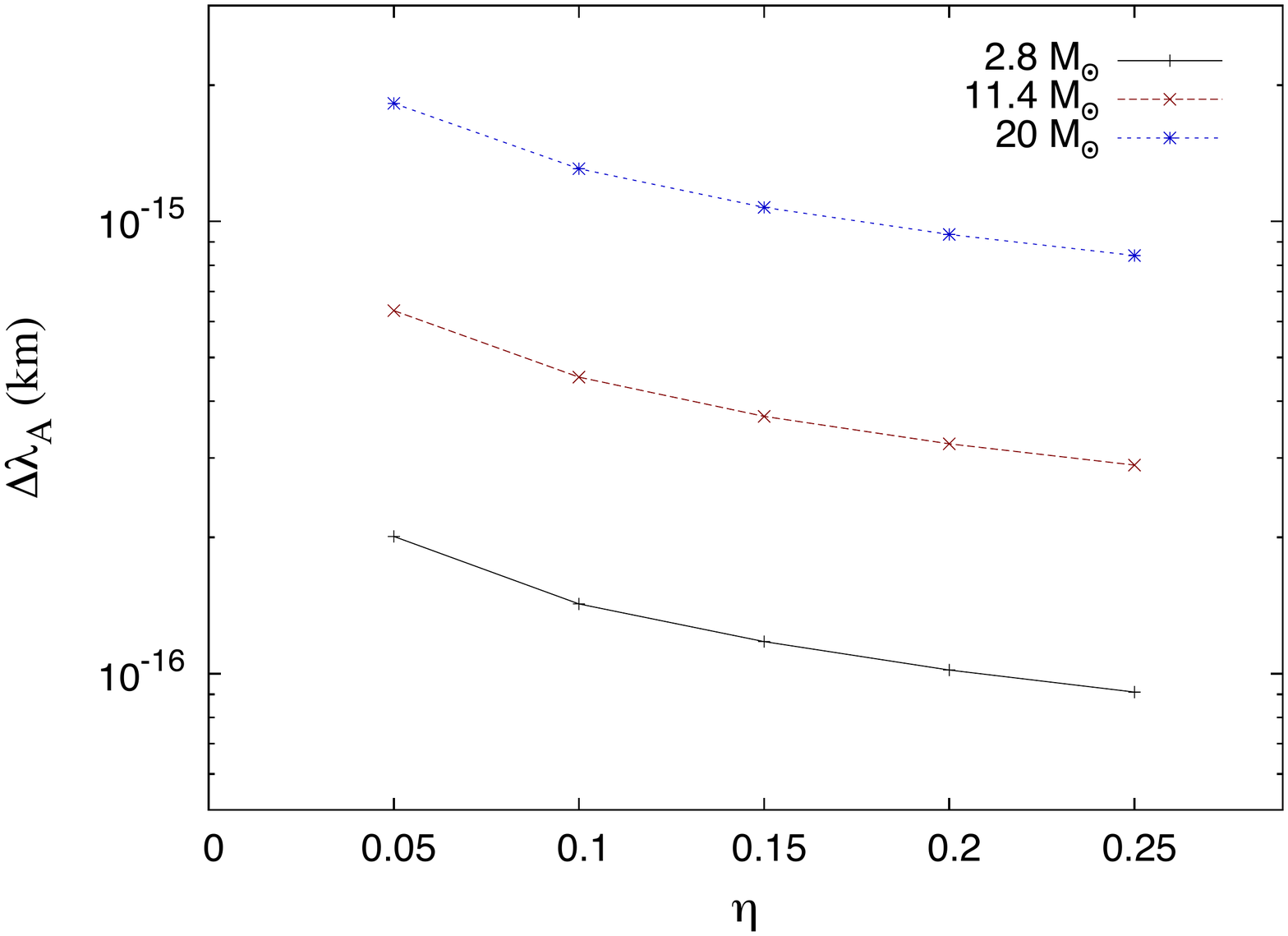} 
\caption{\label{Other-Bounds-figs2} Bounds on $\lambda_{\mathbb A}$ as a function of $\eta$ for different total masses. This is for Ad.~LIGO, with the SNR of $\rho = 10$ and assuming alternative theories in which $\alpha=3$. }
\end{figure}

With all of this information at hand, it seems likely that gravitational-wave detection would provide useful information about Lorentz-violating graviton propagation. For example, if a Bayesian analysis were carried out, once a gravitational wave is detected, and the ppE parameters peaked around $b_{\ppE} = 2$ or $3$, this could possibly indicate the presence of some degree of Lorentz violation. Complementarily, if no deviation from general relativity is observed, then one could constrain the magnitude of $\mathbb{A}$ to interesting levels, considering that no bounds exist to date.
\section{Conclusions and Discussion}\label{conclusions}

We studied whether Lorentz symmetry-breaking in the propagation of gravitational waves could be measured with gravitational waves from non-spinning, compact binary inspirals. We considered modifications to a massive graviton dispersion relation that scale as $\mathbb{A} p^{\alpha}$, where $p$ is the graviton's momentum while $\mathbb{A}$ and $\alpha$ are phenomenological parameters. We found that such a modification introduces new terms in the gravitational-wave phase due to a delay in the propagation: waves emitted at low frequency, early in the inspiral, travel slightly slower than those emitted at high frequency later. This results in an offset in the relative arrival times at a detector, and thus, a frequency-dependent phase correction. We mapped these new gravitational-wave phase terms to the recently proposed ppE scheme, with ppE phase parameters $b_{\ppE} = \alpha-1$.

We then carried out a simple Fisher analysis to get a sense of the accuracy to which such dispersion relation deviations could be measured with different gravitational-wave detectors. We found that indeed, both the mass of the graviton and additional dispersion relation deviations could be constrained. For values of $\alpha>4$, there is not enough information in the waveform to produce an invertible Fisher matrix. Certain values of $\alpha$, like $\alpha = 0$ and $2$, also cannot be measured, as they become $100\%$ correlated with other system parameters. 

In deriving these bounds, we have made several approximations that force us to consider them only as rough indicators that gravitational waves can be used to constrain generic Lorentz-violation in gravitational-wave propagation. For example, we have not accounted for precession or eccentricity in the orbits, the merger phase of the inspiral, the spins of the compact objects or carried out a Bayesian analysis. We expect the inclusion of these effects to modify and possibly worsen the bounds presented above by roughly an order of magnitude, based on previous results for bounds on the mass of the graviton~\cite{Will:1997bb,Will:2004xi,BBW2,stavridis,arunwill09,keppel:2010qu, 2010PhRvD..81f4008Y}. However, the detection of $N$ gravitational waves would lead to a $\sqrt{N}$ improvement in the bounds~\cite{2011arXiv1107.3528B}, while the modeling of only the Lorentz-violating term, without including the mass of the graviton, would also increase the accuracy to which $\lambda_{\mathbb{A}}$ could me measured~\cite{Cornish:2011ys}.

Future work could concentrate on carrying out a more detailed data analysis study, using Bayesian techniques. In particular, it would be interesting to compute the evidence for a general relativity model and a modified dispersion relation model, given a signal consistent with general relativity, to see the betting-odds of the signal favoring GR over the non-GR model. A similar study was already carried out in~\cite{Cornish:2011ys}, but there a single ppE parameter was considered.  Another interesting avenue for future research would be to consider whether there are any theories (quantum-inspired or not) that predict fractional $\alpha$ powers or values of $\alpha$ different from $3$ or $4$. 
 
\blankpage


\makeatletter
\let\orig@chapter\@chapter
\def\@chapter[#1]#2{\ifnum \c@secnumdepth >\m@ne
                       \if@mainmatter
                         \refstepcounter{chapter}%
                         \typeout{\@chapapp\space\thechapter.}%
                         \addcontentsline{toc}{chapter}%
                                   {Appendix~\protect\numberline{\thechapter.} #1}%
                       \else
                         \addcontentsline{toc}{chapter}{#1}%
                       \fi
                    \else
                      \addcontentsline{toc}{chapter}{#1}%
                    \fi
                    \chaptermark{#1}%
                    \addtocontents{lof}{\protect\addvspace{10\p@}}%
                    \addtocontents{lot}{\protect\addvspace{10\p@}}%
                    \if@twocolumn
                      \@topnewpage[\@makechapterhead{#2}]%
                    \else
                      \@makechapterhead{#2}%
                      \@afterheading
                    \fi}
\makeatother


\addtocontents{toc}{\vspace{2em}} 

\myapp{APPENDICES}{
}{}{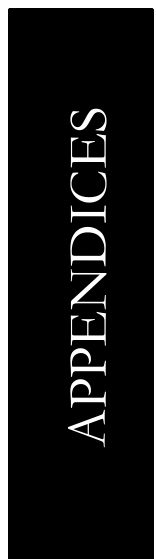} 

\appendix 

\chapter{Evaluations} 

\label{AppendixA} 
\thispagestyle{myplain}
\lhead[\thepage]{Appendix A.}      
\rhead[Appendix A.]{\thepage}
\ClearWallPaper

\def\xone{\mathbf{x}_1}
\def\xtwo{\mathbf{x}_2}
\def\n{\mathbf{n}}
\def\a{\mathbf{a}}
\def\b{\mathbf{b}}
\def\c{\mathbf{c}}
\def\x{\mathbf{x}}
\def\vone{\mathbf{v}_1}
\def\vtwo{\mathbf{v}_2}
\def\aone{\mathbf{a}_1}
\def\atwo{\mathbf{a}_2}
\def\datwo{\dot{\mathbf{a}}_2}
\def\Nvone{(\mathbf{v}_1\cdot\mathbf{n})}
\def\Nvtwo{(\mathbf{v}_2\cdot\mathbf{n})}
\def\vonevtwo{(\mathbf{v}_1\cdot\mathbf{v}_2)}
\def\ddatwo{\ddot{\mathbf{a}}_2}
\def\Naone{(\mathbf{a}_1\cdot\mathbf{n})}
\def\Natwo{(\mathbf{a}_2\cdot\mathbf{n})}

\section{Basic Facts}
\label{basicfacts}
\begin{figure}

    \subfigure [Inspirallig binary system configuration] {
        \includegraphics[width=0.5\textwidth]{./Figures/center.pdf}
        \label{fig:sensor3}
    }
    \subfigure[The location of field points relative to the origin and relative to each other]{
        \includegraphics[width=0.5\textwidth]{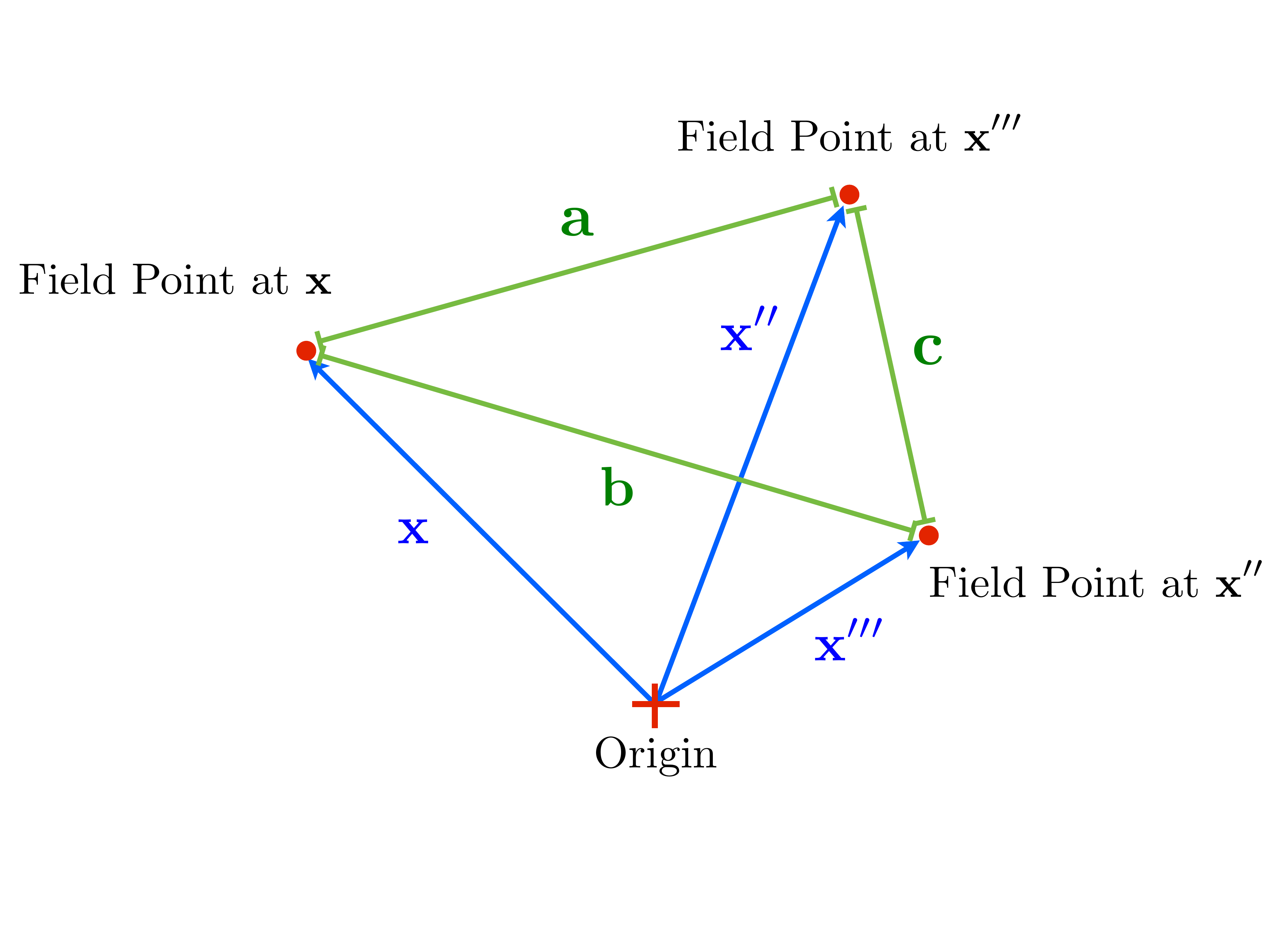}
         \label{fig:tiger}
    }
    \caption{A shematic configuration of an inspiralling compact binary system and the related integral variables}
    \label{fig:Sensor}   

    \end{figure}

\beq
\n^i\equiv \frac{(\xone-\xtwo)^i}{|\xone-\xtwo|}
\eeq

\beq
r\equiv |\xone-\xtwo|
\eeq

\begin{center}
\line(1,0){250}
\end{center}

\bea
\a_2^i=\alpha m_1 \frac{\n^i}{r^2}, &\displaystyle\a_1^i=-\alpha m_2 \frac{\n^i}{r^2}
\eea

\begin{center}
\line(1,0){250}
\end{center}
\beq
\atwo\cdot\n=\alpha \frac{m_1}{r^2}
\eeq
\beq
\aone\cdot\n=-\alpha \frac{m_2}{r^2}
\eeq
\beq
\atwo\cdot\vtwo=\frac{\alpha m_1 \Nvtwo}{r^2}
\eeq
\beq
\frac{d}{dt}\Nvtwo=\alpha \frac{m_1}{r^2}+\frac{\vonevtwo-\vtwo^2-\Nvone\Nvtwo+\Nvtwo^2}{r}
\eeq
\beq
\datwo^i=\frac{\alpha m_1}{r^3} (\vone^i-\vtwo^i)-3 \frac{\alpha m_1}{r^3}\biggl(\Nvone-\Nvtwo\biggr)\n^i
\eeq
\beq
\datwo\cdot\n=-2\frac{\alpha m_1}{r^3} \biggl[\Nvone-\Nvtwo \biggr]
\eeq
\beq
\datwo\cdot\vtwo=\frac{\alpha m_1}{r^3}\biggl(\vonevtwo-v_2^2\biggr)-3\frac{\alpha m_1}{r^3}\biggl[\Nvone \Nvtwo-\Nvtwo^2 \biggr]
\eeq
\beqa\nonumber
\ddatwo^i&=&-\alpha m_1 \biggl[\frac{\atwo^i-\aone^i}{r^3} +\frac{6}{r^4}(\vone^i-\vtwo^i)\biggl(\Nvone-\Nvtwo\biggr)\biggr.\\
&&\biggl.-\frac{15}{r^4}\biggl(\Nvone-\Nvtwo\biggr)^2\n^i+\frac{3}{r^3}\biggl(\Naone-\Natwo\biggr)\n^i+\frac{3}{r^4}(v_1-v_2)^2 \n^i\biggr]\quad
\eeqa
\beqa
\ddatwo\cdot\n&=&-\alpha m_1 \biggl[\frac{2}{r^3}\biggl(\Naone-\Natwo\biggr)-\frac{9}{r^4}\biggl(\Nvone-\Nvtwo\biggr)^2+\frac{3}{r^4}(v_1-v_2)^2\biggr]\qquad
\eeqa

\begin{center}
\line(1,0){250}
\end{center}

\beq
\frac{d}{dt}\biggl(\frac{1}{r}\biggr)=-\frac{1}{r^2}(\vone \cdot \n-\vtwo\cdot\n)
\eeq

\beq
\frac{d}{dt}\biggl(\frac{1}{r^3}\biggr)=-\frac{3}{r^4}(\vone \cdot \n-\vtwo\cdot\n)
\eeq

\beq
\frac{d}{dt}\biggl(\frac{1}{r^5}\biggr)=-\frac{5}{r^6}(\vone \cdot \n-\vtwo\cdot\n)
\eeq

\beq
\frac{d}{dt}\n=\frac{d}{dt}\biggl(\frac{\xone-\xtwo}{|\xone-\xtwo|}\biggr)=\frac{\vone-\vtwo}{r}-\frac{\Nvone-\Nvtwo}{r}\n
\eeq

\begin{center}
\line(1,0){250}
\end{center}

\beqa
\frac{1}{A}&=&\frac{1}{2r} (1-\frac{\epsilon}{2r}+\frac{1}{4r^2}\epsilon^2+\cdots)\\
\frac{1}{A^2}&=&\frac{1}{4r^2} (1-\frac{\epsilon}{r}+\frac{3}{4} \frac{\epsilon^2}{r^2}+\cdots)\\
\frac{1}{A^3}&=&\frac{1}{8r^3} (1-\frac{3}{2}\frac{\epsilon}{r}+\frac{3}{2} \frac{\epsilon^2}{r^2}+\cdots)
\eeqa
\begin{center}
\line(1,0){250}
\end{center}

\beqa\nonumber
\partial_k\; \partial_i{''} \partial_j{'''}\ln{A}&=&-(\frac{\delta^{ik}-\a^i\a^k}{|\a|})(\frac{\b^j+\c^j}{A^2})-(\frac{\delta^{jk}-\b^j\b^k}{|\b|})(\frac{\a^i-\c^i}{A^2})\\
&&+(\frac{\delta^{ij}-\c^i\c^j}{|\c|}) (\frac{\a^k+\b^k}{A^2})+\frac{2}{A^3} (\a^i-\c^i) (\b^j+\c^j) (\a^k+\b^k),
\eeqa

\beqa\nonumber
\partial_k \;\partial_i{'''} \partial_j{'''}\ln{A}&=&-\frac{1}{A}\biggl\{\frac{\b^i}{|\b|}(\frac{\delta^{jk}-\b^j\b^k}{|\b|})+\frac{\b^j}{|\b|}(\frac{\delta^{ik}-\b^i\b^k}{|\b|})+\frac{\b^k}{|\b|}(\frac{\delta^{ij}-\b^i\b^j}{|\b|})\biggr\}\\\nonumber
&&-(\frac{\delta^{ik}-\b^i\b^k}{|\b|}) (\frac{\b^j+\c^j}{A^2})-(\frac{\delta^{ij}-\b^i\b^j}{|\b|}) (\frac{\a^k+\b^k}{A^2})\\\nonumber
&&-(\frac{\delta^{jk}-\b^j\b^k}{|\b|}) (\frac{\b^i+\c^i}{A^2}) -(\frac{\delta^{ij}-\c^i\c^j}{|\c|}) (\frac{\a^k+\b^k}{A^2})\\
&&+\frac{2}{A^3} (\b^i+\c^i) (\b^j+\c^j) (\a^k+\b^k),
\eeqa

\beqa\nonumber
\partial_k \;\partial_i{''} \partial_j{''}\ln{A}&=&-\frac{1}{A}\biggl\{\frac{\a^i}{|\a|}(\frac{\delta^{jk}-\a^j \a^k}{|\a|})+\frac{\a^j}{|\a|}(\frac{\delta^{ik}-\a^i \a^k}{|\a|})+\frac{\a^k}{|\a|}(\frac{\delta^{ij}-\a^i \a^j}{|\a|})\biggr\}\\\nonumber
&&-(\frac{\delta^{ik}-\a^i\a^k}{|\a|}) (\frac{\a^j-\c^j}{A^2})-(\frac{\delta^{ij}-\a^i\a^j}{|\a|}) (\frac{\a^k+\b^k}{A^2})\\\nonumber
&&-(\frac{\delta^{jk}-\a^j\a^k}{|\a|}) (\frac{\a^i-\c^i}{A^2}) -(\frac{\delta^{ij}-\c^i\c^j}{|\c|}) (\frac{\a^k+\b^k}{A^2})\\
&&+\frac{2}{A^3} (\a^i-\c^i) (\a^j-\c^j) (\a^k+\b^k),
\eeqa

\beqa
\partial''_i\partial'''_j\ln{A}&=&-\frac{1}{A}(\frac{\delta^{ij}-\c^i\c^j}{|\c|})-\frac{(\a^i-\c^i) (\b^j+\c^j)}{A^2},
\eeqa

\beqa
\partial_k\partial'''_j\ln{A}&=&-\frac{1}{A}(\frac{\delta^{jk}-\b^j\b^k}{|\b|}+\frac{1}{A^2}(\a^k+\b^k) (\b^j+\c^j),
\eeqa

\beqa\nonumber
\partial'''_k\partial''_i\partial''_j\ln{A}&=&\frac{1}{A}\biggl[\frac{\c^i}{|\c|}(\frac{\delta^{jk}-\c^j\c^k}{|\c|})+\frac{\c^j}{|\c|}(\frac{\delta^{ik}-\c^i\c^k}{|\c|})+\frac{\c^k}{|\c|}(\frac{\delta^{ij}-\c^i\c^j}{|\c|})\biggr]\\\nonumber
&&+(\frac{\delta^{ik}-\c^i\c^k}{|\c|}) (\frac{\c^j-\a^j}{A^2})+(\frac{\delta^{ij}-\c^i\c^j}{|\c|}) (\frac{\b^k+\c^k}{A^2})\\\nonumber
&&+(\frac{\delta^{jk}-\c^j\c^k}{|\c|}) (\frac{\c^i-\a^i}{A^2}) +(\frac{\delta^{ij}-\a^i\a^j}{|\a|}) (\frac{\b^k+\c^k}{A^2})\\
&&-\frac{2}{A^3}(\c^i-\a^i) (\c^j-\a^j) (\b^k+\c^k)
\eeqa

\beqa\nonumber
\partial''_k\partial''_i\partial''_j\ln{A}&=&\frac{1}{A}\biggl[\frac{\a^i}{|\a|}(\frac{\delta^{jk}-\a^j\a^k}{|\a|})+\frac{\a^j}{|\a|}(\frac{\delta^{ik}-\a^i\a^k}{|\a|})+\frac{\a^k}{|\a|}(\frac{\delta^{ij}-\a^i\a^j}{|\a|})\\\nonumber
&&-\frac{\c^i}{|\c|}(\frac{\delta^{jk}-\c^j\c^k}{|\c|})-\frac{\c^j}{|\c|}(\frac{\delta^{ik}-\c^i\c^k}{|\c|})-\frac{\c^k}{|\c|}(\frac{\delta^{ij}-\c^i\c^j}{|\c|})\biggr]\\\nonumber
&&+\frac{1}{A^2}\biggl[(\a^i-\c^i)\biggl(\frac{\delta^{jk}-\a^j\a^k}{|\a|}+\frac{\delta^{jk}-\c^j\c^k}{|\c|}\biggr)+(\a^j-\c^j)\biggl(\frac{\delta^{ik}-\a^i\a^k}{|\a|}+\frac{\delta^{ik}-\c^i\c^k}{|\c|}\biggr)\\
&&+(\a^k-\c^k)\biggl(\frac{\delta^{ij}-\a^i\a^j}{|\a|}+\frac{\delta^{ij}-\c^i\c^j}{|\c|}\biggr)\biggr]+\frac{2}{A^3}(\a^i-\c^i)(\a^j-\c^j)(\a^k+\b^k),
\eeqa

and,
\beqa\nonumber
\partial{''}_i \;\partial_k{'''} \partial_l{'''}\ln{A}&=&-\frac{1}{A}\biggl\{\frac{\c^i}{|\c|}(\frac{\delta^{kl}-\c^k \c^l}{|\c|})+\frac{\c^k}{|\c|}(\frac{\delta^{il}-\c^i \c^l}{|\c|})+\frac{\c^l}{|\c|}(\frac{\delta^{ik}-\c^i \c^k}{|\c|})\biggr\}\\\nonumber
&&+(\frac{\delta^{kl}-\b^k\b^l}{|\b|}) (\frac{\a^i-\c^i}{A^2})-(\frac{\delta^{ik}-\c^i\c^k}{|\c|}) (\frac{\b^l+\c^l}{A^2})\\\nonumber
&&+(\frac{\delta^{kl}-\c^k\c^l}{|\c|}) (\frac{\a^i-\c^i}{A^2}) -(\frac{\delta^{il}-\c^i\c^l}{|\c|}) (\frac{\b^k+\c^k}{A^2})\\
&&+\frac{2}{A^3} (\c^i-\a^i) (\b^k+\c^k) (\b^l+\c^l),
\eeqa
where
\beq
A\equiv |\a|+|\b|+|\c|,
\eeq
and
\beqa\nonumber
\a=\x-\x'',&&\a^i=\frac{(\x-\x'')^i}{|\x-\x''|}\\\nonumber
\b=\x-\x''',&&\b^i=\frac{(\x-\x''')^i}{|\x-\x'''|}\\
\c=\x''-\x''',&&\c^i=\frac{(\x''-\x''')^i}{|\x''-\x'''|}.
\eeqa
Using the following straghtforward relations  help following the procedure to obtain above results.
\beqa
\partial_kA=\a^k+\b^k,&\partial''_kA=-\a^k+\c^k,&\partial_k''' A=-(\b^k+\c^k),\\
\partial_k \frac{1}{A}=-\frac{\a^k+\b^k}{A^2},&\partial_k''\frac{1}{A}=\frac{\a^k-\c^k}{A^2},&\partial_k'''\frac{1}{A}=\frac{\b^k+\c^k}{A^2},\\
\partial_k\frac{1}{A^2}=-2(\frac{\a^k+\b^k}{A^3}),&\partial''_k\frac{1}{A^2}=2(\frac{\a^k-\c^k}{A^3}),&\partial'''_k\frac{1}{A^2}=2(\frac{\b^k+\c^k}{A^3}),
\eeqa
and
\beqa
\partial_i\a^j=\frac{\delta^{ij}-\a^i\a^j}{|\a|},&\partial_i\b^j=\frac{\delta^{ij}-\b^i\b^j}{|\b|},&\partial_i\c^j=0,\\
\partial''_i\a^j=\frac{-\delta^{ij}+\a^i\a^j}{|\a|},&\partial''_i\c^j=\frac{\delta^{ij}-\c^i\c^j}{|\c|},&\partial''_i\b^j=0,\\
\partial_i'''\b^j=\frac{-\delta^{ij}+\b^i\b^j}{|\b|},&\partial_i'''\c^j=\frac{-\delta^{ij}+\c^i\c^j}{|\c|},&\partial'''_i\a^j=0.
\eeqa

\section{Multipole Moments for Two-Body Systems}\ClearWallPaper
\label{sec:moments}

Here we evaluate the multipole moments that appear in the radiation reaction expressions (\ref{15PNeom}) and (\ref{25PNeom}) to the order required to obtain $2.5$PN-accurate contributions.  The scalar dipole moment ${\cal I}_s^i$ in Eq.\ (\ref{15PNeom}) must be evaluated to $1$PN order.   Substituting $\tau_s$ from Eq.\ (\ref{tausPN}) and $\sigma_s$ from Eq.\ (\ref{sigmasPN}) to $1$PN order into Eq.\ (\ref{IsQ}), we obtain
\begin{equation}
{\cal I}_s^i = G \zeta m_1 x_1^i (1-2s_1) \left [ 1 - \frac{1}{2} v_1^2 - \frac{G \alpha m_2}{r} \left ( 1 - 4 \frac{\bar{\beta}_1}{\bar{\gamma}} \right ) \right ]  + ( 1 \rightleftharpoons 2 ) \,.
\end{equation}

Most of the multipole moments that appear in the $2.5$PN expressions (\ref{25PNeom}) can be evaluated to the lowest PN order, so that we may write
\begin{subequations}
\begin{eqnarray}
{\cal I}^{ij} &=& G(1-\zeta) \left (  m_1 x_1^{ij} + m_2 x_2^{ij} \right )  \,,
\\
{\cal I}^{ijk} &=& G(1-\zeta) \left (  m_1 x_1^{ijk} + m_2 x_2^{ijk} \right )  \,,
\\
{\cal J}^{qj} &=& G(1-\zeta) \epsilon^{qab} \left (  m_1 v_1^b x_1^{aj} + m_2 v_2^b x_2^{aj} \right ) \,,
\\
{\cal I}_s^{ij} &=& G\zeta  \left (  m_1 (1-2s_1) x_1^{ij} + m_2 (1-2s_2) x_2^{ij} \right )  \,,
\\
{\cal I}_s^{ijk} &=& G\zeta \left (  m_1 (1-2s_1) x_1^{ijk} + m_2 (1-2s_1) x_2^{ijk} \right )  \,.
\end{eqnarray}
\end{subequations}
The exception to this rule is the scalar monopole moment $M_s = \int_{\cal M} \tau_s d^3x$; formally it contributes at $0.5$PN order, as can be seen in Eq.\ (\ref{bigexpansiondPsi}), but its leading contribution is constant in time, and hence it is the $1$PN correction that matters.  Inserting  $\tau_s$ and $\sigma_s$ from Eqs.~(\ref{tausPN}) and (\ref{sigmasPN}) to $1$PN order, we obtain
\begin{equation}
M_s  = G \zeta m_1 (1-2s_1) \left [ 1 - \frac{1}{2} v_1^2 - \frac{G \alpha m_2}{r} \left ( 1 -4 \frac{\bar{\beta}_1}{\bar{\gamma}} \right ) \right ]  + ( 1 \rightleftharpoons 2 ) \,.
\end{equation}
Since the first term is constant, it can be dropped.

\section{Phase of the Gravitational Waveform to 3.5PN order}
\label{sec:35PNwaveform}

The phasing expression of \eref{7.17} was valid to 2PN order.  Here we quote the full expression, which has been calculated through 3.5PN order, as in  Eq.~(3.18) in ~\cite{2009PhRvD..80h4043B} 
\begin{eqnarray}
\psi^{({\rm F2})}_{3.5}(f) & = & 2\pi f t_c-\phi_c-\frac{\pi}{4}
+\frac{3}{128\,\eta\, v^{5}}\; \biggl[ 1 +\frac{20}{9}
\biggl( \frac{743}{336} + \frac{11}{4}\eta \biggr)v^2 
- 16\pi v^3 \\
&+& 10\,\biggl( \frac{3058673}{1016064} 
+ \frac{5429\, }{1008}\,\eta + \frac{617}{144}\,\eta^2 \biggr)v^4 +  \pi\biggl(\frac{38645 }{756} - \frac{65}{9}\eta\biggr)
\biggl\{1  + 3 \ln \biggl(\frac{v}{v_{\rm lso}}\biggr)\biggr\} v^5
\nonumber\\ 
&+&
\biggl\{ \frac{11583231236531}{4694215680} - \frac{640}{3}\pi^2 -
\frac{6848\,\gamma }{21} -\frac{6848}{ 21} \ln\biggl(4\;{v}\biggr) \nonumber\\ 
&+& \biggl( - \frac{15737765635}{3048192} 
+ \frac{2255\,{\pi }^2}{12} \biggr)\eta 
+\frac{76055}{ 1728}\eta^2-\frac{127825}{ 1296}\eta^3
\biggr \} v^6 \nonumber\\
&+& \pi\biggl(\frac{77096675 }{254016} + \frac{378515}{1512}\,\eta 
- \frac{74045}{756}\,\nu^2\biggr)v^7 \biggr],\nonumber
\label{eq:3.5PN-phasing}
\end{eqnarray}
where $v=(\pi M f)^{1/3}$, and $\gamma=0.577216\cdots$ is the Euler constant.


\chapter{Calculations} 

\label{AppendixB} 
\thispagestyle{myplain}
\lhead[\thepage]{Appendix B. \emph{}}      
\rhead[Appendix B. \emph{}]{\thepage}

\section{Sample Derivations and Evaluations}
\subsection{Integration of 1PN potentials}
\label{sec:1pn-potentials}
At 1PN order, the integration is straightforward. We consider a particular example

\be
\int_1 \rho^* \Phi_{1,i}\; d^3x = -\int_1 \rho^* \int \frac{\rho^{*'}}{\mid \vec{x}-\vec{x'}\mid ^3} (x-x')_j\; d^3x'\; d^3x.
\ee

Since we are interested in integration over body 1, the coordinate $x$ only needs to be integrated over body 1 instead of the entire space beacuase $\rho^*$ is zero everywhere else. There is, however, no such restriction on $\rho^{*'}$, and the $x'$ coordinate is to be evaluated over bodies 1 and 2, as these are the only bodies in the problem. Therefore, the integral splits into two integrals, one over each body.

\bea\label{B3}
 -\int_1 \rho^* \int \frac{\rho^{*'}}{\mid \vec{x}-\vec{x'}\mid ^3} (x-x')_j\; d^3x'\; d^3x &=&\nonumber\\
 -\int_1 \rho^* \int_1 \frac{\rho^{*'}}{\mid \vec{x}-\vec{x'}\mid ^3} (x-x')_j\; d^3x'\; d^3x &-&  \int_1 \rho^* \int_2 \frac{\rho^{*'}}{\mid \vec{x}-\vec{x'}\mid ^3} (x-x')_j\; d^3x'\; d^3x.
\eea

We consider these two pieces indivisually. First, the integral with both the $x$ and $x'$ coordinates evaluated over body 1 is a self-integral over body 1. With $v=v_1+\bar{v}'$, we have
\bea\label{B4}
 &&-\int_1 \rho^* \int_1 \frac{\rho^{*'}}{\mid \vec{x}-\vec{x'}\mid ^3} (x-x')_j\; d^3x'\; d^3x\nonumber\\
 &=&\int_1\frac{\rho^* \rho^{*'} (v_1^2+2 v_1\cdot\bar{v}'+\bar{v}'^2)}{\mid \vec{x}-\vec{x'}\mid ^3} (x-x')_j\; d^3x'\; d^3x\nonumber\\
&=&2 v_1^i H_1^{ij}+t_1^j.
\eea
where 
\bea
t_1^j&=&\displaystyle \int_1  \frac{\rho^* \rho^{*'} \bar{v}'^2 (x-x')^j}{\mid \vec{x}-\vec{x'}\mid ^3} d^3x'\; d^3x,\\
H_1^{ij}&=&\displaystyle \int_1  \frac{\rho^* \rho^{*'} \bar{v}'^i (x-x')^j}{\mid \vec{x}-\vec{x'}\mid ^3} d^3x'\; d^3x.
\eea

The integral that involves $v_1^2$ is zero because by symmetry the integral is automatically zero.

The second integral of \eref{B3} is integrated with $x$ ranging over body 1 and $x'$ over body 2. Let $\vec{v}'=\vec{v}_2+\bar{\vec{v}}'$, so we have

\bea\label{B5}
\int_1 \rho^* \int_2 \frac{\rho^{*'}}{\mid \vec{x}-\vec{x'}\mid ^3} (x-x')_j\; d^3x' &=& \int_1 \rho^* \int_2  \rho^{*'} (v_2^2+2 v_2\cdot\bar{v}'+\bar{v}'^2) \frac{n_j}{r^2} \; d^3x'\; d^3x\nonumber\\
&=& m_1 m_2 v_2^2 \frac{n^j}{r^2} + 2 m_1 \mathcal{I}_2 \frac{n^j}{r^2},
\eea
where we have approximated $\mid \vec{x}-\vec{x}' \mid$ as $r$, the distanse of separation between the two bodies, and $n^i$ is a unit vector in the direction of $\vec{x_1}-\vec{x_2}$. Note that the second term, involving $v_2\cdot \bar{v}'$ integrates to zero because $\int_2 \rho^{*'} \bar{v}' d^3x=0$ by the definition of center of mass. Combining \eref{B4} and \eref{B5}, we have
\be
\int_1 \rho^* \Phi_{1,i}\; d^3x = -2 v_1^2 H_1^{ij}-t_1^j-m_1 m_2 v_2^2 \frac{n^j}{r^2}-2 m_1 \mathcal{I}_2 \frac{n^j}{r^2}.
\ee

If the integral involves an additional coordinate $x''$ and an additional conserved density ${\rho^{*}}''$, there will be 4 integrals from all combinations of permutating ${\rho^{*}}'$ and ${\rho^{*}}''$ between bodies 1 and 2.

\subsection{Integration of 2PN potentials - part I}
At 2PN order, there are many more terms to integrate in the equation of motion than at 1PN order (compare the number of terms between \eref{1pnEOM} and \eref{2pnEOM}), and it is much work to include all possible combinations of permutating the various $\rho^*$ between bodies 1 and 2. Therefore, we neglect terms that involve any self-integrals, and only consider terms that involve the masses and velocities of the bodies, and the distance of separation between them. First we express the integrals in terms of the conserved densities, velocities, and the coordinates. Then, if (a) for each coordinate, a corresponding conserved density also appears, and (b) there are no ``triangle'' terms in any of the denominators, then there is no problem of divergence, and the integration is straightforward. We simply associate x with body 1, then assign the coordinates $x'$ and $x''$ to either body 1 or body 2 in such a way that all distances that appear in denominators of the integral are a difference between a coordinate associated with body 1 and a coordinate associated with body 2. There should be no distance in any denominators of the integral that is a difference between two coordinates associated with the same body. Such terms would be singular in the point mass limit, and our procedure is to discard such terms. For example,

\bea
\frac{1}{m_1}\int_1 \rho^* v^j U V_{j,i} d^3x = -\frac{1}{m_1}\int_1 \rho^* v^j \frac{{\rho^*}'}{\mid\vec{x}-\vec{x}'\mid} \frac{{\rho^*}'' {v_j}'' (x-x'')_i}{\mid\vec{x}-\vec{x}'\mid^3} d^3x''\, d^3x'\, d^3x=-m_2^2 \frac{v_1\cdot v_2}{r^3} n^i\\
\frac{1}{m_1}\int_1 \rho^* v^j \Phi^i_{2,j} d^3x =-\frac{1}{m_1}\int_1 \rho^* v^j {\rho^*}' \frac{(\vec{x}-\vec{x}')_j}{\mid\vec{x}-\vec{x}'\mid^3} \frac{{\rho^*}'' {v_j}''}{\mid\vec{x}-\vec{x}'\mid} d^3x''\, d^3x'\, d^3x=- m_1 m_2 \frac{v_1\cdot n}{r^3} v_1^i,
\eea
where in the first example both $x'$ and $x''$ are associated with body 2, but in the second example, $x'$ is associated with body 2 while $x''$ is associated with body 1.

\subsection{Integration of 2PN potentials - part II}
Not all terms in \eref{2pnEOM} can be integrated by the method described in the last section. Potentials such as $P_2^{ij}$ and related potentials such as $H$, $G_1$, $G_2$ have to be integrated with the use of the integral 
\be\label{B9}
\frac{1}{4\pi}\int \mid \vec{x}-\vec{x_a}\mid^{-1} \mid \vec{x}-\vec{x_b}\mid^{-1} \mid \vec{x}-\vec{x_c}\mid^{-1} d^3 x = -\ln (\mid \vec{x_a}-\vec{x_b}\mid+ \mid \vec{x_b}-\vec{x_c}\mid+ \mid \vec{x_a}-\vec{x_c}\mid)+1.
\ee

Using this integral, integration of ${P^{ij}}_{2,k}$ becomes
\bea\label{B10}
\frac{1}{m_1} \int_1 \rho^* {P^{ij}}_{2,k}&=& \frac{1}{4\pi m_1} \int_1 \rho^* \partial_k \biggl(U'_{,i} U'_{,j} \mid\vec{x}-\vec{x}'\mid^{-1}\biggr) d^2x\; d^3x'\nonumber\\
&=&\frac{1}{4\pi m_1}\int_1 \rho^* \partial_k {\partial}''_i {\partial}'''_j \frac{{\rho^*}'' {\rho^*}'''}{\mid\vec{x}-\vec{x}'\mid \mid\vec{x}'-\vec{x}''\mid \mid\vec{x}''-\vec{x}'''\mid} d^3x\; d^3x'\; d^3x''\;d^3x'''\nonumber\\
&=& -\frac{1}{m_1}\int_1 \rho^* {\rho^*}'' {\rho^*}''' \partial_k {\partial_i}'' {\partial_j}''' \ln ( \mid\vec{x}-\vec{x}''\mid+ \mid\vec{x}-\vec{x}'''\mid+ \mid\vec{x}''-\vec{x}'''\mid) d^3x\, d^3x''\, d^3x'''\nonumber\\
&=&\frac{1}{m_1}\int_1 \rho^* {\rho^*}'' {\rho^*}''' \biggl[-\frac{2}{A^3} (\hat{a}^k+\hat{b}^k) (\hat{a}^i+\hat{c}^i) (\hat{b}^j+\hat{c}^j) +\frac{1}{\mid \vec{a}\mid A^2} (\delta^{ik}-\hat{a}^i \hat{a}^k) (\hat{b}^j+\hat{c}^j)\nonumber\\
&&+\frac{1}{\mid\vec{b}\mid A^2} (\delta^{jk}-\hat{b}^j \hat{b}^k) (\hat{a}^i-\hat{c}^i)-\frac{1}{\mid\vec{c}\mid A^2} (\delta^{ij}-\hat{c}^i \hat{c}^j)(\hat{a}^k +\hat{b}^k)\biggr]\,d^3xd^3x''d^3x''',
\eea
where $\vec{a}=\vec{x}-\vec{x}''$, $\vec{b}=\vec{x}-\vec{x}'''$, $\vec{c}=\vec{x}''-\vec{x}'''$, $A=+\mid a \mid+\mid b \mid\mid c \mid$, and the hat notation denotes a unit vector. We then integrate \eref{B10} over all possibilities by associating the coordinates $x''$ and $x'''$ with bodies 1 and 2 in turn, and keep only finite terms. The coordinate x is associated with body 1, so when the coordinates $x''$ and $x'''$ are both assigned to body 1, the result is a self integral, which we discard. Now, if we assign the coordinate $x''$ to body 1 and $x'''$ to body 2, then $\vec{a}\rightarrow 0$, $\vec{b}=\vec{r}$, and $\vec{c}=\vec{r}$, where $\vec{r}=\vec{x}_1-\vec{x}_2$. Special care is needed in taking the limit of $\mid \vec{a}\mid \rightarrow 0$ as ther emay be finite terms associated with the limit. For example, let $\mid \vec{a} \mid=\varepsilon\ll \mid \vec{r}\mid$ and $r=\mid\vec{r}\mid$ and consider the term $\mid\vec{a}\mid^{-1} A^{-2} (\delta^{ik}-\hat{a}^i \hat{a}^k) (\hat{b}^j+\hat{c}^j)$,

\bea
\mid\vec{a} \mid^{-1} A^{-2} (\delta^{ik}-{\hat{a}}^i\hat{a}^k) (\hat{b}^j+\hat{c}^j)&=& \frac{1}{(2r+\varepsilon)^2 \varepsilon} (\delta^{ik}-\hat{a}^i\hat{a}^k) (2 n^j)\nonumber\\
&\simeq& \frac{1}{4\varepsilon r^2} \biggl(1-\frac{\varepsilon}{r}+\frac{3}{4} \displaystyle(\frac{\varepsilon}{r})^2+\cdots\biggr) (\delta^{ik}-\varepsilon^2) (2 n^j)\nonumber\\
&=&-\frac{n^j}{2 r^3} \delta^{ik},
\eea

where we have discarded terms that diverge as $\varepsilon^{-1}$ as well as terms that tend to zero in the limit of $\varepsilon\rightarrow 0$. Note that although, to the leading order, the term $\mid \vec{a} \mid^{-1} A^{-2} (\delta^{ik}-\hat{a}^i \hat{a}^k) (\hat{b}^j+ \hat{c}^j)$ diverges as $\varepsilon^{-1}$ as $\varepsilon\rightarrow 0$, there is a finit contribution to the integral that we could not have obtained had we naively, and incorrectly, discarded the entire term.

Repeating the process of associating the coordinates $x''$ and $x'''$ to bodies $1$ and $2$, the final result of $\int P_{2,k}^{ij}$ becomes

\bea\label{B12}
\frac{1}{m_1}\int \rho^* P_{2,k}^{ij}d^3x &=& m_1 m_2 \frac{1}{4r^3} (4n_i n_j n_k-\delta^{kj} n_i-\delta^{ij} n_k-2 \delta^{ik} n_j)\nonumber\\
&& +m_1 m_2 \frac{1}{4 r^3} (4n_i n_j n_k-\delta^{ik} n_j-\delta^{ij} n_k-2 \delta^{kj} n_i)\nonumber\\
&&+m_2^2 \frac{1}{4r^3} (-4n_i n_j n_k+\delta^{ik} n_j+\delta^{jk} n_i+2 \delta^{ij} n_k),
\eea

where the first line of \eref{B12} is obtained by associating $x''$ with body $1$ and $x'''$ with body $2$, so that $\vec{a}\rightarrow 0$, $\vec{b}=\vec{r}$, and  $\vec{c}=\vec{r}$. The second line of the equation is obtained by associating $x''$ with body $2$ and $x'''$ with  body $1$, so that $\vec{a}= \vec{r}$, $\vec{b}\rightarrow 0$, and  $\vec{c}=-\vec{r}$. Finally, the third line of the equation is obtained by associating both $x''$ and $x'''$ with body 2, so that $\vec{a}= \vec{r}$, $\vec{b}= \vec{r}$, and  $\vec{c}\rightarrow 0$.

For potentials  $P_{2,k}^{ij}$, the procedure of integration is similar. One difference is that, once \eref{B9} is used to simplify the integral, we may need to take derivatives with respect to different coordinates than the ones taken in $P_{2,k}^{ij}$. Another difference is that many integrals have quantities such as $\vec{v}''$, and so we assign them to the appropriate bodies consistent with the assignment of the associated coordinates, for example, ($\vec{x}''$ for the case with $\vec{v}''$).

\subsection{Integration of 2PN potentials - part III}

Of all the potentials at 2PN order, the potential $H \equiv P(U^{ij} P_2^{ij})$ most difficult to integrate it is ``doubly triangular''. Although the principle of integration is same as that in Appendix \ref{basicfacts}, there is no closed form expression such as \eref{B9} that simplifies the ``triangular'' potentials, and the integration must be carried out on mathematical software. We first simplify $H$ by partial integration
\bea\label{B13}
H&=&\frac{1}{4\pi}\int \frac{d^3x'}{\mid\vec{x}-\vec{x'}\mid} (U_{,jk} P_2^{jk})\nonumber\\
&=& \frac{1}{4\pi} \int{d^3x'} \biggl[\partial'_j \biggl(\frac{1}{\mid\vec{x}-\vec{x'}\mid} U_{,k} P_2^{jk}\biggr)-\partial'_j \biggl(\frac{1}{\mid\vec{x}-\vec{x'}\mid}\biggr) U_{,k} P_2^{jk}\nonumber\\
&&-\frac{1}{\mid\vec{x}-\vec{x'}\mid} U_{,k} \biggl(\frac{1}{2} \Phi_{2,k}-\frac{1}{2} U U_{,k} -\Sigma(U_{,k})\biggr)\biggr],
\eea
where we have used the formula
\be
\partial_j P_2^{ij}=\frac{1}{2} \Phi_{2.i} -\frac{1}{2} U U_{,i}-\Sigma(U_{,i}).
\ee

The first term of \eref{B13} vanishes because it can be converted to surface integral at infinity. So the 2PN potential that we need to integrate becomes
\be\label{B15}
\frac{1}{m_1} \int_1 \rho^* H_{,j} d^3x=\frac{1}{4\pi m_1} \int_1 \rho^* \biggl[\partial_j \partial_i  \biggl(\frac{1}{\mid\vec{x}-\vec{x'}\mid}\biggr)  U_{,k} P_2^{jk}+ \frac{(x-x')_j}{\mid\vec{x}-\vec{x'}\mid^3} U_{,k} \biggl(\frac{1}{2} \Phi_{2,k}-\frac{1}{2} U U_{,k} -\Sigma(U_{,k})\biggr)\biggr]
\ee

The second term of \eref{B15} involving $U_{,k} \Phi_{2,k}$ can be integrated using the methods of Appendix \ref{basicfacts}, but all other terms in the equation have to be integrated using a mathematical software. As an example we consider the third term of \eref{B15} 

\bea\label{B16}
\frac{1}{4\pi m_1} \int_1 \rho^* d^3x \,d^3x' \frac{(x-x')_j}{\mid\vec{x}-\vec{x'}\mid^3} U U^2_{,k} &=& \frac{1}{4\pi m_1} \int_1 \rho^* d^3x \,d^3x' \frac{(x-x')_j}{\mid\vec{x}-\vec{x'}\mid^3}  \biggl( \frac{m_1}{\mid\vec{y_1}\mid} + \frac{m_2}{\mid\vec{y_2}\mid}\biggr)\\
&&\times\biggl(\frac{m_1^2}{\mid\vec{y_1}\mid^4}+ \frac{m_2^2}{\mid\vec{y_1}\mid^4}+2 \frac{m_1 m_2 (\vec{y_1}\cdot\vec{y_2})}{\mid\vec{y}_1\mid^3 \mid\vec{y}_2\mid^3}\biggr)\nonumber\\
&=& \frac{1}{4\pi m_1} \int_1 \rho^* d^3x \,d^3x' \frac{(x-x')_j}{\mid\vec{x}-\vec{x'}\mid^3}  \biggl[ \frac{m_1^3}{\mid\vec{y_1}\mid^5}+ \frac{m_1 m_2^2}{\mid\vec{y_1}\mid \mid\vec{y}_2\mid^4}\nonumber\\
&&+2 \frac{m_1^2 m_2 (\vec{y_1}\cdot\vec{y_2})}{\mid\vec{y}_1\mid^4 \mid\vec{y}_2\mid^3}+ \frac{m1^2 m_2}{\mid\vec{y}_1\mid^4 \mid\vec{y}_2\mid} +\frac{m_2^3}{\mid\vec{y}_2\mid^5}+2\frac{m_1 m_2^2 (\vec{y}_1\cdot\vec{y}_2)}{\mid\vec{y}_1\mid^3 \mid\vec{y}_2\mid^4}\biggr],\nonumber
\eea
where $\vec{y}_1=\vec{x}'-\vec{x}_1$ and $\vec{y}_2=\vec{x}'-\vec{x}_2$, and we use a point mass expression for $U$. Since the coordinate $\vec{x}'$ is not assigned to either body, the potentials $U$ and $U_{,k}$, which are functions of $\vec{x}'$, are then the sum of the potential form each body. Of the six terms in \eref{B16}, the first term involves solely $\vec{y}_1$ and is a self-integral, which we discard. The second term becomes
\bea\label{B17}
\frac{1}{4\pi m_1} \int_1 \rho^* d^3x \,d^3x' \frac{(x-x')_j}{\mid\vec{x}-\vec{x'}\mid^3}  \frac{m_1 m_2^2}{\mid\vec{y}_1\mid \mid\vec{y}_2\mid^4}&=&-\frac{1}{4\pi m_1} \int_1 \rho^* d^3x \,d^3x' m_1 m_2^2 \frac{y_1^j}{\mid\vec{y}_1\mid^4 \mid\vec{y}_2\mid^4}\nonumber\\
&=&+\frac{1}{4\pi m_1} \int_1 \rho^* d^3x \,d^3x' m_1 m_2^2 \frac{y_2^j}{\mid\vec{y}_1\mid^2 \mid\vec{y}_2\mid^6}\nonumber\\
&=&-\frac{m_1 m_2^2}{2} n_j \int_{z=\varepsilon}^\infty \int_{\theta=0}^\pi z^2 dz \sin\theta d\theta \frac{z \cos\theta}{z^6 (z^2-2 z r \cos\theta+r^2)}\nonumber\\
&=&-\frac{m_1 m_2^2}{2} n_j \int_{z=\varepsilon}^\infty \int_{u=-1}^1 \frac{u\, dz\, du}{z^3 (z^2-2 z r u+r^2)},
\eea
where we first note that since $\vec{x}$ is assigned to body $1$, $(x-x')_j/\mid\vec{x}-\vec{x}'\mid^3$ is simply $-y_1^j/\mid\vec{y}_1\mid^3$. We next perform a partial integration and discard the surface term. The partial integration is not always necessary and is done in this case to avoid having Maple crash. We then make the substitution $z=\mid\vec{y}_2\mid$ and express $\mid\vec{y}_1\mid$ in terms of $z$ and the distance of separation between the two bodies $r=\mid\vec{x}_1-\vec{x}_2\mid$. We also choose the $(z, \theta, \phi)$ spherical coordinate system
such that the $\theta=0$ axis is parallel to $\vec{x}_1-\vec{x}_2$, and integrate over the azimuthal angle $\phi$. Since the integral in \eref{B17} must in the end be proportional to $n$ (the only vector in the problem), we need only to evaluate the projection
of the integral onto the $z$-axis. This integral can be done either by hand or using a mathematics software package,
and we used Maple 14 for our calculations. We first integrate over $u$ from $-1$ to $1$; because the result contains terms
proportional to $(z^2-r^2)^{-k}$, we then integrate $z$ from $\varepsilon$ to $r$ and from $r$ to infinity. While one can show that the
integral over $z\sim r$ is non-singular, splitting the integral avoids having Maple crash. We then expand the result in
powers of $\varepsilon$, and discard all terms proportional to $\varepsilon^{-2}$, $\varepsilon^{-1}$, and $\varepsilon^k$, and keep only the terms independent of $\varepsilon$. For the integral in \eref{B17}, we obtain the following result,

\be
-\frac{m_1 m_2^2}{2} n_j \int_{z=\varepsilon}^\infty \int_{u=-1}^1 \frac{u\, dz\, d\theta}{z^3 (z^2-2 z r u+r^2)}=m_1 m_2^2 \biggl(-\frac{2}{3} \frac{1}{r^3 \varepsilon}+\frac{4}{15} \frac{\varepsilon}{r^5}+\frac{2}{35} \frac{\varepsilon^3}{r^7}+\mathcal{O}(\varepsilon^5)\biggr) n^j
\ee

Therefore, the contribution of the integral in \eref{B17} to \eref{B16} is zero.

As another example, we consider the third term of \eref{B16}.
\bea\label{B19}
\frac{  m_1^2 m_2 }{2\pi m_1} \int_1 \rho^* d^3x \,d^3x' \frac{(x-x')_j}{|\vec{x}-\vec{x'}|^3} \frac{\vec{y}_1\cdot\vec{y}_2}{|\vec{y}_1|^4 |\vec{y}_2|^3}&=& -\frac{m_1^2 m_2}{2\pi} \int_1 \rho^* d^3x \,d^3x' \frac{y_1^i\, (\vec{y}_1\cdot\vec{y}_2)}{\mid\vec{y}_1\mid^7 \mid\vec{y}_2\mid^3}\nonumber\\
&=&-m_1^2 m_2  n_j \int_{z=\varepsilon}^\infty \int_{\theta=0}^\pi z^2 dz \sin\theta d\theta \frac{z \cos\theta (z^2+r z \cos\theta)}{z^7 (z^2+2 z r \cos\theta+r^2)^{3/2}}\nonumber\\
&=&-m_1^2 m_2  n_j \int_{z=\varepsilon}^\infty \int_{u=-1}^1 dz\, du \frac{u (z+r\,u)}{z^3 (z^2+2 z r u+r^2)^{3/2}}\nonumber\\
&=& m_1^2 m_2 n_j \biggl(\frac{1}{3 r^2 \varepsilon^2}-\frac{3}{5} \frac{1}{r^4}\biggr).
\eea
where we make similar substitutions as the previous example, except this time we let $z=\mid\vec{y}_1\mid$, and note that $\vec{y}_1\cdot\vec{y}_2=z\cdot (z+r)=z^2=r z \cos\theta$ because $r$ is projected onto the $z$ axis. Note that integration by parts is {\em not} necessary for this example, and that the final result is exact. Since we only keep terms that are independent of $\varepsilon$, the contribution of this terms to \eref{B16} is $-(3/5) m_1^2 m_2 n_j /r^4$. Repeating this procedure for all terms in \eref{B16}, we obtain the final result
\be
\frac{1}{4\pi m_1} \int_1 \rho^* d^3x \frac{(x-x')_j}{\mid\vec{x}-\vec{x}'\mid^3} U U^2_{,k} d^3x'=\biggl(\frac{1}{2} m_2^2 -\frac{1}{2} m_1^2 m_2\biggr) \frac{n_j}{r^4}.
\ee

\section{Results of Integration of 2PN Potentials}
In this section we give the results of integration of 2PN potentials that appear in \cref{chapter9} and \cref{chapter10}. The integrated results of individual terms of \eref{2pnEOM} are:

\be
\frac{4}{m_1}\int_1\rho^* v^i v^k v^j V^{j}_{,k} d^3x=-4 m_2 \frac{\vonevtwo \Nvone}{r^2} v_1^i,
\ee
\be
\frac{1}{m_1}\int_1\rho^* \dot{U} v^2 v^i d^3x = m_2 v_1^2 \frac{\Nvtwo}{r^2} v_1^i,
\ee
\be
-\frac{1}{m_1}\int_1\rho^* \dot{U}_{s} (1-2s) v^2 v^i d^3x = - (1-2s_1) (1-2s_2) m_2 v_1^2 \frac{\Nvtwo}{r^2} v_1^i,
\ee
\be
-\frac{2}{m_1}\int_1\rho^* v^i v^j \Phi_{1,j} d^3x = 2 m_2 v_2^2 \frac{\Nvone}{r^2} v_1^i,
\ee
\be
\frac{4}{m_1}\int_1\rho^* v^i v^j \Phi_{2,j} d^3x = -4 m_1 m_2 \frac{\Nvone}{r^3} v_1^i,
\ee
\be
\frac{4}{m_1}\int_1\rho^* v^i v^j \Phi_{2s,j}^{s} d^3x = -4 (1-2s_1) (1-2s_2) m_1 m_2 \frac{\Nvone}{r^3} v_1^i,
\ee
\be
-\frac{2}{m_1}\int_1\rho^* v^i v^j \ddot{X}_{,j} d^3x = 2 m_2 v_2^2 \frac{\Nvone}{r^2} v_1^i-6 m_2 \frac{\Nvtwo^2 \Nvone}{r^2} v_1^i+4 m_2 \frac{\Nvtwo \vonevtwo}{r^2}v_1^i,
\ee
\be
\frac{2}{m_1}\int_1\rho^* v^i v^k \Phi^{ik}_{1,j}d^3x=-2 m_2 \frac{\vonevtwo^2}{r^2} n^j,
\ee
\be
-\frac{4}{m_1}\int_1\rho^* v^i v^k \Phi^{jk}_{1,i}d^3x = 4 m_2 \frac{\vonevtwo \Nvone}{r^2} v_2^j,
\ee
\bea
\frac{2}{m_1}\int_1\rho^* v^i v^k P_{2,j}^{ik} d^3x=4 m_1 m_2 \frac{\Nvone^2}{r^3} n^j-m_1 m_2 \frac{v_1^2}{r^3} n^j-3m_1 m_2 \frac{\Nvone}{r^3} v_1^j\nonumber\\
-2 m_2^2\frac{\Nvone^2}{r^3}n^j+m_2^2\frac{v_1^2}{r^3} n^j+m_2^2\frac{\Nvone}{r^3}v_1^j,
\eea
\bea
\frac{2}{m_1}\int_1\rho^* v^i v^k P_{2s,j}^{ik} d^3x=\biggl\{4 m_1 m_2 \frac{\Nvone^2}{r^3} n^j-m_1 m_2 \frac{v_1^2}{r^3} n^j-3m_1 m_2 \frac{\Nvone}{r^3} v_1^j\biggr\}\nonumber\\ \times(1-2s_1)(1-2s_2)
+\biggl\{-2 m_2^2\frac{\Nvone^2}{r^3}n^j+m_2^2\frac{v_1^2}{r^3} n^j+m_2^2\frac{\Nvone}{r^3}v_1^j\biggr\}(1-2s_2)^2,
\eea
\bea
-\frac{4}{m_1}\int_1\rho^* v^i v^k P_{2,i}^{jk} d^3x=-8 m_1 m_2 \frac{\Nvone^2}{r^3} n^j+3m_1 m_2 \frac{v_1^2}{r^3} n^j+5m_1 m_2 \frac{\Nvone}{r^3} v_1^j\nonumber\\
+4 m_2^2\frac{\Nvone^2}{r^3}n^j-m_2^2\frac{v_1^2}{r^3} n^j-3m_2^2\frac{\Nvone}{r^3}v_1^j,
\eea
\bea
-\frac{4}{m_1}\int_1\rho^* v^i v^k P_{2s,i}^{jk} d^3x=\biggl\{-8 m_1 m_2 \frac{\Nvone^2}{r^3} n^j+3m_1 m_2 \frac{v_1^2}{r^3} n^j+5m_1 m_2 \frac{\Nvone}{r^3} v_1^j \biggr\} \nonumber\\ \times (1-2s_1)(1-2s_2)
+\biggl\{4 m_2^2\frac{\Nvone^2}{r^3}n^j-m_2^2\frac{v_1^2}{r^3} n^j-3m_2^2\frac{\Nvone}{r^3}v_1^j\biggr\} (1-2s_1)^2,
\eea
\be
-\frac{1}{2m_1}\int_1\rho^* v^2 \Phi_{1,j} d^3x=\frac{1}{2} m_2 v_1^2 v_2^2 \frac{n^j}{r^2},
\ee
\be
\frac{1}{2m_1}\int_1\rho^* (1-2s) v^2 \Phi^s_{1,j} d^3x=-\frac{1}{2} (1-2s_1) (1-2s_2) m_2 v_1^2 v_2^2 \frac{n^j}{r^2},
\ee
\be
-\frac{1}{m_1}\int_1\rho^* v^2 \Phi_{2,j} d^3x=m_1 m_2 v_1^2 \frac{n^j}{r^3},
\ee
\be
\frac{1}{m_1}\int_1\rho^* (1-2s) v^2 \Phi^s_{2,j} d^3x=-  (1-2s_1) (1-2s_2) m_1 m_2 v_1^2 \frac{n^j}{r^3},
\ee
\be
-\frac{1}{m_1}\int_1\rho^* [1-\zeta-(2\lambda_1+\zeta)(1-2s)] v^2 \Phi^s_{2s,j} d^3x= [1-\zeta-(2\lambda_1+\zeta)(1-2s1)] (1-2s_1) (1-2s_2) m_1 m_2 v_1^2 \frac{n^j}{r^3},
\ee
\be
\frac{2}{m_1}\int_1\rho^* [\lambda_1 (1-2s)+2 \zeta s'] v^2 U_s U_{s,j} d^3x= -2 [\lambda_1 (1-2s_1)+2 \zeta s'_1] (1-2s_2)^2 m_2^2 v_1^2 \frac{n^j}{r^3},
\ee
\be
\frac{4}{m_1}\int_1\rho^* (1-2s) v^2 [\Sigma(a_s U_s)]_{,j} d^3x=-4 (1-2s_1)^2 a_{s2}\; m_1 m_2\; v_1^2 \frac{n^j}{r^3},
\ee
\be
\frac{1}{2m_1}\int_1\rho^* v^2 \ddot{X}_{,j} d^3x=-\frac{1}{2} m_2 \frac{v_1^2 v_2^2}{r^2} n^j+\frac{3}{2} m_2 \frac{v_1^2 \Nvtwo^2}{r^2} n^j-m_2 \frac{v_1^2 \Nvtwo}{r^2} v_2^j,
\ee
\bea
-\frac{1}{2m_1}\int_1\rho^* (1-2s) v^2 \ddot{X}_{s,j} d^3x=\biggl\{\frac{1}{2} m_2 \frac{v_1^2 v_2^2}{r^2} n^j-\frac{3}{2} m_2 \frac{v_1^2 \Nvtwo^2}{r^2} n^j+m_2 \frac{v_1^2 \Nvtwo}{r^2} v_2^j\biggr\}\nonumber\\
\times (1-2s_1)(1-2s_2),\quad
\eea
\be
\frac{3}{m_1}\int_1\rho^* v^j \dot{\Phi}_{2} d^3x=-3 m_1 m_2 \frac{\Nvone}{r^3} v_1^j+6 m_1 m_2 \frac{\Nvtwo}{r^3} v_1^j,
\ee
\be
-\frac{1}{m_1}\int_1\rho^* (1-2s) v^j \dot{\Phi}^s_{2} d^3x= \biggl\{m_1 m_2 \frac{\Nvone}{r^3} v_1^j-2 m_1 m_2 \frac{\Nvtwo}{r^3} v_1^j\biggr\}\times (1-2s_1) (1-2s_2),
\ee
\bea
\frac{1}{m_1}\int_1\rho^* [3(1-\zeta)-(2\lambda_1+\zeta)(1-2s)] v^j \dot{\Phi}^s_{2s} d^3x=\biggl\{- m_1 m_2 \frac{\Nvone}{r^3} v_1^j+2 m_1 m_2 \frac{\Nvtwo}{r^3} v_1^j\biggr\}\nonumber\\
\times [3(1-\zeta)-(2\lambda_1+\zeta)(1-2s_1)] (1-2s_1) (1-2s_2),\qquad
\eea
\be
-\frac{4}{m_1}\int_1\rho^* v^j \dot{\Sigma}(a_s\; U_s) d^3x=\biggl\{4 m_1 m_2 \frac{\Nvone}{r^3} v_1^j-8 m_1 m_2 \frac{\Nvtwo}{r^3} v_1^j\biggr\}\times a_{s2}\;(1-2s_1)^2,
\ee
\be
-\frac{2}{m_1}\int_1\rho^* [\lambda_1 (1-2s)+2\zeta s'] v^j U_s \dot{U}_s d^3x=-2  [\lambda_1 (1-2s_1)+2\zeta s'_1]  (1-2s_2)^2 m_2^2 \frac{\Nvtwo}{r^3} v_1^j,
\ee
\be
-\frac{1}{2m_1}\int_1\rho^*v^j \dot{\Phi}_{1}d^3x=-\alpha\;m_1 m_2 \frac{\Nvtwo}{r^3} v_1^j-\frac{1}{2} m_2 v_2^2 \frac{\Nvtwo}{r^2} v_1^j,
\ee
\be
-\frac{1}{2m_1}\int_1\rho^* (1-2s) v^j \dot{\Phi}^s_{1}d^3x=\biggl\{-\alpha\;m_1 m_2 \frac{\Nvtwo}{r^3} v_1^j-\frac{1}{2} m_2 v_2^2 \frac{\Nvtwo}{r^2} v_1^j\biggr\}\times (1-2s_1) (1-2s_2),
\ee
\bea
-\frac{3}{2m_1}\int_1\rho^* v^j \dddot{X} d^3x=3 \alpha\;m_1 m_2 \frac{\Nvtwo}{r^3} v_1^j-3 \alpha\;m_1 m_2 \frac{\Nvone}{r^3} v_1^j\nonumber\\
-\frac{9}{2} m_2 v_2^2 \frac{\Nvtwo}{r^2} v_1^j+\frac{9}{2} m_2 \frac{\Nvtwo^3}{r^2} v_1^j,
\eea
\bea
\frac{1}{2m_1}\int_1\rho^* (1-2s) v^j \dddot{X}_s d^3x=\biggl\{- \alpha\;m_1 m_2 \frac{\Nvtwo}{r^3} v_1^j+ \alpha\;m_1 m_2 \frac{\Nvone}{r^3} v_1^j\nonumber\\
+\frac{3}{2} m_2 v_2^2 \frac{\Nvtwo}{r^2} v_1^j-\frac{3}{2} m_2 \frac{\Nvtwo^3}{r^2} v_1^j\biggr\}\times (1-2s_1)(1-2s_2),
\eea
\be
\frac{4}{m_1}\int_1\rho^* v^j V^i U_{,i} d^3x=-4 m_2^2 \frac{\Nvtwo}{r^3} v_1^j,
\ee
\be
\frac{4}{m_1}\int_1\rho^* v^j V^i U_{s,i} d^3x=-4 (1-2s_2)\; m_2^2 \frac{\Nvtwo}{r^3} v_1^j,
\ee
\be
\frac{8}{m_1}\int_1\rho^* v^i V^j U_{,i} d^3x=-8 m-2^2 \frac{\Nvone}{r^3} v_2^j,
\ee
\be
\frac{8}{m_1}\int_1\rho^* (1-2s) v^i V^j U_{s,i} d^3x=-8 m-2^2 (1-2s_1) (1-2s_2) \frac{\Nvone}{r^3} v_2^j,
\ee
\bea
-\frac{4}{m_1}\int_1\rho^* v_i \dot{P}_{2}^{ij} d^3x=-24 m_1 m_2 \frac{\Nvone^2}{r^3}n^j-8 m_1 m_2 \frac{\vonevtwo}{r^3} n^j+32 m_1 m_2 \frac{\Nvone \Nvtwo}{r^3}n^j\nonumber\\
+5 m_1 m_2 \frac{v_1^2}{r^3} n^j+11 m_1 m_2 \frac{\Nvone}{r^3} v_1^j-8 m_1 m_2 \frac{\Nvtwo}{r^3} v_1^j-8m_1m_2\frac{\Nvone}{r^3}v_2^j\nonumber\\
+m_2^2\frac{\vonevtwo}{r^3}n^j-4 m_2^2 \frac{\Nvone\Nvtwo}{r^3}n^j+2m_2^2 \frac{\Nvtwo}{r^3}v_1^j+m_2^2\frac{\Nvone}{r^3} v_2^j,\qquad
\eea
\bea
-\frac{4}{m_1}\int_1\rho^* v_i \dot{P}_{2s}^{ij} d^3x=\biggl\{-24 m_1 m_2 \frac{\Nvone^2}{r^3}n^j-8 m_1 m_2 \frac{\vonevtwo}{r^3} n^j+32 m_1 m_2 \frac{\Nvone \Nvtwo}{r^3}n^j\nonumber\\
+5 m_1 m_2 \frac{v_1^2}{r^3} n^j+11 m_1 m_2 \frac{\Nvone}{r^3} v_1^j-8 m_1 m_2 \frac{\Nvtwo}{r^3} v_1^j-8m_1m_2\frac{\Nvone}{r^3}v_2^j\biggr\}\times (1-2s_1) (1-2s_2)\nonumber\\
+\biggl\{m_2^2\frac{\vonevtwo}{r^3}n^j-4 m_2^2 \frac{\Nvone\Nvtwo}{r^3}n^j+2m_2^2 \frac{\Nvtwo}{r^3}v_1^j+m_2^2\frac{\Nvone}{r^3} v_2^j\biggr\}\times (1-2s_2)^2,\qquad
\eea
\bea
\frac{2}{m_1}\int_1\rho^* v_i \ddot{X}_{,i}^j=2 \alpha\;m_1 m_2 \frac{\Nvone}{r^3} v_1^j-2\alpha\; m_1 m_2 \frac{\Nvone}{r^3} v_2^j-6\alpha\; m_1 m_2 \frac{\Nvone^2}{r^3}n^j\nonumber\\
+10\alpha\; m_1 m_2\frac{\Nvone\Nvtwo}{r^3} n^j
-4\alpha\; m_1 m_2 \frac{\vonevtwo}{r^3}n^j-2m_2\frac{v_2^2 \Nvone}{r^2}v_2^j \nonumber\\ -4m_2\frac{\vonevtwo \Nvtwo}{r^2} v_2^j+6m_2\frac{\Nvtwo^2 \Nvone}{r^2} v_2^j,
\eea
\bea
-\frac{2}{m_1}\int_1\rho^* v_i \ddot{X}_{,j}^i=-2\alpha\; m_1 m_2 \frac{v_1^2}{r^3} n^j+2 \alpha\;m_1 m_2 \frac{\vonevtwo}{r^3} n^j+6\alpha\; m_1 m_2 \frac{\Nvone^2}{r^3}n^j\nonumber\\
-10\alpha\; m_1 m_2\frac{\Nvone\Nvtwo}{r^3} n^j
+4\alpha\; m_1 m_2 \frac{\Nvone}{r^3}v_2^j+2m_2\frac{v_2^2 \vonevtwo}{r^2}n^j \nonumber\\ +4m_2\frac{\vonevtwo \Nvtwo}{r^2} v_2^j-6m_2\frac{\Nvtwo^2 \vonevtwo}{r^2} n^j,
\eea
\be
\frac{2}{m_1}\int_1\rho^* v^i V^j_{3,i} d^3x=-2m_2 \frac{v_2^2 \Nvone}{r^2} v_2^j,
\ee
\be
-\frac{2}{m_1}\int_1\rho^* v^i V^i_{3,j} d^3x=2m_2 \frac{v_2^2 \vonevtwo}{r^2} n^j,
\ee
\be
-\frac{8}{m_1}\int_1\rho^* v^i \Phi^j_{2,i} d^3x=8 m_1 m_2 \frac{\Nvone}{r^3} v_1^j,
\ee
\be
\frac{8}{m_1}\int_1\rho^* v^i \Phi^i_{2,j} d^3x=-8 m_1 m_2 \frac{v_1^2}{r^3} n^i,
\ee
\bea
\frac{16}{m_1}\int_1\rho^* v^i G^j_{7,i} d^3x=4\alpha\;m_1 m_2 \frac{\Nvone^2}{r^3} n^j-6\alpha\; m_1 m_2 \frac{\Nvone}{r^3} v_1^j+2\alpha\;m_1m_2 \frac{v_1^2}{r^3}n^j\nonumber\\
+4\alpha\;m_1m_2\frac{\Nvone\Nvtwo}{r^3}n^j+2\alpha\;m_1m_2\frac{\Nvtwo}{r^3}v_1^j-\alpha\;m_1m_2\frac{\Nvone}{r^3}v_2^j\nonumber\\
-5\alpha\;m_1m_2 \frac{\vonevtwo}{r^3}n^j-4m_2^2 \frac{\Nvone\Nvtwo}{r^3}n^j+m_2^2\frac{\vonevtwo}{r^3}n^j\nonumber\\
+m_2^2 \frac{\Nvtwo}{r^3}v_1^j+2 m_2^2\frac{\Nvone}{r^3}v_2^j,
\eea
\bea
-\frac{16}{m_1}\int_1\rho^* v^i G^i_{7,j} d^3x=-4\alpha\;m_1 m_2 \frac{\Nvone^2}{r^3} n^j+3\alpha\; m_1 m_2 \frac{\Nvone}{r^3} v_1^j+\alpha\;m_1m_2 \frac{v_1^2}{r^3}n^j\nonumber\\
-4\alpha\;m_1m_2\frac{\Nvone\Nvtwo}{r^3}n^j-2\alpha\;m_1m_2\frac{\Nvtwo}{r^3}v_1^j+\alpha\;m_1m_2\frac{\vonevtwo}{r^3}n^j\nonumber\\
+5\alpha\;m_1m_2 \frac{\Nvone}{r^3}v_2^j+4m_2^2 \frac{\Nvone\Nvtwo}{r^3}n^j-m_2^2\frac{\Nvone}{r^3}v_2^j\nonumber\\
-m_2^2 \frac{\Nvtwo}{r^3}v_1^j-2 m_2^2\frac{\vonevtwo}{r^3}n^j,
\eea
\bea
-\frac{8}{m_1}\int_1\rho^* v^i P(\dot{U}_s U_s^{,[j})^{,i]} d^3x= m_1 m_2 (1-2s_1) (1-2s_2) \biggl\{ \frac{\Nvone}{r^3}v_1^j- \frac{v_1^2}{r^3}n^j\biggr\}\nonumber\\
+ m_2^2 (1-2s_2)^2 \biggl\{\frac{\Nvone}{r^3} v_2^j- \frac{\vonevtwo}{r^3}n^j\biggr\},
\eea
\be
-\frac{8}{m_1}\int_1\rho^* v^i U V_{j,i} d^3x=8 m_2^2 \frac{\Nvone}{r^3} v_2^j,
\ee
\be
\frac{8}{m_1}\int_1\rho^* v^i U V_{i,j} d^3x=-8 m_2^2 \frac{\vonevtwo}{r^3} n^j,
\ee
\be
\frac{4}{m_1}\int_1\rho^* v^i U V^j_{2,i} d^3x=-4m_1 m_2 \frac{\Nvone}{r^3} v_2^j,
\ee
\be
-\frac{4}{m_1}\int_1\rho^* v^i U V^i_{2,j} d^3x=4m_1 m_2 \frac{\vonevtwo}{r^3} n^j,
\ee
\be
\frac{4}{m_1}\int_1\rho^* v^i U V^{s\,j}_{2s,i} d^3x=-4 (1-2s_1) (1-2s_2)\; m_1 m_2 \frac{\Nvone}{r^3} v_2^j,
\ee
\be
-\frac{4}{m_1}\int_1\rho^* v^i U V^{s\,i}_{2s,j} d^3x=4 (1-2s_1) (1-2s_2)\; m_1 m_2 \frac{\vonevtwo}{r^3} n^j,
\ee
\be
-\frac{4}{m_1}\int_1\rho^* v^i \dot{\Phi}_{1}^{ij}d^3x=-4 \alpha\;m_1 m_2 \frac{\vonevtwo}{r^3} n^j-4\alpha\;m_1m_2\frac{\Nvone}{r^3} v_2^j-4m_2\frac{\vonevtwo\Nvtwo}{r^2} v_2^j,
\ee
\be
\frac{4}{m_1}\int_1\rho^* \Phi_2 U_{,j} d^3x=-4 m_1 m_2^2 \frac{n^j}{r^4},
\ee
\be
\frac{4}{m_1}\int_1\rho^* \Phi^s_{2s} U_{,j} d^3x=-4 (1-2s_1) (1-2s_2) m_1 m_2^2 \frac{n^j}{r^4},
\ee
\be
\frac{2}{m_1}\int_1\rho^* [\lambda_1(1-2s)+\zeta(2\a_s-s-2s^2)]\; \Phi^s_2 U_{s,j} d^3x=-2 [\lambda_1(1-2s_1)+\zeta(2\a_{s1}-s_1-2s_1^2)]\; (1-2s_2)^2 m_1 m_2^2 \frac{n^j}{r^4},
\ee
\bea
\frac{2}{m_1}\int_1\rho^* \biggl[2\zeta(1-\zeta)(1-s)+(2\lambda_1+\zeta)[\lambda_1(1-2s)+\zeta(2a_s-s-2s^2)]\biggr] \Phi^s_{2s} U_{s,j} d^3x=\qquad\nonumber\\
-2 \biggl[2\zeta(1-\zeta)(1-s_1)+(2\lambda_1+\zeta)[\lambda_1(1-2s_1)+\zeta(2a_{s1}-s_1-2s_1^2)]\biggr]\nonumber\\
\times (1-2s_1)(1-2s_s)^2 m_1 m_2^2 \frac{n^j}{r^4},\qquad
\eea
\be
\frac{4}{m_1}\int_1\rho^* (1-2s) \Phi_2 U_{s,j} d^3x=-4 (1-2s_1)(1-2s_2)\;m_1 m_2^2 \frac{n^j}{r^4},
\ee
\bea
\frac{8}{m_1}\int_1\rho^* [\lambda_1(1-2s)+\zeta(2\a_s-s-2s^2)]\; \Sigma(a_s\; U_s) U_{s,j} d^3x=\nonumber\\
-8 [\lambda_1(1-2s_1)+\zeta(2\a_{s1}-s_1-2s_1^2)]\; a_{s2}\;(1-2s_1) (1-2s_2) m_1 m_2^2 \frac{n^j}{r^4},
\eea
\bea
-\frac{4}{m_1}\int_1\rho^* G_{6,j} d^3x=16 m_1 m_2 \frac{\Nvone^2}{r^3}-8m_1m_2 \frac{\Nvone}{r^3} v_1^j-4m_1m_2\frac{v_1^2}{r^3}n^j\nonumber\\
+2m_1m_2\frac{\Nvtwo^2}{r^3}n^j
-2m_1m_2\frac{\Nvtwo}{r^3}+12 m_2^2\frac{\Nvtwo^2}{r^3}n^j-6m_2^2\frac{\Nvtwo}{r^3} v_2^j-2m_2^2\frac{v_2^2}{r^3}n^j,\qquad
\eea
\bea
-\frac{4}{m_1}\int_1\rho^* G_{6s,j} d^3x=\biggl\{16 m_1 m_2 \frac{\Nvone^2}{r^3}-8m_1m_2 \frac{\Nvone}{r^3} v_1^j-4m_1m_2\frac{v_1^2}{r^3}n^j\nonumber\\+2m_1m_2\frac{\Nvtwo^2}{r^3}n^j
-2m_1m_2\frac{\Nvtwo}{r^3}+12 m_2^2\frac{\Nvtwo^2}{r^3}n^j-6m_2^2\frac{\Nvtwo}{r^3} v_2^j-2m_2^2\frac{v_2^2}{r^3}n^j\biggr\}\nonumber\\
\times(1-2s_1)(1-2s_2),\qquad
\eea
\bea
\frac{3}{4m_1}\int_1\rho^* \ddot{X}_{1,j} d^3x=\frac{3}{2} m_1^2 m_2 \frac{n^j}{r^4}-\frac{3}{2}m_1 m_2 \frac{v_2^2}{r^3} n^j+\frac{3}{2} m_1 m_2 \frac{\vonevtwo}{r^3}n^j\nonumber\\
+\frac{15}{2} m_1 m_2 \frac{\Nvtwo^2}{r^3}n^j-\frac{9}{2}m_1m_2\frac{\Nvone\Nvtwo}{r^3}n^j-3m_1 m_2 \frac{\Nvtwo}{r^3} v_2^j\nonumber\\
-\frac{3}{4}m_2\frac{v_2^4}{r^2}n^j-\frac{3}{2} m_2 \frac{v_2^2 \Nvtwo}{r^2}v_2^j+\frac{9}{4} m_2 \frac{v_2^2 \Nvtwo^2}{r^2}n^j,
\eea
\bea
-\frac{1}{4m_1}\int_1\rho^* \ddot{X}(v^2)_{s,j} d^3x=-\frac{1}{3}\biggl\{\frac{3}{2} m_1^2 m_2 \frac{n^j}{r^4}-\frac{3}{2}m_1 m_2 \frac{v_2^2}{r^3} n^j+\frac{3}{2} m_1 m_2 \frac{\vonevtwo}{r^3}n^j\nonumber\\
+\frac{15}{2} m_1 m_2 \frac{\Nvtwo^2}{r^3}n^j-\frac{9}{2}m_1m_2\frac{\Nvone\Nvtwo}{r^3}n^j-3m_1 m_2 \frac{\Nvtwo}{r^3} v_2^j\nonumber\\
-\frac{3}{4}m_2\frac{v_2^4}{r^2}n^j-\frac{3}{2} m_2 \frac{v_2^2 \Nvtwo}{r^2}v_2^j+\frac{9}{4} m_2 \frac{v_2^2 \Nvtwo^2}{r^2}n^j\biggr\}\times (1-2s_1)(1-2s_2),
\eea
\be
-\frac{2}{m_1}\int_1\rho^* U \ddot{X}_{,j} d^3x=2m_2^2\frac{v_2^2}{r^3}n^j-6m_2^2\frac{\Nvtwo^2}{r^3}n^j+4m_2^2\frac{\Nvtwo}{r^3} v_2^j,
\ee
\be
-\frac{2}{m_1}\int_1\rho^* (1-2s)\;U \ddot{X}_{s,j} d^3x=\biggl\{2 m_2^2\frac{v_2^2}{r^3}n^j-6m_2^2\frac{\Nvtwo^2}{r^3}n^j+4m_2^2\frac{\Nvtwo}{r^3} v_2^j\biggr\}\times (1-2s_1)(1-2s_2),
\ee
\bea
-\frac{1}{m_1}\int_1\rho^* [\lambda_1(1-2s)+\zeta (2\;a_{s}-s-2s^2)]\;U_s \ddot{X}_{s,j} d^3x=\frac{1}{2}\biggl\{2 m_2^2\frac{v_2^2}{r^3}n^j-6m_2^2\frac{\Nvtwo^2}{r^3}n^j\nonumber\\
+4m_2^2\frac{\Nvtwo}{r^3} v_2^j\biggr\}
\times [\lambda_1(1-2s_1)+\zeta (2\;a_{s1}-s_1-2s_1^2)]\;(1-2s_2)^2,\quad
\eea
\be
\frac{9}{2m_1}\int_1\rho^* [\Sigma(U\,v^2)]_{,j}d^3x=-\frac{9}{2}m_1 m_2 \frac{v_2^2}{r^3} n^j,
\ee
\bea
-\frac{1}{2m_1}\int_1\rho^* [3 (1-\zeta)-(2\lambda_1+\zeta) (1-2s)] [\Sigma_s(U_s\,v^2)]_{,j}d^3x=\nonumber\\
\frac{1}{2}[3 (1-\zeta)-(2\lambda_1+\zeta) (1-2s_1)] (1-2s_1)(1-2s_2)\; m_1 m_2 \frac{v_2^2}{r^3} n^j,
\eea
\be
-\frac{3}{2m_1}\int_1\rho^*(1-2s) [\Sigma_s(U\,v^2)]_{,j}d^3x=\frac{3}{2}(1-2s_1)(1-2s_2)\; m_1 m_2 \frac{v_2^2}{r^3} n^j,
\ee
\be
\frac{2}{m_1}\int_1\rho^* [\Sigma(a_s\;U_s\,v^2)]_{,j}d^3x=-2 (1-2s_1)^2 a_{s2}\; m_1 m_2 \frac{v_2^2}{r^3} n^j,
\ee
\be
-\frac{6}{m_1}\int_1\rho^* U \Phi_{1,j} d^3x=6 m_2^2\frac{v_2^2}{r^3} n^j,
\ee
\be
\frac{2}{m_1}\int_1\rho^* (1-2s)\; U \Phi^s_{1,j} d^3x=-2 (1-2s_1)(1-2s_2)\; m_2^2\frac{v_2^2}{r^3} n^j,
\ee
\be
\frac{1}{m_1}\int_1\rho^* [\lambda_1 (1-2s)+\zeta(2 a_s-s-2 s^2)] U_s \Phi^s_{1,j} d^3x=- [\lambda_1 (1-2s_1)+\zeta(2 a_{s1}-s_1-2 s_1^2)] (1-2s_2)^2 m_2^2 \frac{v_2^2}{r^3} n^j,
\ee
\be
\frac{3}{2m_1}\int_1\rho^* [\Sigma(U^2)]_{,j}d^3x=-\frac{3}{2}m_1^2 m_2 \frac{n^j}{r^4},
\ee
\be
\frac{3}{2m_1}\int_1\rho^* (1-2s) [\Sigma_s(U^2)]_{,j}d^3x=-\frac{3}{2}(1-2s_1)(1-2s_2)\;m_1^2 m_2 \frac{n^j}{r^4},
\ee
\be
\frac{1}{m_1}\int_1\rho^* [\Sigma(U_s^2)]_{,j}d^3x=-(1-2s_1)^2 m_1^2 m_2 \frac{n^j}{r^4},
\ee
\be
\frac{1}{2m_1}\int_1\rho^* (1-2s) [\Sigma_s(U_s^2)]_{,j}d^3x=-\frac{1}{2}(1-2s_1)^3 (1-2s_2)\;m_1^2 m_2 \frac{n^j}{r^4},
\ee
\be
\frac{1}{m_1}\int_1\rho^* (1-2s) [\Sigma_s(U U_s)]_{,j}d^3x=- (1-2s_1)^2 (1-2s_2)\;m_1^2 m_2 \frac{n^j}{r^4},
\ee
\be
\frac{2}{m_1}\int_1\rho^* (1-2s) [\Sigma(a_s\; U_s^2)]_{,j}d^3x=-2 a_{s2}\;(1-2s_1)^3 m_1^2 m_2 \frac{n^j}{r^4},
\ee
\be
-\frac{4}{m_1}\int_1\rho^* (1-2s) [\Sigma(b_s\;U_s^2)]_{,j}d^3x=4 b_{s2}\;(1-2s_1)^3 m_1^2 m_2 \frac{n^j}{r^4},
\ee
\be
\frac{4}{m_1}\int_1\rho^* (1-2s) [\Sigma(a_s\;U U_s)]_{,j}d^3x=-4 a_{s2}\;(1-2s_1)^2 m_1^2 m_2 \frac{n^j}{r^4},
\ee
\be
\frac{4}{m_1}\int_1\rho^* U \Phi_{2,j}d^3x=-4 m_1 m_2^2 \frac{n^j}{r^4},
\ee
\be
\frac{4}{m_1}\int_1\rho^* [1-\zeta+(2\lambda_1+\zeta)(1-2s)]\; U \Phi^s_{2s,j}d^3x=-4 [1-\zeta+(2\lambda_1+\zeta)(1-2s_1)]\; (1-2s_1)(1-2s_2)\; m_1 m_2^2 \frac{n^j}{r^4},
\ee
\be
\frac{4}{m_1}\int_1\rho^* (1-2s)\;U \Phi^s_{2,j}d^3x=-4 (1-2s_1)(1-2s_2)\; m_1 m_2^2 \frac{n^j}{r^4},
\ee
\be
\frac{2}{m_1}\int_1\rho^* [\lambda_1 (1-2s)+\zeta(2a_s-s-2s^2)] U_s \Phi^s_{2,j}d^3x=-2 [\lambda_1 (1-2s_1)+\zeta(2a_{s1}-s_1-2s_1^2)]\;(1-2s_2)^2 m_1 m_2^2 \frac{n^j}{r^4},
\ee
\bea
\frac{2}{m_1}\int_1\rho^* [\lambda_1 (1-2s)+\zeta(2a_s-s-2s^2)] U_s \Phi^s_{2s,j}d^3x=-2 [\lambda_1 (1-2s_1)+\zeta(2a_{s1}-s_1-2s_1^2)]\nonumber\\
\times(1-2s_1) (1-2s_2)^2 m_1 m_2^2 \frac{n^j}{r^4},\qquad
\eea
\be
\frac{16}{m_1}\int_1\rho^* (1-2s) U [\Sigma(a_s\; U_s)]_{,j}d^3x=-16\; a_{s2}\; (1-2s_1)^2 m_1 m_2^2 \frac{n^j}{r^4},
\ee
\bea
\frac{8}{m_1}\int_1\rho^* [\lambda_1 (1-2s)+\zeta(2a_s-s-2s^2)] U_s [\Sigma(a_s\; U_s)]_{,j} d^3x=\nonumber\\
-8 [\lambda_1 (1-2s_1)+\zeta(2a_{s1}-s_1-2s_1^2)]
\times a_{s2}\;(1-2s_1) (1-2s_2) m_1 m_2^2 \frac{n^j}{r^4},\qquad
\eea
\bea
-\frac{1}{2m_1}\int_1\rho^* \ddot{X}_{2,j} d^3x=-\frac{1}{2} m_1^2 m_2 \frac{n^j}{r^4}-\frac{1}{2} m_1 m_2^2 \frac{n^j}{r^4}-\frac{3}{2} m_1 m_2 \frac{\Nvone^2}{r^3}n^j\nonumber\\
+4m_1m_2\frac{\Nvone\Nvtwo}{r^3}n^j-4m_1m_2\frac{\Nvtwo^2}{r^3}n^j+\frac{1}{2}m_1 m_2 \frac{v_1^2}{r^3}n^j\nonumber\\
-m_1m_2\frac{\vonevtwo}{r^3}n^j+m_1m_2\frac{v_2^2}{r^3}n^j-m_1m_2\frac{\Nvone}{r^3} v_2^j+2 m_1 m_2\frac{\Nvtwo}{r^3}v_2^j,
\eea
\bea
-\frac{2}{m_1}\int_1\rho^* (1-2s) \ddot{X}(a_s\; U_s)_{,j} d^3x=4\biggl\{-\frac{1}{2} m_1^2 m_2 \frac{n^j}{r^4}-\frac{1}{2} m_1 m_2^2 \frac{n^j}{r^4}-\frac{3}{2} m_1 m_2 \frac{\Nvone^2}{r^3}n^j\nonumber\\
+4m_1m_2\frac{\Nvone\Nvtwo}{r^3}n^j-4m_1m_2\frac{\Nvtwo^2}{r^3}n^j+\frac{1}{2}m_1 m_2 \frac{v_1^2}{r^3}n^j
-m_1m_2\frac{\vonevtwo}{r^3}n^j\nonumber\\
+m_1m_2\frac{v_2^2}{r^3}n^j-m_1m_2\frac{\Nvone}{r^3} v_2^j+2 m_1 m_2\frac{\Nvtwo}{r^3}v_2^j\biggr\}\times a_{s2}\;(1-2s_1)^2,\quad
\eea
\bea
-\frac{1}{2m_1}\int_1\rho^* (1-2s) \ddot{X}^s_{2,j} d^3x=\biggl\{-\frac{1}{2} m_1^2 m_2 \frac{n^j}{r^4}-\frac{1}{2} m_1 m_2^2 \frac{n^j}{r^4}-\frac{3}{2} m_1 m_2 \frac{\Nvone^2}{r^3}n^j\nonumber\\
+4m_1m_2\frac{\Nvone\Nvtwo}{r^3}n^j-4m_1m_2\frac{\Nvtwo^2}{r^3}n^j+\frac{1}{2}m_1 m_2 \frac{v_1^2}{r^3}n^j
-m_1m_2\frac{\vonevtwo}{r^3}n^j\nonumber\\+m_1m_2\frac{v_2^2}{r^3}n^j-m_1m_2\frac{\Nvone}{r^3} v_2^j+2 m_1 m_2\frac{\Nvtwo}{r^3}v_2^j\biggr\}\times (1-2s_1) (1-2s_2),\quad
\eea
\bea
-\frac{1}{2m_1}\int_1\rho^* [1-\zeta+(2\lambda_1+\zeta)(1-2s)] \ddot{X}^{s}_{2s,j} d^3x=\biggl\{-\frac{1}{2} m_1^2 m_2 \frac{n^j}{r^4}-\frac{1}{2} m_1 m_2^2 \frac{n^j}{r^4}\nonumber\\
-\frac{3}{2} m_1 m_2 \frac{\Nvone^2}{r^3}n^j
+4m_1m_2\frac{\Nvone\Nvtwo}{r^3}n^j-4m_1m_2\frac{\Nvtwo^2}{r^3}n^j+\frac{1}{2}m_1 m_2 \frac{v_1^2}{r^3}n^j\nonumber\\
-m_1m_2\frac{\vonevtwo}{r^3}n^j+m_1m_2\frac{v_2^2}{r^3}n^j-m_1m_2\frac{\Nvone}{r^3} v_2^j+2 m_1 m_2\frac{\Nvtwo}{r^3}v_2^j\biggr\}\nonumber\\
\times [1-\zeta+(2\lambda_1+\zeta)(1-2s_1)]\;(1-2s_1) (1-2s_2) ,\qquad
\eea
\bea
\frac{1}{24m_1}\int_1\rho^* \ddddot{Y}_{,j} d^3x=-\frac{15}{8}m_2\frac{\Nvtwo^4}{r^2}n^j+\frac{9}{4}m_2\frac{v_2^2 \Nvtwo^2}{r^2}n^j-\frac{3}{8}m_2\frac{v_2^4}{r^2}n^j+\frac{3}{2}m_2\frac{\Nvtwo^3}{r^2} v_2^j\nonumber\\
-\frac{3}{2}m_2\frac{v_2^2 \Nvtwo}{r^2} v_2^j+\frac{1}{4}m_1^2 m_2\frac{n^j}{r^4}-\frac{1}{2}m_1 m_2^2 \frac{n^j}{r^4}-3 m_1 m_2 \frac{\Nvone^2}{r^3}n^j-m_1m_2 \frac{\vonevtwo}{r^3}n^j\nonumber\\
+4 m_1 m_2 \frac{\Nvone \Nvtwo}{r^3} n^j-m_1 m_2 \frac{\Nvtwo^2}{r^3}n^j+\frac{3}{4} m_1 m_2 \frac{v_1^2}{r^3} n^j+\frac{1}{4}m_1 m_2 \frac{v_2^2}{r^3} n^j\nonumber\\
+\frac{3}{4} m_1 m_2 \frac{\Nvone}{r^3}v_1^j-\frac{1}{4}m_1m_2\frac{\Nvtwo}{r^3} v_1^j-\frac{7}{4} m_1 m_2 \frac{\Nvone}{r^3}v_2^j+\frac{5}{4}m_1m_2\frac{\Nvtwo}{r^3} v_2^j,\qquad
\eea
\bea
\frac{1}{24m_1}\int_1\rho^* \ddddot{Y}_{s,j} d^3x=\biggl\{-\frac{15}{8}m_2\frac{\Nvtwo^4}{r^2}n^j+\frac{9}{4}m_2\frac{v_2^2 \Nvtwo^2}{r^2}n^j-\frac{3}{8}m_2\frac{v_2^4}{r^2}n^j\nonumber\\
+\frac{3}{2}m_2\frac{\Nvtwo^3}{r^2} v_2^j
-\frac{3}{2}m_2\frac{v_2^2 \Nvtwo}{r^2} v_2^j+\frac{1}{4}m_1^2 m_2\frac{n^j}{r^4}-\frac{1}{2}m_1 m_2^2 \frac{n^j}{r^4}\nonumber\\-3 m_1 m_2 \frac{\Nvone^2}{r^3}n^j-m_1m_2 \frac{\vonevtwo}{r^3}n^j
+4 m_1 m_2 \frac{\Nvone \Nvtwo}{r^3} n^j\nonumber\\
-m_1 m_2 \frac{\Nvtwo^2}{r^3}n^j+\frac{3}{4} m_1 m_2 \frac{v_1^2}{r^3} n^j+\frac{1}{4}m_1 m_2 \frac{v_2^2}{r^3} n^j
+\frac{3}{4} m_1 m_2 \frac{\Nvone}{r^3}v_1^j\nonumber\\
-\frac{1}{4}m_1m_2\frac{\Nvtwo}{r^3} v_1^j-\frac{7}{4} m_1 m_2 \frac{\Nvone}{r^3}v_2^j+\frac{5}{4}m_1m_2\frac{\Nvtwo}{r^3} v_2^j\biggr\}\nonumber\\
\times (1-2s_1)(1-2s_2),
\eea
\be
-\frac{2}{m_1}\int_1\rho^* \ddot{X} U_{,j} d^3x=-2 m_1 m_2^2 \frac{n^j}{r^4}+2 m_2^2 \frac{v_2^2}{r^3}n^j-2m_2^2\frac{\Nvtwo^2}{r^3}n^j,
\ee
\be
-\frac{2}{m_1}\int_1\rho^* (1-2s) \ddot{X} U_{s,j} d^3x=\biggl\{-2 m_1 m_2^2 \frac{n^j}{r^4}+2 m_2^2 \frac{v_2^2}{r^3}n^j-2m_2^2\frac{\Nvtwo^2}{r^3}n^j\biggr\}\times (1-2s_1)(1-2s_2),
\ee
\bea
-\frac{1}{m_1}\int_1\rho^* [\lambda_1\;(1-2s)+\zeta\;(2a_s-s-2s^2)]\; \ddot{X}_s U_{s,j} d^3x=\frac{1}{2}\biggl\{-2 m_1 m_2^2 \frac{n^j}{r^4}+2 m_2^2 \frac{v_2^2}{r^3}n^j\nonumber\\
-2m_2^2\frac{\Nvtwo^2}{r^3}n^j\biggr\}\times [\lambda_1(1-2s_1)+\zeta(2a_{s1}-s_1-2s_1^2)]\; (1-2s_2)^2,\quad
\eea
\bea
\frac{2}{m_1}\int_1\rho^* \dddot{X}^j d^3x=-2m_1^2 m_2\frac{n^j}{r^4}+4 m_1 m_2^2 \frac{n^j}{r^4}+6m_2\frac{v_2^2 \Nvtwo}{r^2} v_2^j-6m_2\frac{\Nvtwo^3}{r^2}v_2^j\nonumber\\
-10 m_1 m_2 \frac{\Nvtwo}{r^3} v_2^j+16 m_1 m_2 \frac{\Nvone}{r^3}v_2^j+6m_1m_2\frac{\Nvtwo^2}{r^3}n^j\nonumber\\+6m_1m_2\frac{\Nvtwo}{r^3}v_1^j
-42m_1m_2\frac{\Nvone\Nvtwo}{r^3}n^j-12m_1m_2\frac{\Nvone}{r^3}v_1^j\nonumber\\
+12m_1m_2\frac{\vonevtwo}{r^3}n^j
-6m_1m_2\frac{v_1^2}{r^3}n^j
+30m_1m_2\frac{\Nvone^2}{r^3}n^j,
\eea
\be
-\frac{4}{m_1}\int_1\rho^*[\Sigma(V_i v^i)]_{,j}d^3x=4m_1m_2\frac{\vonevtwo}{r^3}n^j,
\ee
\be
\frac{4}{m_1}\int_1\rho^*(1-2s) [\Sigma_s(V_i v^i)]_{,j}d^3x=-4(1-2s_1) (1-2s_2)\;m_1 m_2 \frac{\vonevtwo}{r^3}n^j,
\ee
\bea
\frac{16}{m_1}\int_1\rho^* \dot{G}^j_{7}d^3x=-2 m_1^2m_2\frac{n^j}{r^4}+2m_1m_2^2\frac{n^j}{r^4}+12m_1m_2\frac{\Nvone^2}{r^3}n^j+5m_1m_2\frac{\vonevtwo}{r^3}n^j\nonumber\\
-16m_1m_2\frac{\Nvtwo^2}{r^3}n^j-4m_1m_2\frac{\Nvone\Nvtwo}{r^3}n^j-6m_1m_2\frac{v_1^2}{r^3}n^j+4m_1m_2\frac{v_2^2}{r^3}n^j\nonumber\\
-2m_1m_2\frac{\Nvone}{r^3}v_1^j-2m_1m_2\frac{\Nvtwo}{r^3}v_1^j+m_1m_2\frac{\Nvone}{r^3}v_2^j+8m_1m_2\frac{\Nvtwo}{r^3}v_2^j\nonumber\\
+4m_2^2\frac{\Nvtwo^2}{r^3}n^j-m_2^2\frac{v_2^2}{r^3}n^j-3m_2^2\frac{\Nvtwo}{r^3}v_2^j,\;\;\;
\eea
\bea
-\frac{4}{m_1}\int_1\rho^* \dot{P}(\dot{U}_s U_s^{,i}) d^3x=4 \biggl\{-\frac{m_1 m_2}{4r^3} \biggl(-4 v_2^2 n^j-8 \Nvtwo v_2^j+16 \Nvtwo^2 n^j\biggr) (1-2s_1) (1-2s_2)
\nonumber\\
-\frac{m_1 m_2}{4r^3} \biggl(2 v_1^2 n^j+6 \Nvone v_1^j-12 \Nvone^2 n^j\biggr) (1-2s_1)(1-2s_2)
\nonumber\\
-\frac{m_2^2}{4r^3}\biggl(2\Nvtwo v_2^j-2\Nvtwo^2 n^j\biggr) (1-2s_2)^2\-\frac{\alpha m_1^2 m_2}{2r^4}n^j (1-2s_1)(1-2s_2)
 \nonumber\\
+\frac{\alpha m_1 m_2^2}{2 r^4}n^j (1-2s_1)(1-2s_2)
\nonumber\\
-\frac{m_1 m_2}{4r^3} \biggl(3\vonevtwo n^j+3 \Nvone v_2^j+2\Nvtwo v_1^j-12 \Nvone \Nvtwo n^j\biggr) (1-2s_1) (1-2s_2) 
\nonumber\\
+\frac{m_1 m_2}{r^3} \biggl(\vonevtwo n^j+\Nvtwo v_1^j+\Nvone v_2^j-4\Nvone\Nvtwo n^j\biggr) (1-2s_1)(1-2s_2)
\nonumber\\
-\frac{m_2^2}{4r^3}\biggl(v_2^2 n^j+\Nvtwo v_2^j-2\Nvtwo^2 n^j\biggr) (1-2s_2)^2\biggr\},\qquad
\eea
\be
-\frac{1}{m_1}\int_1\rho^* [\Sigma(\Phi_2)]_{,j} d^3x=m_1m_2^2\frac{n^j}{r^4},
\ee
\be
-\frac{1}{m_1}\int_1\rho^* [\Sigma(\Phi^s_{2s})]_{,j} d^3x=(1-2s_1)(1-2s_2)\;m_1m_2^2\frac{n^j}{r^4},
\ee
\be
-\frac{1}{m_1}\int_1\rho^* (1-2s) [\Sigma_s(\Phi_2)]_{,j} d^3x=(1-2s_1)(1-2s_2)\;m_1m_2^2\frac{n^j}{r^4},
\ee
\bea
\frac{1}{m_1}\int_1\rho^* \biggl[(2\lambda_1+\zeta) [1-\zeta+(2\lambda_1+\zeta)(1-2s)]-\zeta (1-\zeta) (1-2s)\biggr] [\Sigma_s(\Phi^s_{2s})]_{,j} d^3x\nonumber\\
=- \biggl[(2\lambda_1+\zeta) [1-\zeta+(2\lambda_1+\zeta)(1-2s_1)]-\zeta (1-\zeta) (1-2s_1)\biggr]\nonumber\\
\times(1-2s_1)(1-2s_2)^2\;m_1m_2^2\frac{n^j}{r^4},\qquad
\eea
\be
\frac{1}{m_1}\int_1\rho^* [1-\zeta+(2\lambda_1+\zeta)(1-2s)]\; [\Sigma_s(\Phi^s_2)]_{,j} d^3x=-[1-\zeta+(2\lambda_1+\zeta)(1-2s_1)]\; (1-2s_1)(1-2s_2) m_1m_2^2\frac{n^j}{r^4},
\ee
\be
4\frac{1}{m_1}\int_1\rho^* (1-2s) [\Sigma(a_s\;\Phi^s_2)]_{,j} d^3x=-4 a_{s2}\;(1-2s_1)(1-2s_2)\;m_1m_2^2\frac{n^j}{r^4},
\ee
\be
4\frac{1}{m_1}\int_1\rho^* (1-2s) [\Sigma(a_s\;\Phi^s_{2s})]_{,j} d^3x=-4 a_{s2}\;(1-2s_1)^2 (1-2s_2)\;m_1m_2^2\frac{n^j}{r^4},
\ee
\be
16\frac{1}{m_1}\int_1\rho^* (1-2s) [\Sigma(a_s\;\Sigma(a_s\; U_s))]_{,j} d^3x=-16 a_{s1}\;a_{s2}\;(1-2s_1)(1-2s_2)\; m_1m_2^2\frac{n^j}{r^4},
\ee
\be
4\frac{1}{m_1}\int_1\rho^* [1-\zeta+(2\lambda_1+\zeta)(1-2s)]\;[\Sigma_s(\Sigma(a_s\; U_s))]_{,j} d^3x=-4 [1-\zeta+(2\lambda_1+\zeta)(1-2s_1)]\; a_{s1} (1-2s_2)^2 m_1m_2^2\frac{n^j}{r^4},
\ee
\be
-\frac{8}{m_1}\int_1\rho^* U \dot{V}_{j} \;d^3x=-8 m_1 m_2^2 \frac{n^j}{r^4}-8m_2^2\frac{\Nvtwo}{r^3} v_2^j,
\ee
\bea
-\frac{6}{m_1}\int_1\rho^* G_{1,j} d^3x=-12 m_1 m_2 \frac{\Nvone\Nvtwo}{r^3} n^j+3m_1m_2\frac{\Nvone}{r^3}v_2^j\nonumber\\+3m_1m_2\frac{\vonevtwo}{r^3} n^j+6m_1m_2\frac{\Nvtwo}{r^3}v_1^j+6m_2^2\frac{\Nvtwo^2}{r^3}n^j\nonumber\\
-3m_2^2\frac{\Nvtwo}{r^3}v_2^j-3m_2^2\frac{v_2^2}{r^3}n^j,
\eea
\bea
\frac{2}{m_1}\int_1\rho^* G_{1s,j} d^3x=-\frac{1}{3}\biggl\{-12 m_1 m_2 \frac{\Nvone\Nvtwo}{r^3} n^j+3m_1m_2\frac{\Nvone}{r^3}v_2^j\nonumber\\+3m_1m_2\frac{\vonevtwo}{r^3} n^j+6m_1m_2\frac{\Nvtwo}{r^3}v_1^j+6m_2^2\frac{\Nvtwo^2}{r^3}n^j\nonumber\\
-3m_2^2\frac{\Nvtwo}{r^3}v_2^j-3m_2^2\frac{v_2^2}{r^3}n^j\biggr\}\times (1-2s_2),
\eea
\be
-\frac{3}{2m_1}\int_1\rho^* [\Sigma(\Phi_1)]_{,j} d^3x=\frac{3}{2} m_1 m_2\frac{v_1^2}{r^3} n^j,
\ee
\be
-\frac{3}{2m_1}\int_1\rho^* (1-2s) [\Sigma_s(\Phi_1)]_{,j} d^3x=\frac{3}{2} (1-2s_1)(1-2s_2)\; m_1 m_2\frac{v_1^2}{r^3} n^j,
\ee
\be
\frac{2}{m_1}\int_1\rho^* [\Sigma(a_s\;\Phi_1^s)]_{,j} d^3x=-2 (1-2s_1)^2\;m_1 m_2\frac{v_1^2}{r^3} n^j,
\ee
\be
\frac{1}{2m_1}\int_1\rho^* [1-\zeta+(2\lambda_1+\zeta)(1-2s)] [\Sigma_s(\Phi_1^s)]_{,j} d^3x=-\frac{1}{2}(1-2s_1)(1-2s_2)\; m_1 m_2\frac{v_1^2}{r^3} n^j,
\ee
\be
\frac{7}{8m_1}\int_1\rho^* [\Sigma(v^4)]_{,j} d^3x=-\frac{7}{8}m_2\frac{v_2^4}{r^2} n^j,
\ee
\be
-\frac{1}{8m_1}\int_1\rho^* (1-2s) [\Sigma_s (v^4)]_{,j} d^3x=\frac{1}{8}(1-2s_1) (1-2s_2) \;m_2\frac{v_2^4}{r^2} n^j,
\ee
\be
-\frac{8}{m_1}\int_1\rho^* \dot{\Phi}^j_{2} d^3x=8m_1m_2^2\frac{n^j}{r^4}+8m_1m_2\frac{\Nvone}{r^3}v_1^j-16m_1m_2\frac{\Nvtwo}{r^3}v_1^j,
\ee
\bea
\frac{8}{m_1}\int_1\rho^* G_{3,j} d^3x=-36 m_1 m_2 \frac{\Nvone\Nvtwo}{r^3}n^j+10 m_1m_2\frac{\Nvone}{r^3}v_2^j+10m_1m_2\frac{\Nvtwo}{r^3}v_1^j\nonumber\\
+8m_1m_2\frac{\vonevtwo}{r^3}n^j-24m_2^2\frac{\Nvtwo^2}{r^3}+12m_2^2\frac{\Nvtwo}{r^3} v_2^j+4m_2^2\frac{v_2^2}{r^3}n^j,\qquad
\eea
\bea
\frac{8}{m_1}\int_1\rho^* (1-2s) G_{3s,j} d^3x=\biggl\{-36 m_1 m_2 \frac{\Nvone\Nvtwo}{r^3}n^j+10 m_1m_2\frac{\Nvone}{r^3}v_2^j\nonumber\\
+10m_1m_2\frac{\Nvtwo}{r^3}v_1^j
+8m_1m_2\frac{\vonevtwo}{r^3}n^j-24m_2^2\frac{\Nvtwo^2}{r^3}+12m_2^2\frac{\Nvtwo}{r^3} v_2^j+4m_2^2\frac{v_2^2}{r^3}n^j\biggr\}\nonumber\\
\times (1-2s_1)(1-2s_2),\qquad
\eea
\be
\frac{8}{m_1}\int_1\rho^* V_i V_{i,j}d^3x=-8m_2^2\frac{v_2^2}{r^3}n^j,
\ee
\be
-\frac{4}{m_1}\int_1\rho^* H_{,j} d^3x=-8 m_1 m_2^2\frac{n^j}{r^4}-m_2^3 \frac{n^j}{r^4},
\ee
\be
-\frac{4}{m_1}\int_1\rho^* H_{s,j} d^3x=-4 \biggl\{\frac{1}{5} (1-2s_1) (s_1-s_2) m_1^2 m_2  \frac{n^j}{r^4}+2 (1-2s_1)(1-2s_2) m_1 m_2^2  \frac{n^j}{r^4}+\frac{1}{4} (1-2s_2)^2 m_2^3\frac{n^j}{r^4} \biggr\},
\ee
\be
-\frac{4}{m_1}\int_1\rho^* (1-2s) H^s_{,j} d^3x=-4 (1-2s_1) \biggl\{-\frac{1}{5} (s_1-s_2) m_1^2 m_2  \frac{n^j}{r^4}+2 (1-2s_2) m_1 m_2^2  \frac{n^j}{r^4}+\frac{1}{4} (1-2s_2) m_2^3\frac{n^j}{r^4} \biggr\},
\ee
\be
-\frac{4}{m_1}\int_1\rho^* (1-2s) H^s_{s,j} d^3x=-4 (1-2s_1) \biggl\{2 (1-2s_1)(1-2s_2)^2 m_1 m_2^2 \frac{n^j}{r^4} +\frac{1}{4} (1-2s_2)^3 m_2^3\frac{n^j}{r^4} \biggr\},
\ee
\bea
-\frac{4}{m_1}\int_1\rho^* G_{2,j} d^3x=2m_1^2m_2 \frac{n^j}{r^4}+2m_1m_2^2\frac{n^j}{r^4}+16 m_1 m_2\frac{\Nvtwo^2}{r^3}n^j-8m_1m_2\frac{\Nvtwo}{r^3}v_2^j\nonumber\\
-4m_1m_2\frac{v_2^2}{r^3}n^j+2m_1m_2\frac{\Nvone^2}{r^3}n^j-2m_1m_2\frac{\Nvone}{r^3}v_1^j+12m_2^2\frac{\Nvtwo^2}{r^3}n^j\nonumber\\
-6m_2^2\frac{\Nvtwo}{r^3}v_2^j-2m_2^2\frac{v_2^2}{r^3}n^j,\qquad
\eea
\bea
-\frac{4}{m_1}\int_1\rho^* (1-2s)\;G_{2s,j} d^3x=\biggl\{2m_1^2m_2 \frac{n^j}{r^4}+2m_1m_2^2\frac{n^j}{r^4}+16 m_1 m_2\frac{\Nvtwo^2}{r^3}n^j\nonumber\\-8m_1m_2\frac{\Nvtwo}{r^3}v_2^j
-4m_1m_2\frac{v_2^2}{r^3}n^j+2m_1m_2\frac{\Nvone^2}{r^3}n^j-2m_1m_2\frac{\Nvone}{r^3}v_1^j\nonumber\\+12m_2^2\frac{\Nvtwo^2}{r^3}n^j
-6m_2^2\frac{\Nvtwo}{r^3}v_2^j-2m_2^2\frac{v_2^2}{r^3}n^j\biggr\}\times (1-2s_1) (1-2s_2),
\eea
\bea
\frac{8}{m_1}\int_1\rho^* G_{4,j} d^3x=16m_1m_2\frac{\Nvone\Nvtwo}{r^3}n^j-4m_1m_2\frac{\Nvtwo}{r^3}v_1^j-4m_1m_2\frac{\vonevtwo}{r^3}n^j\nonumber\\
-8m_1m_2\frac{\Nvone}{r^3}v_2^j-8m_2^2\frac{\Nvtwo^2}{r^3}n^j+4m_2^2\frac{\Nvtwo}{r^3}v_2^j+4m_2^2\frac{v_2^2}{r^3}n^j,\quad
\eea
\be
\frac{4}{m_1}\int_1\rho^* V_j \dot{U} d^3x=4m_2^2\frac{\Nvtwo}{r^3}v_2^j,
\ee
\be
-\frac{4}{m_1}\int_1\rho^* (1-2s) V_j \dot{U}_s d^3x=-4 (1-2s_1)(1-2s_2)\;m_2^2\frac{\Nvtwo}{r^3}v_2^j,
\ee
\be
\frac{8}{m_1}\int_1\rho^* U^2 U_{,j} d^3x=-8m_2^3\frac{n^j}{r^4},
\ee
\be
\frac{8}{m_1}\int_1\rho^* (1-2s) U^2 U_{s,j} d^3x=-8 (1-2s_1)(1-2s_2)\;m_2^3\frac{n^j}{r^4},
\ee
\be
\frac{4}{m_1}\int_1\rho^* [2\lambda_1(1-2s)+\zeta(2a_s-s-2s^2)] U U_s U_{s,j} d^3x=-4 [2\lambda_1(1-2s_1)+\zeta(2a_{s1}-s_1-2s_1^2)]\; (1-2s_2)^2\; m_2^3\frac{n^j}{r^4},
\ee
\bea
-\frac{1}{m_1}\int_1\rho^* [(8\lambda_1^2-2\zeta\lambda_1-2\lambda_2)(1-2s)+6\lambda_1\zeta(2a_s-s-2s^2)\nonumber\\
+\zeta^2(2a_s+12 s\; a_s-4b_s-s-2s^2-8s^3)] U_s^2 U_{s,j} d^3x= [(8\lambda_1^2-2\zeta\lambda_1-2\lambda_2)(1-2s_1)\nonumber\\+6\lambda_1\zeta(2a_{s1}-s_1-2s_1^2)+\zeta^2(2a_{s1}+12s_1\;a_{s1}-4b_{s1}-s_1-2s_1^2-8s_1^3)] \;(1-2s_2)^3\; m_2^3\frac{n^j}{r^4},\qquad
\eea
\be
-\frac{4}{m_1}\int_1\rho^* P_2^{ij} U_{,i} d^3x=-4m_1m_2^2\frac{n^j}{r^4},
\ee
\be
-\frac{4}{m_1}\int_1\rho^* P_{2s}^{ij} U_{,i} d^3x=-4 (1-2s_1)(1-2s_2)\;m_1m_2^2\frac{n^j}{r^4},
\ee
\be
-\frac{4}{m_1}\int_1\rho^* (1-2s) P_2^{ij} U_{s,i} d^3x=-4 (1-2s_1)(1-2s_2) m_1m_2^2\frac{n^j}{r^4},
\ee
\be
-\frac{4}{m_1}\int_1\rho^* (1-2s) P_{2s}^{ij} U_{s,i} d^3x=-4 (1-2s_1)^2 (1-2s_2)^2 m_1m_2^2\frac{n^j}{r^4},
\ee
\be
-\frac{4}{m_1}\int_1\rho^* \Phi_1^{ij} U_{,i} d^3x=4m_2^2\frac{\Nvtwo}{r^3} v_2^j,
\ee
\be
-\frac{4}{m_1}\int_1\rho^* (1-2s) \Phi_1^{ij} U_{s,i} d^3x=4(1-2s_1) (1-2s_2)\;m_2^2\frac{\Nvtwo}{r^3} v_2^j,
\ee
\be
\frac{4}{m_1}\int_1\rho^* \dot{V}^j_{2} d^3x=4m_1^2m_2\frac{n^j}{r^4}-4m_1m_2\frac{\Nvone}{r^3} v_2^j+8m_1m_2\frac{\Nvtwo}{r^3}v_2^j,
\ee
\be
-\frac{4}{m_1}\int_1\rho^* \dot{V}^{s\;j}_{2s} d^3x=-\biggl\{4m_1^2m_2\frac{n^j}{r^4}-4m_1m_2\frac{\Nvone}{r^3} v_2^j+8m_1m_2\frac{\Nvtwo}{r^3}v_2^j\biggr\} (1-2s_1) (1-2s_2),
\ee
\be
\frac{2}{m_1}\int_1\rho^* \dot{V}^j_{3} d^3x=4m_1m_2\frac{\Nvtwo}{r^3} v_2^j+2m_1m_2\frac{v_2^2}{r^3}n^j+2m_2\frac{v_2^2 \Nvtwo}{r^2}v_2^j,
\ee
\be
-\frac{1}{2m_1}\int_1\rho^* [\Sigma(\ddot{X})]_{,j}d^3x=-\frac{1}{2}m_1m_2^2\frac{n^j}{r^4}+\frac{1}{2}m_1m_2\frac{v_1^2}{r^3}n^j-\frac{1}{2}m_1m_2\frac{\Nvone^2}{r^3}n^j,
\ee
\bea
-\frac{1}{2m_1}\int_1\rho^* (1-2s) [\Sigma_s(\ddot{X})]_{,j}d^3x=\biggl\{-\frac{1}{2}m_1m_2^2\frac{n^j}{r^4}+\frac{1}{2}m_1m_2\frac{v_1^2}{r^3}n^j\nonumber\\
-\frac{1}{2}m_1m_2\frac{\Nvone^2}{r^3}n^j\biggr\}
\times (1-2s_1)(1-2s_2),\qquad
\eea
\bea
-\frac{1}{2m_1}\int_1\rho^* [1-\zeta+(2\lambda_1+\zeta) (1-2s)] [\Sigma_s(\ddot{X}_s)]_{,j}d^3x=\biggl\{-\frac{1}{2}m_1m_2^2\frac{n^j}{r^4}+\frac{1}{2}m_1m_2\frac{v_1^2}{r^3}n^j
\nonumber\\
-\frac{1}{2}m_1m_2\frac{\Nvone^2}{r^3}n^j\biggr\}
\times [1-\zeta+(2\lambda_1+\zeta) (1-2s_1)]\;(1-2s_1)(1-2s_2),\qquad
\eea
\bea
-\frac{2}{m_1}\int_1\rho^* (1-2s) [\Sigma(a_s\;\ddot{X}_s)]_{,j}d^3x=4\biggl\{-\frac{1}{2}m_1m_2^2\frac{n^j}{r^4}+\frac{1}{2}m_1m_2\frac{v_1^2}{r^3}n^j\nonumber\\-\frac{1}{2}m_1m_2\frac{\Nvone^2}{r^3}n^j\biggr\}
\times a_{s2}\;(1-2s_1)^2,\qquad
\eea
\be
-\frac{2}{m_1}\int_1\rho^* \Phi_1 U_{,j} d^3x=2m_2^2\frac{v_2^2}{r^3} n^j,
\ee
\be
-\frac{2}{m_1}\int_1\rho^* (1-2s)\; \Phi_1 U_{s,j} d^3x=2 (1-2s_1)(1-2s_2)\;m_2^2\frac{v_2^2}{r^3} n^j,
\ee
\bea
\frac{1}{m_1}\int_1\rho^* [\lambda_1(1-2s)+\zeta(2\;a_s-s-2 s^2)]\; (1-2s)\; \Phi^s_1 U_{s,j} d^3x=\nonumber\\
- [\lambda_1(1-2s_1)+\zeta(2\;a_{s1}-s_1-2 s_1^2)]\; (1-2s_2)^2\;m_2^2\frac{v_2^2}{r^3} n^j.
\eea

\addtocontents{toc}{} 

\backmatter

\lhead[\thepage]{Biblography \emph{}}      
\rhead[Biblography \emph{}]{\thepage}

\bibliographystyle{acm}
\bibliography{master} 


\end{document}